\documentclass [ 11pt, twoside] {rauls}[2012/06/19]
\usepackage{amssymb,latexsym,mathbbol}
\usepackage{amsmath,amsbsy}
\usepackage{epsfig,bm}
\usepackage{graphicx,comment}
\usepackage[figuresright]{rotating}
 
\usepackage{alltt}  %

 \usepackage{caption}

\usepackage{subfigure}
\newcommand{\nn}{\nonumber}
\newcommand{\benn}{\begin{displaymath}}
\newcommand{\eenn}{\end{displaymath}}

\usepackage{epsfig}
\usepackage{amssymb}
\usepackage{amsthm}
\usepackage{amsmath}
\usepackage{amsmath}
\usepackage{hyperref}
\usepackage{multirow}
\newcommand{\singlet}{{^1}\hspace{-.06 cm}{S_0}}
\newcommand{\triplet}{{^3}\hspace{-.06 cm}{S_1}}
\newcommand{\pinot}{\not\hspace{-.08cm}{\pi}}

\usepackage{amsmath}
\begin{document}
 
%
%

\prelimpages
 
%
%
\Title{On the Determination of  Elastic and Inelastic \\
Nuclear Observables from Lattice QCD}
\Author{Ra\'ul A. Brice\~no}
\Year{2013}
\Program{Department of Physics}
 
\Chair{Martin Savage}{Professort}{Department of Physics}

\Signature{Stephen Sharpe}

\Signature{Huey-Wen Lin}

{\Degreetext{A dissertation%
  \\
  submitted in partial fulfillment of the\\ requirements for the degree of}
 \def\thefootnote{\fnsymbol{footnote}}
 \let\footnoterule\relax
 \titlepage
 }
\setcounter{footnote}{0}

%
%

\setcounter{page}{-1}
\abstract{
One of the overarching goals of nuclear physics is to rigorously compute properties of hadronic systems directly from the fundamental theory of the strong interaction, Quantum Chromodynamics (QCD). In particular, the hope is to perform reliable calculations of nuclear processes which would impact our understanding of environments ranging from big bang nucleosynthesis, stars and supernovae, to nuclear reactors and high-energy density facilities. Such calculations, being truly \emph{ab-initio}, would include all two-nucleon and three-nucleon (and higher) interactions in a consistent manner.  

Currently, lattice QCD (LQCD) provides the only reliable option for performing calculations of low-energy hadronic observables. LQCD calculations are necessarily performed in a finite Euclidean spacetime. As a result, it is necessary to construct formalism that maps the finite-volume observables determined via LQCD to the infinite-volume quantities of interest. For $2\rightarrow2$ bosonic elastic scattering processes, Martin L\"uscher \cite{luscher1,luscher2} first showed that one can obtain the physical scattering phase
shifts from the \textit{finite volume} (FV)  two-particle spectrum (for lattices with spatial extents that are much larger than the range of interactions). This thesis discusses the extension of this formalism for three important classes of systems. 

Chapter~\ref{intro} discusses key aspects of the standard model, paying close attention to QCD at  low-energies and the necessity of effective field theories (EFTs) and LQCD. Chapter~\ref{mmsys} reviews the result by L\"uscher for two bosons with arbitrary momentum. After a detailed derivation of the quantization condition for two bosons below the inelastic threshold, it is straightforward to determine the spectrum of a system with arbitrary number of channels composed of two hadrons with nonzero total momentum. In Section \ref{scalardimer}, L\"uscher's result is re-derived using the \emph{auxilary field formalism}, also known as the ``\emph{dimer formalism}".

Chapter~\ref{NNsys} briefly reviews the complexity of the nuclear sector, as compared to the scalar sector, and it shown that this rich structure can be recovered by the generalization of the \emph{auxilary field formalism} for the two nucleon system. Using this formalism, the quantization condition for two non-relativistic nucleons\footnote{Assuming the isospin limit where the proton and the neutron are degenerate and can be thought of as components of an isospin doublet, $N^{T}=(p~n)$.} in a finite volume is derived. The result presented hold for a two nucleon system with arbitrary partial-waves, spin and parity.  Provided are the explicit relations among scattering parameters and their corresponding point group symmetry class eigenenergies with orbital angular momentum $l\le3$.

Finally, Chapter~\ref{mmmsys} presents the quantization condition for the spectrum of three identical bosons in a finite volume. Unlike the two-body analogue, the quantization condition of the three-body sector is not algebraic and in general requires numerically solving an integral equation. However, for systems with an attractive two-body force that supports a two-body bound-state, a \emph{diboson}, and for energies below the diboson breakup, the quantization condition reduces to the well-known L\"uscher formula with exponential corrections in volume that scale with the diboson binding momentum. To accurately determine infinite volume phase shifts, it is necessary to extrapolate the phase shifts obtained from the L\"uscher formula for the boson-diboson system to the infinite volume limit. For energies above the breakup threshold, or for systems with no two-body bound-state (with only scattering states and resonances) the L\"uscher formula gets power-law volume corrections and consequently fails to describe the three-particle system. These corrections are nonperturbatively included in the quantization condition presented.

}
 
%
%
\tableofcontents
\listoffigures

%
%
\acknowledgments{

I would like to thank my advisor, Martin Savage. Writing my first paper was probably the most challenging thing I have ever done. In those two years of writing I learnt many important lessons, both about myself and physics. For one, I can confidently say today that I do want to pursue a career in science. I learnt that the stories we tell about our work are just as important as the work itself. None of this would have ever been possible if you had not repeatedly kept a straight face and let me know that it was ``\emph{not good enough}" and that I had ``\emph{more work to do}". These are just the words that I needed to hear. So thank you for that, that and all the support, patience, and insight. It is not any faculty that would encourage two graduate students to organize a workshop on ``Nuclear Reactions from Lattice QCD".  
 
To my collaborators, Daniel Bolton, Huey-Wen Lin, Tom Luu, and Zohreh Davoudi, I am infinitely grateful. I have learnt so much from each one of you in the past few years. Zohreh, the past couple of years would not have been half as productive without you. I have enjoyed every minute of our endless discussions/arguments both scientific and non-scientific. I am particularly thankful for Steve Sharpe's and Huey-Wen Lin's support and for always having theirs doors open for physics and non-physics questions. Other people I would like to thank are Alan Jamison, Brian Mattern, Max Hansen, Amy Nicholson, Joe Wasem, John Vinson, and Brian Smigielski for the endless number of discussions throughout the years.

}

%
%
%
\dedication{
\begin{center}
\emph{Estos a\~nos de trabajo y esfuerzo se los dedico a mi madre y a mi abuela.} 
\end{center}

These years of hard work and effort I dedicate to my mother and grandmother. Despite myself and others, the two of them never doubted my ability to succeed and gave everything they had so I could be the best man possible. Of course the rest of my immediate family (mi papa, Ely, Ernesto, Erika, y Titi) have been nothing but loving and supportive, even when they are thousands of miles away.  
 \\

 Then, there are my friends and loved ones, too many to list here. Some I want to mention are those who made sure that I am not a not just a physics machine. These include the \emph{Book Club for Music}/climbing crew/etc: Naomi, Tyler, Amy, Chris, Jared, Alan, Phil, Megan, Sam, Elisa, Roy, Ashely, etc. You have made the last few years in Seattle more than just tolerable. I will you miss you all.

\begin{center}``Un ser sin estudios, es un ser incompleto." \\-Sim\'on Bol\'ivar \end{center}
}

%

%
%

\textpages
 
 
\chapter {Introduction}{\label{intro}}

One of the most important achievements of modern science is the unification of three of the four known fundamental forces of nature into what is known as the \emph{Standard Model of particle physics}. The Standard Model describes the electroweak \cite{Glashow1961579, Salam:1964ry, Weinberg:1967tq, Glashow:1970gm} and strong nuclear interactions \cite{Fritzsch:1972jv, Fritzsch:1973pi, Politzer:1973fx, Gross:1973id} of all subatomic particles observed to this date. One of the goals of modern day particle physics is to find extensions of the Standard Model that would incorporate gross features of our universe that are currently missed by the Standard Model. Some key examples include the fact that neutrinos are massive, as well as the fact that the standard model does not incorporate gravity nor a fundamental description of dark matter. 

An equally important scientific program is to utilize the Standard Model to obtain further insight to physical systems that are either experimentally inaccessible or whose experimental programs are plagued by systematic errors. In particular, it would be desirable to understand the implications of the Standard Model to the evolution of stars, Big Bang/supernovae nucleosynthesis, the composition of neutron stars, as well as reactions occurring in nuclear reactors and high-energy density facilities. This would require a tight control of the strong nuclear sector for a wide range of energy regimes. This sector of the Standard Model is described by a \emph{quantum field theory} (QFT) known as \emph{Quantum Chromodynamics} (QCD) \cite{Fritzsch:1972jv, Fritzsch:1973pi}. As will be discussed in the subsections to come, at low energies QCD, whose fundamental degrees of freedom are \emph{quarks and gluons}, is non-perturbative \cite{Politzer:1973fx, Gross:1973id}. The non-perturbative nature of QCD has historically introduced uncontrolled systematic errors to theoretical calculations of nuclear physics phenomena directly from QCD. 

At low energies QCD is not only non-perturbative but it is also \emph{confining}. This means that despite quarks and gluons being the fundamental degrees of freedom, they are always bound together to form \emph{hadrons}. As a consequence, it is convenient to describe low-energy nuclear systems using {effective field theories} (EFTs) whose fundamental degrees of freedoms are the hadrons themselves rather than the quarks and gluons. Fermionic hadrons composed of three constituent quarks (or three antiquarks) are known as \emph{baryons}. The proton and the neutron are two important examples of baryons, as they are the building blocks of all nuclei. Bosonic hadrons composed of a quark-antiquark pair are known as \emph{mesons}. Pions, kaons and etas are examples of mesons. They are identified as the approximate pseudoGoldstone bosons of QCD and mediate the nuclear force at long distances. Section \ref{efts_sec} reviews the two most important low-energy EFTs for nuclear physics. The first is Chiral Perturbation Theory \cite{Weinberg:1978kz, Gasser:1983yg, Gasser:1984gg}, which describes the interactions of pions, kaons and etas.  The second is ``\emph{pionless EFT}" or EFT$\left(\pinot\right)$ \cite{pds, pds2}, which has been shown to accurately describes the two- and three-nucleon interactions at low energies. 

State of the art nuclear physics calculation, such as no core shell model (see Refs.~\cite{Zheng:1993qx, Jaqua:1993zz, Zheng:1994zza, Navratil:1996vm,  Navratil:1998uf}) with continuum or  lattice effective field theory, use chiral interactions derived from EFT$\left(\pinot\right)$+$\chi PT$.  Due to poor determination of the low energy coefficients appearing in these EFTs, present day nuclear physics calculations are plagued by systematic errors. The only known way to circumvent the predictive limitations of such EFTs is to perform non-perturbative numerical calculations of physical obsevables directly from QCD. This is program referred to Lattice QCD (LQCD)~\cite{Wilson:1974sk}. LQCD calculations are performed in a discretized, finite Euclidean spacetime. As a result, it is necessary to develop formalism that connects quantities evaluated via LQCD to the physical observables of interest. This thesis addresses two major obstacles towards the study of nuclear reactions directly from QCD: scattering processes above inelastic thresholds \cite{Briceno:2012rv, Briceno:2012yi} and the determination phases and mixing angles for arbitrary channels in the two nucleon sector \cite{Briceno:2013lba}. The work discussed in Chapters~\ref{mmsys}-\ref{mmmsys} were previously presented in 

\begin{itemize}
\item Ra\'ul A. Brice\~no and Zohreh Davoudi. Moving multichannel systems in a finite volume with application to proton-proton fusion, \href{http://prd.aps.org/abstract/PRD/v88/i9/e094507}{\emph{Phys. Rev.} {\bf D88} 094507 (2012)}, \href{http://arxiv.org/abs/arXiv:1204.1110}{arXiv:1204.1110 [hep-lat].}
\item Ra\'ul A. Brice\~no, Zohreh Davoudi, and Thomas C. Luu. Two-Nucleon Systems in a Finite Volume: (I) Quantization Conditions, \href{http://prd.aps.org/abstract/PRD/v88/i3/e034502}{{\em Phys. Rev. \/} {\bf D88} 034502 (2013)}, \href{http://arxiv.org/abs/arXiv:1305.4903}{arXiv:1305.4903 [hep-lat].} 
\item Ra\'ul A. Brice\~no and Zohreh Davoudi. Three-particle scattering amplitudes from a finite volume formalism,  \href{http://prd.aps.org/abstract/PRD/v87/i9/e094507}{{\em Phys.~Rev.~\/}~{\bf D87}~094507 (2012)}, \href{http://arxiv.org/abs/arXiv:1212.3398}{arXiv:1212.3398 [hep-lat].}
\end{itemize}
To understand the context of the work presented, it is necessary to first review the basics of the Standard Model, QCD, EFTs, and LQCD.

\section{QCD and the Standard Model of Particle Physics \label{qcdsec}}
The standard model is a relativistic quantum field theory that describes the strong, weak and electromagnetic interaction of quarks and leptons. The strong sector, which is the main focus of this thesis, is described by a non-Abelian gauge theory \cite{Yang:1954ek} with gauge group $SU(3)_C$, whose fundamental degrees of freedom are quarks and gluons. The electroweak weak sector is described by a $SU(2)_L\times U(1)_Y$ gauge group \cite{Glashow1961579, Salam:1964ry, Weinberg:1967tq, Glashow:1970gm}. As it stands, the standard model has three generations of spin-1/2 quarks and leptons, a spin-zero Higgs boson \cite{Higgs:1964ia, Higgs:1964pj, Higgs:1966ev}, and four mediating spin-1 gauge bosons (the photon, gluon, $W^{\pm}$ and $Z^0$). Although the electroweak Lagrangian has $SU(2)_L\times U(1)_Y$ symmetry, this symmetry is spontaneous broken down to $U(1)_Q$ by the vacuum expectation value of the Higgs field. As a consequence, $W^{\pm}$ and $Z^0$ are massive particles, with masses of $m_W=80.385(15)$~GeV and $m_Z=98.1876(21)$~GeV \cite{pdg}, while the photon remains massless.

Two defining characteristics of low energy QCD are \emph{confinement} and  \emph{asymptotic freedom}. Confinement refers to the fact that the low-energy degrees of freedom are color singlet bound states (hadrons) of quarks and gluons. The fact that QCD is non-perturbative distinguishes it from the spontaneously broken $SU(2)_L\times U(1)_Y$ sector, where perturbation theory can be implemented in a controlled way. As a result, the dominant contribution to a large subset of hadronic physics is encapsulated by QCD, which must be treated non-perturbatively. At high energies QCD becomes perturbative and quarks become asymptotically free, this is commonly referred to as \emph{asymptotic freedom}.  

There are currently six quark flavors: \emph{up, down, strange, charm, bottom,} and \emph{top}. The six fields associated with the quark flavors can be compactly written as components in a six-dimensional vector in ``flavor space", $q^T=\left(u,d,s,c,b,t\right)$. With this, one can write down the QCD Lagrange density \cite{Fritzsch:1972jv, Fritzsch:1973pi} \footnote{Although \emph{Lagrange density} is the correct nomenclature, it is more common for this object to be referred to as a \emph{Lagrangian}. These will be used interchangeably throughout this work.} as
\begin{eqnarray}
\label{qcdlag}
\mathcal{L}_{QCD}
=\bar{q}\left(i~\slash\hspace{-.24cm} {\rm{D}} -{\rm{m}}_q\right)q-\frac{1}{4}\text{Tr}~[G_{\mu\nu}G^{\mu\nu}],\end{eqnarray}
where $\slash\hspace{-.24cm} {\rm{D}}=D^0\gamma^0-{\bf{D}\cdot{\boldsymbol\gamma}},$ and $\gamma^{\mu}$ are the standard Dirac matrices satisfying the Clifford algebra $\{\gamma^\mu,\gamma^\nu\}=2g^{\mu\nu}$. Above $D^\mu=\partial_\mu+ig A_{\mu}^at^a$ denotes the SU(3) color covariant derivative\footnote{Repeated indices are summed over, unless explicitly mentioned.}, and $G^a_{\mu\nu}$ is the gluon field strength tensor, 
\begin{eqnarray}
G^a_{\mu\nu} = \partial_{\mu}A_\nu^a-\partial_{\nu}A_\mu^a-gf^{abc}A^b_{\mu}A^c_{\nu},
\end{eqnarray}
where the structure constant $f^{abc}$ is defined by $[t^a,t^b]=f^{abc}t^c$, $\text{Tr}~[t^at^b]=\delta^{ab}/2$. With eight $A_{\mu}^a$ in color space (denoted here by the index ``$a$"), gluons are described in the adjoint representation of $SU(3)$ \cite{Fritzsch:1973pi}. Quarks transform in the fundamental representation of $SU(3)$, and as a result are described by a three-dimensional vector field in color. $\rm{m}_q$ denotes the six-dimensional diagonal mass matrix with ${\rm{m}}_q=diag(m_u,m_d,m_s,m_c,m_b,m_t)$. Due to the fact that quarks are not asymptotic states, their masses are renormalization scheme dependent. In the $\overline{MS}$ scheme, the masses can be determined using a self-consistent definition $m_q(\mu=m_q)=m_q$, where $\mu$ denotes the renormalization scale of the theory. Using this definition, the charm, bottom and top have masses of $m_c=1.275(25)~{\rm{GeV}},~m_b=4.18(3)~{\rm{GeV}},$ and $m_t=173.5(6)(8)~{\rm{GeV}}$, respectively \cite{pdg}. For sufficiently low energies ($E\ll m_c$), these particles can not go \emph{on-shell} ($E^2=m^2+p^2$). Therefore their contribution to low-energy observables are kinematically suppressed and will be neglected from the remainder of this discussion.  
 
The QCD Lagrangian, Eq. (\ref{qcdlag}), is necessarily invariant under $SU(3)$ gauge transformations. Under this transformation, the quark field transforms as 
\begin{eqnarray}
\label{gaugetr_q}
q(x)\rightarrow \Omega(x)q(x)
\end{eqnarray}
 where $\Omega(x)\in SU(3)$, while the gauge field transforms as 
 \begin{eqnarray}
 \label{gaugetr_A}
 A_\mu\rightarrow \Omega(x) A_\mu \Omega(x)^{-1}+\frac{i}{g}(\partial_\mu \Omega(x)) \Omega(x)^{-1}.
 \end{eqnarray}
It is easy to show that both the Dirac term and theYang-Mills term in the QCD Lagrangian remain invariant under gauge transformations. 

\begin{figure}[t]
\begin{center} 
\subfigure[]{
\label{alpha}
\includegraphics[scale=0.7]{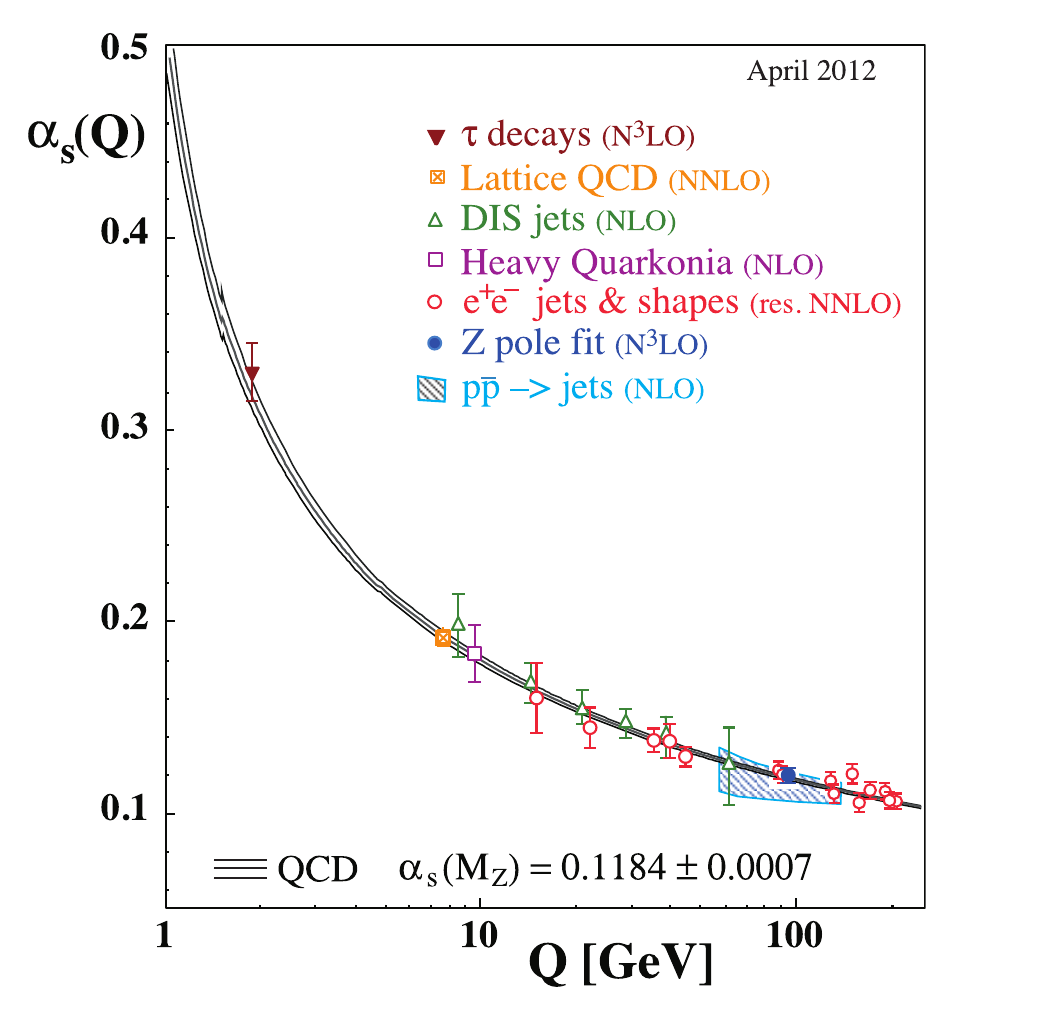}}
\subfigure[]{
\label{alpha_LQCD}
\includegraphics[scale=0.2]{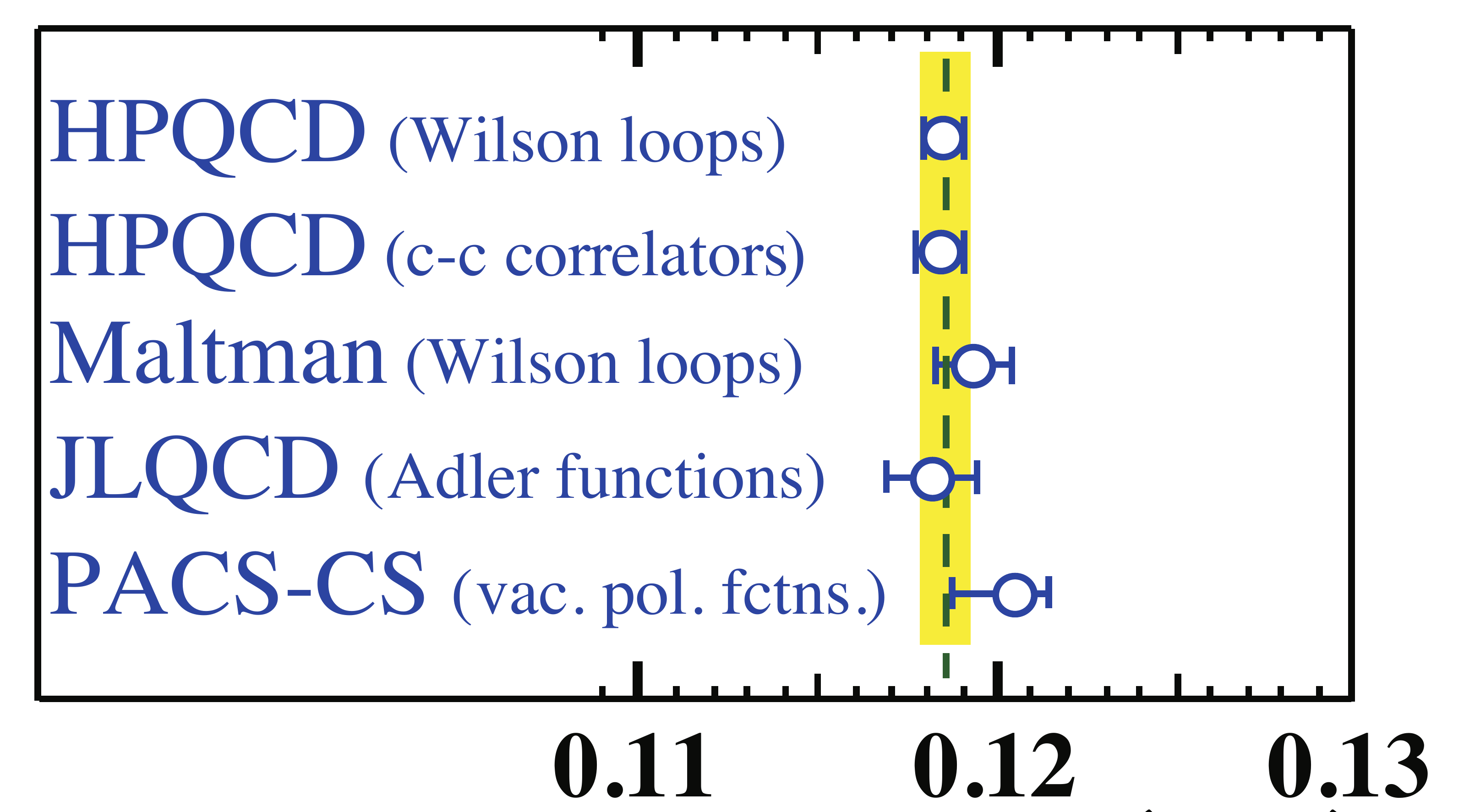}}
\caption[$\alpha_s(Q)$]{{\small a) Shown is a summary of $\alpha_s$ as a function of energy energy scale $\mu=Q$ using a variety of different techniques. In parenthesis is shown the order in perturbation theory used (NLO: next-to-leading order; NNLO: next-to-next-to leading order; res. NNLO: NNLO matched with rsummed next-to-leading logarithms; N3LO: next-to-NNLO) \cite{pdg}. b) Currently the most precise determination of the running coupling constant is obtained via LQCD, $\alpha_s^{\text{latt}}(m_Z) = 0.1185 \pm 0.0007$, which is obtained as a world average of Refs.~\cite{Maltman:2008nf, Aoki:2009tf, Shintani:2010ph, McNeile:2010ji,Davies:2003ik}.
The figure is reproduced with permission from the authors of the Particle Data Group's review on quantum chromodynamics.}}\label{running_alpha}
\end{center}
\end{figure}

Having written down the Lagrangian, one may proceed to construct the QCD Feynman rules and calculate observables. The quantity that best illustrates the challenges associated with understanding QCD is $\alpha_s(\mu)=g^2(\mu)/4\pi$, the QCD analogue of the fine-structure constant. $g(\mu)$ is the running coupling between quarks and gluons at the renormalization scale $\mu$. At next-to-leading order (NLO) in the perturbative expansion, the $\beta$-function of QCD is~\cite{Politzer:1973fx, Gross:1973id}
\begin{eqnarray}
\label{betaf}
\beta{(g)}\equiv \mu \frac{\partial}{\partial \mu}g(\mu)=-\frac{g^3}{16\pi^2}\left(11-\frac{2N_f}{3}\right)+\mathcal{O}{(g^5)},
\end{eqnarray}
where $N_f$ is the number of dynamical quarks, typically between three to six. Since the $\beta$-function is always negative for $N_f\leq 16$, the interactions between quarks and gluons asymptotically vanishes at large energies. This phenomenon is known as \emph{asymptotic freedom}~\cite{Politzer:1973fx, Gross:1973id}. Equation (\ref{betaf}) can be solved for $\alpha_s(\mu)$ in terms of a subtraction point $\Lambda_{QCD}$
\begin{eqnarray}
\label{alphas}
\alpha_s{(\mu)}=\frac{12\pi}{(33-2N_q)\log(\mu^2/\Lambda_{QCD}^2)}.
\end{eqnarray}
As $\mu$ approaches $\Lambda_{QCD}$, the strength of the interactions quickly diverges, illustrating the fact that standard perturbative tools fail and non-perturbative effects become important. From experiments and LQCD calculations, it is observed that $\Lambda_{QCD}\sim200~\rm{MeV}$, which sets the scale for non-perturbatively strong effects. Figure~\ref{running_alpha} shows a summary of $\alpha_s$ as a function of the energy scale $\mu=Q$ determined from a variety of physical processes. 

Although it may seem as an insurmountable challenge to study low-energy hadronic physics, this obstacle can be overcome by the implementation of two ideas. The first is to construct a low-energy EFT with hadrons as degrees of freedom, where the high-energy degrees of freedom, quarks and gluons, have been integrated out yet the underlying symmetries of QCD are still manifested. EFTs allow to have analytic control over the low-energy physics, yet, as will be discussed in the subsequent subsection, it has limited predictability. This is due to the fact that the EFT Lagrangian includes an infinite tower of operators with low energy coefficients (LECs) that are a priori undetermined.  
The second approach is to discretized the QCD action in order numerically evaluate observable in a non-perturbative way, known as latttice QCD (LQCD)~\cite{Wilson:1974sk}. The advantage of LQCD is that it is in principle fully predictable at the cost of loosing analyticity. Therefore, these are complimentary programs, that together allow one to have complete control of low-energy phenomena. 

In order to understand the most prominent EFTs in nuclear physics, it is key to first understand the symmetries of QCD. By definition, QCD has exact $SU(3)$ gauge symmetry. QCD also exhibits an approximate \emph{chiral} symmetry. With $m_u^{\overline{MS}}(\mu=2{~\rm{GeV}})=2.3^{+0.7}_{-0.5}~{\rm{MeV}},~m_d^{\overline{MS}}(\mu=2{~\rm{GeV}})=4.0^{+0.7}_{-0.3}~{\rm{MeV}},~m_s^{\overline{MS}}(\mu=2{~\rm{GeV}})=95(5)~{\rm{MeV}}$~\cite{pdg}, the $u$, $d$, and $s$ quarks are light compared to the nonpertubative scale $\Lambda_{QCD}$. Therefore, one can consider performing perturbations about the chiral point, where the ``light" quarks are exactly massless. In this limit, the part of the QCD Lagragian that includes quark fields, Eq. (\ref{qcdlag}), reduces to
\begin{eqnarray}
\label{chiqcd}
\mathcal{L}_{\chi QCD}
=\bar{q}i~\slash\hspace{-.24cm} {\rm{D}} q
=\bar{q}_Li~\slash\hspace{-.24cm} {\rm{D}} q_L
+\bar{q}_Ri~\slash\hspace{-.24cm} {\rm{D}} q_R
,\end{eqnarray}
where $q_R=\frac{1+\gamma_5}2 q, ~q_L=\frac{1-\gamma_5}2 q$. In this limit the QCD Lagrangian has an \emph{accidental} $SU(3)_L\times SU(3)_R$ chiral symmetry
\begin{eqnarray}
\label{chiralR}
q_L\rightarrow L~q_L,\hspace{2cm}q_R\rightarrow R~q_R,
\end{eqnarray}
where $L\in SU(3)_L,~ R\in SU(3)_R$ mix the three light flavors. The Lagrangian also has $U(1)_V\times U(1)_A$ symmetries. The vector $U(1)$ rotates both left- and right-handed fields by the same phase. The conserved charge associated with this symmetry is known as \emph{baryon number}. The axial $U(1)$ rotates the left-handed quarks by a phase and the right-handed quarks by the opposite phase. Although the axial $U(1)$ is a symmetry of the QCD Lagrangian, it is not a symmetry of the QCD path integral. The symmetry breaking of the axial $U(1)$ by quantum corrections is known as the \emph{axial anomaly} and it dominates for the $\pi^0\longrightarrow \gamma\gamma$ decay~\cite{Adler:1969gk, Adler:1969er, Bardeen:1969md, Bell:1969ts}.

Applying Noether's theorem~\cite{Noether:1918zz} to the chiral QCD Lagrangian, Eq. (\ref{chiqcd}), after an infinitesimal transformation of the form of Eq. (\ref{chiralR}), one obtains the currents
associated with the transformations of the left-handed or right-handed quarks
\begin{eqnarray}
L^{\mu,a}&=&\bar{q}_L\gamma^\mu\frac{t^a}{2}q_L,\hspace{1cm}\partial_\mu L^{\mu,a}=0\\
R^{\mu,a}&=&\bar{q}_R\gamma^\mu\frac{t^a}{2}q_R,\hspace{1cm}\partial_\mu R^{\mu,a}=0.
\end{eqnarray}
It is convenient to consider linear combination of these currents that do not mix under parity transformations. In particular, one can construct the vector and axial currents 
\begin{eqnarray}
V^{\mu,a}&=&R^{\mu,a}+L^{\mu,a}=\bar{q}\gamma^\mu\frac{t^a}{2}q\\
A^{\mu,a}&=&R^{\mu,a}-L^{\mu,a}=\bar{q}\gamma^\mu\gamma_5\frac{t^a}{2}q.
\end{eqnarray}
The $V^{\mu}$ and $A^{\mu}$ currents associate with the $U(1)_V$ and $U(1)_A$ symmetries are similarly defined by replacing ${t^a}$ with the identity. 

The light quark masses are in fact nonzero, therefore one should expect the divergences of the $SU(3)_L\times SU(3)_R$ currents to be nonzero and proportional to the quark masses. By performing Noether's theorem to the full Lagrangian, Eq. (\ref{qcdlag}), one observes that this is indeed the case
\begin{eqnarray}
\partial_\mu V^{\mu,a}&=&i\bar{q}~[m_q,\frac{t^a}{2}]~q\\
\partial_\mu A^{\mu,a}&=&i\bar{q}~\{m_q,\frac{t^a}{2}\}~\gamma_5q\\
\partial_\mu V^{\mu}&=&0\\
\label{anomaly}
\partial_\mu A^{\mu}&=&2i\bar{q}~m_q\gamma_5q +\frac{3g^2}{32\pi^2}\epsilon_{\mu\nu\sigma\rho}~\text{Tr}~[G^{\mu\nu}G^{\sigma\rho}],
\end{eqnarray}
where the second term in Eq. (\ref{anomaly}) corresponds to the axial anomaly~\cite{Adler:1969gk, Adler:1969er, Bardeen:1969md, Bell:1969ts}, which makes the divergence of the current nonzero even in the chiral limit. 

Although $SU(3)_L\times SU(3)_R$ is an approximate symmetry of the QCD Lagrangian, it is not a symmetry of the QCD vacuum (even in the chiral limit). This is most clearly observed by considering the implications of this symmetry onto the hadronic spectrum. For instance, under parity the vector and axial currents do not mix,
 \begin{eqnarray}
 A^{\mu,a}(t,{\bf{x}})&\rightarrow&  A_{\mu}^a(t,-{\bf{x}})\\
  V^{\mu,a}(t,{\bf{x}})&\rightarrow& - V_{\mu}^a(t,-{\bf{x}}),
 \end{eqnarray}
but under chiral transformations these currents do in fact mix. As a result, one should expect parity partners to be degenerate, or nearly degenerate considering that the light quark masses are indeed nonzero. Yet, this is not observed experimentally. For example, the proton, with quantum numbers\footnote{This is standard notation for labeling the quantum numbers of a particle, where $I$=isospin, $J$=angular momentum and $P$=parity.} of $I(J^{P})=\frac{1}{2}(\frac{1}{2}^{+})$ and a mass of $m_{p}=938.2727046(21)$~MeV, is approximately 600~MeV lighter than its parity partner $N(1535)$, which has $I(J^{P})=\frac{1}{2}(\frac{1}{2}^{-})$~\cite{pdg}. This observations leads us to conclude that $SU(3)_L\times SU(3)_R$ is spontaneously broken by the vacuum expectation value (VEV) of the mass operator 
\begin{eqnarray}
\label{qcdvev}
\langle \bar{q}_R^a q_L^b\rangle= v~\delta^{ab},
\end{eqnarray}
where $v$ is the VEV and it is of the order $\Lambda_{QCD}^3$. The presence of a mass in the Lagrangian explicitly breaks chiral symmetry, therefore one should expect a dynamically generated mass to spontaneously break it. Under chiral transformations Eq.~(\ref{qcdvev}) goes to
\begin{eqnarray}
\label{qcdvev2}
\langle \bar{q}_R^a q_L^b\rangle\rightarrow v~(LR^\dag)^{ba}\equiv v \Sigma^{ab}.
\end{eqnarray}
Transformations under which the left-handed and right-handed quarks are simultaneously rotated with $L=R$ leave the condensate invariant. For transformations where $L\neq R$, $\Sigma^{ab}$ parametrizes a different vacuum of QCD than the one shown in Eq. (\ref{qcdvev}) with the same energy. Therefore, the QCD vacuum spontaneously breaks $SU(3)_L\times SU(3)_R$ and reduces it to the diagonal subgroup $SU(3)_V$. 

Goldstone's theorem \cite{Goldstone:1961eq, Goldstone:1962es} dictates that for each generator of the symmetry that is spontaneously broken there is a massless, spinless boson in the spectrum of the theory, commonly referred to as \emph{Goldstone boson}.  In this case, the symmetry that is broken is an axial vector current, therefore the Goldstone boson are in fact pseudoscalars. Furthermore, since the chiral symmetry was not an exact symmetry to begin with, these \emph{pseudo-Goldstone bosons} (pGB) are not expected to be massless but rather much lighter than other particles in the spectrum. This fact is most evident when considering the special case where only the up and down masses are treated as light. In this case, the chiral symmetry reduces to $SU(2)_V$, commonly known as \emph{isospin symmetry}. Since the symmetry broken is a two-dimensional special unitary symmetry, there are three generators and therefore three pGB. These are the pions, $\{\pi^+,\pi^0,\pi^-\}$, with masses that are about three times smaller than any other particle in the hadronic spectrum. The quantum numbers of the mesons can be identified by the quark model  $\{\pi^+:u\bar{d},~\pi^0:\frac{u\bar{u}-d\bar{d}}{\sqrt{2}},~\pi^-:d\bar{u}\}$. With $m_{\pi^{\pm}}=139.57018(35)~\rm{MeV}$~\cite{pdg}, this isospin triplet is nearly degenerate. The splitting can be understood by the fact that the up and down quarks are not degenerate and the inclusion of electromagnetic effects. The identification of the pions as the pGBs of QCD is the foundation of SU(2) chiral perturbation theory ($\chi$PT), which is the subject of section~\ref{chipt}.

Although the strange quark is significantly more massive than the up and down quarks, one can still consider the implications of a spontaneously broken $SU(3)_L\times SU(3)_R$ chiral symmetry. Since $SU(3)$ has eight generators, one should expect there to be eight pGB, three of the which the pions listed above. The remaining five can be identified as $\{K^+,K^0,\bar{K}^0,K^-,\eta\}$. With masses in the range of $490$-$550$~MeV, it is clear that explicit symmetry breaking effects associated with a non-zero strange quark mass are significant. As will be discussed in section~\ref{chipt}, these effects can be addressed in a controlled fashion using SU(3)~$\chi$PT. 

Isospin symmetry plays an important role in nuclear physics, since the proton and neutron are isospin partners. With a quark content of $\{p:uud,~n:ddu\}$ and masses $m_p=938.2727046(21) $~MeV and $m_n=939.565379(21)$~MeV~\cite{pdg}, the proton and neutron can be represented as components of an isospin doublet $N^T=(p,~n)$. This fact will be used extensively through out this thesis, and it will play an key role when constructing an effective field theory for nucleons in chapter \ref{NNsys}.


\section{Effective Field Theory \label{efts_sec}}

The guiding principle of EFTs is that low energy phenomena should be largely insensitive to the details of the fundamental high energy theory. As a result, EFTs are a versatile and extremely powerful tool. There are two classic examples of EFTs. The first corresponds to a theory with light degrees of freedom, $\psi_l$, with mass $m_l$, and heavy excitations $\Psi_h$, with mass $M_h$. For energies well below $M_h$, the heavy modes cannot go on-shell and can be systematically integrated out. This is principal notion behind Fermi's effective theory of weak interactions \cite{Fermi:1934hr, Fermi:1934sk}. Weak processes are mediated by $W^{\pm}$ and $Z^0$ bosons\footnote{$m_W=80.385(15)$~GeV, $m_Z=98.1876(21)$~GeV \cite{pdg}}. For energies in the order of $M_h\sim \{m_W,~m_Z\}$ the intermediate particle propagator, $1/(p^2-M_h^2)$, has a pole; therefore correlation functions involving these particles exhibit complicated non-analytic behavior. For energies well below $M_h$, one can Taylor expand the propagator $1/(p^2-M_h^2)\approx -1/M_h^2-p^2/M_h^4+\cdots$, effectively integrating out effects from the $W^{\pm}$ and $Z^0$ bosons. In this energy regime, the analytic behavior of the correlation functions can be reproduced by contact operators in terms of the asymptotic states of the theory. In fact, section \ref{pinot1} discusses in great detail an analogous EFT for nucleons for interaction energies well below the pion production. In this case the pions can be integrated out of the theory, resulting in a EFT without pions as degrees of freedom and is commonly referred to as EFT$\left(\pinot\right)$ \cite{pds, pds2}. 

The second example of EFTs is one where the relationship between the low-energy and high-energy degrees of freedom is a non-perturbative one. This is the case of the strong sector, where at high energies it is manifested as a fundamental interaction between quarks and gluons, while at low-energies only hadrons are observed. Of course, QCD is not the only theory where such phenomena is observed. In fact, this is a common practice for studying condense matter systems, where collective excitations can be typically described via a low-energy EFT. Probably the best well known example is Landau's theory of phase transitions~\cite{landau1936theory}, which set the foundation of our understanding of the manifestation of spontaneous symmetry breaking in the universe. 

By integrating out high-energy excitations, an infinite tower of operators that are consistent with the symmetries of the theory are generated~\cite{Wilson:1973jj}. To have predictive power, it is essential to define a hierarchy between the operators. For low-energy systems, the typical expansion parameter is $p/\Lambda$, where $p$ momentum of the interactions and $\Lambda$ is the energy scale at which the EFT breaks down. 

\subsection{Chiral Perturbation Theory \label{chipt}}

To get a deeper understanding of the power of EFTs, it is necessary to look at several examples. In this section, we will review probably the most widely used low-energy EFT for hadronic systems, $\chi$PT, which describes the dynamics of pions, kaons and etas. In section \ref{qcdsec}, it was stated that the QCD vacuum spontaneously breaks the approximate chiral $SU(3)_L\times SU(3)_R$ down to $SU(3)_V$, and as a consequence there are eight nearly massless pGB, one for each broken $SU(3)$ generator. It is convenient to parametrize these long-wavelength excitations by upgrading $\Sigma$ in Eq. (\ref{qcdvev2}) to a local operator
\begin{eqnarray}
\label{sigma_matrix}
{\Sigma}(x)=\exp\left(\frac{2i{\boldsymbol{\pi}(x)}}{f}\right), \hspace{1cm}\boldsymbol{\pi}(x)=\boldsymbol{\pi}^a(x)~t^a,
\end{eqnarray} 
where $f$ is the meson decay constants in the chiral limit\footnote{In this convention $f_\pi^{\rm{phys}}=130.41(20)~\rm{MeV}$ and $f_K^{\rm{phys}}=156.1(9)~\rm{MeV}$ at the physical point~\cite{pdg}}, and $\boldsymbol{\pi}^a$ paramatrices the eight pGB bosons transforming as an octet in $SU(3)_V$: $\boldsymbol{\pi}(x)\rightarrow V\boldsymbol{\pi}(x)V^{\dag}, ~V\in SU(3)_V$. In terms of the particle basis, $\boldsymbol{\pi}(x)$ can be written as $\boldsymbol{\pi}(x)$
\begin{eqnarray}
\boldsymbol{\pi}=
\begin{pmatrix} 
\frac{{\pi}^0}{\sqrt{2}} +\frac{{\eta}}{\sqrt{6}} &{\pi}^+ &{K}^+\\
{\pi}^- & -\frac{{\pi}^0}{\sqrt{2}}+\frac{{\eta}}{\sqrt{6}}&{K}^0 \\
{K}^- &{\bar{K}}^0&-{\eta}\sqrt{\frac{2}{3}} \\
\end{pmatrix}.  \end{eqnarray} 
One can verify this representation by evaluating the azimuthal component of isospin ($I_3$), the hypercharge ($Y$), and charge ($Q$) of $\boldsymbol{\pi}$:

\begin{eqnarray}
I_3=\frac{1}{2}\begin{pmatrix} 
1&0 &0\\
0&-1 &0\\
0&0 &0\\
\end{pmatrix},\hspace{1cm}
\left[I_3,\boldsymbol{\pi}\right]=
\begin{pmatrix} 
0&{\pi}^+ &\frac{{K}^+}{2}\\
-{\pi}^- &0&-\frac{{K}^0}{2} \\
-\frac{{K}^-}{2} &\frac{{\bar{K}}^0}{2}&0 \\
\end{pmatrix}\\ 
Y=\frac{1}{3}\begin{pmatrix} 
1&0 &0\\
0&1 &0\\
0&0 &-2\\
\end{pmatrix},\hspace{1cm}
\left[Y,\boldsymbol{\pi}\right]=
\begin{pmatrix} 
0&0&{{K}^+}\\
0 &0&{{K}^0} \\
-{{K}^-} &-{{\bar{K}}^0}&0 \\
\end{pmatrix}\\ 
Q=\frac{1}{3}\begin{pmatrix} 
2&0 &0\\
0&-1 &0\\
0&0 &-1\\
\end{pmatrix},\hspace{1cm}
\left[Q,\boldsymbol{\pi}\right]=
\begin{pmatrix} 
0&{\pi}^+ &{{K}^+}\\
-{\pi}^- & 0&0\\
-{{K}^-}&0&0 \\
\end{pmatrix}.  \end{eqnarray}

With this, we are ready to construct the $\chi$PT Lagrangian. As mentioned earlier, the low-energy EFT must have the same symmetries as QCD. Namely, it should it must be Lorentz invariant, conserve charge and parity. Furthermore, in the chiral limit, $m_{q}\rightarrow 0$, the $\chi$PT Lagrangian must be invariant under chiral transformations 
\begin{eqnarray}
\label{chiraltrans}
\Sigma(x)\rightarrow L\Sigma(x) R^\dag.
\end{eqnarray}
The simplest operator that can be constructed that satisfies all of this criteria is $\text{Tr}~[\Sigma^\dag \Sigma]=N_f$, where $N_f$ is the number of light flavors. This operator adds an overall constant to the Lagrangian that can be absorbed into the vacuum state energy. 

The first nontrivial operator to consider is $\rm{Tr}\left[\partial^\mu \Sigma^{\dag}\partial_\mu\Sigma\right]$. Similarly, one can construct operators with more than two derivatives, but in order to have a proper expansion parameter it is necessary to define the expansion parameter of the theory. With a mass of $m_\rho=775.49(34)$~MeV and a strong overlap with the $\pi\pi$ P-wave state, is natural to set the $\rho$ mass as the chiral symmetry breaking scale $\Lambda_\chi\sim800$~MeV. It will be demonstrated later that the chiral expansion is in terms of $\Lambda_\chi=4\pi f_\pi\sim1.6$~GeV, improving the convergence of the theory. Therefore at $\mathcal{O}(p^2/\Lambda_\chi^2)$, $\rm{Tr}\left[\partial^\mu \Sigma^{\dag}\partial_\mu\Sigma\right]$ is the only chirally symmetric operator.

As has already been pointed out, QCD is in fact not chirally symmetric, and chiral symmetry breaking corrections must be included which scale with the light quark masses. Therefore a natural operator to consider is $ \rm{Tr}\left[m_q\Sigma^{\dag}\right]$, where $m_q=diag(m_u,m_d,m_s)$. Note that this operator is only invariant under $SU(3)_V$ when all the quark masses are degenerate. As this is approximately true for the $SU(2)$, it is convenient to write the quark matrix in the isospin limit $m_q\rightarrow diag(m_l,m_l,m_s)$. Alternatively, one can write this operator in terms of the light meson masses $\mathbb{M}= \frac{1}{2}diag(m_{\pi}^2,m_{\pi}^2, 2m_{K}^2 -m_{\pi}^2)$, $\sim \rm{Tr}\left[\mathbb{M}\Sigma^{\dag}\right]$, and as will be shown below the difference is an overall constant. Treating $m_l^2/\Lambda_\chi^2$ to be in the order of  $\mathcal{O}(p^2/\Lambda_\chi^2)$, this operator comes in at order $\mathcal{O}(p^2/\Lambda_\chi^2)$. 

At LO the light mesonic  $\chi$PT Lagrangian is~\cite{GellMann:1968rz, Weinberg:1968de}
\begin{eqnarray} 
    \label{LO_chipt}
  {\mathcal L}_{\chi PT}^{LO}&=& 
  \frac{f^2}{8}\rm{Tr}\left[\partial^\mu \Sigma^{\dag}\partial_\mu\Sigma\right]
    +\frac{f^2}{4}\rm{Tr}\left[ \mathbb{M}\Sigma^{\dag}+\rm{h.c.}\right]+\mathcal{O}\left(\frac{p^4}{\Lambda_\chi^2}\right)\\
    &=& \frac{f^2}{8}\rm{Tr}\left[\partial^\mu \Sigma^{\dag}\partial_\mu\Sigma\right]
    +\frac{B_0 f^2}{4}\rm{Tr}\left[ {m_q}\Sigma^{\dag}+\rm{h.c.}\right]+\mathcal{O}\left(\frac{p^4}{\Lambda_\chi^2}\right).  \nn    
      \end{eqnarray}     
By expanding $\Sigma=1+\frac{2i{\boldsymbol{\pi}}}{f}+\frac{1}{2!}\left(\frac{2i{\boldsymbol{\pi}}}{f}\right)^2+\cdots$, we obtain the isospin limit of the Gell-Mann-Okubo formula $m_\eta^2={(4m_K^2-m_\pi^2)}/{3}$
\cite{GellMann:1968rz}. Also, one finds the relationship between the bare quark masses and the bare meson masses, $\{m_\pi^2=2B_0~m_l,m_K^2=B_0~(m_l+m_s)\}.$ Of course both of these relations will receive quantum corrections. 

\begin{figure}[t]
\begin{center}
\subfigure[]{
\label{chiptdiag1}
\label{chiptLO}
\includegraphics[scale=0.3]{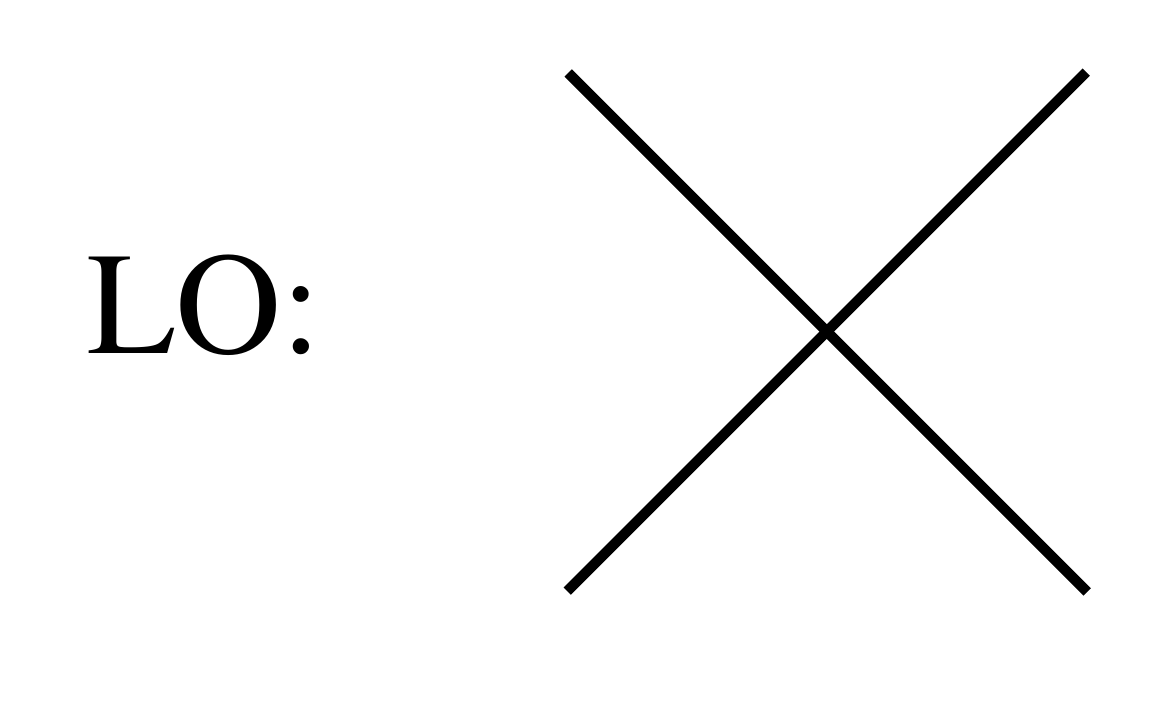}}
\subfigure[]{
\label{chiptdiag2}
\includegraphics[scale=0.3]{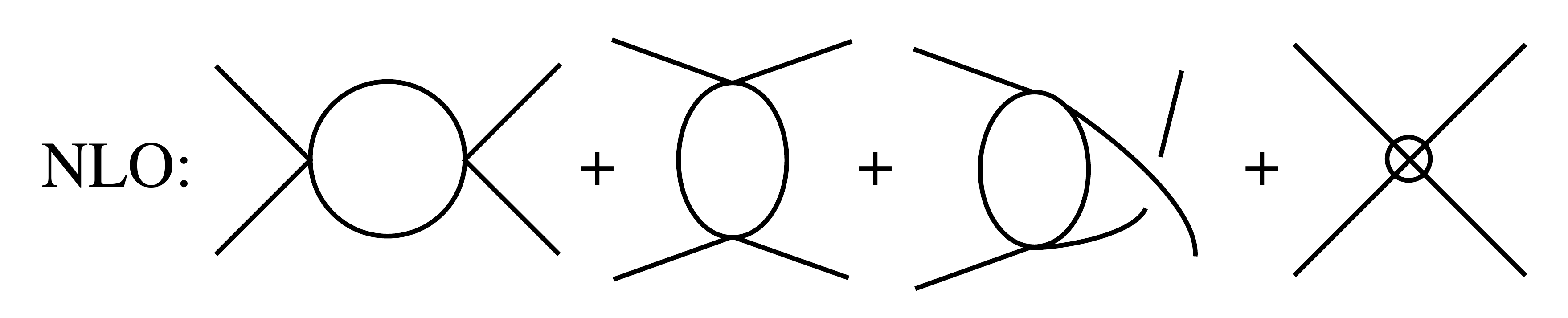}
}
\caption[LO and NLO scattering from $\chi$PT]{a) Leading order contribution to the meson-meson scattering amplitude arising from contact interactions in Lagrangian, Eq. (\ref{LO_chipt}). b) NLO corrections to the scattering amplitude. The vertices appearing in the first loops are determined from Eq. (\ref{LO_chipt}), while the fourth diagram denotes the counter terms appearing in Eq. (\ref{NLO_chipt}).} 
\end{center}
\end{figure}

A remarkable feature of this Lagrangian is that by expanding $\Sigma(x)$ to fourth-order in $\boldsymbol{\pi}(x)$, one obtains a prediction for the S-wave scattering amplitude of $\{\pi\pi\rightarrow \pi\pi,\pi K\rightarrow \pi K,K K\rightarrow K K,\ldots\}$ in terms of just the masses and decay constants. As discussed in appendix \ref{scattheory}, the scattering amplitude can be extracted from the sum of $2\rightarrow 2$ Feynman diagrams. For example, the $\pi^+\pi^+$ and $\pi^+ K^+$ scattering at LO in SU(3) $\chi$PT (depicted in Fig.~\ref{chiptLO}) are ~\cite{Weinberg:1966kf}
\begin{eqnarray}
\label{su3LO}
m_{ \pi}a_{\pi^+\pi^+}=\frac{m^2_{ \pi}}{8\pi f^2_\pi},\hspace{2cm}
\mu_{\pi K} a_{\pi^+ K^+}=\frac{\mu^2_{\pi K}}{4\pi f_\pi f_K},
\end{eqnarray}
where $\mu_{\pi K}$ is the reduced mass of the pion-kaon systems, satisfying $\mu_{\pi K}^{-1}=m^{-1}_\pi+m^{-1}_K$. There is currently no experimental determination of $\mu_{\pi K}a_{\pi^+ K^+}$, but the LO theoretical prediction of $m_{ \pi}a^{LO}_{\pi^+ \pi^+}=0.04558(13)$\footnote{Note, the uncertainty quoted only includes propagated uncertainties due to the experimental uncertainties of the masses and decay constant but not systematic uncertainties due to the chiral expansion.}~\cite{Weinberg:1966kf} compares well with experimental determinations \cite{Pislak:2003sv, Pislak:2001bf}
\begin{eqnarray}
\label{apipi_exp}
m_{ \pi}a^{\rm{exp}}_{\pi^+ \pi^+}=0.0454(31).
\end{eqnarray}
To properly theoretically determine the scattering lengths, it is necessary to perform the calculation to NLO to assure that there is convergence. At $\mathcal{O}(p^4/\Lambda_\chi^4)$, the scattering lengths get contributions from s-,t-,u-channel loops as depicted in Fig. \ref{chiptdiag2}, but also contributions from LECs appearing in the $\mathcal{O}(p^4/\Lambda_\chi^4)$ Lagrangian \cite{Gasser:1984gg}
\begin{eqnarray} 
    \label{NLO_chipt}
  {\mathcal L}_{\chi PT}^{NLO}&=& 
  L_1\rm{Tr}\left[\partial^\mu \Sigma^{\dag}\partial_\mu\Sigma\right]^2
+  L_2\rm{Tr}\left[\partial^\mu \Sigma^{\dag}\partial^\nu\Sigma\right] 
\rm{Tr}\left[\partial_\mu \Sigma^{\dag}\partial_\nu\Sigma\right]
+  L_3\rm{Tr}\left[\partial^\mu \Sigma^{\dag}\partial_\mu\Sigma
~\partial^\nu \Sigma^{\dag}\partial_\nu\Sigma\right]\nn\\
&+&  L_4\rm{Tr}\left[\partial^\mu \Sigma^{\dag}\partial_\mu\Sigma\right]\rm{Tr}\left[ \mathbb{M}\Sigma^{\dag}+\rm{h.c.}\right]
+ L_5\rm{Tr}\left[\partial^\mu \Sigma^{\dag}\partial_\mu\Sigma\left( \mathbb{M}\Sigma^{\dag}+\rm{h.c.}\right)\right]\nn\\
&+& L_6\rm{Tr}\left[ \mathbb{M}\Sigma^{\dag}+\rm{h.c.}\right]^2
+ L_7\rm{Tr}\left[ \mathbb{M}\Sigma^{\dag}-\rm{h.c.}\right]^2
+ L_8\rm{Tr}\left[ \mathbb{M}\Sigma^{\dag}\mathbb{M}\Sigma^{\dag}+\rm{h.c.}\right],     
      \end{eqnarray}     
where UV effects are encapsulated in the low-energy coefficients (LECs) $L_i$. Note that the convention in the literature is to use $\chi\equiv 2B_0m_q$, but at this order in the perturbation theory, this is equal to $\mathbb{M}$. 

Corrections to the scattering lengths from the NLO Lagrangian are depicted in Fig. \ref{chiptdiag2}. 
The loops appearing in Fig. \ref{chiptdiag2} are UV divergent and must be regulated. These divergences can be absorbed by the LECs, at the cost of introducing a renormalization scale, $\mu$, into the problem, and therefore the $L_i$'s become $\mu$-dependent. For example, at NLO, $m_{ \pi}a_{\pi^+ \pi^+}$ is equal to~\cite{Gasser:1984gg}
\begin{eqnarray}
m_{\pi} a_{\pi^+\pi^+}=	-\frac{m^2_{\pi}}{8\pi f_\pi^2}\left\{1+\frac{m_\pi^2}{(4\pi f_\pi)^2}
\left[3\log\left(\frac{m_\pi^2}{\mu^2}\right)
+\frac{1}{9}\log\left(\frac{m_\eta^2}{\mu^2}\right)
-\frac{8}{9}-32(4\pi)^2L^{I=2}_{\pi\pi}(\mu)\right]\right\}\nn\\
\label{apipi3}
\end{eqnarray}
where $L^{I=2}_{\pi\pi}=2L_1+2L_2+L_3-2L_4-L_5+2L_6+L_8$. Note that subleading effects are kinematically suppressed by $m_\pi^2/(4\pi f_\pi)^2\sim m_\pi^2/\Lambda_\chi^2$, justifying the assertion at the beginning of the section that $\Lambda_\chi\sim 4\pi f_\pi$. Since scattering lengths, masses, and decay constants are physical observables, they necessarily cannot depend on the renormalization scale. Requiring Eq.~(\ref{apipi3}) to be $\mu$-independent leads to the evolution of $L_{\pi\pi}^{I=2}(\mu)$ with the renormalization scale
\begin{eqnarray}
\label{evolution}
32(4\pi)^2L_{\pi\pi}^{I=2}(\mu)&=&32(4\pi)^2L_{\pi\pi}^{I=2}(\mu_0)+\frac{28}{9}\log\left(\frac{\mu^2_0}{\mu^2}\right).
\end{eqnarray}

The LECs are not fixed by the symmetries of the theory and must be determined by matching to experiments or evaluated directly from QCD. In fact, LQCD has been extremely successful in evaluating low-energy phase shifts for meson-meson scattering, superseding other theoretical tools and obtaining higher precision than experiments in the ${\pi^+\pi^+}\rightarrow{\pi^+\pi^+}$ channel. Currently the most precise determination of the scattering length for this channel is by the NPLQCD Collaboration~\cite{beane:2007xs}
\begin{eqnarray}
\label{nplqcdapipi}
m_{\pi} a_{\pi^+\pi^+}^{\rm{latt}}=0.04330(42)
\end{eqnarray} 
where the standard deviation includes statistical and systematic uncertainties added in quadrature. A compete list of references that have studied $\pi^+\pi^+$ scattering via LQCD include \cite{Aoki:2002in, Yamazaki:2004qb, Du:2004ib, Beane:2005rj, Li:2007ey, Dudek:2010ew, Yagi:2011jn,  Dudek:2012gj, Fu:2013ffa}. To understand the sources of error and the great success of LQCD in performing these calculations will require a thorough introduction to LQCD and the L\"uscher formalism, which will be done in sections  \ref{lqcd}\&\ref{mmsys}. In section \ref{pipiscat}, this problem will be revisited and it will be described in detailed how the LECs and consequently the scattering lengths are determined. \footnote{For a detailed introduction to $\chi PT$ see Ref.~\cite{Scherer:2002tk}.}

\subsection{Pionless EFT: EFT$\left(\pinot\right)$  \label{pinot1}}
As approximate pGB of QCD, pions are the mediators of the nuclear force at long-distances. Therefore, it would desirable to generalize the $\chi$PT formalism to incorporate baryons. An appropriate way to this was outlined by Jenkins and Manohar and it is known as Heavy Baryon $\chi$PT (HB$\chi$PT) \cite{Jenkins:1990jv, Jenkins:1991es}. HB$\chi$PT presents a natural way to describe the interactions between nucleons and pions that is consistent with the approximate chiral symmetry of QCD. Weinberg proposed using this formalism to describe $NN$ scattering \cite{Weinberg:1990rz, Weinberg:1991um}. For reasons that will be outlined below, this framework led to an ill-defined power counting scheme for strongly interactive systems, which is referred to as \emph{Weinberg's power counting scheme}. This issue will be discussed in section \ref{weinbergpc}.   

Alternatively, for energies well below pion-production, the pions can be effectively integrated out of the theory, akin to the $Z^0$ and $W^\pm$ in Fermi's effective theory of weak interactions. In this limit, the nucleon can be treated non-relativistically and NN-interactions can be accurately described in terms of an infinite series of operators with only contact and derivative interactions, known as EFT$\left(\pinot\right)$ \cite{pds, pds2}.

Being spin-1/2 fermions and forming an isospin doublet, two-nucleon systems have a rather rich structure. Requiring the nucleon state to be antisymmetric under  the interchange of the two-nucleons, there are four possible allowed channels: $\rm{(I,S)}^{\rm{P}}$: $(0,1)^{+}$, $(1,0)^{+}$, $(1,1)^{-}$, $(0,0)^{-}$. Each of these can be further decomposed into an infinite number of partial waves in orbital angular momentum, and some of these may even mix onto each other. For example, the deuteron channel is defined by $J^P=0^+$ and $I=0$ with ${^3S_1}$-${^3D_1}$ mixing. Chapter \ref{NNsys} discusses in great detail the generalization of EFT$\left(\pinot\right)$ for arbitrary two-nucleon channels. Therefore in this section it is sufficient to study a scalar analogue to two-nucleons. Furthermore, only S-wave scattering of identical bosons will be discussed in this section. 
 \begin{center}
\begin{table}
\label{tab:param1}
\resizebox{15cm}{!}{
\begin{tabular}{|cc|ccccccccccc|}
\hline
&$J^P~$&$0^+$&$0^-$&$1^+$&$1^-$&$2^+$&$2^-$&$3^+$&$3^-$&$4^+$&$4^-$&$\cdots$ \\\hline \hline
\multirow{2}{*}{$(L,S)$} &I=0&---&---&\{(0,1),(2,1)\}&(1,0)&(2,1)&---&\{(2,1),(4,1)\}&(3,0)&(4,1)&---&$\cdots$\\
&I=1&(0,0)&(1,1)&---&(1,1)& (2,0)&\{(1,1),(3,1)\}&---&(3,1)&(4,0)&\{(3,1),(5,1)\}&$\cdots$
\\\hline\hline
\end{tabular}}
\caption{All possible two-nucleon states with $J\leq 4$, where I=isospin, S=spin, and L=orbital angular momentum.}
\label{JP}
\end{table} 
\end{center}
To begin, it is important to be reminded of the non-relativistic reduction of a relativistic field theory. Consider the free Lagrangian of a complex scalar field $\Phi$, 
\begin{eqnarray}
 \label{rellag}
 \mathcal{L}_{free}&=&\Phi^\dag\left(-\partial^2-m^2\right)\Phi.
\end{eqnarray}
In $\chi PT$, the range of validity of the theory was defined by $\Lambda_\chi\sim800$~MeV. So it was natural to treat $m_\pi/\Lambda\sim p/\Lambda$ as a perturbation. For a theory where dynamical pions have been integrated out and we are interested in low-energy scattering of two-scalar ``nucleons", then $\Lambda\sim m_\pi$ and $m\gg m_\pi$. Therefore, in order to introduce a well-defined low-energy expansion it is necessary to perform perturbations about $E\sim m$. This is easily done by performing a field redefinition 
\begin{eqnarray}
\label{NRreduc}
\Phi\equiv\frac{1}{\sqrt{2m_b}}e^{-imt}{\phi}.
\end{eqnarray}
In doing so Eq. (\ref{rellag}) reduces to
\begin{eqnarray} 
\label{lagNR}
 \mathcal{L}_{free}&=&\phi^\dag\left(i\partial_t+\frac{\bf{\nabla}^2}{2m_b}-\frac{\partial_t^2}{2m_b}\right)\phi
 =\phi^\dag\left(i\partial_t+\frac{\bf{\nabla}^2}{2m_b}+\frac{\bf{\nabla}^4}{8m^3_b}+\cdots\right)\phi,
\end{eqnarray}
where the equations of motion [$-i\partial_t=\frac{\bf{\nabla}^2}{2m_b}+\cdots$] have been used in the second equality. Note that relativistic corrections to the NR theory appear naturally. That being said, in the remainder of this discussion these corrections will be neglected. Having removed the rest mass, the residual energy is the interaction energy, which scales like $Em\sim p^2$ and has a definitive low-energy expansion. 

Having introduced the free NR Lagrangian we need to introduce interactions. For sufficiently low energies the relevant symmetry is invariance under Lorentz transformations of small velocities, known as velocity reparametrization invariance \cite{Luke:1996hj}, most commonly known as Galilean invariance. For a scalar theory this is most easily done by constructing two-body operators of the form $\phi(x) \overleftrightarrow{\bf{\nabla}}^{2n} \phi(x)$. In momentum space this operator can be written as $({\bf{P-2k}})^{2n}\phi_{\bf{P-k}}\phi_{\bf{k}}$, which is cleraly invariant under boosts of the whole systems ${\bf{P}}\rightarrow {\bf{P}}+\delta {\bf{P}}$. With that it is straightforward to construct the interactive Lagrangian describing S-wave scattering,
\begin{eqnarray} 
\label{lagNR_int}
\mathcal{L}_{int}=
-\frac{C_0}{(2!)^2} \phi^{\dag2} \phi^2
-\frac{C_2}{(2!)^4} \left( \phi^{\dag2}(\phi \overleftrightarrow{\bf{\nabla}}^2 \phi)+\text{h.c.}\right)
-\frac{C_4}{(2!)^6} \left(\phi^\dag \overleftrightarrow{\bf{\nabla}}^2 \phi^\dag\right)\left( \phi \overleftrightarrow{\bf{\nabla}}^2 \phi\right)+\cdots.
\end{eqnarray}
Unlike the relativistic analogue, in a NR field theory particles are neither created nor destroyed. As a result only s-channel scattering diagrams contribute and single particle propagators are not renormalized. Of course, one can attempt to calculate self-energy, t- and u-channel diagrams using Eq. (\ref{lagNR_int}), and they would exactly vanish. 

\begin{figure}[t]
\begin{center}
\subfigure[]{
\label{PDS_LO}
\includegraphics[scale=0.5]{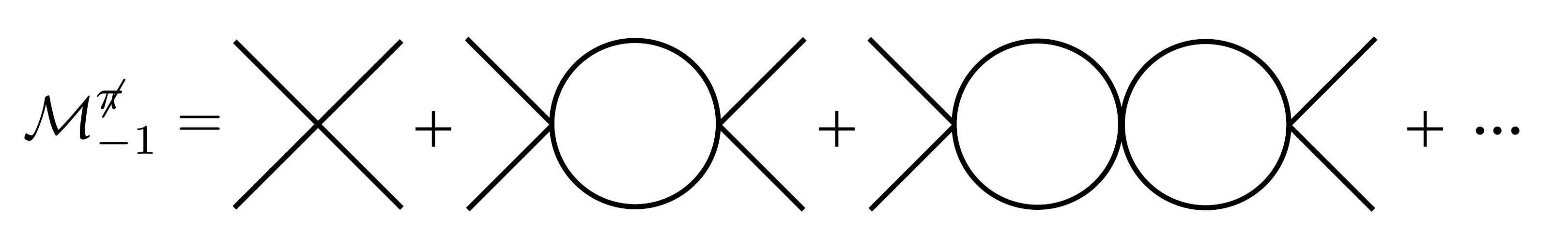}}
\subfigure[]{
\label{PDS_NLO}
\includegraphics[scale=0.5]{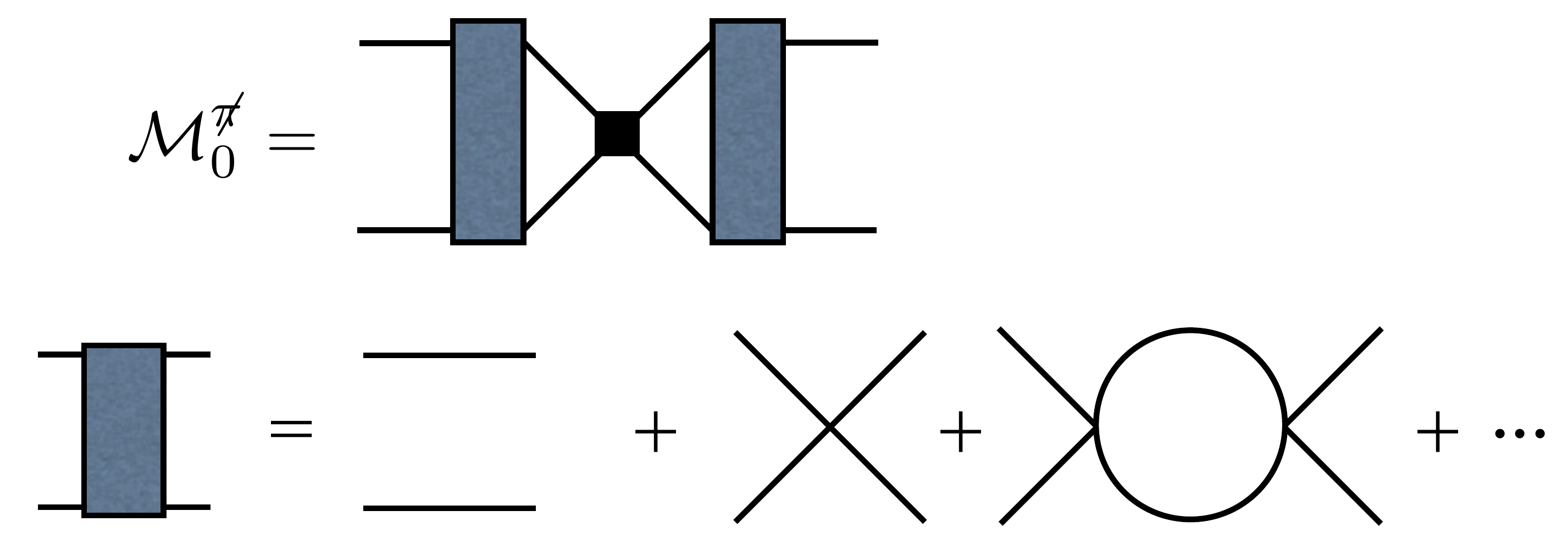}}
\caption[Scattering in EFT($\pinot$)]{a) The sum over ``bubble diagrams" constitutes the leading order EFT$\left(\pinot\right)$ scattering amplitude. The bare vertex denotes the contribution from $C_0(\mu)$. b) NLO corrections to the EFT$\left(\pinot\right)$ scattering amplitude. The square vertex denotes a single insertion of the contact interaction proportional to $C_2(\mu)$, while the two-particle propagator with boxes around them denotes the LO dressed two-particle propagator.}\label{PDS_diag}
\end{center}
\end{figure}

For weakly interacting systems, it would be natural to consider $C_0$ in Eq. (\ref{lagNR_int}) as the LO contribution to the scattering amplitude. Using the NR expression for the scattering amplitude, Eq. (\ref{scatNR}), and the effective range expansion, Eq. (\ref{EFE}), this leads to
\begin{eqnarray}
\label{naive}
C_0\stackrel{.}{=}\frac{8\pi a}{m}.
\end{eqnarray}
Therefore if $a\ll 1/p$ then it is natural to expect this to be a the LO contribution to the scattering amplitude. In nuclear physics, there are two possible S-wave scattering channels, the spin-triplet (${^3S_1}$) and the spin-singlet (${^1S_0}$). Both of these have unnaturally large scattering lengths
\begin{eqnarray}
a^{{^3S_1}}=5.425~{\rm{fm}}, \hspace{1cm}a^{{^1S_0}}=-23.714~\rm{fm}.
\end{eqnarray}
By unnaturally large it is meant that it is larger than the natural low-energy length scale of the theory $m_\pi^{-1}\sim1.4~\rm{fm}$. In fact, since at long distances the nuclear force is mediated by the pion, the range of the interaction should be approximately determined by the inverse of the pion mass. It is reassuring to see that the effective range of both of these channels are of natural size,
\begin{eqnarray}
r^{{^3S_1}}=1.75~{\rm{fm}}, \hspace{1cm}r^{{^1S_0}}=2.734~\rm{fm}.
\end{eqnarray}
Considering that nuclear physics is near the \emph{unitary limit} $(a^{-1}=0)$, it is important to properly understand the the power counting scheme when $a\gg 1/p$. A good place to start is to study the expansion of the NR scattering amplitude, Eq.~(\ref{scatNR}), about the unitary limit. This can be done by treating $rq^{*}$ and $\rho_0 q^{*3}$ to be small, where $\rho_0$ is the shape parameter, while keeping $aq^{*}$ to all orders,
\begin{eqnarray}
\label{unitary}
\mathcal{M}&=&\frac{8\pi}{m  }\frac{1}{-{a^{-1}}+\frac{rq^{*2}}{2}+\frac{\rho_0 q^{*4}}{4!}+\cdots-iq^{*}}\\
&=&
\frac{8\pi}{m  }\frac{-1}{a^{-1}+iq^{*}}\left[1+\frac{rq^{*2}/2}{({a^{-1}+iq^{*}})}+\frac{\left(rq^{*2}/2\right)^2}{({a^{-1}+iq^{*}})^2}+\frac{\rho_0 q^{*4}/4!}{({a^{-1}+iq^{*}})}\cdots\right],\\
&\equiv&\mathcal{M}_{-1}+\mathcal{M}_{0}+\mathcal{M}_{1}+\cdots
\end{eqnarray}
where the subscript denotes the scaling with the relative momentum $q^{*}=Em-P^2/4$, where $E(P)$ is the total energy(momentum) of the system. Finally one arrives at the conclusion that Eq.~(\ref{naive}) would fail to reproduce the leading order behavior of the scattering amplitude near the unitary limit. It turns out that in order to properly recover the leading order term of the scattering amplitude, $\mathcal{M}_{-1}$, one needs to evaluate an infinite series of ``bubble diagrams" shown in Fig. \ref{PDS_LO}.

In general, one needs to evaluate a loop of the form 
\begin{eqnarray}
\label{Idimreg}
I_n&=&\left(\frac{\mu}{2}\right)^{4-D}\int\frac{d^{D-1}{\bf{k}}}{(2\pi)^{D-1}}\frac{k^{*2n}}{\frac{q^{*2}}{m}-\frac{k^{*2}}{m}+i\epsilon}\\
&=&-mq^{*2n}(-q^{*2}-i\epsilon)^{(D-3)/2}\Gamma\left(\frac{D-3}{2}\right)\frac{(\mu/2)^{4-D}}{(4\pi)^{(D-1)/2}}.
\end{eqnarray}
Note that using dimensional regularization and setting $D=4$, this integral is finite. This is a consequence of the fact that dimensional regularization removes all power-law UV divergences and only logarithmic divergences survive. Using the standard \emph{minimal subtraction} (MS) scheme, where only $D=4$ singularities are removed, this integral reduces to
\begin{eqnarray}
I_n^{MS}=-i\frac{m}{4\pi}q^{*2n+1}.
\end{eqnarray}
Therefore, the $n^{th}$ term in the series of bubble diagrams scales like $C_0(C_0mp)^n\sim \frac{8\pi a}{m}({8\pi ap})^n$. For the case where $a>1/p$, this would imply that higher order terms in the series would give higher contributions, yet the sum of all terms would give a finite result, $\mathcal{M}_{-1}$.
This is a highly undesirable power-counting scheme. Kaplan, Savage and Wise  \cite{pds, pds2} argued that a simple solution to this power-counting conundrum can be achieved by modifying the subtraction scheme applied to Eq.~(\ref{Idimreg}). As an alternative scheme, they proposed subtracting the $D=3$ pole  appearing in Eq.~(\ref{Idimreg}).  This ``pole" can be removed by introducing a counterterm
\begin{eqnarray}
\delta I_n=-\frac{mq^{*2n}\mu}{4\pi(D-3)}.
\end{eqnarray} 
Since this poles arises from power-law divergences in $D=4$, this subtraction scheme is known as \emph{power divergence subtraction} (PDS). The PDS renormalized integral becomes
\begin{eqnarray}
\label{pdsint}
I_n^{PDS}=I_n+\delta I_n=-\frac{m}{4\pi}q^{*2n}(\mu+iq^{*})=q^{*2n}I_0^{PDS}.
\end{eqnarray}
Using this subtraction scheme, the LO contribution to the scattering amplitude (depicted in Fig. \ref{PDS_LO}) that includes no vertices with derivative couplings can be written as 
\begin{eqnarray}
i\mathcal{M}_{-1}=-iC_0(\mu)\sum_{n=0}^{\infty}\left(-\frac{1}{2}C_0(\mu)I_0^{PDS}\right)^n=\frac{-iC_0(\mu)}{1+mC_0(\mu)(\mu+iq^{*})/8\pi} .
\end{eqnarray}
Requiring this to reproduce $\mathcal{M}_{-1}$ appearing in Eq. (\ref{unitary}), leads to
\begin{eqnarray}
C_0(\mu)=\frac{8\pi}{m}\frac{1}{a^{-1}-\mu}.
\end{eqnarray}
With all the pieces at hand, it is straightforward to see that, in fact, the power-counting issue is resolved. From Eq.~(\ref{pdsint}) one sees that the integral scales like $I_0^{PDS}\sim \mu$, while the low-energy coefficient $C_0(\mu)$ scales like $\mu^{-1}$, therefore the product  $C_0(\mu)I_0^{PDS}$\footnote{All NR integrals will be evaluated using PDS, and from here on the superscript will be omitted.} is order 1 and must be summed non-perturbatively. 

Having defined the LO piece of the scattering amplitude  in terms of the $C_0$, it is straightforward to calculate subleading terms in a consistent fashion. For example, the NLO contribution comes with a single insertion of $C_2 q^{*2}$. This term can be treated perturbatively, while the infinite set of bubble diagrams dressing the outgoing two-particle propagator (depicted in Fig.~\ref{PDS_NLO}) must be summed non-perturbatively. Doing so, one obtains 
\begin{eqnarray}
i\mathcal{M}_{0}&=&\frac{-iC_2(\mu)q^{*2}}{(1-mC_0(\mu)(\mu+iq^{*})/8\pi)^2} \equiv-\frac{8\pi}{m}
\frac{rq^{*2}/2}{({a^{-1}+iq^{*}})^2}\\
\Rightarrow C_2(\mu)&=&\frac{8\pi}{m}\frac{r}{(a^{-1}-\mu)^2}.
\end{eqnarray} 
Therefore we find that $C_2(\mu)\sim \mu^{-2}$; in fact, one can show that in general $C_{2n}(\mu)\sim \mu^{-(n+1)}$.  By asserting that $\mu\sim q^{*}$, then the power-counting scheme can be summarized by:

1.~The loop measure $\int d^4k$ scales as $q^{*5}$, since $E\sim q^{*2}$

2.~Single particle propagators scale as $\sim 1/q^2$

3.~Vertices with $C_{2n}q^{*2n}$ scale as $q^{*n-1}$.\\

The KSW expansion demonstrates how to reliably reproduce  effective range expansion of the two-body strongly interacting system, which is well experimentally constrained. There are two major advantages of having parametrized the two-body strong interaction. The first is that it allows for the evaluation of two-body matrix elements of electroweak operators. Section \ref{ppfusion} will discuss the incorporation of weak currents responsible for proton-proton fusion ($pp\rightarrow d e^+\nu_e$), and a new method will be proposed for evaluating matrix elements of these currents directly from LQCD. The second advantage is that having constrained the two-body force, one can proceed to evaluate few-body observables. In principle, for systems involving three or more particles there will be contributions from three-and potentially fourth-body forces. Depending on the system, these contributions may be large and cannot be neglected a priori. Section \ref{dimersec1} will discuss the incorporation of three-body forces into this formalism, which will then be utilized in Section \ref{mmmsys} to determine the quantization condition for the spectrum for three-particles in a finite volume.

\subsubsection{Including pions  \label{weinbergpc}}
The toy model presented above is a great model for describing low-energy scalar bosons near unitarity. Generalizing this formalism for nuclear systems is complicated by the spin/isospin structure of the nucleons. Furthermore, for sufficiently large energies contributions from pions can no longer be neglected. This section briefly discusses the framework proposed by Kaplan, Savage and Wise  \cite{pds, pds2} to include pions in the spin singlet channel and demonstrate that in fact the contribution of pions is subleading. Discussion regarding the spin/isospin structure of contact operators will be delayed to section \ref{mmsys}. 

\begin{figure}[t]
\begin{center}

\subfigure[]{
\label{PDS_pi_NLO}
\includegraphics[scale=0.3]{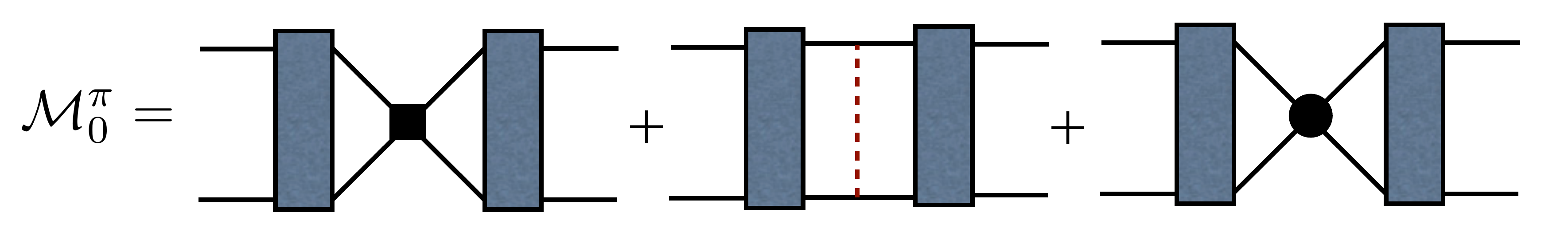}}
\subfigure[]{
\label{C_0_mpi}
\includegraphics[scale=0.090]{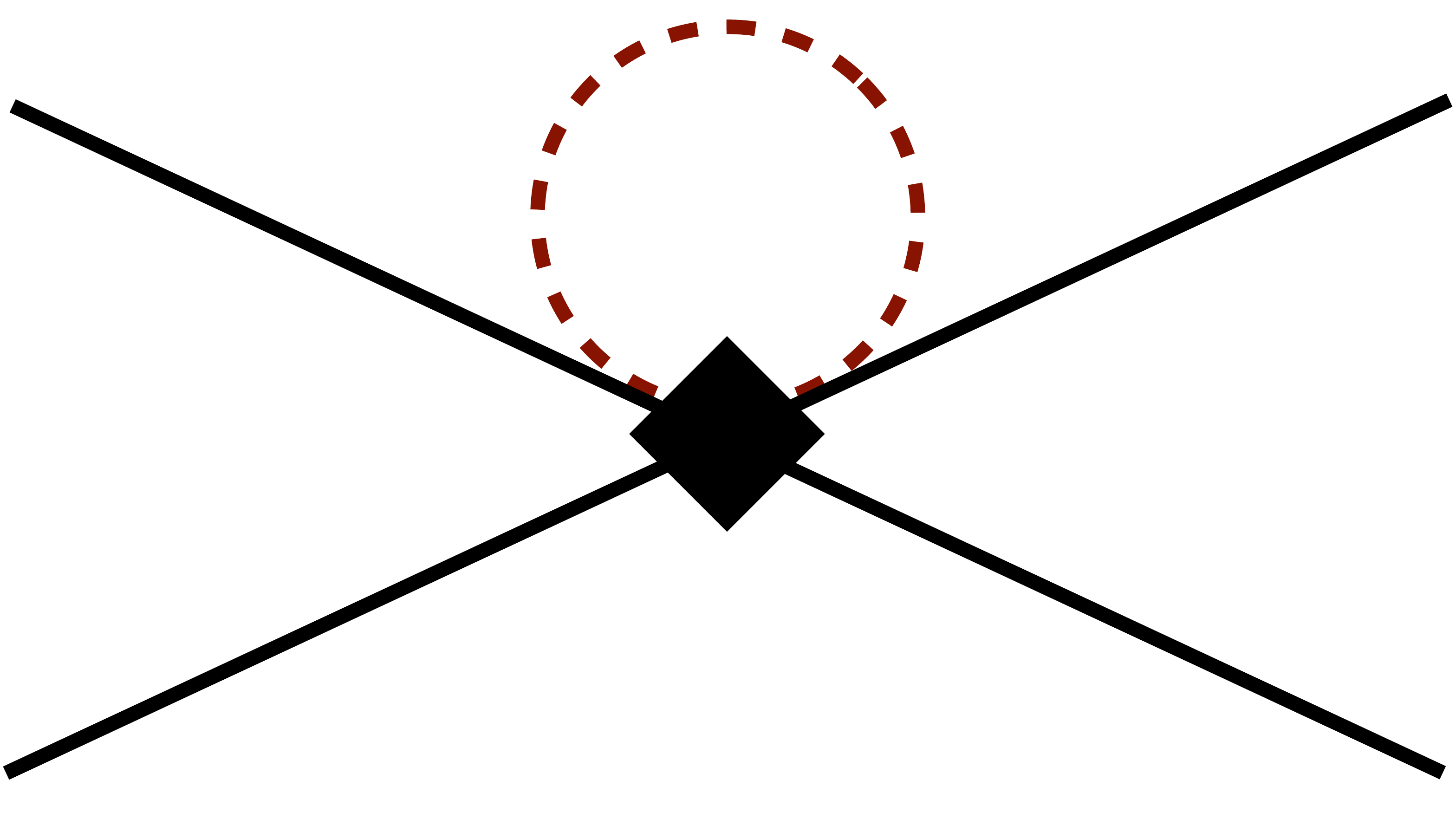}} 
\caption[LO $m_\pi$-dependence in EFT($\pinot$)]{a) In the presence of pions, the NLO scattering amplitudes, depicted in Fig~\ref{PDS_NLO} gets two corrections. The first arising from pion-exchange diagrams, as depicted by the dashed line. The second comes from the LO $m_\pi$-dependence of the momentum independent two-body operator, $m_\pi^2~C_{0,2}(\mu)$ [circle vertex]. Note that the dressed two-particle propagator defined in Fig~\ref{PDS_NLO} has been used. b) An example of the diagrams contributing to the $m_\pi$-dependence of the LECs of EFT($\pinot$). }\label{PDS_pi}
\end{center}
\end{figure}

In the $SU(2)$ limit of $\chi PT$ Eq.~(\ref{sigma_matrix}) reduces to
\begin{eqnarray}
\label{sigma2}
{\Sigma}(x)=\exp\left(\frac{2i{\boldsymbol{\pi}(x)}}{f}\right)\equiv\xi^{2}, \hspace{1cm}\boldsymbol{\pi}=
\begin{pmatrix}
\frac{{\pi}^0}{\sqrt{2}} &{\pi}^+ \\
{\pi}^- & -\frac{{\pi}^0}{\sqrt{2}} \\
\end{pmatrix},  \end{eqnarray} 
where $\xi=\sqrt{\Sigma}$ plays an important role in the construction of the HB$\chi$PT Langrangian~\cite{Jenkins:1990jv, Jenkins:1991es}. The nucleon as an isospin doublet, has a well defined transformation under $SU(2)_V$, but its transformation under $SU(2)_L\times SU(2)_R$ is ambiguously defined. Whatever it may be, the chiral transformation of the nucleon needs to respect the fact that the nucleon has positive parity.  Therefore, it is convenient to require the nucleon field $N$ to transform as 
\begin{eqnarray}
\label{Ntrans}
N(x)\rightarrow U(x) N(x)
\end{eqnarray} 
where $U$ is a symmetric combination of $L$ and $R$, since these are mixed under parity.  In order to construct a Lagrangian that is invariant under such transformation, it is important to think of the transformation of the $\xi(x)=\sqrt{\Sigma(x)}$ field appearing in Eq.~(\ref{sigma2}). Given the transformation of $\Sigma$ field, Eq.~(\ref{chiraltrans}), under $SU(2)_L\times SU(2)_R$ $\xi(x)$ can be chosen to transform as 
\begin{eqnarray}
\label{xitrans}
\xi(x)\rightarrow L~\xi(x)~U^\dag(x)=U(x)~\xi(x)~R^\dag.
\end{eqnarray} 
Note that despite chiral transformations being global, the presence of a square root relating the fields $\xi$ and $\Sigma$ makes the chiral transformation of $\xi$ a local one. It is not an accident that $U(x)$ appears in both equations above. If alternatively, one chooses a nucleon field with chiral transformation of the form $\tilde{N}: \tilde{N}\rightarrow L\tilde{N}$. Then one could perform a field redefinition of $\tilde{N}$ to arrive at $N$ satisfying Eq.(\ref{Ntrans}),
\begin{eqnarray}
N=\xi^{\dag}\tilde{N}.
\end{eqnarray}
Such field-redefinition would impact off-shell quantities but not S-matrix elements. Having made chiral transformations local,  Eq.~(\ref{Ntrans}), has effectively ``\emph{gauged}" interactions between pions and nucleons. Enforcing $SU(2)_L\times SU(2)_R$ chiral symmetry, requires all terms coupling nucleons and pions to be of the form $N^\dag F[\xi,\xi^\dag] N$, where the function $F$ must satisfy $F[\xi,\xi^\dag]\rightarrow U^\dag F[\xi,\xi^\dag]U$ under chiral transformations. This requirement rules out terms of the form $N^\dag\Sigma N$. Since $\xi^\dag\xi=1$, the first non-trivial terms should involve the vector and axial current
\begin{eqnarray} 
\mathbb{V}^{\mu}=\frac{1}{2}\left({\xi}\partial^{\mu}{\xi}^{\dag}+{\xi}^{\dag}\partial^{\mu}{\xi}\right),
&\hspace{1cm}&
\mathbb{A}^{\mu} =\frac{i}{2}\left({\xi}\partial^{\mu}{\xi}^{\dag}-{\xi}^{\dag}\partial^{\mu}{\xi}\right).
 \end{eqnarray} 
It is convenient to introduce the covariant derivative ${D}_\mu =d_\mu+\mathbb{V}_{\mu}$, which has the same transformation properties as the axial current 
\begin{eqnarray} 
{D}^{\mu}\rightarrow U {D}^{\mu} U^\dag,
&\hspace{1cm}&
\mathbb{A}^{\mu} =U \mathbb{A}^{\mu} U^\dag.	
 \end{eqnarray} 
With these pieces, it is straightfoward to construct the NR Lagrangian coupling the nucleons to pions 
\begin{eqnarray}
 \label{rellag}
 \mathcal{L}_{N\pi}&=&N^\dag\left(i{D}_{t}+\frac{{ \bf {D}}^2}{2m}+{g_A}{ \bf {A}}\cdot { \bf {\sigma}}\right)N
+\mathcal{O}(p^2/\Lambda_\chi^2),
\end{eqnarray}
where $\sigma$ denotes the spin of the nucleon in the rest frame, and $g_A = 1.2701(25)$~\cite{pdg}  is the nucleon axial charge. Therefore in the presence of pions, the scattering amplitude would get corrections from pion-exchange diagrams, depicted in Fig.~\ref{PDS_pi_NLO}, that scale like
\begin{eqnarray}
\mathcal{M}_0^{\pi,1}\sim \frac{g_A^2}{f_\pi^2}\frac{{q}^2}{q^2+m_\pi^2}\sim\mathcal{O}(q^{*0}),
\end{eqnarray}
where $q$ denotes the momentum carried by the intermediate pion, while $q^{*}$ is the relative momentum of the two-nucleon system. Note, contributions from pions are important when $q^{*}\sim m_\pi$, therefore in the presence of pions the power counting is modified such that $\mu\sim m_\pi$, which assures that $\mathcal{M}_0^{\pi,1}$ contributes at $\mathcal{O}(q^{*0})$. 

In the presence of pions, there will also be higher-dimensional operators coupling pions to two-nucleon states which are consistent with the symmetries of theory. In absence of pions in the asymptotic states, these operators give rise to diagrams involving pion loops such as the one shown in Fig.~\ref{C_0_mpi}, which can be absorbed by the $m_\pi$-dependence of the LECs appearing in Eq.~(\ref{lagNR_int}),
\begin{eqnarray}
\label{Cmpi}
C_i\rightarrow C_i(m_\pi^2)=C_{i,0}+C_{i,2}~m_\pi^2+C_{i,4}~m_\pi^4+\cdots.
\end{eqnarray}
In the power-counting scheme $m_\pi^2~C_{0,2}$ would come in at the same order as $C_{2,0}$. Therefore at $\mathcal{O}(q^{*0})$, the scattering amplitude will receive a contribution from the LO $m_\pi$-dependence of $C_0(m_\pi)$, as shown in Fig.~\ref{PDS_pi_NLO}. Because the contribution to the scattering amplitude proportional $C_{0,2}$ is momentum independent, it will contribute to the LO $m_\pi$-dependence of the scattering length. In the physical world the pion mass is fixed, but as will be discussed section~\ref{lqcd}, in LQCD calculations one can vary the quark mass. Therefore, in principle LQCD calculations will be able to more precisely constrain such dependences than experiments. 

\subsubsection{Auxiliary-Field Formalism  \label{dimersec1}}
\begin{figure}[t]
\begin{center}
\subfigure[]{
\label{dimer1}
\includegraphics[scale=0.3]{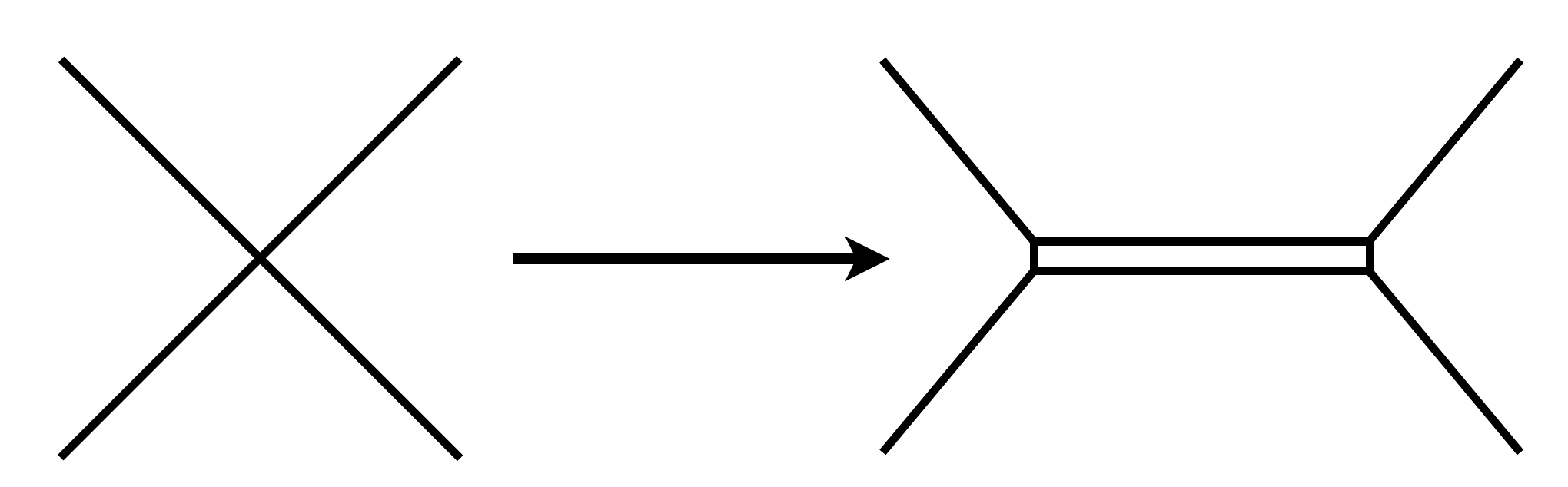}}
\subfigure[]{
\label{dimer2}
\includegraphics[scale=0.3]{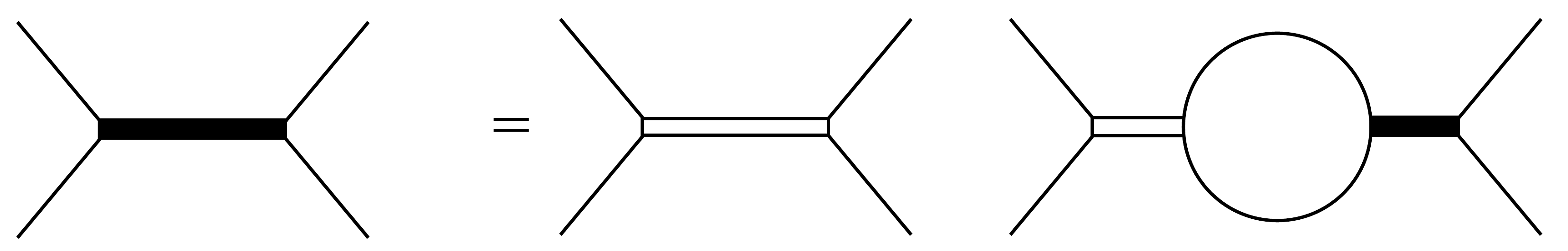}} 
\caption[S-wave scattering mediated by S-wave dimer field]{a) Schematic representation of the replacement of $2\rightarrow 2$ contact interactions by the $2\rightarrow 2$ interactions mediated by an auxiliary-field, whose bare propagator is depicted as a double lines. b) Self-consistent equation defining the the $2\rightarrow 2$ scattering amplitude up to NLO in the effective range expansion, with a fully dressed dimer propagator (thick black line).}\label{dimer}
\end{center}
\end{figure}

The EFT($\pinot$) Lagrangian, Eqs.~(\ref{lagNR}\&\ref{lagNR_int}), can be rewritten using an auxiliary-field \cite{pionless2, pionless3}. This formalism, referred to in the literature as the \emph{dimer formalism}, has proved to be useful for studying three-body physics \cite{Bedaque:1997qi, Bedaque:1998mb, Bedaque:1998kg, Bedaque:1998km, Gabbiani:1999yv, Bedaque:1999vb, Bedaque:1999ve, Bedaque:2000ft, Griesshammer:2004pe}. The idea is to introduce an auxiliary (or dimer) field, $d$, that mediates the two-body interactions (as is schematically shown in Fig.~\ref{dimer1}). In practice, one constructs a Lagrangian in terms of $d$ and $\phi$, that after integrating the dimer field out reduces to Eqs.~(\ref{lagNR}\&\ref{lagNR_int}). In the two-body S-wave sector, it is straightforward to write down such Lagrangian,
\begin{eqnarray}
\label{lag_dimer}
\mathcal{L}_{\phi d}&=&\phi^\dag\left(i\partial_t+\frac{\nabla^2}{2m}\right)\phi
-d^{\dag}\left(i\partial_t+\frac{\nabla^2}{4m}-\Delta\right)d  
-\frac{g_2}{2}\left(d^{\dag}\phi^2+\text{h.c.}\right)+\cdots
\end{eqnarray}
where $g_2$ and $\Delta$ are bare LECs that must be tuned to reproduce two-body scattering amplitude. To see that indeed this is consistent with Eqs.~(\ref{lagNR}\&\ref{lagNR_int}), one observes that the Lagrangian is quadratic in terms of the dimer field and therefore can be exactly integrated out. Equivalently, one can solve for the equations of motions of the dimer field 
\begin{eqnarray}
\label{lag_dimer}
d&=&\frac{g_2}{2}\frac{-1}{i\partial_t+\frac{\nabla^2}{4m}-\Delta}\phi^2+\cdots
=\frac{g_2}{2}\frac{1}{\Delta}\phi^2+\mathcal{O}(p^2).
\label{lag_3}
\end{eqnarray}
Inserting this expressing in the Lagrangian one observed the recovery of a four-body contact interaction along with derivative insertion that define EFT($\pinot$). 
\begin{figure}[t]
\begin{center}
\subfigure[]{
\label{dimer3}
\includegraphics[scale=0.2]{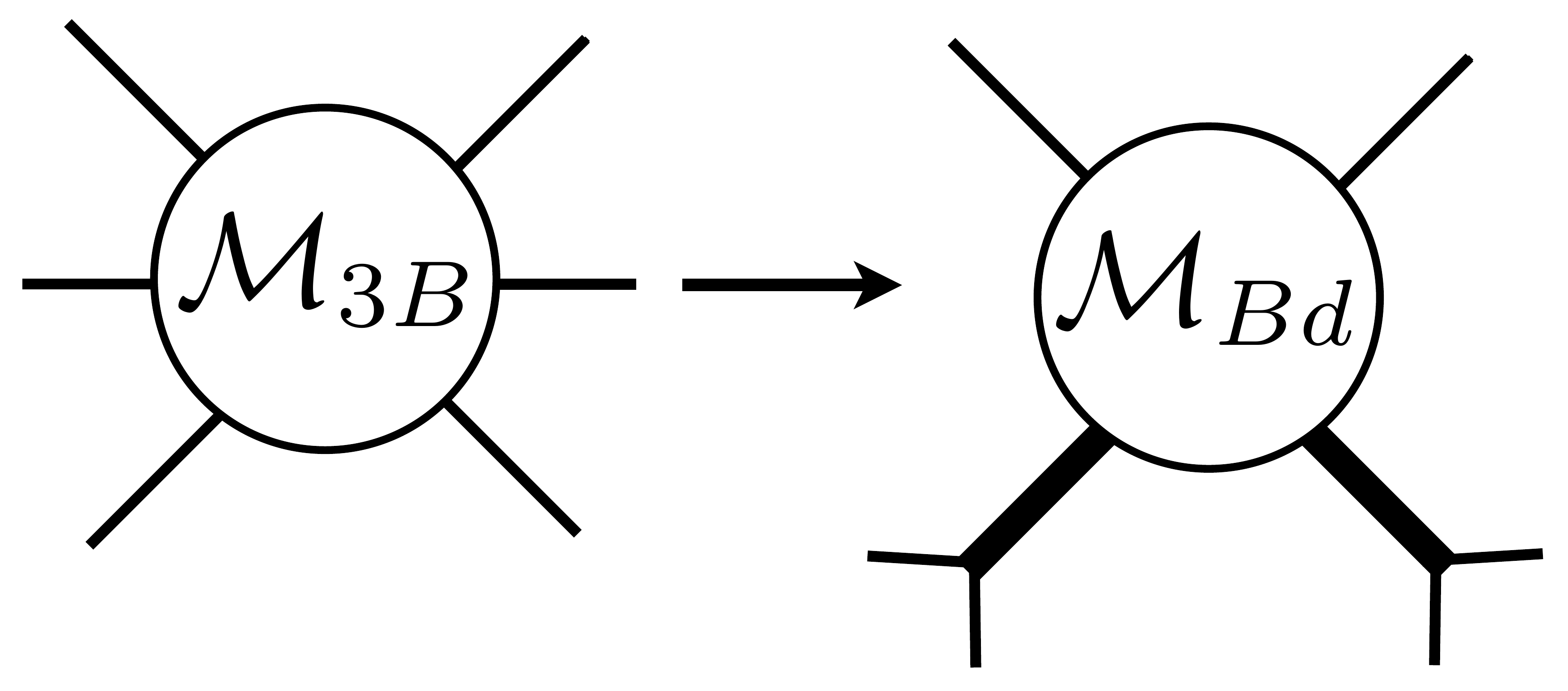}}
\subfigure[]{
\label{dimer4}
\includegraphics[scale=0.3]{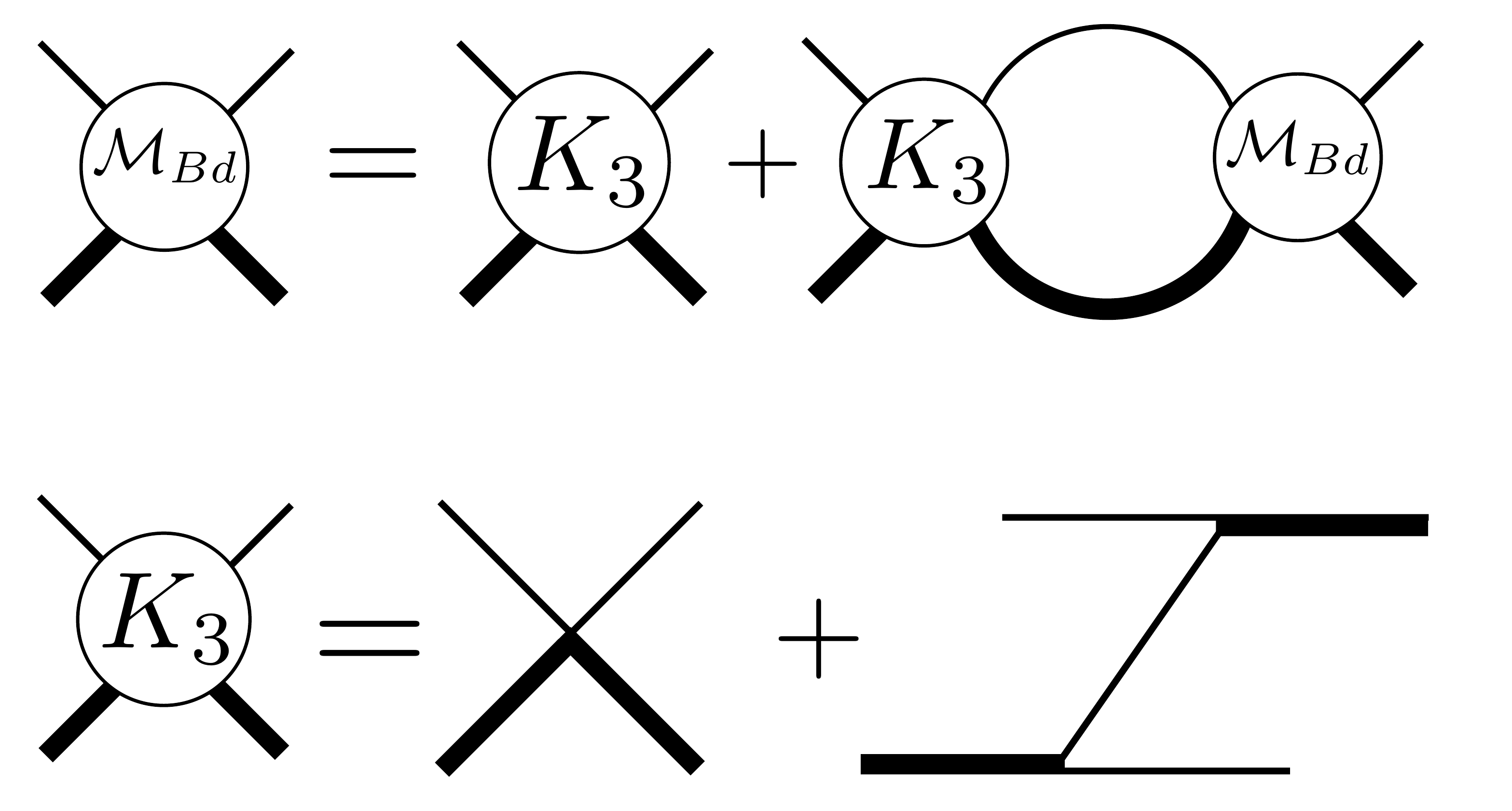}} 
\caption[Three-particle scattering in terms of dimer-boson scattering]{a) Three-particle scattering amplitudes can be reduced to dimer-boson scattering amplitudes where the external dimer legs couple to two-boson states. b) Shown is a diagrammatic representation of the STM equation, \ref{STM}, in terms of the there-body kernel, $K_3$, which is composed of a contact interaction and a boson exchange.}\label{dimer}
\end{center}
\end{figure} 

Having defined the Lagrangian, it is straightforward to calculate boson-boson scattering amplitude by summing over all bubble diagrams. These bubble diagrams dress the dimer propagator, giving the fully dressed dimer propagator. The full scattering amplitude at NLO in the effective range expansion can be written in a self-consistent fashion and it is diagrammatically shown in Fig.~\ref{dimer2}. Using the PDS definition of the two-particle loops, Eq.~(\ref{pdsint}), and matching the solution to the scattering amplitude, one finds that the solutions for the LECs
 \begin{eqnarray}
 g_{2}^2={16\pi}/{m^2r} , \hspace{1cm}
 \Delta=\frac{ g_{2}^2m}{8\pi}\left(\frac{1}{a}-\mu\right),
 \end{eqnarray}
 which leads to the fully-dressed dimer propagator 
\begin{eqnarray}
\label{dimerprop}
i\mathcal{D}^{\infty}(E,P)=-\frac{mr}{2}~
\frac{i}{-a^{-1} +r{q}^{*2}/2-i{q}^{*}+i\epsilon}\Rightarrow
i\mathcal{M}=-g_{2}^2~\mathcal{D}^{\infty}(E,P),
\end{eqnarray}
where $q^{*2}=Em-P^2/4$. 

Thus far, the dimer field is nothing more than an S-wave two-particle interpolating operator, and in reproducing the EFT($\pinot$) result a truncation in the effective range expansion has been made. Section~\ref{scalardimer} will present the generalization of this formalism for arbitrary partial waves and the truncation in the effective range expansion will also be removed. 

As mentioned above, this formalism has been extensively used in the literature for studying three-body physics. Since all $2\rightarrow 2$ are now mediated by the dimer field, three-body scattering amplitude are redefined in this formalism as well. Figure~\ref{dimer3} demonstrates that three-body scattering amplitudes can be treated as boson-dimer scattering where the external legs of the dimers couple to the two bosons\footnote{Despite the fact that the auxiliary-field is referred to as a ``dimer", the two-particle state that it couples to does not need to be bound. This is simply a field redefinition in the Lagrangian.}. The coupling between the dimer and the external legs is completely determined from the two-body sector. Therefore, the only non-trivial part to be determined is the boson-dimer scattering amplitude, $\widetilde{\mathcal{M}}_{Bd}$. To do this, it is necessary to first introduce a three-body force. Since the dimer is a two-particle interpolating operator, this can be achieved by adding a term to the Lagrangian of the form \cite{Bedaque:1998kg, Bedaque:1998km}
\begin{eqnarray}
\label{lag_dimer}
\mathcal{L}_{3}&=&-{g_3}~(d\phi)^\dag~d\phi,
\end{eqnarray}
where $g_3$ is a renormalization scale dependent LEC that must be tuned to assure that the three-body scattering amplitude is renormalization scale independent~\cite{Bedaque:1998kg}. This tuning can be done by requiring the scattering amplitude to have a pole at the three-body bound state energy.

The LO interactions between dimer and boson can be incorporated in an effective three-body Bethe-Salpeter Kernel, $K_3$, Fig.~\ref{dimer4},
\begin{eqnarray}
iK_3(\textbf{p},\textbf{k};{\mathbf{P}},E)&\equiv&-{ig_3}-\frac{i{g_2^2}}{E-\frac{\textbf{p}^2}{2m}-\frac{\textbf{k}^2}{2m}-\frac{(\textbf{P}-\textbf{p}-\textbf{k})^2}{2m}+i\epsilon} \ ,
\label{Kernel}
\end{eqnarray}
where the incoming (outgoing) boson has momentum $\mathbf{p}$ $(\mathbf{k})$ and the incoming (outgoing) dimer has momentum $\mathbf{P}-\mathbf{p}$ $(\mathbf{P}-\mathbf{k})$, and $(E,\mathbf{P})$ denote the total energy and momentum of the three-particle system as before. Note that the incoming/outgoing boson has been put on-shell. The first term in the Kernel, Eq. (\ref{Kernel}), is the three-body contact interaction, while the second term describes the interaction of three particles via exchange of an intermediate particle through two-body contact interactions. It is important to remember that for a given a total energy ($E$) and total momentum ($P$), the kinematics of the three-particle system are not fixed. In the infinite volume, where the momenta are continuous, this leads to an infinite number of possible configurations between a two-particle subsystem and an spectator particle. Formally speaking this is manifested by a cut in the exchange diagrams that leads to loops in Fig.~\ref{dimer4} to be coupled. This is why in the infinite volume the boson-dimer scattering amplitude satisfies an integral equation, known as the Skorniakov-Ter-Martirosian (STM) equation \cite{Skorniakov}, 
\begin{eqnarray}
\widetilde{\mathcal{M}}_{Bd}\left(\mathbf{p},\mathbf{k};\mathbf{P},E\right)= K_3(\textbf{p},\textbf{k};{\mathbf{P}},E) -
\int\frac{d^3q}{(2\pi)^3}K_3(\textbf{p},\textbf{q};{\mathbf{P}},E){\mathcal{D}^{\infty}(E-\frac{q^2}{2m},|\mathbf{P}-\mathbf{q}|)} \widetilde{\mathcal{M}}_{Bd}\left(\mathbf{q},\mathbf{k};\mathbf{P},E\right). \nn\\
\label{STM}
\end{eqnarray}
This formalism has been used extensively to study properties of Bose gases, both in infinite volume   \cite{Bedaque:1998kg, Bedaque:1998km, Bedaque:2000ft} and finite volume \cite{Kreuzer:2009jp}. In particular, it is has allowed to frame Efimov physics \cite{Efimov:1970zz, Efimov:1978pk} in a modern day language of renormalization group flow \cite{Bedaque:1998kg}, which has opened a new class of problems to study, e.g. four-body systems~\cite{Platter:2004he}. The nuclear analogue of this EFT has proven to be useful for studying deuteron-nucleon scattering, below and above the break-up of the deuteron \cite{Bedaque:1997qi, Bedaque:1998mb, Bedaque:1999ve}. Incorporating pions can be done by including the coupling between the nucleons and pions, Eq.~(\ref{rellag}). In Ref.~\cite{Bedaque:1999vb} this was done to determine the real and imaginary part of the quartet S-wave phase shift in deuteron-neutron scattering (${^4S_{3/2}}$) for centre-of-mass momenta of up to 300 MeV, and higher partial waves for deuteron-neutron scattering have also been studied~\cite{Gabbiani:1999yv}. Recently, this formalism has implemented for studying three-body problems with resonant P-wave interactions \cite{Braaten:2011vf}. Sections \ref{mmsys} \& \ref{NNsys} discusses the generalization of the dimer field with arbitrary partial waves, both in the scalar and nuclear sector. In section \ref{mmmsys} this formalism will be used to study three-body systems in a finite volume.

\section{Lattice QCD \label{lqcd}} 
Having thoroughly discussed challenges associated with having an analytic control of QCD phenomena at low-energies, we finally discuss the numerical evaluation of QCD observables in a finite, discretized Euclidean spacetime, referred to as Lattice QCD (LQCD). This program, that was initiated in 1974 by Wilson's seminal paper on ``\emph{Confinement of quarks}"~\cite{Wilson:1974sk}, relies on several key features, but at the core of it are two fundamental of equations. Firstly, the QCD spectrum, as well as matrix elements of operators can be determined from correlation functions in a Euclidean spacetime. For example, consider a creation operator $\hat{O}^{\dag}$, and its correlation function
\begin{eqnarray}
\label{greenfunction}
\langle O(t) O^{\dag}(0) \rangle_T&\equiv&\frac{1}{Z_T}\text{Tr}~[e^{-(T-t)\hat{H}}\hat{O}e^{-t\hat{H}}\hat{O}^{\dag}]\\
&=&\sum_{n,m}\langle m|\hat{O}|n\rangle \langle n|\hat{O}^{\dag}|m\rangle e^{-(T-t)E_m} e^{-tE_n}\nn\\
\label{groundstate}
\Rightarrow\lim_{T\rightarrow \infty}\langle O(t) O^{\dag}(0)\rangle_T&=&\sum_{n}\langle \Omega|\hat{O}|n\rangle \langle n|\hat{O}^{\dag}|\Omega\rangle  e^{-tE_n},
\end{eqnarray}
where $\hat{H}\equiv \hat{H}_{QCD}$ is the QCD Hamiltonian, $T$ is \emph{imaginary temporal} extent of the Euclidean spacetime, $\Omega$ is the groundstate of the system, and $Z_T\equiv \text{Tr}~[e^{-T\hat{H}}]$ is the partition function of the system\footnote{Euclidean QFT in (3+1)-dimensional with a finite temporal extent $T$ is equivalent to a 3-dimensional quantum statistical theory in a finite temperature, $1/T$.}. For sufficiently large $T$ and $t$, only the lowest energy state that has that has the same quantum numbers as $\hat{O}^{\dag}$ would survive.  

The second equation relates the correlation function to the path integral 
\begin{eqnarray}
\label{pathint}
\frac{1}{Z_T}\text{Tr}~[e^{-(T-t)\hat{H}}\hat{O}e^{-t\hat{H}}\hat{O}^{\dag}]&=&\frac{1}{Z_T}
~\int[\mathcal{D}A,\mathcal{D}q,\mathcal{D}\bar{q}]~{O}(t)O^{\dag}(0)~
e^{-S_E}
\end{eqnarray}
where $S_E\equiv\int_T dt\int_{V} d^3x\mathcal{L}^{E}_{QCD}$ is the action in a finite Euclidean spacetime, where $V$ is the volume which will be assumed to be cubic with length $L$ in each one of the sides, and 
\begin{eqnarray}
\label{qcdlag_E}
\mathcal{L}^{E}_{QCD}=\bar{q}\left(\slash\hspace{-.24cm} {\rm{D}}+{\rm{m}}_q\right)q+\frac{1}{4}\text{Tr}~[G^{\mu\nu}G^{\mu\nu}]\equiv\bar{q}~\mathbb{M}~q+\frac{1}{4}\text{Tr}~[G^{\mu\nu}G^{\mu\nu}],\end{eqnarray}
where the metric is now the identity and $\mathbb{M}\equiv{\rm{D}}+{\rm{m}}_q$. 

These two equations allow one to access the QCD spectrum by performing a path integral. The power of this simple observation lies on two facts. The first is that the Lagrangian is quadratic in terms of the quarks, and therefore part of the integral can be done analytically. For instance, the partition function 
\begin{eqnarray}
\label{partition}
{Z_T}=\text{Tr}~[e^{-T\hat{H}}]=
~\int[\mathcal{D}A,\mathcal{D}q,\mathcal{D}\bar{q}]~
e^{-S_E}\nn
=~\int~\mathcal{D}A~{\rm{Det}}(\mathbb{M})~
e^{-\int d^4x~\frac{1}{4}\text{Tr}~[G^{\mu\nu}G^{\mu\nu}]}.
\end{eqnarray}
Similarly, the fermionic integral in Eq.~(\ref{pathint}) can be performed, and can be schematically written as 
\begin{eqnarray}
\label{pathint2}
\frac{1}{Z_T}
~\int[\mathcal{D}A,\mathcal{D}q,\mathcal{D}\bar{q}]~{O}(t)O^{\dag}(0)~
e^{-S_E}=\frac{1}{Z_T}\int~\mathcal{D}A~{\rm{F}}[A]~{\rm{Det}}(\mathbb{M})~
e^{-S_E},
\end{eqnarray}
where ${\rm{F}}[A]$ denotes the functional form of the product of ${O}(t)O^{\dag}(0)$ after having performed the Wick contraction of the quark fields present in the operators. In section~ \ref{contract}, we will discuss the correlation functions of the pions, which will allow us to explicit write down the functional ${\rm{F}[A]}$. 

The second key fact is that the remainder of the path integral integral can be performed using Monte Carlo techniques. This is done by sampling the phase space of the gauge field, $A$, using 
\begin{eqnarray}
\label{propden}
{\rm{Det}}(\slash\hspace{-.24cm} {\rm{D}}+{\rm{m}}_q)
e^{-\int d^4x~\frac{1}{4}\text{Tr}~[G^{\mu\nu}G^{\mu\nu}]}
\end{eqnarray}
 as a probability distribution. In practice, one would obtain a finite sample of \emph{gauge field configurations} $\{A_1,A_2,\ldots,A_{N_G}\}$ of size $N_G$, and the green function above would be approximated as
\begin{eqnarray}
\frac{1}{Z_T}
~\int[\mathcal{D}A,\mathcal{D}q,\mathcal{D}\bar{q}]~({O}O^{\dag})(A,q,\bar{q})~
e^{-S_E}\approx\frac{1}{N_G}\sum_i^{N_{G}}{\rm{F}}[A_i].
\end{eqnarray}
Although an approximation has been made in order to evaluate the path integral, this is a controlled approximation which can be systematically corrected by increasing the number of gauge configurations in the ensemble.  

\subsection{Discretization of the QCD action}
In order to numerically evaluate the QCD partition function, it is necessary to first discretize spacetime. Furthermore, due to limited computational power, the spacetime is necessarily truncated. In this section we will review the basics of the discretization of the action, and in section \ref{FVEuc} finite volume physics will be introduced. 

\begin{figure}[t]
\begin{center} 
\includegraphics[scale=0.3]{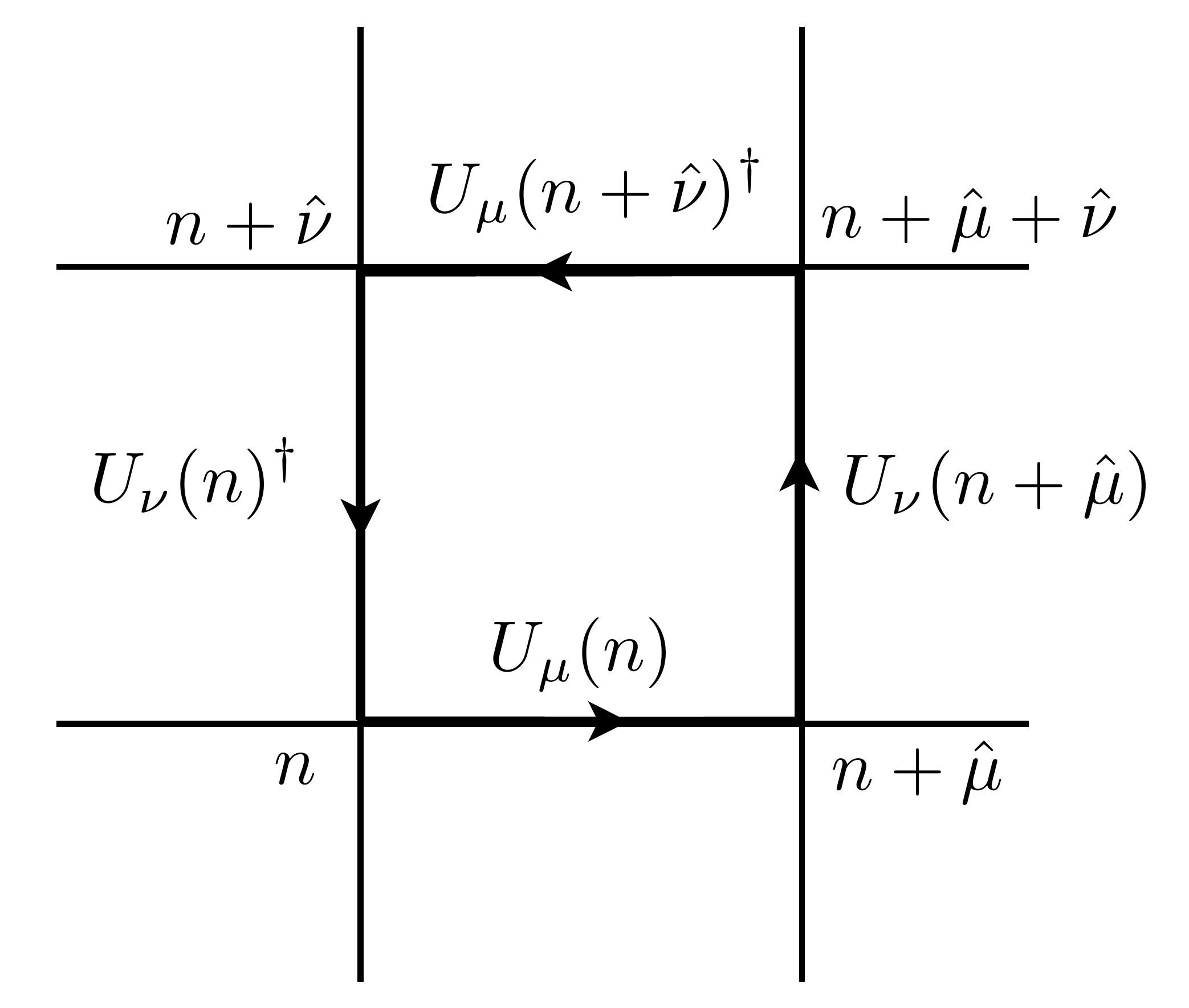}
\caption[Single plaquette]{Pictorial representation of the plaquette in Eq.~(\ref{plaq}).}\label{plaquette}
\end{center}
\end{figure}

Discretizing spacetime implies introducing a finite, minimum separation between point, known as ``\emph{lattice spacing}" and will be denoted\footnote{The lattice spacing is most commonly denoted by $a$ but we want to avoid confusion with the scattering length.} here as $b$. The lattice spacing need not be the same in all direction, but here it will be assumed that spacetime is isotropic. Having introduced an intrinsic length separation between \emph{lattice sites}, one should expect the QCD Lagrangian, Eq.~(\ref{qcdlag_E}), to no longer be local. In fact the discretized Lagrangian need only be local in the continuum limit ($b\rightarrow0$). That being said, it is important to require that gauge invariance is preserved even for nonzero lattice spacing. Quarks fields reside only on the lattice sites, and their gauge transformation is still described by Eq.~(\ref{gaugetr_q}), and the covariant derivative appearing in Eq.~(\ref{qcdlag_E}) now couples quarks in different lattice sites. Therefore, one should not expect the gauge field $A_\mu$ to be the object appearing in the LQCD action, but rather the \emph{gauge transformer}
\begin{eqnarray}
G(x,y)=P\exp\left(ig\int_{\mathcal{C}_{xy}}A\cdot ds\right),
\end{eqnarray}
where $P$ denotes the path ordered integral connecting $x$ and $y$ via the path ${\mathcal{C}_{xy}}$. From the transformation properties of the gauge field, Eq.~(\ref{gaugetr_A}), one finds that $G(x,y)$ has the following transformation properties
\begin{eqnarray}
G(x,y)\rightarrow \Omega(x)G(x,y)\Omega^\dag(y).
\end{eqnarray}
In the limiting case that $x=nb$ and $y=(n+\hat{\mu})b$, i.e. the two points are separated by a single lattice spacing, the integral above can be approximated as 
\begin{eqnarray}
\label{gaugelink}
G(x,y)\equiv U_\mu(n)=\exp\left(igbA_\mu\right)+\mathcal{O}(b^2).
\end{eqnarray}
Note that the argument of the ``\emph{gauge link}", $U_\mu(n)$, denotes the spacetime location of the first point in lattice units, and the subscript depicts the direction of the unit step. Having defined this object, it is straightforward to construct gauge invariant quantities. In particular, the gauge action in Eq.~(\ref{qcdlag_E}) can be built in terms of the shortest, nontrivial closed loop on the lattice, the \emph{plaquette}. The plaquette $U_{\mu\nu}(n)$ is defined as the product over four gauge links [see Fig.~\ref{plaquette}]
\begin{eqnarray}
U_{\mu\nu}(n)&=&U_\mu(n)~U_\nu(n+\hat{\mu})~U_{-\mu}(n+\hat{\mu}+\hat{\nu})~U_{-\nu}(n+\hat{\nu})\nn\\
\label{plaq}
&=&U_\mu(n)~U_\nu(n+\hat{\mu})~U_{\mu}(n+\hat{\nu})^{\dag}~U_{\nu}(n)^{\dag},
\end{eqnarray}
where we have used the relation $U_{-\mu}(n)\equiv U_{\mu}(n-\hat{\mu})^{\dag}$, which follows from Eq.~(\ref{gaugelink}). Under gauge transformations 
\begin{eqnarray}
U_{\mu\nu}(n)\rightarrow\Omega(n)~U_{\mu\nu}(n)~\Omega^\dag(n),
\end{eqnarray}
and as a result its trace is invariant. With this one can write down the gauge action in terms of the plaquettes that reduces to the second term in Eq.~(\ref{qcdlag_E}) in the continuum limit~\cite{Wilson:1974sk}, 
\begin{eqnarray}
S_G[U]=\frac{1}{g^2}\sum_{n,\mu<\nu}\text{Re}~\text{Tr}\left[1-U_{\mu\nu}(n)\right]=\frac{b^4}{4}\sum_{n,\mu<\nu}\text{Tr}~[G^{\mu\nu}G^{\mu\nu}]+\mathcal{O}(b^6).
\end{eqnarray}
Having determined the lattice gauge action, one would imagine that it is straightforward to write a well-behaved discretized fermion action. The challenges associated with constructing a discretized fermionic action is best summarized by the Nielsen-Ninomiya theorem \cite{Nielsen:1980rz}. Which states that for a fermionic action of the form 
\begin{eqnarray}
S_F[q,\bar{q},U]=\int_{-\pi/b}^{\pi/b}\frac{d^4p}{(2\pi)^4}~\bar{q}_{-\textbf{p}}~\widetilde{D}(\textbf{p})~q_{\textbf{p}},
\end{eqnarray}
the Fourier transform of the Dirac operator, $\widetilde{D}(\textbf{p})$, cannot simultaneously satisfy all of the following criteria:

1. $\widetilde{D}(\textbf{p})$ is a periodic and analytic function of $\textbf{p}$ everywhere except at $\textbf{p}=0$,

2. $\widetilde{D}(\textbf{p})\propto \gamma_\mu p_\mu$ as $a p\ll1$,

3. $\{\gamma_5,\widetilde{D}(\textbf{p})\}=0.$

The first two are conditions required to assure that the Dirac operator in coordinate space is local and that there there is only a single pole present for every continuum state. The third is the condition that the Dirac operator must satisfy to have definitive chiral symmetry in the continuum. In fact, it is the last condition that is too stringent for a lattice action. Ginsparg and Wilson showed that in for a finite lattice spacing the Dirac operator should satisfy \cite{Ginsparg:1981bj}
\begin{eqnarray}
\{\gamma_5,{D}\}=b~{D}\gamma_5{D}.
\end{eqnarray}
This is known as the Ginsparg-Wilson equation, and operators that satisfy it have an exact symmetry that is the discretized analogue of the continuum chiral symmetry \footnote{I direct the readers to Ref.~\cite{Kaplan:2009yg} for a review of chiral fermions on the lattice, as well as a instructive derivation of the Ginsparg-Wilson equation.}.

Two solutions to the Ginsparg-Wilson equation are known as \emph{domain wall fermions} \cite{Kaplan:1992bt,  Kaplan:1992sg} and \emph{overlap fermions} \cite{Narayanan:1993ss, Narayanan:1994gw, Narayanan:1995ft}. The domain wall fermion actions is a five dimensional theory of massive fermions without any chiral symmetry, which was shown to have massless fermion bound to the four-dimensional edge of the lattice. Domain wall fermions only satisfy the Ginsparg-Wilson Equation in the limit that the fifth-dimension is taken to infinity, and in that limit these two formalism are equivalent. The general overlap operator 
\begin{eqnarray}
\label{overlapop}
D_{ov}=\frac{1}{b}\left(1+\gamma_5~\text{sign}[H]\right)\equiv\frac{1}{b}\left(1+\gamma_5\frac{H}{\sqrt{H^{\dag}H}}\right),\hspace{1cm}H=\gamma_5 K
\end{eqnarray}
is the only known explicit solution to the Ginsparg-Wilson equation, where $K$ is a $\gamma_5$-Hermitian ``kernel" Dirac operator, i.e. $K^{\dag}=\gamma_5 K \gamma_5$. Since $K$ is $\gamma_5$-Hermitian so is $D_{ov}$. For $\gamma_5$-Hermitian operators, the Ginsparg-Wilson equation can be written as
\begin{eqnarray}
bDD^\dag=D+D^\dag,
\end{eqnarray}
which after a couple of lines of algebra is clearly satisfied by the overlap operator. 

The final piece to have chiral lattice fermions is to find a suitable Dirac kernel, $K$. The simplest choice is to use the massless Wilson Dirac operator, $D_W$ \cite{Wilson:1974sk},
\begin{eqnarray}
K&=&bD_W,
\\
D_W(n,m)&=&\frac{4}{b}\delta_{n,m}-\frac{1}{2b}\sum_{\mu=\pm1}^{\pm4}(1-\gamma_\mu)U_\mu(n)\delta_{n+\hat{\mu},m},
\end{eqnarray}
 where $\gamma_{-\mu}=-\gamma_\mu$. Note that in the absence of a quark mass, this operator has an effective mass term of $4/b$. This was introduced by Wilson to circumvent the ``\emph{doubling problem}". To get further insight consider the Fourier transform of the Wilson Dirac operator in the limit where the gauge link is set to one,
 \begin{eqnarray}
\widetilde{D}_{W}(\textbf{p})=\frac{i}{b}\sum_{\mu=1}^{4}\gamma_\mu~\sin(p_\mu b)+\frac{1}{b}\sum_{\mu=1}^{4}\gamma_\mu~(1-\cos(p_\mu b)).
\end{eqnarray}
For momenta $p\in[0,\pi/b]$, this operator only has one zero, $\textbf{p}=0$. In absence of the second term above, there would have be 15 other zeros corresponding to 
\begin{eqnarray}
p=(\pi/b,0,0,0),(0,\pi/b,0,0),\ldots,(\pi/b,\pi/b,\pi/b,\pi/b)
\end{eqnarray}
the so-called \emph{doublers}\footnote{The name doublers refers to the fact that for $D$-dimensions the naive Dirac operator would have $D^{2}$ zeros}. With the Wilson action, these modes have a mass that is inversely proportional to the lattice spacing and can be safely ignore. Showing that Wilson operator is $\gamma_5$-Hermitian is straightforward using $\gamma_\mu^\dag=\gamma_\mu$, $\{\gamma_\mu,\gamma_5\}=0$, and  $U_{-\mu}(n)=U_\mu(n-\hat{\mu})^{\dag}$. Since the Wilson operator is proportional to $\gamma_\mu p_\mu$ in the continuum limit, so is the overlap operator. 

Inverting large matrices is computationally very expensive, specially if these have large condition numbers.  To obtain the overlap operator, Eq.(\ref{overlapop}), one needs to first determine ${({H^{\dag}H})^{-1/2}}$. Furthermore, one needs to also evaluate the inverse of $D_{ov}+m_q$ to obtain the quark propagators, which has nearly vanishing eigenvalues at the physical point. These two facts make LQCD calculations with chiral, light-quarks technically challenging. Although some calculations have been performed with physical light-quark masses~[e.g. see Refs.~\cite{Durr:2010aw, Bazavov:2012xda}], most calculations continue to be performed at unphysical light-quark masses. With increasingly faster algorithms and more computational power, it is not unrealistic to expect most calculations to be performed at or near the physical point with appropriate chiral symmetry in the near future. 

Although calculations being performed at unphysical quark masses might seem to be a short coming of present day calculation, the ability to perform calculations at different values of the the light-quark masses can give us a great deal of insight. As mentioned in the previous section, the nuclear force is analytically understood in terms of a low-energy EFT with an infinite number of operators with non-trivial $m_\pi$-dependence, e.g. see Eq.~(\ref{Cmpi}). As a result, obtaining the chiral nuclear forces requires performing calculations with many different values of the light-quark masses. It is for this reason that LQCD calculations will complement experiments, by not just giving access to on-shell quantities but also off-shell quantities that are experimentally challenging to determine. 
 
\subsubsection{On the construction of particle correlation functions \label{contract} }
As discussed above, the first part of any LQCD calculations is to generate a set of $N_G$ configurations using the Eq.~(\ref{propden}) as a probability density. With the gauge configurations, one can then proceed to evaluate correlation functions, Eq.~(\ref{greenfunction}). Most operators of interest in nuclear physics will be entirely composed of quark and antiquark operators. From Wick's theorem, we know that Green's functions for such operators can be written in terms of contracted quark and antiquark propagators. The full propagator for a quark with flavor ``f", $D_f^{-1}$, is non-local with matrix, and its elements $D_f^{-1}({x}_f,{x}_i)^{\beta\alpha}_{ba}$ denotes the propagator of a quark created at $x_i$ with (spin,color)=$(\alpha,a)$ and annihilated at $x_f$ with (spin,color)=$(\beta,b)$. The full Dirac operator is a square matrix of $(3\times 4\times L^3\times T)^2$ in size, making it extremely computationally challenging to invert.  With the up and down quarks having the lightest masses, their propagators are the most computationally expensive to calculate. Most modern day calculations are performed in the isospin limit. Another reduction in computational cost can be obtained by noting that most actions are $\gamma_5$-Hermitian, which related forward-moving (quark) propagators and backward-moving (antiquark) propagators,
\begin{eqnarray}
\gamma_5D^{-1}\gamma_5\Leftrightarrow [\gamma_5D^{-1}_f(x_i,x_f)\gamma_5]^{\alpha\beta}_{ab}=[D^{-1}_f(x_f,x_i)^*]^{\beta\alpha}_{ba}.
\end{eqnarray}
Having calculated the propagators from the light-quarks, one needs to \emph{build operators} that have good overlap with the states of interest and then proceed to perform the propagator contractions. For example, the interpolating operators for the pions must have the right isospin structure, must be a pseudoscalar, and must transform correctly under charge conjugation. It is easy to see that the following satisfy all of these criteria
\begin{eqnarray}
\label{pioninter}
\pi^{+}=\bar{d}\gamma_5u,\hspace{1cm}
\pi^{0}=\frac{1}{\sqrt{2}}(\bar{u}\gamma_5u-\bar{d}\gamma_5d),\hspace{1cm}
\pi^{-}=\bar{u}\gamma_5d.
\end{eqnarray}
Under parity $\pi^{+}(\textbf{x},t)\rightarrow -\pi^{+}(-\textbf{x},t)$ as do the other two pions. Under charge conjugations $\pi^{+}(\textbf{x},t)\rightarrow \pi^{-}(\textbf{x},t)$ and $\pi^{0}(\textbf{x},t)\rightarrow \pi^{0}(\textbf{x},t)$. Having the interpolating operators at our disposal, the last step is to perform the Wick contractions. The correlation functions for $\pi^+$ and $\pi^0$ with total momentum $P$ are
\begin{figure}[t]
\begin{center}
\subfigure[]{
\label{connected}
\includegraphics[scale=0.4]{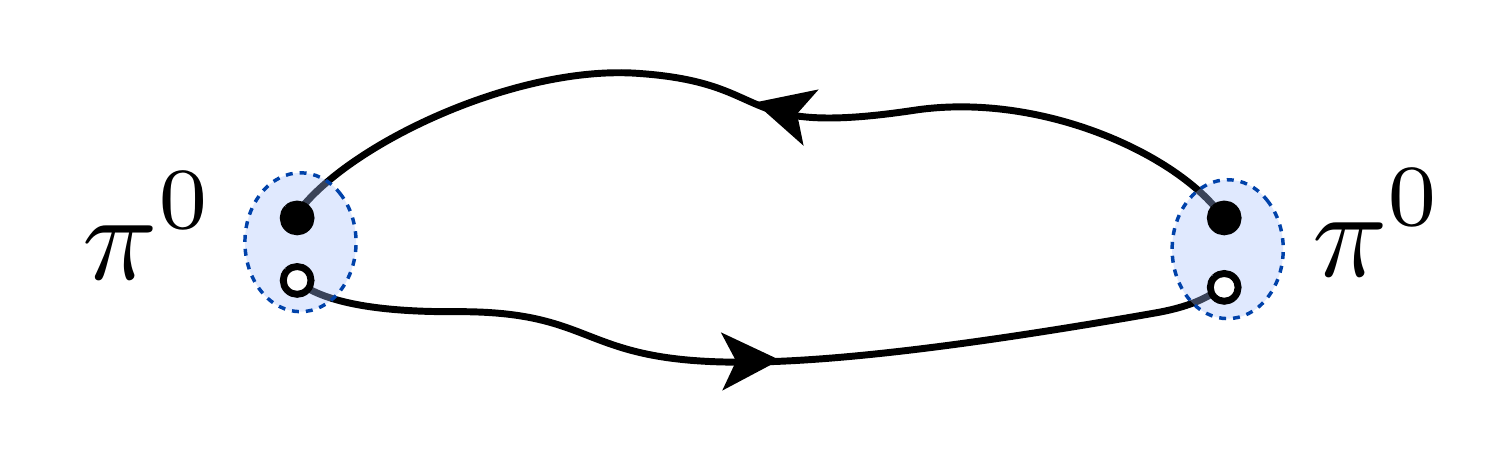}}
\subfigure[]{
\label{disconnected}
\includegraphics[scale=0.4]{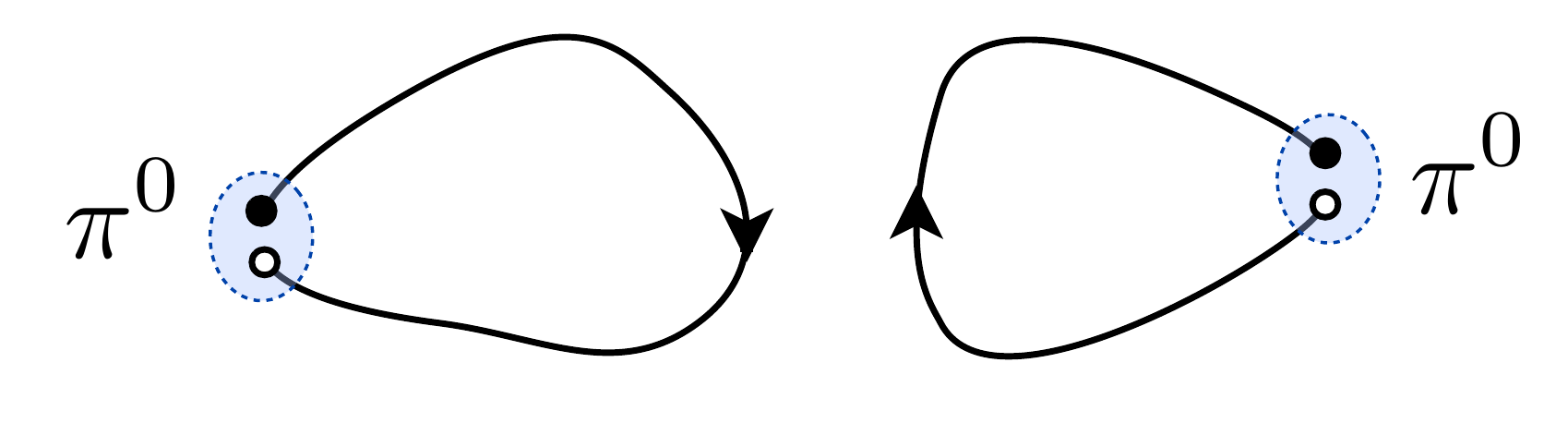}}
\caption[$\pi^0$ correlation function]{a) Connected and b) disconnected diagrams that contribute to the $\pi^0$ correlation function. The latter involve all-to-all propagators.}\label{pi0}
\end{center}
\end{figure}
\begin{eqnarray}
C_{\pi^{+}}(t,\textbf{P})&=&\sum_{\textbf{x}}e^{i\textbf{x}\cdot \textbf{P}}\langle\pi^{+}(x)\bar{\pi}^{+}(0)\rangle=\sum_{\textbf{x}}e^{i\textbf{x}\cdot \textbf{p}}\langle\bar{d}(x)\gamma_{5}u(x)\bar{u}(0)\gamma_{5}d(0)\rangle\nn\\
\label{pipCORR}
&=&-\sum_{\textbf{x}}e^{i\textbf{x}\cdot \textbf{p}}~\text{Tr}\left(D_u^{-1}(x,0)\gamma_5D_d^{-1}(0,x)\gamma_5\right)\longrightarrow  2Z_{\pi^+}e^{-E_\pi T/2}\cosh\left(E_\pi  t_T\right),\\
C_{\pi^{0}}(t,\textbf{P})&=&-\frac{1}{2}\sum_{\textbf{x}}e^{i\textbf{x}\cdot \textbf{p}}~\text{Tr}\left(D_u^{-1}(x,0)\gamma_5D_u^{-1}(0,x)\gamma_5\right)-\frac{1}{2}\sum_{\textbf{x}}e^{i\textbf{x}\cdot \textbf{p}}~\text{Tr}\left(D_d^{-1}(x,0)\gamma_5D_d^{-1}(0,x)\gamma_5\right)\nn\\
&+&\sum_{\textbf{x}}e^{i\textbf{x}\cdot \textbf{p}}~\text{Tr}\left(D_u^{-1}(x,x)\gamma_5)~~\text{Tr}(D_u^{-1}(0,0)\gamma_5\right)\nn\\
&+&\sum_{\textbf{x}}e^{i\textbf{x}\cdot \textbf{p}}~\text{Tr}\left(D_d^{-1}(x,x)\gamma_5)~~\text{Tr}(D_d^{-1}(0,0)\gamma_5\right)
\longrightarrow  2Z_{\pi^0}e^{-E_\pi T/2}\cosh\left(E_\pi  t_T\right)
\label{pi0CORR}
\end{eqnarray}
where the asymptotic behavior has been deduced from Eq.~(\ref{greenfunction}) and $ t_T=t-T/2$. Although the ``\emph{sources}" for the correlation functions above were placed at the origin, this need not be the case; the correlation function only depends on the time separation between the source and the ``\emph{sink}". By utilizing the $\gamma_5$-Hermiticity and considering the isospin limit, the $\pi^+$ correlation function, Eq.~(\ref{pipCORR}), can be written in terms of a single quark propagator going from the one point to all other points on the lattice. These are known as ``\emph{point-to-all}" propagators. These propagators are a factor of volume smaller than the full-Dirac operator. The first two terms in the $\pi^0$ correlation function, Eq.~(\ref{pi0CORR}), involve point-to-all propagators, whose contribution to the correlation function can be portrayed by \emph{``connected"} diagrams, Fig.~\ref{connected}. This correlation function also has terms with propagators starting and ending at the same point $x$, where $x$ is being summed over the lattice volume. These propagators, depicted by the \emph{``disconnected"} diagram in Fig.~\ref{disconnected}, are known as ``\emph{all-to-all}" propagators and require inverting the full Dirac operator. Historically, these contributions have made the study of unflavored hadronic systems nearly impossible. With increasingly faster inversion algorithms, these calculations are now starting to be possible. With this progress in mind, in section \ref{coudpledchannels} we will pay close attention to $\pi\pi-K\bar{K}$ mixing in the isosinglet channel.

\subsection{Finite Euclidian Spacetime \label{FVEuc}}

Since calculation are necessarily performed in a finite volume, it is important to have finite volume effects under control. Chapters \ref{mmsys}, \ref{NNsys}, \& \ref{mmmsys} discuss finite volume artifacts for two-particle and three-particle systems. In this section the one-particle sector is discussed and the challenges associated two-particle sector will be introduced.

As illustrated by Eq.~(\ref{groundstate}), LQCD 	can reliably determine QCD spectrum in a finite Euclidean spacetime. Although there has been much success in extracting excited states, ground-state energies are always determined with higher precision and their implication for the infinite volume spectrum are better understood. The ground states energy extracted from a correlation function of a single particle with total momentum equal to zero can be interpreted as the particle's mass in a finite spacetime volume and finite lattice spacing, $m_h(L,T,a)$. Hadrons composed of quarks with light masses $m_l$ satisfying $m_l b\ll1$, their discretization effects are suppressed. For heavy-quark masses satisfying $bm_Q{>}{1}$, sytematic errors due to discretization effects are sizable and need to be properly addressed. To circumvent this issue, modern day calculations use non-relativistic QCD (NRQCD) \cite{NRQCD} actions for the bottom sector and relativistic heavy-quark actions \cite{RHQ0, RHQ1, RHQ2, RHQ3,RHQ4}, where all $\mathcal{O}((m_Q a)^n)$ corrections are systematically removed, for the charm sector. These calculations must be performed at multiple lattice spacing and results must be extrapolated to the continuum.  

Assuming that discretization effects are under control, we proceed to discuss the finite volume dependence of the hadron masses. In this sector, finite volume effects arise from interactions of a particle with its neighboring mirror images. For large $T$ and $L$, this interaction is mediated by pion exchange, therefore it is natural to compare the correlation length of the pion, $1/m_\pi$, to the spatial $(L)$ and temporal extents $(T)$ of the lattice. For $T/m_\pi\sim L/m_\pi\gg1$ it will be shown that these finite volume effects are exponentially suppressed~\cite{luscher0}; this regime is known as the p-regime \cite{pregime, pregime2} and will be the focus of this work.

In general the volume dependence of  the spectrum of $N$-particles can be obtained by solving for the poles of the $N$-particle propagator in a finite volume. For single-particle systems, the bare propagator contains no finite volume (FV) or finite temperature (FT) dependence and consequently all FV and FT effects are encoded in the self-energy corrections, $\Sigma(L,T,p^2)$, 
\begin{eqnarray}
\frac{i~Z_h(L,T)}{p^2-m_h(L,T)^2+i~\epsilon}&\equiv&\frac{i}{p^2-m_0^2+i~\epsilon}\left(1-\frac{\Sigma(L,T,p^2)}{p^2-m_0^2+i~\epsilon}+\cdots\right)\\
&=&\frac{i}{p^2-m_0^2+~\Sigma(L,T,p^2)+i\epsilon}.
\end{eqnarray}
Therefore the poles of the propagator are found by 
\begin{eqnarray}
p^2&=&m_0^2-~\Sigma(L,T,p^2)= m_0^2-~\Sigma(L,T,m_0^2)+\cdots\\
&=& \underbrace{m_0^2-~\Sigma(\infty,\infty,m_0^2)}_{(m_h^{\rm{phys}})^2}-~\delta\Sigma(L,T,m_0^2)\\
&=&(m_h^{\rm{phys}})^2-~\delta\Sigma(L,T,(m_h^{\rm{phys}})^2)+\cdots\equiv(m_h(L,T))^2,
\end{eqnarray}
where $\delta\Sigma(L,T,m^2)\equiv\Sigma(L,T,m^2)-\Sigma(\infty,\infty,m^2)$, and it has been assumed that the interaction leading to self-energies is perturbative. Leading order self-energy corrections for the light-meson sector can be calculated using Eq.~(\ref{LO_chipt}) \cite{Gasser:1986vb, Colangelo:2003hf, Colangelo:2005gd}; the nucleon sector can be studied using Eq.~(\ref{rellag}) \cite{pregime}. Assuming $SU(2)_L\times SU(2)_R$ chiral symmetry and ignoring the contribution of the Delta resonances for the nucleon, one finds the LO FV contribution to the pion and nucleon masses are
\begin{eqnarray}
\label{mpiL}
\delta m_\pi^2(L)=\frac{m_\pi}{2 f_\pi^2}\sum_{\textbf{n}\neq0}\left(\frac{m_\pi}{2 \pi L~n}\right)^{3/2}{e^{-n m_\pi L}}+\cdots\\
\label{mNL}
\delta m_N(L)=\frac{3g_A^2}{16\pi f_\pi^2}\sum_{\textbf{n}\neq0}\frac{1}{L~n}e^{-n m_\pi L}+\cdots.
\end{eqnarray}
Instead of rederiving this result, we will use $\lambda\phi^4$-theory as a toy model for the meson sector, as it illustrates all the key-features of FV physics,
\begin{eqnarray}
 \label{rellag}
 \mathcal{L}_{toy}&=&\frac{1}2\phi\left(-\partial^2-m^2\right)\phi-\frac{\lambda}{4!}\phi^4.
\end{eqnarray}
\begin{figure}[t]
\begin{center}
\includegraphics[scale=0.4]{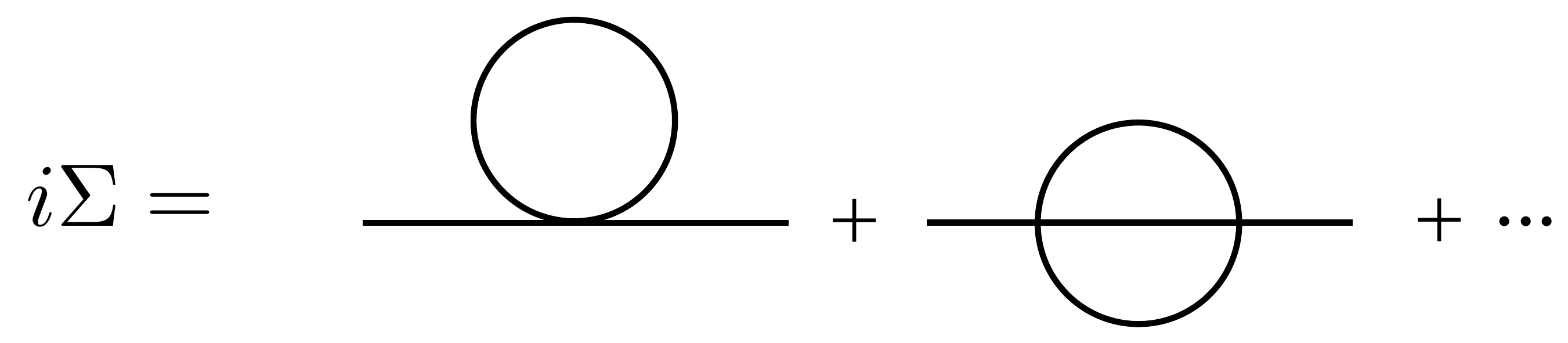}
\caption[Finite volume self-energy]{Contributions to the self-energy in the $\lambda\phi^4$-theory. In a finite volume, the momentum flowing through the loops is discretized.}\label{self_energy}
\end{center}
\end{figure}
Having periodic boundary conditions in a the spatial extent leads to the momenta to be discretized, $\textbf{p}_\textbf{n}=2\pi \textbf{n}/L$. Although the boundary conditions in the spatial extent are a choice, quarks necessarily have antiperiodic boundary conditions in the temporal extent of the lattice. This is a consequence of the fermionic nature of the quarks and the definition of the partition function, Eq.~(\ref{partition}). This means that mesons(baryons) have periodic(antiperiodic) boundary condition in the temporal extent. Therefore energies are also discretized
\begin{eqnarray}
\omega_{n_0}&=&\frac{2\pi n_0}{T}~~\text{(for mesons)}\\
\omega_{n_0}&=&\frac{\pi~(2n_0+1)}{T}~~\text{(for baryons)},
\end{eqnarray}
these are known as the Matsubara frequencies. With this, one finds that the LO contribution to self-energy, first diagram shown in Fig.~\ref{self_energy}, is equal to 
\begin{eqnarray}
i\Sigma(L,T)&=&\frac{\lambda}{L^3 T}\sum_{\textbf{n},n_0}\frac{1}{(\frac{2\pi n_0}{T})^2-(\frac{2\pi \textbf{n}}{L})^2-m^2+i\epsilon}\nn\\
&\rightarrow&
\label{self1}
-\frac{\lambda}{L^3 T}\sum_{\textbf{n},n_0}\frac{1}{(\frac{2\pi n_0}{T})^2+(\frac{2\pi \textbf{n}}{L})^2+m^2},
\end{eqnarray}
where the Matsubara frequencies have been Wick-rotated. This is equivalent to starting with a Lagrangian in Euclidean spacetime. 
Summing over the Matsubara frequencies using the Abel-Plana formula, Eq.~(\ref{AbelPlana}), 
{\small
\begin{eqnarray}
  \label{AbelPlana} \frac{1}{T}\sum_{n} f(\frac{2\pi n}{T}) =
\int_{-\infty}^\infty \frac{dz}{2\pi} f(z) - i \text{Res}
(\frac{f(z)}{e^{iT z}-1})|_{\rm lower plane} + i \text{Res}
(\frac{f(z)}{e^{-iT z}-1})|_{\rm upper plane}\nn\\
\label{abel}
\end{eqnarray}}
we obtain
{\small
 \begin{eqnarray}
 \label{self2}
i\Sigma(L,T)&=&-i\frac{\lambda}{L^3}\sum_{\textbf{n}}\left[\frac{1}{2\sqrt{(\frac{2\pi \textbf{n}}{L})^2+m^2}}
+\frac{1}{\sqrt{(\frac{2\pi \textbf{n}}{L})^2+m^2}}\frac{1}{e^{T{\sqrt{(\frac{2\pi \textbf{n}}{L})^2+m^2}}}-1}\right].
\end{eqnarray}}
 The sum over the $\textbf{n}$ can be done using the Poisson Resummation formula, 
 \begin{eqnarray}
\label{poisson}
  \frac{1}{L^3}\sum_{\bf{k}=\frac{2\pi \textbf{n}}L} f(\textbf{k})=  \int \frac{d^3k}{(2\pi)^3}  f(\textbf{k}) +
\sum_{\textbf{n}\neq 0}    \int \frac{d^3k}{(2\pi)^3} f(\textbf{k})~{e^{-L \textbf{n}\cdot \textbf{k}}}. 
\end{eqnarray}
With this tools we arrive at 
 \begin{eqnarray}
 \label{self3}
 \delta\Sigma(L,T)
&=&
-\frac{\lambda}{2 m}\sum_{\textbf{n}\neq0}\left(\frac{m}{2 \pi L~n}\right)^{3/2}{e^{-n m L}}\nn\\
&-&
\frac{\lambda}{2\pi^2}\int_0^\infty{dk}\frac{k^2}{\sqrt{k^2+m^2}}\frac{1}{e^{T{\sqrt{k^2+m^2}}}-1}+
\cdots,
\end{eqnarray}
where terms that are exponentially suppressed in both $L$ and $T$ have been neglected. It is easy to numerically show that the $T$-dependent piece is two orders of magnitude smaller than the $L$-dependent piece where $T=L$. Most calculations are performed using $T>L$, therefore the temperature-dependent piece can be safely neglected. We conclude that for this toy model, the LO FV correction to the mass is  
\begin{eqnarray}
\delta m^2(L)=\frac{\lambda}{2 m}\sum_{\textbf{n}\neq0}\left(\frac{m}{2 \pi L~n}\right)^{3/2}{e^{-n m L}},
\end{eqnarray}
exactly reproducing the result by L\"uscher~\cite{luscher0}. 

Therefore, we conclude by reiterating that, in the one-body sector, finite volume effects are exponentially suppressed. As will be discussed in great detail for the remainder of this work, this is not the case for systems with two or more particles. From here on, it will be assume that exponential corrections scaling with the pion mass are negligible (i.e. $m_\pi L\sim  m_\pi T\gg1$ must hold) and consequently will be ignored.

Ultimately, we are interested in determining $S$-matrix elements directly from LQCD. The definition of the $S$-matrix strongly relies on the notion of asymptotic states, which are nonexistent in periodic-finite volumes with spatial extents of the order of a few fermi. One could naively argue that by studying two-particle Green's functions at different values of $L$ one could extrapolate to $L=\infty$ and in doing so obtaining information regarding ``asymptotic states". This speculation is trumped by the fact that LQCD calculations of Green's functions are performed in Euclidean spacetime. Furthermore, these are numerical approximations of the Green's functions, therefore they cannot be analytically continued to Minkowski space. Maiani and Testa studied the infinite volume limit of a Euclidean theory \cite{Maiani:1990ca}, and found that infinite volume Euclidean GreenÕs functions for $\pi(q)+\pi(-q)\rightarrow n$ depend on the desired scattering amplitude $\mathcal{M}_{\pi\pi\rightarrow n}\propto{_{out}\langle }n| \pi,q,\pi,-q\rangle_{in}-{_{out}\langle} n| \pi,q,\pi,-q\rangle_{out}$ as well as the average $({_{out}\langle n| \pi,q,\pi,-q\rangle_{in}+ {_{out}\langle} n| \pi,q,\pi,-q\rangle_{out}})/2$. It is only after analytically continuing to Minkowski space that one recovers the LSZ reduction formula. They showed that only in the kinematic threshold, the $\pi +\pi\rightarrow \pi  +\pi$ Green's function reduces to 
\begin{eqnarray}
G(t_1,t_2)=\langle \pi(t_1)\pi(t_2)J(0)^\dag \rangle\stackrel{\small{t_1\gg t_2\gg0}}{\longrightarrow}\frac{Z_\pi }{(2m_\pi)^2}~f(4m_\pi^2)~\left(1-a\sqrt{\frac{m_\pi}{4\pi~t_2}}+\cdots\right),\nn\\
\end{eqnarray}
where $J$ is a current operator that couples to two pions, $f(q^2)$ is the form factor of $J$, and $a$ is the S-wave scattering length.

This naive approach is rather limiting, and it is necessary to circumvent this limitation to be able to extract $S$-matrix elements with arbitrary momentum. The solution to this problem was first postulated by Martin L\"uscher \cite{luscher1,luscher2}, who found a one-to-one mapping between the finite volume two-particle spectrum and the infinite volume scattering phase shifts using a field theoretical approach. L\"uscher derived this relation for two-scalar bosons with non-relativistic momentum below inelastic thresholds. In the following chapter we will re-derive the relativistic analogue of this problem \cite{movingframe, sharpe1, sharpe2, movingframe2}.  The remainder of this work will present new results regarding the generalization of this formalism for coupled-channels, baryonic systems and three-particle systems.

 
\chapter{Meson-Meson Systems in a Finite Volume}{\label{mmsys}}

Scattering processes in hadronic physics provide useful information
about the properties of particles and their interactions. Some of
these processes are well investigated in experiments with reliable
precision. 
However, there are interesting two-body hadronic processes whose experimental determinations continue to pose challenges. They mainly include two-body
hadronic scatterings near or above the kinematic threshold with the
possibility of the occurrence of resonances. In this section we discuss  scalar sector of QCD, whose nature is still puzzling (See for example
\cite{coup3} and references therein). While some phenomenological models
suggest the scalar resonances to be tetraquark states (as first proposed by Jaffe \cite{fourquark}), others
propose these to be weakly bound mesonic molecular states. The
most famous of which are the flavorless $a_{0}(980)$ and $f_{0}(980)$, which
are considered to be candidates for a $K\bar{K}$ molecular states
\cite{kkmol, uchipt, kkmol2}. 
In order to shine a light on the nature of these resonances, it would be necessary to perform model-independent multi-channel calculations including the  $\{\pi\pi,\pi\pi\pi\pi, K\bar K, \eta\eta\}$ scattering states directly from the underlying theory of QCD. 

As briefly discussed in the previous section , L\"uscher showed
how to obtain the infinite volume scattering phase shifts below inelastic thresholds from 
calculating energy levels of two-body scattering states in the finite
volume \cite{luscher1,luscher2}. In order to present the generalization of this formalism above inelastic thresholds, it is key to first understand L\"uscher's result. The non-relativistic, center of mass reference frame result by L\"uscher has been generalized to the moving frames in Refs. \cite{movingframe, sharpe1, movingframe2}. In following section we follow in detail the generalization by Kim, Sachrajda, and Sharpe, and discuss the implications for $\pi^+\pi^+$ scattering. In section \ref{coudpledchannels} we present the generalization of this formalism for N arbitrarily strongly coupled two-body channels in a moving frame and discuss the implications for the $\pi\pi-K\bar{K}$ isosinglet spectrum. In section \ref{scalardimer} we see the generalization of the dimer formalism for arbitrary partial wave and observe that indeed the dimer formalism recover L\"uscher's well known result. Finally in section \ref{EW2B} we observe the implications of the coupled-channel formalism for electroweak processes involve two particles both in the initial and final states, e.g.~$\pi K\rightarrow \pi\pi$.

\section{Below inelastic thresholds}
\begin{figure}[t]
\begin{center} 
\subfigure[]{
\label{twopar_FV1}
\includegraphics[scale=0.4]{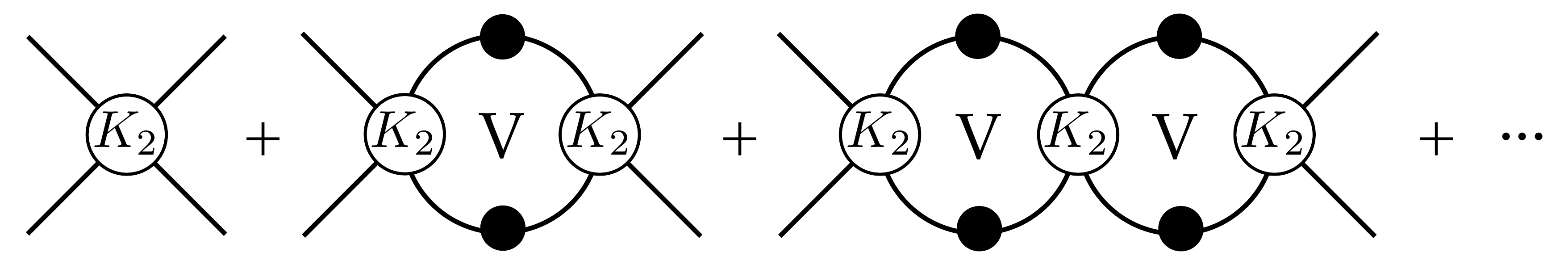}}
\subfigure[]{
\includegraphics[scale=0.25]{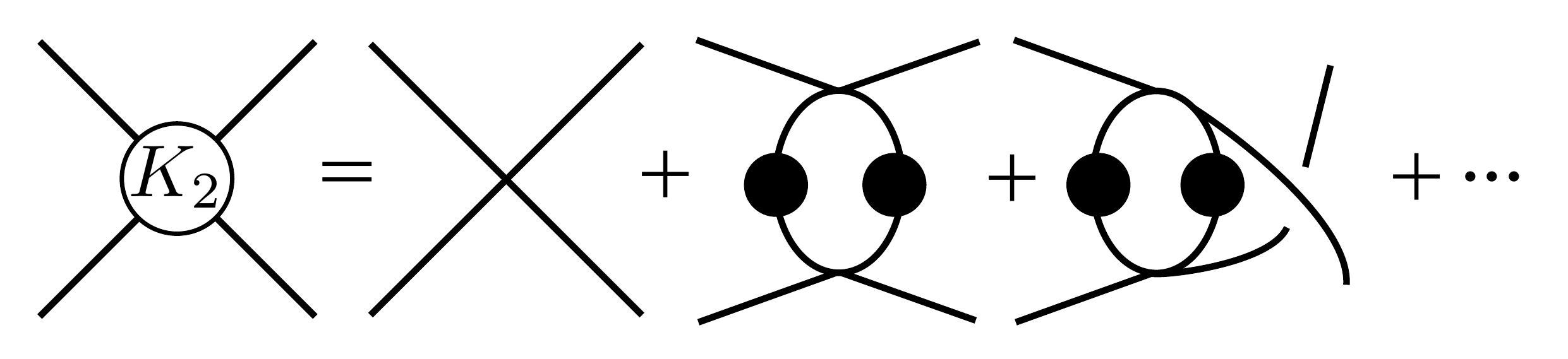}
\label{twopar_FV2}}
\subfigure[]{
\includegraphics[scale=0.25]{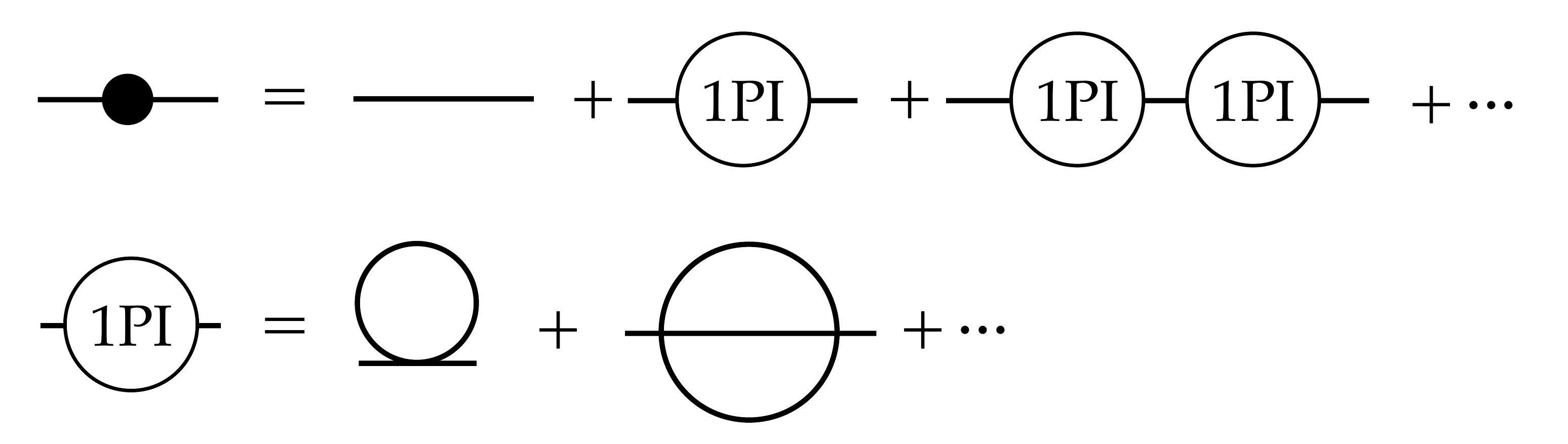}
\label{twopar_FV3}}
\caption[Relativistic finite-volume two-particle propagator]{a) Sum of all $2\rightarrow 2$ amputated diagrams written in terms of the $K_2$ and the fully-dressed one-particle propagator. b) $K_2$ is two particle Bethe-Salpeter kernel, i.e. the sum of all s-channel two-particle irreducible diagrams. c) The fully dressed one-particle propagator is the sum of all one-particle irreducible diagrams.}
\label{twopar_FV}\end{center}
\end{figure}

The goal is to obtain a relationship between the finite volume two-particle spectrum and infinite volume scattering amplitudes. As mentioned in section~\ref{FVEuc} the N-particle spectrum can be obtained from the pole condition of the N-particle propagator. Therefore we need to evaluate the full two-particle propagator in a finite volume, which is equal to the sum of all $2\rightarrow 2$ amputated diagrams, shown in Fig.~\ref{twopar_FV}. For the time being, we will consider systems composed of two particle with mass $m_1$ and $m_2$ with $m_1\leq m_2$. The system has a total momentum $\textbf{P}$ and energy E satisfying $m_1+m_2\leq \sqrt{E^2-P^2} \ll 3m_1+m_2$. In section~\ref{FVEuc} we observed that the self-energy diagrams appearing in the one-particle propagator, Fig.~\ref{twopar_FV3}, are exponentially close to their infinite volume counterpart. From here on all $\mathcal{O}(e^{-m_\pi L})$ correction will be neglected, and as a result the fully dressed one-particle propagator has a pole at the physical mass, $m^{\rm{phys}}_i$, with a residue of one (LO exponential corrections for $\pi^+\pi^+$ \cite{Bedaque:2006yi} and $NN$ \cite{Sato:2007ms} have been previously calculated). The finite volume two-particle propagator, as is shown in Fig.~\ref{twopar_FV1}, can be written in terms of the Bethe-Salpeter kernel, $K_2$, which  is also exponentially close to its infinite volume counterpart. In fact, only diagrams where all intermediate particles can be simultaneously put on-shell are power-law in volume, all other diagrams exponentially suppressed. For energies below the particle production threshold, this corresponds solely to s-channel diagrams. 

The first non-trivial diagram to consider is the first loop appearing in Fig.~\ref{twopar_FV1}. Having shown in section~\ref{FVEuc} that finite temperature effects can be safely neglected, the zero-momentum appearing in the loops will be assumed to be continuous, and the spatial momentum will be discretized. If the two particle appearing in the loop are identical, there is a symmetry factor of 2, this will be encapsulated in an overall constant $n$ which is equal to 1 for distinguishable particles and 1/2 for indistinguishable particles. With this, the second diagram in Fig.~\ref{twopar_FV1} is equal to

\begin{eqnarray}
\label{loop1}
iG^{V}(\mathbf{p}_i,\mathbf{p}_f)&\equiv&\frac{n}{L^3}\sum_{\mathbf{k}}\int\frac{dk^0}{2\pi}\frac{K_2(\mathbf{p}_i,\mathbf{k})~K_2(\mathbf{k},\mathbf{p}_f)}{[(k-P)^2-m_{1}^2+i\epsilon][k^2-m_{2}^2+i\epsilon]}\\
&=&\frac{-in}{L^3}\sum_{\mathbf{k}}
\left\{
\frac{K_2(\mathbf{p}_i,\mathbf{k})~K_2(\mathbf{k},\mathbf{p}_f)}{[(\omega_{k,2}-E)^2-\omega_{Pk,1}^2
+i\epsilon][2\omega_{k,2}+i\epsilon]}
\right.\nn\\
\label{loop2}
&&\left.\hspace{3cm}+\frac{K_2(\mathbf{p}_i,\mathbf{k})~K_2(\mathbf{k},\mathbf{p}_f)}{[2\omega_{Pk,1}+i\epsilon][(\omega_{Pk,1}+E)^2-\omega_{k,2}^2+i\epsilon]}\right\},
\end{eqnarray}
where the integral over $k^0$ has been performed and the following kinematic functions have been defined
\begin{eqnarray}
\omega^2_{k,i}=|\textbf{k}|^2+m_i^2,\hspace{1cm}
\omega^2_{Pk,i}=|\textbf{P}-\textbf{k}|^2+m_i^2.
\end{eqnarray}
The ingoing and outgoing momentum are denoted as $\mathbf{p}_i$ and $,\mathbf{p}_f$, where the momentum is necessarily fixed by the on-shell condition but the direction is left unfixed. It is easy to see that for $E\leq 3m_1+m_2$, the second term appearing in Eq.~(\ref{loop2}) is finite for all values of $\textbf{k}$. From Poisson's 
Resummation formula Eq.~(\ref{poisson}) it is easy to see that this term can be replaced with its infinite volume counterpart up to exponential corrections scaling like $\mathcal{O}(e^{-m_1 L})\leq \mathcal{O}(e^{-m_\pi L})$ and can be safely neglected. Therefore the finite-volume effects are encoded in the first term appearing above
\begin{eqnarray}
\label{sum_rel1}
S_{rel}
&=&-i\frac{n}{L^3}\sum_{\mathbf{k}}\frac{1}{2\omega_{k,2}}
\frac{K_2(\mathbf{p}_i,\mathbf{k})K_2(\mathbf{k},\mathbf{p}_f)}{[(\omega_{k,2}-E)^2-\omega_{Pk,1}^2+i\epsilon]}.
\end{eqnarray}
It is convenient to write this sum in terms of center of mass (CM) frame coordinates $(E^*, q^*)$, where  $E^*=\sqrt{E^2-P^2}$ is the CM energy and $q^*$ is the CM relative momentum 
\begin{eqnarray}
\label{momentum}
q^{*2}=\frac{1}{4}\left(E^{*2}-2(m_{1}^2+m_{2}^2)+\frac{
(m_{1}^2-m_{2}^2)^2}{E^{*2}}\right),
\end{eqnarray}
which simplifies to $\frac{E^{*2}}{4}-m^2$ when $m_{1}=m_{2}=m$. The Lab frame coordinates $\textbf{k}=({k}_{||},{k}_\perp)$ and $\omega_{k,i}$ appearing in the summand can be transformed to CM coordinates $\textbf{k}^*=({k}^*_{||},{k}^*_\perp)$ and $\omega_{k,i}^*=\sqrt{{k}^{*2}+m_i^2}$ using the standard Lorentz transformations 
\begin{eqnarray}
{k}^*_{||}&=&\gamma({k}_{||}-\beta \omega_{k,i}), \hspace{1cm}
{k}^*_{\perp}={k}_{\perp}, \hspace{1cm}
\omega_{k,i}^*=\gamma(\omega_{k,i}-\beta {k}_{||}),
\end{eqnarray}
where $\gamma = \frac{E^*}{E},~\beta=\frac{P}{E}$. Using these relations and schematically writing the functional form of the kernels in the CM frame as $K_2^{*}$, it takes a few lines of algebra to rewrite Eq.~(\ref{sum_rel1}) as 
\begin{eqnarray}
\label{sum_rel2}
S_{rel}
&=&-i\frac{n}{L^3}\sum_{\mathbf{k}}\frac{1}{2E^*}\frac{\omega_{k,2}^*}{\omega_{k,2}}
\frac{K_2^*(\mathbf{p}_i^*,\mathbf{k}^*)K_2^*(\mathbf{k}^*,\mathbf{p}_f^*)}{q^{*2}-k^{*2}+i\epsilon}\left(\frac{E^*+\frac{m_1^2-m_2^2}{E^*}-2\omega^*_{k,2}}{4\omega^*_{k,2}}\right).
\end{eqnarray}
Notice that at the pole, the term in the parenthesis is exactly equal to one. The on-shell condition fixes the overall momentum of the kernel leaving the direction of the momentum unfixed. This leads us to decompose the kernels and the product of them into spherical harmonics
\begin{eqnarray}
K_2^*(\mathbf{p}_i^*,\mathbf{k}^*)&=&\sum_{l',m'}[K_2^*(\mathbf{p}_i^*,{k}^*)]_{l'm'}\sqrt{4\pi}Y^*_{l'm'}(\hat{k}^*),\\
K_2^*(\mathbf{k}^*,\mathbf{p}_f^*)&=&\sum_{l'',m''}[K_2^*({k}^*,\mathbf{p}_f^*)]_{l''m''}\sqrt{4\pi}Y_{l''m''}(\hat{k}^*),\\
K_2^*(\mathbf{p}_i^*,\mathbf{k}^*)K_2^*(\mathbf{k}^*,\mathbf{p}_f^*)&=&\sum_{l,m}
f_{lm}(k^*)k^{*l}\sqrt{4\pi}Y_{lm}(\hat{k}^*),\\
\Rightarrow f_{lm}(k^*)&=& \frac{\sqrt{4\pi}}{k^{*l}}\sum_{l',l'',m',m''} [K_2^*(\mathbf{p}_i^*,{k}^*)]_{l',m'}[K_2^*({k}^*,\mathbf{p}_f^*)]_{l'',m''}\nn\\
\label{flm}
&&\hspace{2.5cm}\times\int d\Omega_{\textbf{k}}~Y^*_{l'm'}(\hat{k}^*)Y^*_{lm}(\hat{k}^*) Y_{l''m''}(\hat{k}^*).
\end{eqnarray}
The final piece of mathematical artillery that we need relies on the observation that Eq.~(\ref{sum_rel2}) has a single pole. By subtraction the singular contribution, the summation can be replaced by an integral up to exponential corrections that will be neglected,
\begin{eqnarray}
\label{sum_relr}
S_{rel}+i\frac{f_{lm}(q^*)}{2E^*}\frac{n}{L^3}~\sum_{\mathbf{k}}\frac{\omega_{k,2}^*}{\omega_{k,2}}
\frac{k^{*l}~\sqrt{4\pi}Y_{lm}(\hat{k}^*)}{q^{*2}-k^{*2}+i\epsilon}&&\hspace{6cm}\nn\\&&\hspace{-4cm}
=I_{rel}+i\frac{f_{lm}(q^*)}{2E^*}~{n}\int\frac{d^3{\mathbf{k}}}{(2 \pi)^3}\frac{\omega_{k,2}^*}{\omega_{k,2}}
\frac{k^{*l}~\sqrt{4\pi}Y_{lm}(\hat{k}^*)}{q^{*2}-k^{*2}+i\epsilon},
\end{eqnarray}
where $I_{rel}$ denotes the infinite volume counterpart of Eq.~(\ref{sum_rel2}) and repeated indices are summed. By transforming the  variable of integration transformation from ${\mathbf{k}}$ to ${\mathbf{k}}^*$ we introduce a Jacobian that exactly cancels the overall factor of ${\omega_{k,2}^*}/{\omega_{k,2}}$ appearing in the last integral.  The $i\epsilon$ appearing in the denominator of the summation can be safely ignored, for the integral it is convenient to replace the denominator appearing into two parts, $\frac{1}{q^{*2}-k^{*2}+i\epsilon}=-i\pi \delta(q^{*2}-k^{*2})+\mathcal{P}\frac{1}{q^{*2}-k^{*2}}$, where $\mathcal{P}$ denotes the principal value. Equation Eq.(\ref{sum_relr}) can be rewritten as
\begin{eqnarray}
S_{rel}-I_{rel}
&=&n\left(\frac{q^*f_{00}(q^*)}{8\pi E^*}+\frac{i}{2 E^*}\sum_{l,m}f^*_{lm}(q^*)c^{\textbf{P}}_{lm}(q^{*2})\right),
\end{eqnarray}
where 
\begin{eqnarray}
\label{clm}
c^{\textbf{P}}_{lm}(x)&=&\frac{1}{\gamma}\left[\frac{1}{L^3}\sum_{\textbf{k}}-\mathcal{P}\int\frac{d^3\mathbf{k}}{(2\pi)^3}\right]\frac{\sqrt{4\pi}Y_{lm}(\hat{k}^*)~r^{l}}{{r}^{*2}-x} 
\end{eqnarray}
where $r={\gamma}^{-1}(\mathbf k_{||}-\alpha \mathbf P)+\mathbf k_{\perp}$, and $\alpha=\frac{1}{2}\left[1+\frac{m_1^2-m_2^2}{E^{*2}}\right]$~\cite{ Leskovec:2012gb,Davoudi, Fu:2011xz}. This reduces to the non-relativistic value of $\alpha=\frac{m_1}{m_1+m_2}$ as is presented in Ref.~\cite{Bour:2011ef}. Note that this result is equivalent to the result obtained in Refs. \cite{movingframe, sharpe1,movingframe2} for the boosted systems of particles with identical masses.\footnote{The kinematic function $c^{\textbf{P}}_{lm}(q^{*2})$ can also be written in terms of the three-dimensional Zeta function, $\mathcal{Z}^\textbf{d}_{lm}$,
 \begin{eqnarray}
 \nonumber
 	c^{\textbf{P}}_{lm}(q^{*2})=\frac{\sqrt{4\pi}}{\gamma L^3}\left(\frac{2\pi}{L}\right)^{l-2}\mathcal{Z}^\textbf{d}_{lm}[1;(q^*L/2\pi)^2],\hspace{1cm} 
\mathcal{Z}^\textbf{d}_{lm}[s;x^2]=\sum_{\mathbf r \in P_d}\frac{Y_{l,m}(\mathbf{r})}{(r^2-x^2)^s},
\end{eqnarray}
where the sum is performed over $P_d=\left\{\mathbf{r}\in \textbf{R}^3\hspace{.1cm} | \hspace{.1cm}\mathbf{r}={\gamma}^{-1}(\mathbf m_{||}-\alpha \mathbf d)+\mathbf m_{\perp} \text{,}m\in \textbf{Z}\right\}$, $\mathbf d$ is the normalized boost vector $\mathbf d=\mathbf{P}L/2\pi$,  and $\alpha$ is defined above~\cite{Davoudi, Fu:2011xz, Leskovec:2012gb}.} Note that the definition of the function $c_{lm}^{\textbf{P}}$, Eq.~(\ref{clm}), differs to that of Ref. \cite{sharpe1} by an overall sign.
Using Eq.~(\ref{flm}), the difference between the infinite volume and finite volume loops can be represented as a product over infinite-dimensional matrices in orbital angular momentum space,
\begin{eqnarray} 
S_{rel}-I_{rel}&=&iG^{V}(\mathbf{p}_i,\mathbf{p}_f)-iG^{\infty}(\mathbf{p}_i,\mathbf{p}_f)\\
&=&[-iK_2^*(\mathbf{p}_i^*,{q}^*)]_{l',m'}  (i\delta \mathcal{G}^{V}_{i})_{l',m';l'',m''} [-iK_2^*({q}^*,\mathbf{p}_f^*)]_{l'',m''}\\
&\equiv&-iK_2^*~ \delta \mathcal{G}^{V}~K_2^*
\end{eqnarray}
Note that we have used the fact that the only finite volume piece appearing in  Eq.(~\ref{loop2}) is $S_{rel}$, and the finite volume contribution to the single loop is
\begin{eqnarray}
(\delta \mathcal{G}^{V})_{l_1,m_1;l_2,m_2}&=&i	\frac{q^*n}{8\pi E^*}\left(\delta_{l_1,l_2}\delta_{m_1,m_2}+i\frac{4\pi}{q^*}\sum_{l,m}\frac{\sqrt{4\pi}}{q^{*l}}c^{\textbf{P}}_{lm}(q^{*2})\int d\Omega^*Y^*_{l_1m_1}Y^*_{lm}Y_{l_2m_2}	\right).\nn\\
\label{def0}
\end{eqnarray}
Denoting $\mathcal{G}^{\infty}$ and $\mathcal{G}^{V}$ as the infinite and finite volume loops respectively, we can determine the full two-particle propagator $i\mathcal{M}^{V}$ as follows
\begin{eqnarray}
\label{FV_prop}
i\mathcal{M}^V&=&-i{K}_2
-i{K}_2\mathcal{G}^{V} {K}_2
-i{K}_2\mathcal{G}^{V} {K}_2\mathcal{G}^{V}{K}_2+\cdots=-i{K}_2\frac{1}{1-\mathcal{G}^{V}{K}_2}.\end{eqnarray}
Currently the poles of the propagator are written in terms of the kernel. This can be circumvented by writing the kernel in terms of the two-particle scattering amplitude. This is defined as the sum of all $2\rightarrow 2$ amputates scattering amplitudes, where the loops are evaluated with continuous momenta,
\begin{eqnarray}
\label{IV_prop}
i\mathcal{M}^\infty&=&-i{K}_2\frac{1}{1-\mathcal{G^{\infty}}{K}_2}
\Rightarrow
{K}_2=-\mathcal{M}\frac{1}{1-\mathcal{G^{\infty}}\mathcal{M}}.\end{eqnarray}
Inserting this definition for $K_2$ into Eq.~(\ref{FV_prop}) one finds that the poles of the finite volume propagator are defined by
\begin{eqnarray}
\label{det0}
\det \left(\left(\mathcal{M}^{\infty}\right)^{-1}+\delta\mathcal{G}^V\right)=0, 
\end{eqnarray} 
where the determinant is over angular momentum space. While the scattering amplitude is diagonal in angular momentum with the diagonal component defined in Eq.~(\ref{scatrel}), $\delta\mathcal{G}^V$ as defined in Eq.~(\ref{def0}), is clearly not. This is a consequence of the fact that angular momentum is not a good quantum number in a cubic finite volume.  As a result, the energy eigenstates in the CM frame are identified with the irreducible representations (irreps) of the octahedral (O) group~\cite{luscher1}. Since the phase shifts are characterized according to the irreps of the SO(3) group, the energy eigenvalues of the system in a given irreps of the octahedral group  in general depends on the phase shifts of more than one partial-wave. For example, if  a interpolating operator for two degenerate mesons is in the $A_1$ irrep of the cubic group, the energy eigenstates of the system have overlap with the $l=0,4,6,\ldots$ angular momentum states at CM frame. When $\textbf{P}\neq 0$, the symmetry group is reduced and the symmetries of the system are defined by the tetragonal ($D_{4}$) group. Consequently at low energies the $l=0$ state will mix with $l=2$ partial wave as well as with higher partial waves \cite{movingframe}. For two mesons with different masses, the symmetry group is even further reduced in the boosted frame, making the mixing to occur between $l=0$ and $l=1$ states as well as with higher angular momentum states \cite{Fu:2011xz}. An easy way to see the latter is to note that in contrast to the case of degenerate masses, the kinematic function $c_{lm}^{\textbf{P}}$ as defined in Eq. (\ref{clm}) does not vanish for odd $l$ when the masses are different. As a result even and odd angular momenta can mix as seen in the quantization condition. This does not indicate that the spectrum of the system is not invariant under parity. As long as all  interactions between the particles are parity conserving, the spectrum of the system and its parity transformed counterpart are the same. One should note that the determinant condition, Eq. (\ref{det0}), guarantees this invariance: any  mechanism, for example, which takes an S-wave scattering state to an intermediate P-wave two-body state, would take it back to the final S-wave scattering state, and the system ends up in the same parity state. \footnote{Note that under parity $\mathcal{Z}^\textbf{d}_{lm}\rightarrow\left(-1\right)^{l}\mathcal{Z}^\textbf{d}_{lm}$. Note  also that under the interchange of particles $\mathcal{Z}^\textbf{d}_{lm}\rightarrow\left(-1\right)^{l}\mathcal{Z}^\textbf{d}_{lm}$, so that for degenerate masses the $c_{lm}^{\textbf{P}}$-function vanishes for odd $l$. This is expected since the parity transformation in the CM frame is equivalent to the interchange of particles. However, as is explained above for the case of parity transformation, despite the fact that $\delta \mathcal{G}^{V}$ is not symmetric with respect to the particle masses, the quantization condition is invariant under the interchange of the particles.}
 
\subsection{Implications for $\pi^+\pi^+$ scattering \label{pipiscat}} 
In practice, it is necessary to truncate the partial waves contributing to Eq.~(\ref{det0}) to some $l_{max}$. For sufficiently small energies one can set $l_{max}=0$. In doing so, the rather complicated quantization condition, Eq.~(\ref{det0}), reduces to just
\begin{eqnarray}
\label{QC_swave}
q^*\cot({\delta}^{(0)})=4\pi c_{00}^{P}(q^{*2}),
\end{eqnarray}
where ${\delta}^{(0)}$ is the S-wave phase shift. Note that the imaginary pieces appearing in Eq.~(\ref{det0}) have exactly canceled. In fact it is easy to show this is true for all partial waves.  

The quantization condition simplifies even further, when introducing a pseudo-phase $\phi$, defined by  
\begin{eqnarray}
 \label{pseudophase0}
{q^*}\cot({\phi})\equiv -4\pi{ c_{00}^P(q^{*2})}\Longrightarrow \delta^{(0)}+\phi=N\pi,
\end{eqnarray}
where $N$ is an integer.  

Having all the pieces in place, we can revisit the determination of the $\pi^+\pi^+$ scattering length from LQCD. First, it is necessary to extract the lab frame energies. To do this is necessary to construct appropriate interpolating operators with the right quantum numbers and good overlap with the cubic irreps. For example, for two pions with $\textbf{P}=0$, $E_{\pi^+\pi^+}^{A_1}$ can be extracted using the following correlation function
\begin{eqnarray}
C_{\pi^{+}\pi^{+}}(t,\textbf{q})&=&\sum_{\textbf{x}}e^{i(\textbf{x}-\textbf{y})\cdot \textbf{q}}\langle\pi^{+}(x)\pi^{+}(y)\bar{\pi}^{+}(0)\bar{\pi}^{+}(0)\rangle\\
&\longrightarrow  &
2~Z_{\pi^+\pi^+}~e^{-E^{A_1}_{\pi^+\pi^+}T/2}~\cosh\left(E^{A_1}_{\pi^+\pi^+}  t_T\right)+
Z^T_{\pi^+\pi^+},
\end{eqnarray}
where $ t_T=t-T/2$, $T$ is the temporal extent of the lattice, $Z_{\pi^+\pi^+}$ and $Z^T_{\pi^+\pi^+}$ are constants.
The relative momentum can then be determined from Eq.~(\ref{momentum}), by setting $m_1$ and $m_2$ equal to the pion mass calculated on the lattice $m_\pi^{latt}$. Using Eq.~(\ref{QC_swave}), for  $q^{*2}\ll m_\pi^{latt}$ one obtains the scattering length [${a^{-1}_{\pi^+\pi^+}} \approx-4\pi c_{00}^{P}(q^{*2})$] for the given value of the pion mass. By performing calculations at multiple values of the light-quark masses one can then obtain the scattering length as a function of $m_\pi$. 

\begin{figure}[t]
\begin{center} 
\includegraphics[scale=0.5]{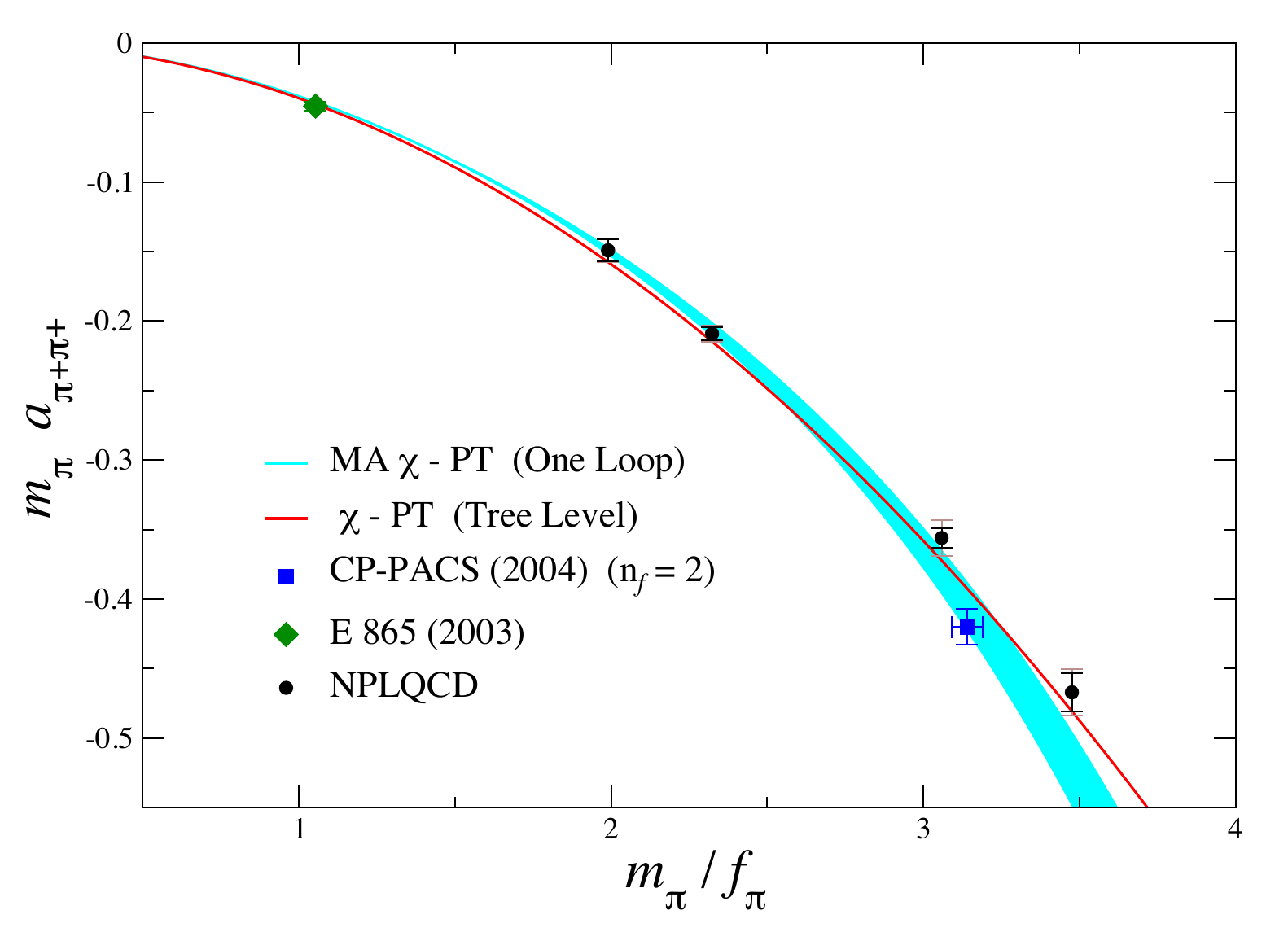}
\caption[$m_\pi a_{\pi^+\pi^+}$ as a function of $m_\pi/f_\pi$ as determined by the NPLQCD Collaboration~\cite{beane:2007xs}]{\small{This figure is from the work of the NPLQCD Collaboration~\cite{beane:2007xs}. Shown are their determination of $m_\pi a_{\pi^+\pi^+}$ (black circles)  as a function of $m_\pi /{f_\pi}$ for four different ensembles. The statistical uncertainty is shown as a dark bar, while the light bars denote systematic and statistical uncertainties added in quadrature. The blue band corresponds to the fit of the four values of $m_\pi a_{\pi^+\pi^+}$ using Eq.~(\ref{apipi3}). They also compare their results to those of the CP-PACS Collaboration~ \cite{Yamazaki:2004qb} (square), the LO $\chi$PT results Eq.~(\ref{su3LO})~\cite{Weinberg:1966kf} (red line), and the experimental measurement Eq.~(\ref{apipi_exp})~\cite{Pislak:2003sv, Pislak:2001bf} (diamond). The difference in sign between the figure and the value quoted in Eq.~(\ref{nplqcdapipi}) is due to the fact that in Ref.~\cite{beane:2007xs} the scattering length is defined as $a^{-1}=\lim_{p\rightarrow0}~p\cot\delta$, which differs by an overall sign from the convention used in this work, Eq.~(\ref{EFE}), which is the customary nuclear physics definition. The figure is reproduced with permission from the NPLQCD Collaboration.
}}
\label{nplqcd_apipi}
\end{center}
\end{figure}
As mentioned before, the most precise determination of the scattering length is by the NPLQCD Collaboration~\cite{beane:2007xs}. They performed the procedure discussed above for four different light quark masses, corresponding to $m_\pi=293.1(1.5)-591.8(1.0)$~MeV. Figure~\ref{nplqcd_apipi} shows the four values of $m_\pi {a_{\pi^+\pi^+}}$ obtained by NPLQCD as a function of $f_\pi/m_\pi$. The ``MA $\chi$-PT labeling makes reference to the fact that discretization effects have been removed using expression from Mixed Action $\chi$PT~\cite{Chen:2006wf}. Also shown is their fit of the scattering lengths using Eq.~(\ref{apipi3}), which allows them to make a prediction at the physical point, Eq.~(\ref{nplqcdapipi}). For other examples of the application of this formalism for extraction of scattering phase shifts and binding energies of two-hadron systems from LQCD the reader is redirected to Refs.~\cite{Durr:2008zz, Beane:2010hg, Beane:2011xf, Beane:2012vq, Beane:2012ey,Yamazaki:2012hi, Beane:2011iw, Beane:2013br, Beane:2011sc, Lang:2011mn, Pelissier:2012pi,  Dudek:2012xn, Dudek:2012gj, Mohler:2013rwa}.

\section{Above inelastic thresholds \label{coudpledchannels}}
The L\"uscher formalism relies on the assumption that the energy of the systems lies sufficiently below any inelastic thresholds. A direct calculation of the near threshold scattering quantities using LQCD can lead to the identification of resonances in QCD such
as $a_{0}(980)$ and $f_{0}(980)$ discussed earlier, and provide reliable predictions for their
masses and their decay widths. One such generalization was developed by
Liu $et\; al.$ in the context of quantum mechanical two-body scattering
\cite{coup02, coup2}. There, the authors have been able to deduce the relation between
the infinite volume coupled channel S-matrix elements and the energy
shifts of the scattering particles in the finite volume by solving
the coupled Schrodinger equation both in infinite volume and on a
torus. The idea is that as long as the exponential volume corrections
are sufficiently small, the polarization effects, as well as other
field theory effects, are negligible. Therefore after replacing the non-relativistic
dispersion relations with their relativistic counterparts, the quantum
mechanical result of Liu $et\; al.$ \cite{coup02, coup2} is speculated to be
applicable to the massive field theory. In another approach, Lage
$et\; al.$ considered a two-channel Lippman-Schwinger equation in
a non-relativistic effective field theory (NR EFT). They presented the mechanism for obtaining the $\bar{K}N$ scattering length, and studying the nature of the $\Lambda\left(1405\right)$ resonance from LQCD \cite{coup1}. Later on, Bernard $et\; al.$ generalized
this method to the relativistic EFT which would be applicable for coupled
$KK-\pi\pi$ channels \cite{coup3}. Unitarized chiral perturbation theory (UCHPT) provides another tool to study a variety of resonances in the coupled channel scatterings. This method uses the Bethe-Salpeter equation for a coupled-channel system to dynamically generate the resonances in both light meson sector and meson-baryon sector in infinite volume, see for example Refs. \cite{Oller:1997ti, Kaiser, Locher, OsetII}. When applied in the finite volume, the volume-dependent discrete energy spectrum can be produced, and by fitting the parameters of the chiral potential to the measured energy spectrum on the lattice, the resonances can be located by solving the scattering equations in infinite volume. This method has been recently used to study the resonances $f_0(600)$, $f_0(980)$ and $a_0(980)$ in Refs. \cite{coup2011, Oset}, $\Lambda(1405)$ in Ref. \cite{MartinezTorres}, $a_1(1260)$ in Ref. \cite{Roca:2012rx}, $\Lambda_c(2595)$ in Ref. \cite{Xie}, and $D^*_{s 0}(2317)$ in Ref. \cite{MartinezTorresII} in the finite volume. One should note that in contrast to the single
channel scattering system, coupled-channel scattering
requires determining a minimum of three independent scattering parameters
which would require at least three measurements of the energy
levels in the finite volume. As proposed in Refs. \cite{coup3, coup2011}, one
can impose twisted boundary condition in the lattice calculation
to increase the number of measurements by varying the twist angle and further constrain the scattering parameters. Another tool
to circumvent this problem is the use of asymmetric lattices as
investigated in Refs. \cite{coup3,coup2011, MartinezTorres}. Alternatively, one can perform calculations with different boost momenta \cite{MartinezTorres,Roca:2012rx}.

Generalizing Eq.~(\ref{det0}) to $N$ arbitrarily strongly coupled two-body channels with arbitrary momentum is straightforward. Consider first the case where there are two open channels. For example, when the pion mass is approximately 300~MeV, the four pion threshold lies above the two-kaon threshold. Therefore the following formalism can be used for studying $\pi\pi-K\bar{K}$ isosingle channel for energies below the four-pion threshold. For a given partial wave, the $S$-matrix is a two-dimensional matrix over the two-open channels, Eq.~(\ref{smatrix2}), where $\delta_{I}$ and $\delta_{II}$ are the phase shifts corresponding to the scattering in channels  $I$ and $II$ respectively, and $\bar{\epsilon}$ is a parameter which characterizes the mixing between the channels.

Again, let $E$ and $P$ denote the total energy and momentum in the laboratory frame and the CM energy is $E^*=\sqrt{E^2-P^2}$. For the $j^{th}$ channel with two mesons each having masses $m_{j,1}$ and $m_{j,2}$, the CM relative momentum is 
\begin{eqnarray}
\label{momentumcc}
q^{*2}_j=\frac{1}{4}\left(E^{*2}-2(m_{j,1}^2+m_{j,2}^2)+\frac{
(m_{j,1}^2-m_{j,2}^2)^2}{E^{*2}}\right),
\end{eqnarray}
\begin{figure}[t] 
\begin{center}
\includegraphics[scale=0.45]{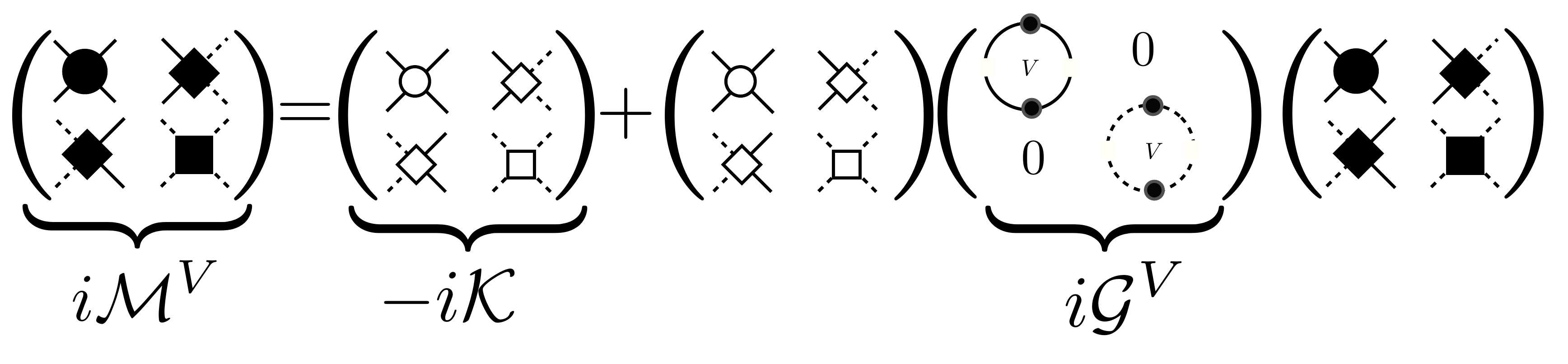}
\caption[Coupled-channel finite-volume two-particle propagators]{The fully-dressed finite volume two-particle propagator, $\mathcal{M}^V$ can be written in a self-consistent way in terms of the Bethe-Salpeter Kernel, $\mathcal{K}$ and the finite-volume function $\mathcal{G}^V$. Note that the only difference between these and their single-channel counterparts is that these are now matrices over the open channels.  }\label{twoparFV4}\end{center}
\end{figure} 
which simplifies to $\frac{E^{*2}}{4}-m_{j}^2$ when $m_{j,1}=m_{j,2}=m_{j}$. Because the $S$-matrix for the $l^{th}$ partial wave is a two-dimensional matrix, the scattering amplitude is also necessarily a two-dimensional matrix with matrix elements defined in Eq.~(\ref{scatrel2}). Similarly, the Kernel and the two-particle propagators get upgraded to matrices in the space of open channels. The full finite volume two-particle propagator is pictorially defined in a self-consistent way in Fig.~\ref{twoparFV4}. It important to note that the channels only mix by off-diagonal terms in the Kernel. That is to say that in the absence of interactions a two-pion state continues to propagate as a two-pion state. With this in mind, in the presence of multiple channels Eq.~(\ref{loop1}) is replaced by 
\begin{eqnarray}
\label{loop3} 
\left[iG^{V}(\mathbf{p}_i,\mathbf{p}_f)\right]_{ab}&\equiv&\frac{n_j}{L^3}\sum_{\mathbf{k}}\int\frac{dk^0}{2\pi}\frac{[\mathcal{K}(\mathbf{p}_i,\mathbf{k})]_{aj}~[\mathcal{K}(\mathbf{k},\mathbf{p}_f)]_{jb}}{[(k-P)^2-m_{j,1}^2+i\epsilon][k^2-m_{j,2}^2+i\epsilon]},\\
&\equiv&-i[\mathcal{K} \mathcal{G}^{V}\mathcal{K}]_{ab}
\end{eqnarray}
where the $\mathcal{G}^{V}$ is now not just an infinite-dimensional matrix in angular momentum but also a two-dimensional diagonal matrix in the number of channels. The subscripts $a, j, b$ denote the initial, intermediate and final states, respectively. The difference between $\mathcal{G}^{V}$  and its infinite volume counterpart has matrix elements defined by 
\begin{eqnarray}
\label{defi}
(\delta \mathcal{G}^{V}_{j})_{l_1,m_1;l_2,m_2}&=&i	\frac{q^*_jn_j}{8\pi E^*}\left(\delta_{l_1,l_2}\delta_{m_1,m_2}+i\frac{4\pi}{q_j^*}\sum_{l,m}\frac{\sqrt{4\pi}}{q_j^{*l}}c^{\textbf{P}}_{lm}(q_j^{*2})\int d\Omega^*Y^*_{l_1m_1}Y^*_{lm}Y_{l_2m_2}	\right),\nn\\
\end{eqnarray}
where $n_j=\frac{1}{2}$ if the particles in the $j^{th}$ loop are identical and $n_j=1$ otherwise. 

Having upgraded all the objects present to matrices in the open channels, one can go through the derivation of the poles of the finite volume propagator in the same fashion as was done in Eqs.~(\ref{FV_prop}-\ref{det0}) to find
\begin{eqnarray}
\label{detcc}
{\rm{Det}}(1+\delta \mathcal{G}^{V}\mathcal{M})={\rm{det}}_{\rm{oc}}\left[\rm{det}_{\rm{pw}}\left[1+\delta \mathcal{G}^{V}\mathcal{M}\right]\right]=0 ,
\end{eqnarray} 
where the determinant $\rm{det}_{\rm{oc}}$ is over the N open channels and the determinant $\rm{det}_{\rm{pw}}$ is over the partial waves. This result was independently derived by \cite{Briceno:2012yi, Hansen:2012tf}. For N=1 Eq.~(\ref{detcc}) reduces to Eq.~(\ref{det0}). 

For N=2, one obtains 
\begin{eqnarray}
\label{allorders}
{\rm{Det}}\begin{pmatrix} 
1+\delta \mathcal{G}^{V}_I\mathcal{M}_{I,I}&\delta \mathcal{G}^{V}_I\mathcal{M}_{I,II}\\
\delta \mathcal{G}^{V}_{II}\mathcal{M} _{I,II}&1+\delta \mathcal{G}^{V}_{II}\mathcal{M}_{II,II}\\
\end{pmatrix}=0.
\end{eqnarray} 
For $l_{max}=0$ it is convenient to introduce a pseudo-phase analogous to Eq. (\ref{pseudophase0})
\begin{eqnarray}
 \label{pseudophase}
{q^*}\cot({\phi}_j)\equiv -4\pi{ c_{00}^P(q^{*2}_j)}.
\end{eqnarray}
Using this, the quantization condition can be written in a manifestly real form,
 \begin{eqnarray}
\label{allorders2}
\cos{2\bar{\epsilon}}\cos{\left(\phi_1+\delta_1-\phi_2-\delta_2\right)}=\cos{\left(\phi_1+\delta_1+\phi_2+\delta_2\right)},
 \end{eqnarray} 
which is equivalent to the result given in Refs. \cite{coup02, coup2} in the CM frame \footnote{The equivalence between Eq. (\ref{allorders2}) and Eq. (37) of Ref. \cite{coup02} can be understood by noting that the pseudo-phase  $\phi_{i}$ as defined in Eq. (\ref{pseudophase}) is equivalent to the negative $\Delta_{i}$ as defined in Eq. (36) of Ref. \cite{coup02}. On the other hand, the mixing parameter $\overline{\epsilon}$ as defined in Eq. (\ref{smatrix2}) is related to the mixing parameter $\eta_{0}$ defined in Eq. (14) of Ref. \cite{coup02} through $\eta_{0}=\cos2\bar{\epsilon}$.}. It is easy to see that in the $\bar \epsilon\rightarrow 0$ limit, one recovers the decoupled quantization conditions for both channels $I$ and $II$, Eq. (\ref{pseudophase0}).

Consider the N=3 case. Unitarity and time-reversal invariance allow us to parametrize the S-matrix using three phases shifts $\{\delta_I, \delta_{II}, \delta_{II}\}$ and three mixing angles  $\{\bar\epsilon_1, \bar\epsilon_2, \bar\epsilon_3\}$ 
\begin{eqnarray}
\label{smatrix3}
S_3=\begin{pmatrix} 
e^{i2\delta_I}c_1
&ie^{i(\delta_I+\delta_{II})}s_1c_3
&ie^{i(\delta_I+\delta_{III})}s_1s_3\\
ie^{i(\delta_I+\delta_{II})}s_1c_2&
e^{i2\delta_{II}}\left(c_1c_2c_3-s_2s_3\right)
&
ie^{i(\delta_I+\delta_{III})}
\left(c_1c_2s_3+s_2c_3\right)\\
ie^{i(\delta_I+\delta_{III})}s_1s_2&
ie^{i(\delta_{II}+\delta_{III})}\left(c_1s_2c_3+c_2s_3\right)&
ie^{i2\delta_{III}}\left(c_1s_2s_3-c_2c_3\right)\\
\end{pmatrix},
\end{eqnarray}
where $c_i=\cos(2\bar{\epsilon}_i)$, $s_i=\sin(2\bar{\epsilon}_i)$. Note that in the limit ${\epsilon}_2={\epsilon}_3=0$ the third channel decouples, and a block diagonal matrix composed of $S_2$ is obtained, corresponding to the $I-II$ coupled channel, as well as a single element corresponding to the scattering in the uncoupled channel $III$. The spectrum of three-coupled channel is defined by
\begin{eqnarray}
\label{det3}
\det\begin{pmatrix} 
1+\delta \mathcal{G}^{V}_I\mathcal{M}_{I,I}&
\delta \mathcal{G}^{V}_I\mathcal{M}_{I,II}
&\delta \mathcal{G}^{V}_I\mathcal{M}_{I,III}
\\
\delta \mathcal{G}^{V}_{II}\mathcal{M}_{II,I}&1+\delta \mathcal{G}^{V}_{II}\mathcal{M}_{II,II}&
\delta \mathcal{G}^{V}_{II}\mathcal{M}_{II,III}
\\
\delta \mathcal{G}^{V}_{III}\mathcal{M}_{III,I}&
\delta \mathcal{G}^{V}_{III}\mathcal{M}_{III,II}
&1+\delta \mathcal{G}^{V}_{III}\mathcal{M}_{III,III}
\end{pmatrix}=0,
\end{eqnarray} 
where the scattering matrix elements can be  determined from Eq. (\ref{smatrix3}) using the relationship between the scattering amplitudes and the S-matrix elements, Eq.~(\ref{scatrel2}).

\subsubsection{Implication for the $\pi\pi-K\bar{K}$ spectrum}
\begin{figure}[t]
\begin{center}
\subfigure[]{
\label{connected2}
\includegraphics[scale=0.4]{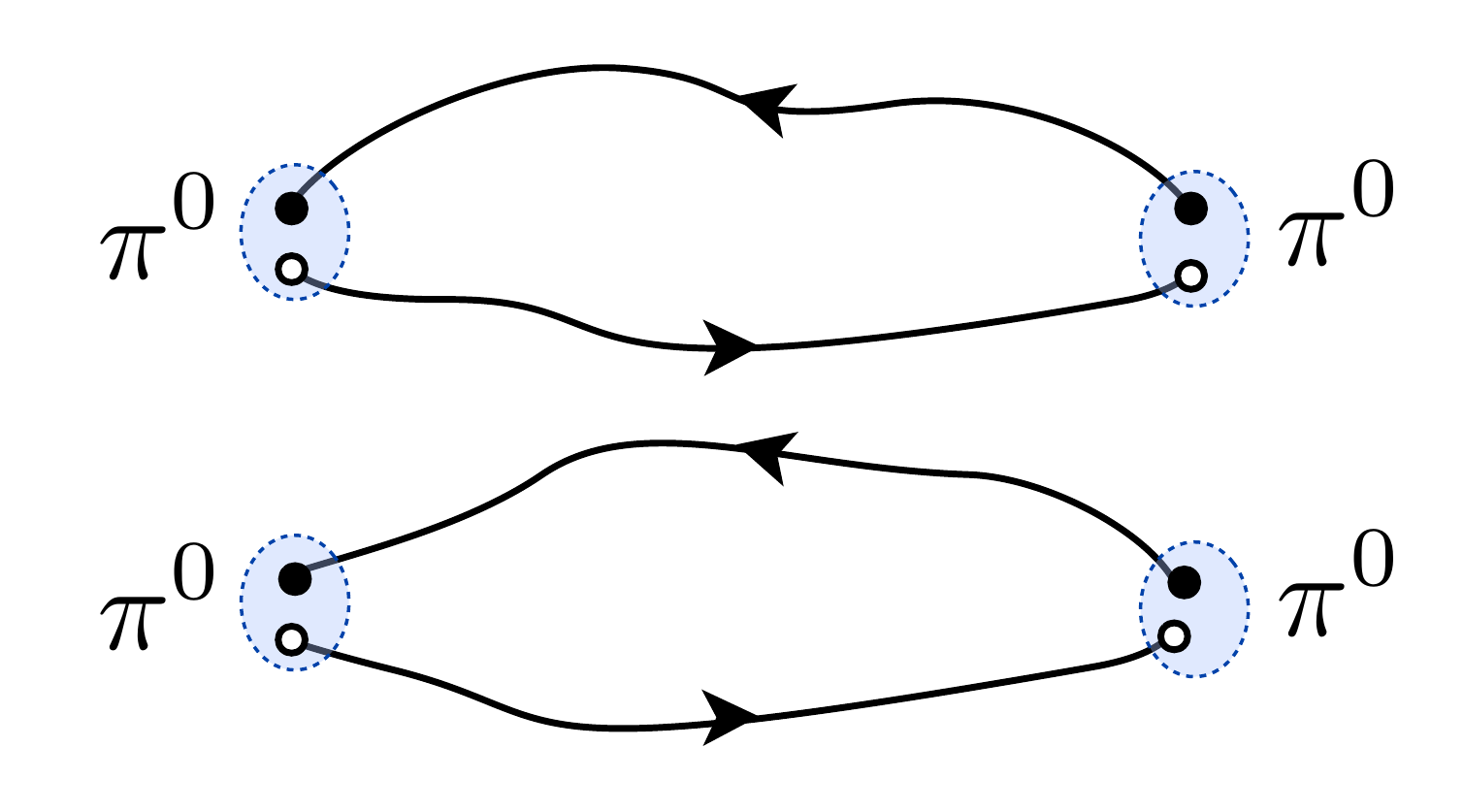}}
\subfigure[]{
\label{disconnected2}
\includegraphics[scale=0.4]{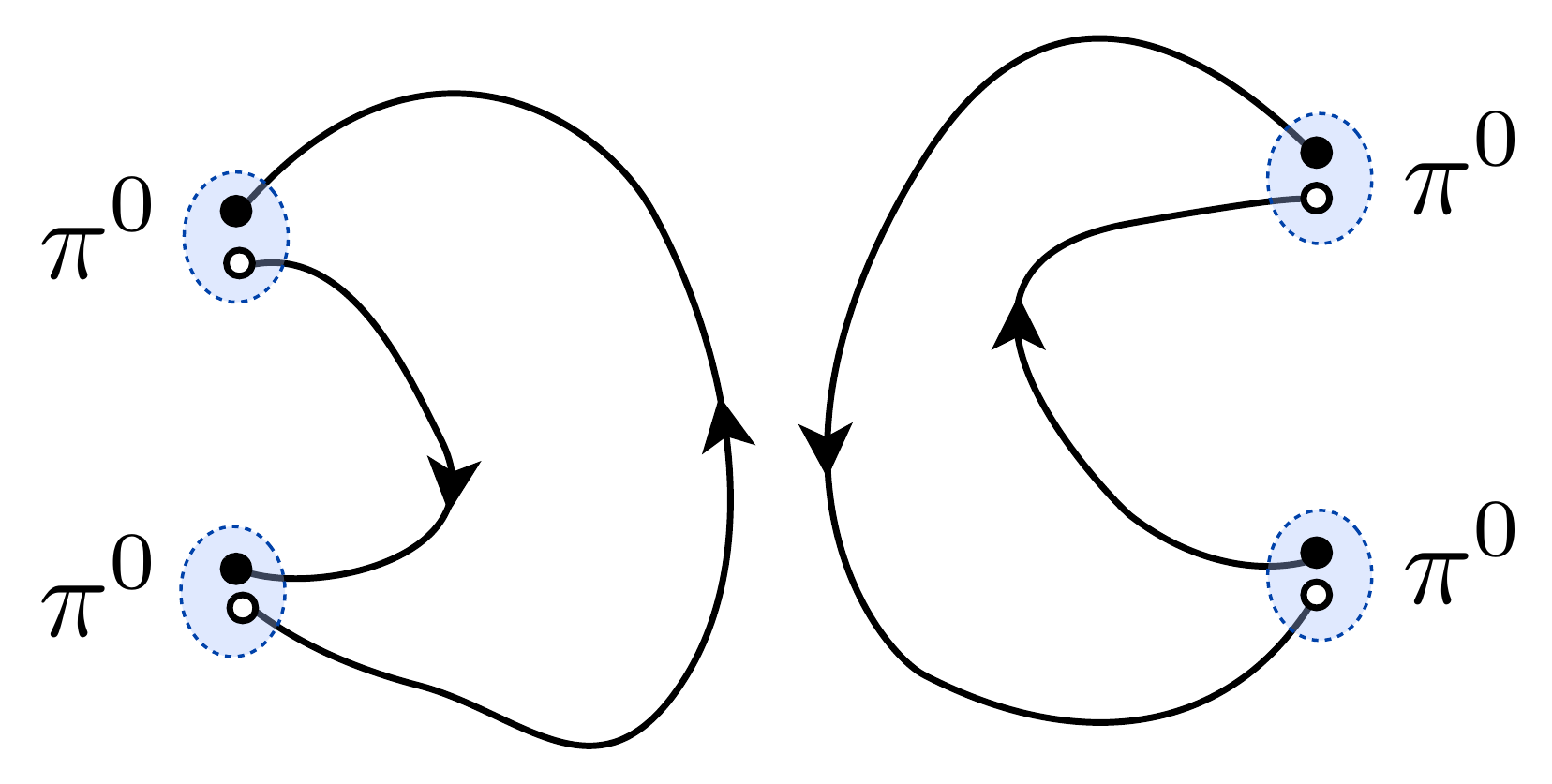}}
\caption[$\pi^0\pi^0$ correlation function]{a) Example of connected and b) disconnected diagrams that contribute to the $\pi^0\pi^0$ correlation function. }\label{pi0pi0}
\end{center}
\end{figure}

To this day the coupled-channel formalism has not be implemented in any LQCD calculations, so it is still preliminary to comment on the successes of this formalism. To understand challenges associated with calculations involving coupled-channels it suffices to consider the $\pi\pi-K\bar{K}$ isosinglet channel. For energies well below the $K\bar{K}$-threshold, the two kaon system cannot go on-shell, therefore up to exponentially small corrections this system can be treated as a single-channel system composed of two pions. A good interpolating operator for two pions in an isosinglet state, can be constructed as a linear combination of the operators appearing in Eq.~(\ref{pioninter})~\cite{Kuramashi:1993ka, Fukugita:1994ve}
\begin{eqnarray}
\label{pipiisosinglet}
\mathcal{O}(t_1,t_2)_{\pi\pi^{I=0}}=\frac{1}{\sqrt{3}}\left(\pi^{+}(t_1)\pi^{-}(t_2)+\pi^{+}(t_2)\pi^{-}(t_1)-\pi^{0}(t_1)\pi^{0}(t_2)\right).
\end{eqnarray}
The correlation function would have contributions from connected [e.g. Fig.~\ref{connected2}] and disconnected diagrams [e.g. Fig.~\ref{disconnected2}]. As discussed earlier, the latter pose a significant computational challenge, and as a result only two calculations of $a^{I=0}_{\pi\pi}$ using full QCD have been done to this day~\cite{Fu:2011bz, Fu:2013ffa}. This is in fact a common feature among most coupled-channel systems\footnote{In section~\ref{NNsys}, we will discuss coupled channels in the two-nucleon sector where disconnected diagrams are not a source of computational limitation}. 

We can utilize the prediction of the scattering amplitudes for the $\pi\pi-K\bar{K}$ system from LO SU(3)~$\chi PT$ to infer what the finite volume spectrum looks like. The nice feature of $\chi PT$ is that it allows us to dial the light-quark masses to unphysical values is a systematic fashion. For $m_\pi\approx 310$~MeV, the kaon has a mass of approximately $m_K\approx 527$~MeV~\cite{Briceno:2012wt}, where the strange quark mass is fixed at its physical value. Therefore for this set of parameters, the four-pion threshold is well above the two-kaon threshold and we can investigate the spectrum using Eq.~(\ref{allorders}) up to energies around 1240~MeV. 

By investigating the analytic structure of Eq.~(\ref{allorders}) one observes that the LO scattering amplitudes obtained using Eq.~(\ref{LO_chipt}) lead to a complex quantization condition. This is in fact not a issue with the quantization condition, but rather an issue regarding $\chi$PT as it does not reproduce the analytic structure of the fully relativistic scattering amplitude~\ref{scatrel}. The solution is to promote the LO scattering amplitude to the Bethe-Salpeter Kernel, $\mathcal{K}_{\chi PT}$, and resume the two-particle irreducible s-channel diagrams non-perturbatively. This formalism is refered to as a Unitarized $\chi$PT~(U$\chi$PT)~\cite{Oller:1997ng, Oller:1998hw}. For the $\pi\pi-K\bar{K}$, $\mathcal{K}_{\chi PT}$ can be written as~\cite{Oller:1997ng, Oller:1998hw}
\begin{eqnarray}
\label{chiPTkernel}
\mathcal{K}_{\chi PT}&=&
\begin{pmatrix} 
\mathcal{K}_{\pi\pi}&\mathcal{K}_{\pi K}\\
\mathcal{K}_{\pi K}&\mathcal{K}_{K\bar{K}}\\
\end{pmatrix},\hspace{1cm}
\mathcal{K}_{\pi\pi}=\frac{2E^{*2}-m_\pi^2}{8\pi E^{*}f_\pi^2}\\
\mathcal{K}_{\pi K}&=&\frac{\sqrt{3}E^{*2}}{16\pi E^{*}f_\pi^2},\hspace{2cm}
\mathcal{K}_{K\bar{K}}=\frac{3E^{*2}}{16\pi E^{*}f_\pi^2}.
\end{eqnarray} 
\begin{figure}[t]
\begin{center} 
\includegraphics[scale=0.30]{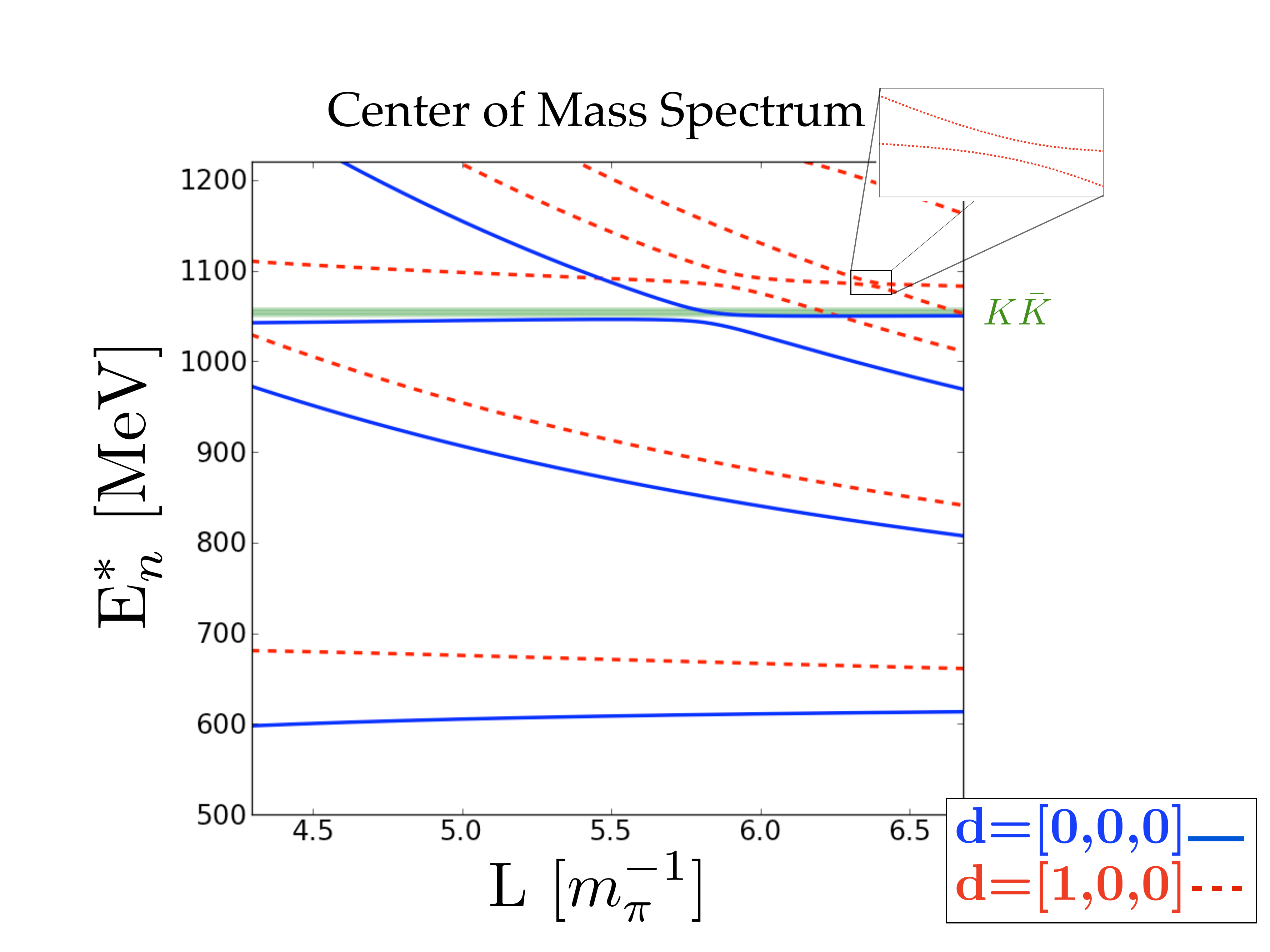}
\caption[Prediction for the $\pi\pi-K\bar{K}$ finite-volume spectrum for $m_\pi\approx 310$~MeV]{Shown is finite volume CM energy for $\pi\pi-K\bar{K}$ as predicted from the quantization condition, Eq.~(\ref{allorders}), using the input from U$\chi$PT~Eq.(\ref{chiPTkernel})~\cite{Oller:1997ng, Oller:1998hw}. The spectrum is generated for two-different total momentum $\textbf{P}$ and are labeled by $\textbf{d}=\textbf{P}L/2\pi$. Also shown is the $K\bar{K}$ threshold (green dashed line).}
\label{cc_spectrum}
\end{center}
\end{figure}
Using this expression for the Kernel and truncating $l_{max}$=0, one can then proceed to generated the $A^+_1$ spectrum of the $\pi\pi-K\bar{K}$ system as a function of volume, shown in Fig.~\ref{cc_spectrum}. The spectrum shown was generated for two different total momenta $\textbf{P}\equiv 2\pi \textbf{d}/L$, corresponding ${d}=0$ and ${d}=1$. Note that the spectrum is generated for $L\leq4/m_\pi$, which is the minimum criteria needed to be able to neglect contributions from exponential corrections to the quantization condition.  

There are a couple of things that can be observed. First, we see a generic feature of coupled-channel systems, which is the fact that energy levels do not cross. In particular, it would be tempting to identify the blue line that is just below the $K\bar{K}$ threshold for $m_\pi L<6$ as a two-kaon sate, but as the volume is chosen so that $m_\pi L\sim6$ it is clear that this state is an admixture of $K\bar{K}$ and $\pi\pi$. In fact, if the off-diagonal terms in Eq.(\ref{chiPTkernel}) were zero, these levels would indeed cross and one could clearly identify one as a two-kaon state and the other as a two-pion state. 

A second and more important observation is that for $m_\pi L\sim6-6.5$, the gap between the two states is  small. Consequently, low statistics calculations will not be able to resolve the spectrum near these points in the parameter space. Therefore to be able to reliably disentangle the spectrum, while simultaneously having FV artifacts well under control, it is preferable to design the future LQCD calculations to be performed in the $m_\pi L\sim4-6$ range. 

Finally, before it was stated that one needs to have three-independent measurements at the same CM energy to be able to extract the two phase shifts and the one mixing angle of the two-channel system. Although this is formally true, one could alternatively use the NLO expressions of U$\chi$PT~\cite{Oller:1997ng, Oller:1998hw} to parametrize the scattering amplitudes, perform calculations at different volumes, boost, and pion masses, and simultaneously fit all of the LQCD calculations to obtain the LECs appearing at NLO in U$\chi$PT.

\section{Auxiliary-Field Formalism for Arbitrary partial waves in the scalar sector \label{scalardimer}}

In section \ref{dimersec1}, we discussed the auxiliary-field formalism and its importance for studying three-body physics. This formalism has only been previously developed for S-wave \cite{pionless2, pionless3} and P-wave \cite{Braaten:2011vf} scattering processes, but as discussed in the previous sections orbital angular momentum is not a good quantum number of the finite volume system. Therefore to be able to study finite volume physics it is necessary to generalize this formalism for higher partial waves. 

\begin{figure}[t]
\begin{center}
\subfigure[]{
\label{DimerIV}
\includegraphics[scale=0.425]{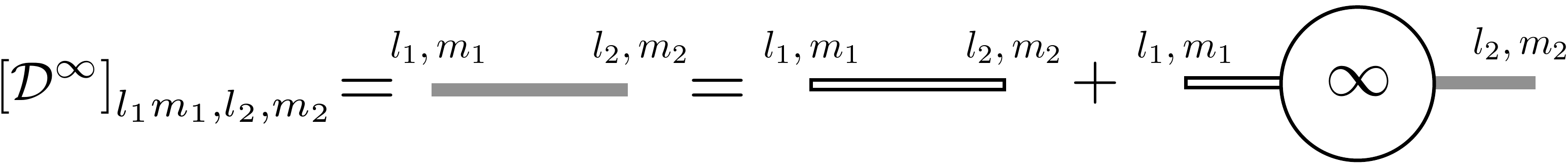}}
\subfigure[]{
\label{DimerFV}
\includegraphics[scale=0.425]{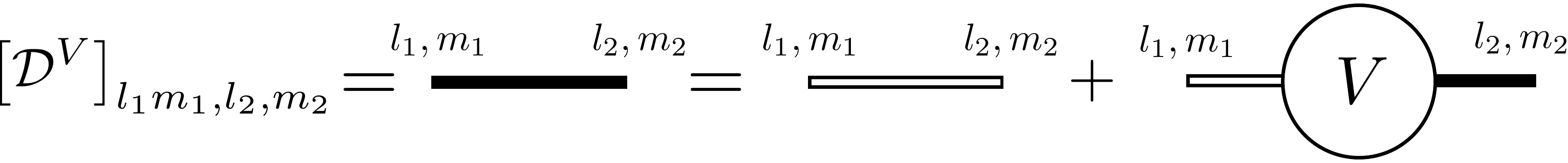}}
\caption[Scalar dimer for arbitrary partial waves]{a) Diagrammatic equation satisfied by the matrix elements of the full dimer propagator in a) infinite volume and b) finite volume. The grey (black) band represents the full infinite (finite) volume propagator, $\mathcal{D}^{\infty}$ ($\mathcal{D}^V$), while the double lines represent the bare propagator, $\mathcal{D}^B$.}\label{fig:dimer}
\end{center}
\end{figure}

Consider two identical bosons with mass $M$ that interact in an arbitrary partial-wave channel $(l,m)$ via a short-range interaction that can be effectively described by derivative couplings to the fields. Let $\phi_{k}$ and  $d_{lm,P}$ denote the interpolating operators that annihilate a boson with four-momentum $k$, and a dimer (with quantum numbers of two bosons) with four-momentum $P$ and angular momentum $(l,m)$, respectively. Then if the total energy and momentum of the CM of the two-boson system is $(E,\mathbf{P})$, one can write a Galilean-invariant action that describes such system in the infinite volume in terms of a Lagrange density in the momentum space,
{\small
\begin{eqnarray}
\label{action}
{S}^{\infty}&=&\int\frac{d^{4}P}{(2\pi)^{4}}\left[\phi_{P}^{\dagger}\left(E-\frac{\textbf{P}^{2}}{2M}\right)\phi_{P}-\sum_{l,m}d_{lm,P}^{\dagger}\left(E-\frac{\textbf{P}^{2}}{4M}-\Delta_{l}+\sum_{n=2}^{\infty}c_{n,l}\left(E-\frac{\textbf{P}^{2}}{4M}\right)^{n}\right)d_{lm,P}\right]
\nonumber\\
&~&\qquad \qquad \qquad  -\int\frac{d^{4}P}{(2\pi)^{4}}~\frac{d^{4}k}{(2\pi)^{4}}\sum_{l,m}~\frac{g_{2,l}}{2}\left[d_{lm,P}^{\dagger}~\sqrt{4\pi}~Y_{lm}(\hat{\textbf{k}}^{*})~|\mathbf{k}^{*}|^{l}\phi_{{k}}\phi_{P-{k}}+h.c.\right],
\end{eqnarray}}
where ${\textbf{k}}^*=\textbf{k}-\textbf{P}/2$ denotes the relative momentum of two bosons in the interaction term. Note that the interactions between bosons in partial-wave channel $(l,m)$ is mediated by a corresponding dimer field, $d_{lm}$. As is evident, upon integrating out such auxiliary-field, one recovers the four-boson interaction term in a Lagrangian with only $\phi$-field degrees of freedom. Since this is a theory of identical bosons, all couplings of the dimer field to a the two-boson state with an odd partial-wave vanish. Eq. (\ref{action}) clearly reduces to the S-wave result of Refs. \cite{pionless2, pionless3, Griesshammer:2004pe} discussed in section \ref{dimersec1}. For systems involving distinguishable scalar bosons this can be easily generalized (e.g. for P-wave scattering see Ref. \cite{Braaten:2011vf}). As usual, the low-energy coefficients (LECs) $\{\Delta_{l},c_{l,n}, g_{2,l}\}$ in the effective Lagrangian must be tuned to reproduce the ERE of the $l^{th}$-partial-wave, Eq.~(\ref{EFE}). The fully dressed dimer propagator can be obtained by summing up the self-energy bubble diagrams to all orders, Fig.~\ref{DimerIV},  
\begin{eqnarray}
\mathcal{D}^{\infty}(E,\mathbf{P})=\frac{1}{(\mathcal{D}^{B})^{-1}-I^{\infty}(E,\mathbf{P})},
\label{D-infinity}
\end{eqnarray}
where $\mathcal{D}^B$ denotes the bare dimer propagator, 
\begin{eqnarray}
\left[\mathcal{D}^{B}(E,\mathbf{P})\right]_{l_1m_1,l_2m_2}=\frac{-i~\delta_{l_1l_2}\delta_{m_1m_2}}{E-\frac{\mathbf{P}^2}{4M}-\Delta_{l}+\sum_{n=2}^{\infty}c_{n,l}(E-\frac{\textbf{P}^{2}}{4M})^{n}+i\epsilon},
\label{D-bare}
\end{eqnarray}
and $I^{\infty}$ denotes the value of the bubble diagram when evaluated using the power divergence subtraction scheme \cite{pds, pds2, Beane:2003da},
\begin{eqnarray}
\left[I^{\infty}(E,\mathbf{P})\right]_{l_1m_1,l_2,m_2}=\frac{iM}{8\pi}g_{2,l_1}^2k^{*2l_1}(\mu+ik^*)\delta_{l_1l_2}\delta_{m_1m_2}.
\label{I-infinity}
\end{eqnarray}
 $\mu$ is the renormalization scale. By requiring the full dimer propagator, $\mathcal{D}^{\infty}$, in the infinite volume to reproduce the full scattering amplitude at any given partial-wave,
\begin{eqnarray} 
\mathcal{M}^{\infty}_{l_1m_1,l_2m_2}&=&
\frac{8\pi}{M}~
\frac{1}{k^{*}\cot{\delta^{(l_1)}_d}-ik^{*}}\delta_{l_1l_2}\delta_{m_1m_2},
\end{eqnarray}
one arrives at
\begin{eqnarray}
g_{2,l}^2=\frac{16\pi}{M^2r_{l}} ~~\forall~~l=2n, ~~ \Delta_l=\frac{2}{Mr_l}\left(\frac{1}{a_l}-\mu k^{*2l}\right), ~~ c_{n,l}=\frac{2}{Mr_l}\frac{\rho_{n,l}M^n}{n!}.
\label{g2l}
\end{eqnarray}
%
%
%
%
%
%

In a finite volume, the two-boson system can still be described by the action in Eq. (\ref{action}) except the periodic boundary conditions constrain the momenta to be discretized. In particular, the integral over momenta in Eq. (\ref{action}) is replaced by a sum over discrete momenta, $P=\frac{2\pi}{L}\mathbf{n}$, where $\mathbf{n}$ is a triplet integer. Then it is straightforward to evaluate the corresponding bubble diagram in the finite volume,
\begin{eqnarray}
\left[I^{V}\right]_{l_1m_1,l_2,m_2}&=&\frac{iM}{8\pi}g_{2,l_1}g_{2,l_2}k^{*l_1+l_2}
\nonumber\\
&~&\times\left[\mu~\delta_{l_1l_2}\delta_{m_1m_2}+\sum_{l,m}\frac{(4\pi)^{3/2}}{k^{*l}}c^{\textbf{P}}_{lm}(k^{*2})\int d\Omega~Y^*_{l_1,m_1}Y^*_{l,m}Y_{l_2,m_2}\right],\nn\\
\label{I-V}
\end{eqnarray}
where $c^{\textbf{P}}_{lm}(x)$ is the non-relativistic limit of Eq.~(\ref{clm}), where $\gamma=1$ and $\alpha=1/2$ for two degenerate particles. The full dimer propagator, $\mathcal{D}^V$, can then be obtained by summing up the infinite series of bubble diagrams in Fig.~\ref{DimerIV}, where the LEC of the theory are matched with the the physical quantities, Eq. (\ref{g2l}),
\begin{eqnarray}
\mathcal{D}^{V}(E,\mathbf{P})=\frac{1}{(\mathcal{D}^{B})^{-1}-(\mathcal{D}^{B})^{-1}I^{V}(E,\mathbf{P})\mathcal{D}^{B}},
\label{D-FV}
\end{eqnarray}
Note that $\mathcal{D}^V$ is a matrix in the angular momentum space as are $\mathcal{D}^B$ and $I^V$, Eqs. (\ref{D-bare}, \ref{I-V}). The poles of the FV dimer propagator give the spectrum of two-boson system in a finite volume in terms of the scattering parameters. These energy eigenvalues satisfy the following determinant condition
\begin{eqnarray}
\det \left[k^*\cot \delta+\mathcal{F}^{FV}\right]=0,
\label{FullQCboson}
\end{eqnarray}
where both $\cot \delta$ and $\mathcal{F}^{FV}$ are matrices in the angular momentum space,
\begin{eqnarray}
\cot \delta \equiv \cot \delta_{l_1}\delta_{l_1l_2}\delta_{m_1m_2},
\label{cot}
\end{eqnarray}
\begin{eqnarray}
\left[\mathcal{F}^{FV}\right]_{l_1m_1,l_2m_2}=-\sum_{l,m}\frac{(4\pi)^{3/2}}{k^{*l}}c^{\textbf{P}}_{lm}(k^{*2})\int d\Omega~Y^*_{l_1,m_1}Y^*_{l,m}Y_{l_2,m_2}.
\label{F}
\end{eqnarray}

This quantization condition agrees with the non-relativistic limit of Eq.~(\ref{det0})~\cite{movingframe, sharpe1, Christ:2005gi}. This derivation shows that upon incorporating the higher partial-waves in the construction of the dimer Lagrangian, as well as accounting for higher order terms in the EFR expansion, all the two-body physics is fully encapsulated in this formalism. As a result the systematic errors of FV multi-particle calculations that have used an S-wave dimer field up to next-to-leading order in ERE (see Refs. \cite{Briceno:2012rv, Kreuzer:2008bi, Kreuzer:2009jp, Kreuzer:2010ti, Kreuzer:2012sr}), can be easily avoided.  

\section{Electroweak mixing in the two-body scalar sector \label{EW2B}}

Electroweak processes in the two-hadron sector of QCD encompass a variety of interesting processes, so it would be desirable to calculate the electroweak matrix elements directly from LQCD. One of the very first attempts to develop a formalism for such processes from a finite volume Euclidean calculation is due to Lellouch and L\"uscher. In  their seminal work \cite{LL1}, they restricted their analysis to $K\rightarrow \pi\pi$ decay in the kaon's rest frame, and showed that the absolute value of the transition matrix element in an Euclidian FV is proportional to the physical transition matrix element. This proportionality factor is known as the LL-factor. This formalism was then generalized to moving frames in Refs. \cite{sharpe1, movingframe2}. Here we present the generalization of Lellouch and L\"uscher formalism to processes where the initial and final states are composed of two-hadrons S-wave states and are coupled solely by two-body currents\footnote{In section~\ref{ppfusion} we will discuss the FV corrections to the axial-vector current in the NN-sector. This is particularly important for studying proton-proton fusion ($pp\rightarrow d e^+\nu_e$) directly from LQCD. There we find that one-body currents lead to large FV corrections to the formalism discussed here.}. In the relativistic case, the coupled-channel result, Eq. (\ref{allorders}), is used to derived the $2\rightarrow 2$ LL-factor for boosted systems.   

In order to derive the relativistic two-body LL-factor, one first notes that in the absence of the weak interaction, the two states decouple, and as a result the $S$-matrix becomes diagonal. As is pointed out by Lellouch and L\"uscher, there is a simple trick to obtain the desired relation by assuming the initial and final states to be nearly degenerate with energy $E^*_0$ (each satisfying Eq. (\ref{pseudophase0})) when there is no weak interaction. Once the perturbative weak interaction is turned on, the degeneracy is lifted, and the energy eigenvalues are
\begin{eqnarray}
\label{deg}
E^*_{\pm}=E^*_0\pm V|\mathcal{M}^V_{I,II}|\equiv E^*_0\pm\Delta E^*, 
\end{eqnarray}
where $\mathcal{M}^V_{I,II}$ is the FV matrix element of the weak Hamiltonian density. 
Consequently, the CM momenta and the scattering phase shifts acquire perturbative corrections of the form
\begin{eqnarray}
\label{pseudomom}
\Delta q^*_i=\frac{1}{4q_i^*}\left(E^*_0-\frac{(m_{j,1}^2-m_{j,2}^2)^2}{ E_0^{*3}}\right)V|\mathcal{M}^V_{I,II}|\equiv \Delta \tilde{q}^*_i \hspace{.1cm}V|\mathcal{M}^V_{I,II}|,
\end{eqnarray}
and
\begin{eqnarray}
\label{pseudomom2}
\Delta \delta_i({q}^*_i )=\delta'_i({q}^*_i ) \Delta \tilde{q}^*_i \hspace{.1cm}V|\mathcal{M}^V_{I,II}|,
\end{eqnarray}
where $\delta'_i({q}^*_i )$ denotes the derivative of the phase shift with respect to the momentum evaluated at the free CM momentum. The perturbed energy necessarily satisfies the quantization condition, Eq. (\ref{allorders}). The generalized LL-factor for $2\rightarrow 2$ scattering is then obtained by Taylor expanding Eq. (\ref{allorders}) to leading order in the weak matrix element about the free energy solution,{\small 
\begin{eqnarray}
\frac{|\mathcal{M}^\infty_{I,II}|^2}{|\mathcal{M}^V_{I,II}|^2}
=V^2 \left\{\Delta \tilde{q}^*_I \Delta \tilde{q}^*_{II}\left(\frac{8 \pi E^*_0}{n_Iq^*_I}\right)\left(\frac{8\pi E^*_{0}}{n_{II}q^*_{II}}\right)
\left(\phi_{I}'(q^*_{I})+\delta_{I}'(q^*_{I})\right)\left(\phi_{II}'(q^*_{II})+\delta_{II}'(q^*_{II})\right)\right\},
\label{mesonLL}
\end{eqnarray}}
where $\phi'_i({q}^*_i )$ denotes the derivative of the pseudo-phase  with respect to the momentum evaluated at the free CM momentum.

Note that we arrived at the generalization of the LL factor for two-body matrix elements using the degeneracy of states argument.  
Lin $et\; al.$ \cite{LL2} showed that  the LL-factor for $K\rightarrow \pi\pi$ can also be derived using the density of states in the large volume limit, and this argument was then generalized by Kim $et\; al.$ \cite{sharpe1} to boosted systems. Here it will be shown that the results in Eq. (\ref{mesonLL}) is also consistent with the work by Kim $et\; al.$ Let $\sigma_i\left(\mathbf{x},t\right)$ be the two-particle annihilation operator for the $i^{th}$ channel. Then the two particle correlation function in FV can be written as
\begin{eqnarray}
\label{fincor}
C_{\mathbf{P},i}^{V}\left(t\right)&\equiv&\int d^3x\hspace{.1cm}e^{i\mathbf{P}\cdot \mathbf{x}}\left\langle 0\right|\sigma_i\left(\mathbf{x},t\right)\sigma^\dag_i\left(\mathbf{0},0\right)\left|0\right\rangle_V= V\sum_m   e^{-E_mt}\left|\left\langle 0\right|\sigma\left(\mathbf{0},0\right)\left|i;\mathbf{P},m\right\rangle_V \right|^{2}\nn\\
&\stackrel{L\rightarrow \infty }\longrightarrow&
V \int dE\rho_{V,i}(E)e^{-Et}\left|\left\langle 0\right|\sigma\left(\mathbf{0},0\right)\left|i;\mathbf{P},E\right\rangle_V \right|^{2}.
\end{eqnarray}
In the first equality we have inserted a complete set of states. In the second equality, we have introduced the density of states for the $i^{th}$ channel, $\rho_{V,i}(E)$, which is defined as $\rho_{V,i}(E)=dm_i/dE$. Using Eqs. (\ref{pseudophase0}), (\ref{pseudomom}) the density of states can be written as $\rho_{V,i}(E^*)=\left(\phi_{i}'(q^*_{i})+\delta_{i}'(q^*_{i})\right)\Delta\tilde{q}^*_{i}/\pi$. In the infinite volume the two-particle correlation function is \cite{LL2}
\begin{equation}
\label{inftycor}
C_{\mathbf{P},i}^{\infty}\left(t\right)=\frac{n_i}{8\pi^{2}}\int dE\frac{q^{*}_i}{E^{*}}e^{-Et}\left|\left\langle 0\right|\sigma\left(\mathbf{0},0\right)\left|i;\mathbf{P},E\right\rangle_\infty \right|^{2}, 
\end{equation}
where the factor of $n_i$ has been introduced to account for the double counting of the phase space when the particles are identical. It is straightforward to show that this relation still holds when the two particles have different masses. From Eqs. (\ref{fincor}), (\ref{inftycor}) the relationship between the states of infinite and asymptotically large (yet finite) volume can be deduced, 
\begin{equation}
\label{FVstates}
\left|i;\mathbf{P},E\right\rangle_\infty\Leftrightarrow
2\pi\sqrt{\frac{2V\rho_{V,i}E_0^*}{n_iq_i^*}}
\left|i;\mathbf{P},E\right\rangle_V.
\end{equation}
This relation therefore recovers the LL-factor as given in Eq. (\ref{mesonLL}). It also demonstrates that the LL-factor accounts for different normalizations of the states in the finite volume and infinite volume in presence of interaction.

\chapter{Nucleon-Nucleon Systems in a Finite Volume}{\label{NNsys}}

 Being able to make reliable predictions for few-body and many-body nuclear systems requires truly \textit{ab initio} methods with quantifiable uncertainties is an important goal of nuclear physics. In the two-body sector empirical models are sufficiently precise to provide a reliable estimation of the two-body nuclear force and accurately reproduce experiments. However, as discussed in previous chapters they do not provide an analytic handle of the nuclear force. Furthermore, they do not give much insight into the nature of such systems at extreme energies and densities where experiments are not available or when more exotic nuclear systems involving hyper-nucleons -- such as those in astrophysical environments -- become relevant. The nature of the nuclear forces in connection to the parameters of the Standard Model of particle physics is unknown and further insights into this problem require first-principle calculations that use these fundamental parameters as input. Currently the only rigorous method with which one can study nuclear systems directly from the fundamental theory of strong interactions is LQCD. Although both analytically and computationally demanding, this approach has been successfully implemented for studying nuclear systems in recent years. With constant developments in formalism and algorithms, as well as ever-increasing computational resources, the precision needed for these calculations will be within reach in the upcoming years.

As has been discussed extensively through this work, Lattice QCD (LQCD) calculations are performed in a finite, discretized Euclidean spacetime volume. L\"uscher presented a master formula, Eq.~(\ref{det0}), for the scattering phase shifts of two \textit{scalar} particles in \textit{arbitrary} partial-waves in connection to the FV energy levels of the two-meson system \cite{luscher1, luscher2}. Although this master formula is self-contained and incorporates all the necessary details to be implemented in practice, deducing the relations among phase shifts in different partial-waves and the energy levels of a specific LQCD calculation requires multiple non-trivial steps. The corresponding procedure is sometimes called the reduction of the L\"uscher formula. The difficulty associated with this procedure is due to the fact that the LQCD energy eigenvalues is determined according to the irreps of the cubic group in the CM frame. Since the phase shifts are characterized according to the irreps of the SO(3) rotational group, the energy eigenvalues of the system in a given irrep of the cubic group  in general depend on the phase shifts of more than one partial-wave channel. Performing  LQCD calculations of energy levels in different irreps of the cubic group would provide multiple QCs depending on different linear combinations of the scattering phase shifts, leading to better constraints on these quantities. Therefore it is necessary to identify all the QCs satisfied by a given scattering parameter in a partial-wave channel. While L\"uscher's original work presents the reduction of the master formula to a QC for the cubic $A_1$ irrep, Ref. \cite{Luu:2011ep} provides the full quantization conditions for the energy eigenvalues of different irreps of the cubic group, in both positive and negative parity sectors for orbital angular momentum $l\leq6$ as well as $l=9$ in the scalar sector. For scattering involving a spin-$\frac{1}{2}$ particle and a scalar particle, the L\"uscher formula can be generalized such that the energy eigenvalues of the meson-baryon system in a given irrep of the double-cover of the cubic group is related to the corresponding phase shifts \cite{Bernard:2008ax}. Here we discuss and present the generalization of this formalism for nucleon-nucleon systems\footnote{The L\"uscher formula to study two-nucleon systems were first presented in Ref. \cite{Beane:2003da}, although due to constraining the calculation to S-wave scattering, the complexity of the two-nucleon systems has not been dealt with. The only previous attempt to address this problem, including the spin, isospin and angular momentum degrees of freedom, is the work by N. Ishizuka \cite{Ishizuka:2009bx}, where the quantization conditions for energy eigenvalues of a two-nucleon system at rest in the positive and negative parity isosinglet channels were obtained for $J\leq 4$.}, where due to the the possibility of physical mixing among different partial-wave channels, more complexities arise. This is an important problem as it provides the formalism needed for a first-principle extraction of the S-D mixing parameter in the deuteron channel, and will eventually shed light on the nature of the tensor interaction in nuclear physics.\footnote{For a different approach in studying tensor nuclear force using LQCD calculations, see Ref. \cite{Aoki:2009ji, Murano:2013xxa}. These calculations rely on the assumption that the potential density $U_E(\textbf{x},\textbf{y})$, which is the Fourier transform of the Bethe-Salpeter kernel, is both energy-independent and ``approximately local". As it clearly stated in L\"uscher seminal work \cite{luscher1, luscher2} and the work of the CP-PACS Collaboration~\cite{Aoki:2005uf} this is not the case and making these assumptions leads to model-dependent potentials whose systematic errors cannot be properly quantified. } The master formula presented here, Eq.~(\ref{NNQC}), is valid below the pion production threshold for all spin and isospin channels in both positive and negative parity sectors and is derived using the generalization of the auxiliary-field formalism presented in section~\ref{scalardimer} for nucleon-nucleon interactions.  

Performing LQCD for systems with different CM momenta gives access to a different spectrum at a given volume and provides additional QCs for the energy eigenvalues of the system in terms of scattering parameters. Boosting the two-particle system however reduces the symmetry of the problem further and introduces more FV-induced mixings among different partial-waves \cite{Moore:2005dw, Luu:2009}. By investigating the symmetry group of the boosted systems along one and two Cartesian axes as well as that of the unboosted system, we have identify all the QCs satisfied by the phase shifts and mixing parameters in channels with total angular momentum $J\leq4$; ignoring scattering in partial-wave channels with $l\geq4$. Different QCs correspond to different irreps of the cubic ($O$), tetragonal ($D_{4}$) and orthorhombic ($D_{2}$) point groups that represent the symmetry group of systems with CM momentum $\mathbf{P}=0$, $\mathbf{P}=\frac{2\pi}{L}(0,0,1)$ and $\mathbf{P}=\frac{2\pi}{L}(1,1,0)$ respectively, where $L$ denotes the spatial extent of the cubic volume. As will be discussed later, these QCs can be also utilized for boost vectors of the form $\frac{2\pi}{L}(2n_1,2n_2,2n_3)$, $\frac{2\pi}{L}(2n_1,2n_2,2n_3+1)$ and $\frac{2\pi}{L}(2n_1+1,2n_2+1,2n_3)$ and all cubic rotations of these vectors where $n_1,n_2,n_3$ are integers.  Although the master formula presented in this chapter at zero CM momentum has been already derived in Ref. \cite{Ishizuka:2009bx} for nucleon-nucleon systems using a relativistic quantum field theory approach, the full classifications of different QCs for all the spin and isospin channels and for two non-zero CM momenta are  presented and tabulated for the first time in the following sections. These relations make the implementation of the generalized L\"uscher formula for nucleon-nucleon systems straightforward for future LQCD calculations of the nucleon-nucleon (NN) system.  

It is important to reiterate that despite the tight empirical constraints on the two-body nuclear force, the investigation of the two-nucleon sector within LQCD is still warranted.  As was discussed in section~\ref{EW2B}, understanding the energy-dependence of the scattering phase-shifts of the two-body hadronic states \cite{nnd, Meyer:2012wk, Briceno:2012yi, Bernard:2012bi, Meyer:2013dxa}, for example, is essential to obtaining physical matrix elements of current operators in the two-body sector. In section~\ref{ppfusion} we will see that in order to study proton-proton fusion directly from LQCD we need to first obtain the axial charge and derivative of the phase shifts at the given value of the pion mass where calculations are performed. Similarly, the study of multi-nucleon systems from LQCD also requires not only the knowledge of two-nucleon phase shifts,  but the $m_\pi$-dependence of the two-nucleon phase shifts \cite{Briceno:2012rv, Kreuzer:2008bi, Kreuzer:2009jp, Kreuzer:2010ti, Kreuzer:2012sr}. In chapter~\ref{mmmsys} we will see this fact explicitly for the scalar analogue of three-nucleon systems.  
 
 \section{Auxiliary-Field Formalism for Arbitrary partial waves in the nuclear sector \label{nucleardimer}}

In section~\ref{scalardimer} we observed that both infinite- and finite-volume physics in the scalar sector can be well described by introducing a dimer that couples to two bosons in an arbitrary partial wave. Due to spin and isospin degrees of freedom, the two-nucleon system exhibits some specific features. In particular, the anti-symmetry of the two-nucleon state constrains the allowed spin and isospin channels for a given parity state. Additionally, any spin-triplet two-nucleon state is an admixture of two different orbital-angular momentum states. For example, as is well-known, the two-nucleon state in the deuteron channel with $J^{P}=1^{+}$ is an admixture of S-wave and D-wave states. In general, a positive parity two-nucleon state with total angular momentum $J$ is a linear combination of states with\footnote{The $L$ that is introduced here and elsewhere as the partial-wave label of quantities should not be confused with the spatial extent of the lattice $L$ that appears in the definition of the $c_{lm}^{\mathbf{P}}$ functions.}
\begin{eqnarray}
\left(L=J\pm\frac{1}{2}(1-(-1)^J), S=\frac{1}{2}(1-(-1)^J)\right),
\label{positive}
\end{eqnarray}
while in the negative parity sector, the states that are being mixed have\footnote{Note, however, that for a $J$-even state in the first case and a $J$-odd state in the second case, there is only one angular momentum state allowed and no mixing occurs.}
\begin{eqnarray}
\left(L=J\pm\frac{1}{2}(1+(-1)^J), S=\frac{1}{2}(1+(-1)^J)\right).
\label{negative}
\end{eqnarray}
Table (\ref{JP}) shows the allowed spin and angular momentum of NN states in both isosinglet and isotriplet channels with $J\leq4$.

In order to write the most general Lagrangian describing nucleon-nucleon scattering in all spin, isospin and angular momentum channels, let us introduce an operator that creates an NN-state with total four-momentum $P$ and the relative momentum ${\textbf{k}}^*={\textbf{k}}-\frac{{\textbf{P}}}{2}$ in an arbitrary partial-wave $(L,M_L)$ in the following way
\begin{eqnarray}
|NN;P,k^*\rangle _{LM_L,SM_S,IM_I}=\mathcal{N}_L\int d\Omega_{\textbf{k}^*}~Y^*_{LM_L}(\hat{\textbf{k}}^{*})k^{*L}\left[N^T_{P-k}~\hat{\mathcal{P}}_{(SM_S,IM_I)}~N_k\right]^\dag|0\rangle,
\end{eqnarray}
where $k^*=\left|\mathbf{k}^*\right|$. $\hat{\mathcal{P}}_{(SM_S,IM_I)}$ is an operator which projects onto a two-nucleon state with spin $(S,M_S)$ and isospin $(I,M_I)$, and $\mathcal{N}_L$ is a normalization factor. By requiring such state to have a non-zero norm, and given the anti-commutating nature of nucleon fields, one can infer that for positive parity states the operator $\hat{\mathcal{P}}_{(SM_S,IM_I)}$ must be necessarily antisymmetric, while for negative parity states it must be symmetric. Since this operator is a direct product of two projection operators in the space of spin and isospin, these requirements can be fulfilled by constructing the corresponding operators using the appropriate combinations of Pauli matrices, $\sigma_j$ ($\tau _j$), that act on the spin (isospin) components of the nucleon field. To proceed with such construction, let us define the following operators
\begin{eqnarray}
\alpha_j^I=\tau_y\tau_j,\hspace{1cm}
\alpha_j^S=\sigma_y\sigma_j,\hspace{1cm}
\beta^I=\tau_y,\hspace{1cm}
\beta^S=\sigma_y. 
\end{eqnarray}
Note that the matrices that are named as $\alpha$ are symmetric while those that are named as $\beta$ are antisymmetric. Superscript $I$ ($S$) implies that the operator is acting on the spin (isospin) space, and index $j=1,2,3$ stands for the Cartesian components of the operators. Alternatively one can form linear combinations of $\alpha^S_j$ ($\alpha^I_j$) that transform as a rank one spherical tensor.\footnote{A Cartesian vector $\mathbf{r}$ can be brought into a spherical vector according to 
\begin{eqnarray}
\label{spherical}
r^{(0)}\equiv r_{z},\hspace{1cm}
r^{(\pm 1)}\equiv\mp \frac{\left(r_{x} \pm ir_{y}\right)}{\sqrt{2}}.
\nonumber
\end{eqnarray}
} Using these matrices, it is straightforward to see that an antisymmetric $\hat{\mathcal{P}}_{(SM_S,IM_I)}$ can have one of the following forms 
\begin{eqnarray}
{\hat{\mathcal{P}}}_{(00,1M_I)} \equiv \frac{\alpha^{(M_I)}_I\otimes \beta_S}{\sqrt{8}},\hspace{.25cm}
{\hat{\mathcal{P}}}_{(1M_S,00)} \equiv \frac{\beta_I\otimes \alpha^{(M_s)}_S}{\sqrt{8}},
\end{eqnarray}
which can project onto two-nucleon states with $\left(S=0,I=1\right)$ and $(S=1,I=0)$ respectively. Note that these are the conventional isotriplet and isosinglet projection operators in the positive parity sector that are used frequently in literature \cite{pds, pds2, Savage:1998ae, Chen:1999tn}.
On the other hand, a symmetric $\hat{\mathcal{P}}_{(SM_S,IM_I)}$ can project onto two-nucleon states with $(S=0,I=0)$ and $(S=1,I=1)$ and should have one of the following forms,
\begin{eqnarray}
{\hat{\mathcal{P}}}_{(00,00)} \equiv \frac{\beta_I\otimes \beta_S}{\sqrt{8}},\hspace{.25cm}
{\hat{\mathcal{P}}}_{(1M_S,1M_I)} \equiv \frac{\alpha^{(M_I)}_I\otimes \alpha^{(M_S)}_S}{\sqrt{8}},
\end{eqnarray}
respectively.

As it is the total angular momentum $J$ that is conserved in a two-nucleon scattering process, as opposed to the orbital angular momentum $L$, it is convenient to project a two-nucleon state in the $|LM_L,SM_S\rangle$ basis into a state in the $|JM_J,LS\rangle$ basis using the Clebsch-Gordan coefficients,
\begin{eqnarray} 
|NN;P,k^*\rangle _{JM_J,LS,IM_I}&=&\sum_{M_L,M_S}\langle JM_J|LM_L,SM_S\rangle~ |NN;P,k^*\rangle _{LM_L,SM_S,IM_I}.
\label{NNstate}
\end{eqnarray} 

Note that isospin remains a conserved quantum number up to small isospin breaking effects that we ignore for the nucleon systems.
In order to describe nucleon-nucleon interactions, we introduce an auxiliary dimer filed, similar to the scalar theory.\footnote{The S-wave dimer field in the nuclear sector is commonly referred to as a di-baryon field.} This field, that will be labeled $d^{LS}_{JM_J,IM_I;P}$, has the quantum numbers of two-nucleon states with total angular momentum $(J,M_J)$ and isospin quantum number of ${(I,M_I)}$ with orbital angular momentum $L$ and spin $S$. Now the action corresponding to the Lagrangian density of free nucleon and dimer fields in the momentum space can be written as
{\small
\begin{eqnarray} 
S^{\infty}_{kin}&=&\int\frac{d^4P}{(2\pi)^4}\left[N^\dag_P(E-\frac{\textbf{P}^2}{2m})N_P \right .
\nonumber\\
&~& ~~~ \left . -\sum_{\substack{J,M_J, I,M_I}}\sum_{L,S}~\left(d^{LS}_{JM_J,IM_I;P}\right)^\dag\left(E-\frac{\textbf{P}^2}{4m}-\Delta^{LS}_{JI}+\sum_{n=2}^{\infty} c^{LS}_{JI,n}(E-\frac{\textbf{P}^2}{4m})^{n}\right)d^{LS}_{JM_J,IM_I;P}\right].
\nonumber\\
\label{Skin}
\end{eqnarray}} 

In order to write the interaction Lagrangian, one should note that, while the total angular momentum, parity, isospin and spin are conserved in a strongly interacting nucleon-nucleon process, the orbital angular momentum can change due to the action of tensor forces in nuclear physics. This is easy to implement in this formalism, as the two-nucleon states that are formed, Eq. (\ref{NNstate}), are compatible with the symmetries of the two-nucleon states. The interacting part of the action that does not mix angular momentum states, ${S}^{\infty}_{int,1}$, can then be written as
{\small
\begin{eqnarray} 
{S}^{\infty}_{int,1}&=&-\int\frac{d^4P}{(2\pi)^4}~\frac{d^4k}{(2\pi)^4}
\sum_{\substack{J,M_J, I,M_I}} ~ \sum_{L,M_L,S,M_S}
~{g^{LS}_{JI}}~
\langle JM_J|LM_L,SM_S\rangle
\nonumber\\
&~& \qquad \qquad \qquad \qquad \times \left[\left(d^{LS}_{JM_J,IM_I;P}\right)^\dag~\sqrt{4\pi}~Y_{LM_L}(\hat{\textbf{k}}^*)~{k}^{*L}~N^T_{k}~\hat{\mathcal{P}}_{(SM_S,IM_I)}~N_{P-k}+h.c. \right],
\nonumber\\
\label{S1}
\end{eqnarray} }
where ${g^{LS}_{JI}}$ denotes the coupling of a dimer field to the two-nucleon state with quantum numbers $\{J,I,L,S\}$. Note that the interactions must be azimuthally symmetric and so the reason the couplings are independent of azimuthal quantum numbers. Eqs. (\ref{positive}, \ref{negative}) now guide us to write the most general form of the interacting part of the action that is not diagonal in the angular momentum space, ${S}^{\infty}_{int,2}$, as follows
{\small
\begin{eqnarray} 
{S}^{\infty}_{int,2}&=&-\int\frac{d^4P}{(2\pi)^4}\frac{d^4k}{(2\pi)^4}
\sum_{\substack{J,M_J, I,M_I}} ~ \sum_{L,M_L,L',M_L',S,M_S}{h_{JI}} ~ \delta_{I,\frac{1+(-1)^J}{2}}\delta_{S,1}(\delta_{L,J+1}\delta_{L',J-1}+\delta_{L,J-1}\delta_{L',J+1})
\nonumber\\
&~& ~ \times
\langle JM_J|L'M_L',SM_S\rangle~\left[\left(d^{LS}_{JM_J,IM_I;P}\right)^\dag~\sqrt{4\pi}~Y_{L'M_{L}'}(\hat{\textbf{k}}^*)~{k}^{*{L'}}~N^T_{k}~\hat{\mathcal{P}}_{(SM_S,IM_I)}~N_{P-{k}}+h.c.\right].
\nonumber\\
\label{S2}
\end{eqnarray} }
Note that in this interacting term, spin, isospin and the initial and final angular momenta are all fixed for any given total angular momentum $J$. As a result we have only specified the $(JI)$ quantum numbers corresponding to coupling $h$. As in the scalar case, all the LECs of this effective Lagrangian, $\{\Delta^{LS}_{JI},c^{LS}_{JI,n}, g^{LS}_{JI}, h_{JI}\}$, can be tuned to reproduce the low-energy expansion of the scattering amplitudes in the $J^{th}$ angular momentum channel with a given spin and isospin. As discussed in Sec. \ref{scalardimer}), in the scalar sector the LECs can be easily determined in terms of the ERE parameters and the renormalization scale. For coupled-channel systems, obtaining the LECs in terms of the scattering parameters requires solving a set of coupled equations. The tuning of the LECs is only an intermediate step in obtaining the relationship between the FV spectrum and the scattering amplitude, which can be easily circumvented by introducing the Bethe-Salpeter kernel.  

\begin{figure}
\begin{centering}
\includegraphics[scale=0.425]{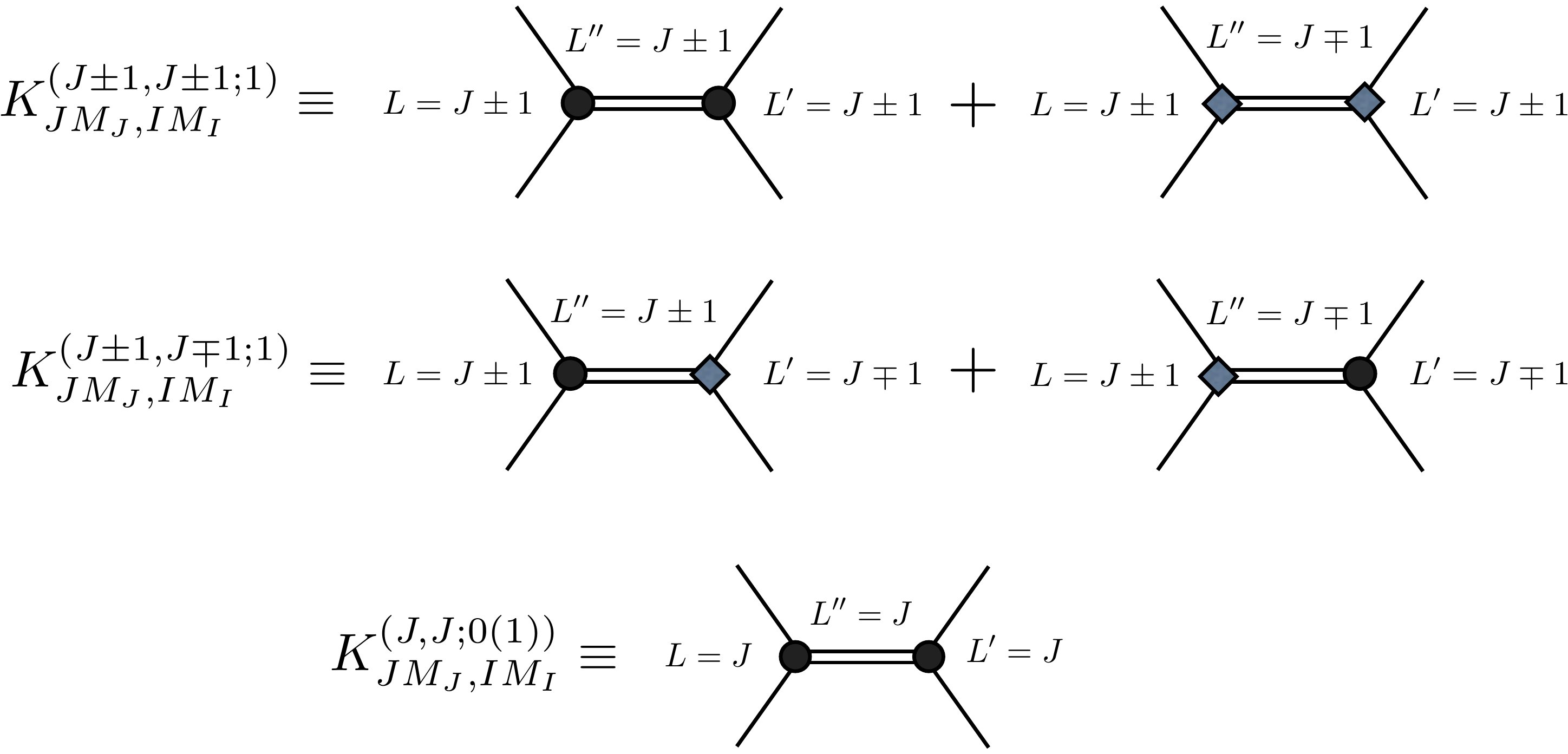}
\par
\caption[Leading $2\rightarrow2$ transition amplitudes in the sector with total angular momentum $J$ in the nuclear sector]{The leading $2\rightarrow2$ transition amplitudes in the sector with total angular momentum $J$, Eq. (\ref{KernelJ}). The superscripts in the kernels denote the initial angular momentum, $L$, final angular momentum, $L'$ and the conserved spin of the channels, $S$, respectively. The black circle represents the interaction vertex that conserves the partial-wave of the channel, and whose strength is parametrized by the coupling $g^{LS}_{JI}$, Eq. (\ref{S1}). The grey diamond denotes the vertex that mixes partial-waves, and whose strength is given by $h_{JI}$, Eq. (\ref{S2}). the double lines are the bare propagators corresponding to a dimer field with angular momentum $L''$.}\label{fig: Kernels}
\end{centering}
\end{figure}

Let us encapsulate the leading $2\rightarrow2$ transition amplitude between a two-nucleon state with $(JM_J,IM_I,LS)$ quantum numbers and a two-nucleon state with $(JM_J,IM_I,L'S')$ quantum numbers in the Bethe-Salpeter kernel, $K$. Note that since total angular momentum, spin and isospin are conserved in each $2\rightarrow 2$ transition, the kernel can be fully specified by $K_{JM_J;IM_I}^{(LL';S)}$. Since $J$ is conserved, the full kernel in the space of total angular momentum can be expressed as a block-diagonal matrix. In fact, it is straightforward to see that for each $J$-sector, the corresponding sub-block of the full matrix has the following form
\begin{eqnarray}
\left(\begin{array}{cccc}
K_{JM_{J};IM_{I}}^{(J-1,J-1;1)} & 0 & 0 & K_{JM_{J};IM_{I}}^{(J-1,J+1;1)}\\
0 & K_{JM_{J};IM_{I}}^{(J,J;0)} & 0 & 0\\
0 & 0 & K_{JM_{J};I'M_{I'}}^{(J,J;1)} & 0\\
K_{JM_{J};IM_{I}}^{(J+1,J-1;1)} & 0 & 0 & K_{JM_{J};IM_{I}}^{(J+1,J+1;1)}
\end{array}\right).
\label{KernelJ}
\end{eqnarray}
Note that for any given $J$, $I$, $L$ and $S$, there are $(2J+1)^2\times(2I+1)^2$ elements accounting for different values of $M_J$ and $M_I$ quantum numbers. Note also that the value of the isospin is fixed for each transition kernel. Explicitly, one finds that $I=\frac{1+(-1)^J}{2}$ and $I'=\frac{1+(-1)^{J+1}}{2}$.\footnote{Note that there is no $(I=0,S=0)$ channel for scattering in an even $J$ sector. Also there is no $(I=1,S=0)$ channel for scattering in an odd $J$ sector.} For the special case of $J=0$, the corresponding sub-sector is 
\begin{eqnarray}
\left(\begin{array}{cc}
K_{00;1M_{I}}^{(0,0;0)} & 0\\
0 & K_{00;1M_{I}}^{(1,1;1)}
\end{array}\right).
\label{KernelJ0}
\end{eqnarray}
These kernels, that correspond to leading transitions in all spin and isospin channels, are depicted in Fig. \ref{fig: Kernels}. Although one can read off the Feynman rules corresponding to these kernels from the  Lagrangian, Eqs. (\ref{Skin}, \ref{S1}, \ref{S2}), the FV energy eigenvalues can be determined without reference to the explicit form, as will become evident shortly.

The scattering amplitude can be calculated by summing up all the $2\rightarrow2$ diagrams which can be obtained by any number of insertions of the transition kernels and the two-particle propagator loops. It can be easily seen that the infinite-volume two-particle loops, $\mathcal{G}^{\infty}$, are diagonal in total angular momentum, spin, isospin and orbital angular momentum. It is easy to show that $\mathcal{G}^{\infty}=2~I^{\infty}$, where $I^{\infty}$ is the infinite-volume loop for two identical bosons, Eq. (\ref{I-infinity}), hence the overall factor of two.
As a result, the scattering amplitude can be expressed as
\begin{eqnarray}
\mathcal{M}^{\infty}=-\mathcal{K}\frac{1}{1-\mathcal{G}^{\infty}\mathcal{K}},
\end{eqnarray}
where $\mathcal{K}$ is a matrix whose $J^{th}$-sub-block is given by Eq. (\ref{KernelJ}). Since $\mathcal{G}^{\infty}$ is diagonal, the $J^{th}$-sub-block of the infinite-volume scattering amplitude reads 
\begin{eqnarray}
\left(\begin{array}{cccc}
\mathcal{M}_{JM_{J};IM_{I}}^{(J-1,J-1;1)} & 0 & 0 & \mathcal{M}_{JM_{J};IM_{I}}^{(J-1,J+1;1)}\\
0 & \mathcal{M}_{JM_{J};IM_{I}}^{(J,J;0)} & 0 & 0\\
0 & 0 & \mathcal{M}_{JM_{J};I'M_{I'}}^{(J,J;1)} & 0\\
\mathcal{M}_{JM_{J};IM_{I}}^{(J+1,J-1;1)} & 0 & 0 & \mathcal{M}_{JM_{J};IM_{I}}^{(J+1,J+1;1)}
\end{array}\right),
\label{amplitude}
\end{eqnarray}
for any non-zero $J$ and
\begin{eqnarray}
\left(\begin{array}{cc}
\mathcal{M}_{00;00}^{(0,0;0)} & 0\\
0 & \mathcal{M}_{00;1M_{I}}^{(1,1;1)}
\end{array}\right),
\label{amplitude}
\end{eqnarray}
for $J=0$. As is conventional, the scattering amplitude in channels with no partial-wave mixing can be parametrized by a scattering phase shift, $\delta_{JI}^{LS}$, according to
\begin{eqnarray}
\mathcal{M}_{JM_{J};IM_{I}}^{(JJ;S)}=\frac{4\pi}{Mk^*}\frac{e^{2i\delta_{JI}^{LS}}-1}{2i}\delta_{L,J}=\frac{4\pi}{Mk^*}\frac{1}{\cot{\delta_{JI}^{LS}}-i}\delta_{L,J},
\label{M-single}
\end{eqnarray}
while in channels where there is a mixing between the partial-waves, it can be characterized by two phase-shifts and one mixing angle, $\bar{\epsilon}_J$, \cite{Smatrix},
\begin{eqnarray}
\mathcal{M}_{JM_{J};IM_{I}}^{(J\pm1,J\pm1;S)}=\frac{4\pi}{Mk^*}\frac{\cos{2\bar{\epsilon}_J}e^{2i\delta_{JI}^{LS}}-1}{2i}\delta_{L,J\pm1},
\label{M-coupled1}
\\
\mathcal{M}_{JM_{J};IM_{I}}^{(J\pm1,J\mp1;S)}=\frac{4\pi}{Mk^*}\sin{2\bar{\epsilon}_J}\frac{e^{2i(\delta_{JI}^{LS}+\delta_{JI}^{L'S})}}{2}\delta_{L,J\pm1}\delta_{L',J\mp1}.
\label{M-coupled2}
\end{eqnarray}
These relations are independent of $M_J$ and $M_I$ as the scatterings are azimuthally symmetric. We emphasize again that Kronecker deltas used to specify the $L$ quantum numbers should not be confused with the phase shifts. Note that for each $J$ sector, there is only one mixing parameter and as result no further labeling other than the $J$ label is necessary for $\bar{\epsilon}_J$.

The FV kernels are equal to the infinite-volume kernels (up to exponentially suppressed terms in volume below the pion production threshold), and in particular the $J^{th}$-sub-block of such kernel is given by Eq. (\ref{KernelJ}). As in the scalar case, the only difference between the finite volume and infinite volume shows up in the s-channel bubble diagrams, where the two particles running in the loops can go on-shell and give rise to power-law volume corrections. It is straightforward to show that the two-nucleon propagator in the finite volume, $\mathcal{G}^V$, can be written as
\begin{eqnarray}
\mathcal{G}^V=\mathcal{G}^{\infty}+\delta\mathcal{G}^V,
\label{M-infinity}
\end{eqnarray}
where $\delta\mathcal{G}^V$ is a matrix in the $(JM_J,IM_I,LS)$ basis whose matrix elements are given by
{\small
\begin{eqnarray}
&& \left[\delta\mathcal{G}^V\right]_{JM_J,IM_I,LS;J'M_J',I'M_I',L'S'}=\frac{iMk^*}{4\pi}\delta_{II'}\delta_{M_IM_I'}\delta_{SS'}\left[\delta_{JJ'}\delta_{M_JM_J'}\delta_{LL'} +i\sum_{l,m}\frac{(4\pi)^{3/2}}{k^{*l+1}}c_{lm}^{\mathbf{P}}(k^{*2}) \right.
\nonumber\\
&& \qquad \qquad \qquad \qquad ~ \left .  \times \sum_{M_L,M_L',M_S}\langle JM_J|LM_L,SM_S\rangle \langle L'M_L',SM_S|J'M_J'\rangle \int d\Omega~Y^*_{L,M_L}Y^*_{l,m}Y_{L',M_L'}\right],
\nonumber\\
\label{deltaG}
\end{eqnarray}}
and, as is evident, is \emph{neither} diagonal in the $J$-basis nor in the $L$-basis. The kinematic function $c_{lm}^{\mathbf{P}}(k^{*2})$ is defined in Eq. (\ref{clm}) and is evaluated at the on-shell relative momentum of two nucleons in the CM frame. The full FV two-nucleon scattering amplitude can be evaluated by summing up all $2\rightarrow2$ FV diagrams,\footnote{One should note that using the notion of FV scattering amplitude is merely for the mathematical convenience. As there is no asymptotic state by which one could define the scattering amplitude in a finite volume, one should in principle look at the pole locations of the two-body correlation function. However, it can be easily shown that both correlation function and the so-called FV scattering amplitude have the same pole structure, so we use the latter for the sake of simpler representation.}
\begin{eqnarray}
\mathcal{M}^{V}=-\mathcal{K}\frac{1}{1-\mathcal{G}^{V}\mathcal{K}}=\frac{1}{(\mathcal{M}^{\infty})^{-1}+\delta\mathcal{G}^{V}},
\label{M-V}
\end{eqnarray}
where in the second equality the kernel is eliminated in favor of $\mathcal{M}^{\infty}$ and $\mathcal{G}^{\infty}$ using Eq. (\ref{M-infinity}). The energy eigenvalues of the two-nucleon system arise from the poles of $\mathcal{M}^V$ which satisfy the following determinant condition
\begin{eqnarray}
\det\left[{(\mathcal{M}^{\infty})^{-1}+\delta\mathcal{G}^{V}}\right]=0.
\label{NNQC}
\end{eqnarray}
This quantization condition clearly reduces to Eq. (\ref{FullQCboson}) for two-boson systems when setting $S=0$ \footnote{The symmetry factor in both scattering amplitude and the FV function will cancel out in the determinant condition, leaving the FV QC, Eq.~(\ref{FullQCboson}), insensitive to the distinguishability of the particles.}, and is in agreement with the result of Ref. \cite{Bernard:2008ax} for meson-baryon scattering after setting $S=1/2$. This result also extends the result of Ref. \cite{Ishizuka:2009bx} for two-nucleon systems to moving frames. The determinant is defined in the basis of $(JM_J,IM_I,LS)$ quantum numbers and is over an infinite dimensional matrix. To be practical, this determinant should be truncated in the space of total angular momentum and orbital angular momentum. Such truncation is justified since in the low-momentum limit the scattering phase shift of higher partial-waves $L$  scales as $k^{*2L+1}$. In the next section, by truncating the partial-waves to $L\leq3$, we unfold this determinant condition further, and present the reduction of this master formula to separate QCs for energy eigenvalues in different irreps of the corresponding symmetry group of the two-nucleon system. The first trivial reduction in the QC clearly takes place among different spin-isospin channels. In particular, it is straightforward to see that the QC in Eq. (\ref{NNQC}) does not mix $(S=0,I=1)$, $(S=1,I=0)$, $(S=0,I=0)$ and $(S=1,I=1)$ sectors, and automatically breaks into four independent determinant conditions that correspond to different spin-isospin sectors,
\begin{eqnarray}
\textrm{Det}\left[{(\mathcal{M}^{\infty})^{-1}+\delta\mathcal{G}^{V}}\right]=\prod_{I=0}^{1}\prod_{S=0}^{1} \det \left[(\mathcal{M}^{\infty}_{(I,S)})^{-1}+\delta \mathcal{G}^{V}_{(I,S)} \right]=0 .
\label{NNQC-IS}
\end{eqnarray}
This is due to the fact that each J-sub block of the scattering amplitude matrix can be separated into three independent sectors as following
\begin{eqnarray}
\mathcal{M}^{\infty}_{(I,1)}\equiv\left(\begin{array}{ccc}
\mathcal{M}_{J;I}^{(J-1,J-1;1)} &  & \mathcal{M}_{J;I}^{(J-1,J+1;1)}\\
\\
\mathcal{M}_{J;I}^{(J+1,J-1;1)} &  & \mathcal{M}_{J;I}^{(J+1,J+1;1)}
\end{array}\right), ~\mathcal{M}^{\infty}_{(I,0)}\equiv\begin{array}{c}
\mathcal{M}_{J;I}^{(J,J;0)}\end{array}, ~\mathcal{M}^{\infty}_{(I',1)}\equiv
\begin{array}{c}
\mathcal{M}_{J;I'}^{(J,J;1)}\end{array},
\nonumber\\
\label{amplitude-IS}
\end{eqnarray}
where $I$ and $I'$ are defined after Eq. (\ref{KernelJ}). Since the $M_J$ and $M_I$ indices are being suppressed, one should keep in mind that each block is still a $(2J+1)^2\times(2I+1)^2$ diagonal matrix. If $J$ is even, these amplitudes describe scattering in the negative parity isotriplet, positive parity isotriplet and positive parity isosinglet channels, respectively. For an odd $J$, these amplitudes correspond to scattering in the positive parity isosinglet, negative parity isosinglet and negative parity isotriplet channels, respectively.
Due to the reduced symmetry of the FV, $\delta \mathcal{G}^V$ has off-diagonal terms in the basis of total angular momentum $J$. So although the QC in Eq. (\ref{NNQC}) fully breaks down in the $(I,S)$-basis, it remains coupled in the $(J,L)$-basis. In order to further reduce the determinant conditions in Eq. (\ref{NNQC-IS}), the symmetries of the FV functions must be studied in more detail. This will be the topic of the next section, Sec. \ref{sec: Reduction}.

\section{Symmetry Considerations and Quantization Conditions \label{sec: Reduction}}
Lattice QCD calculations are performed in cubic volumes with periodic boundary conditions on the fields in spatial directions. As a result, the energy eigenstates of the two-particle systems at rest transform according to various irreps of the cubic group, depending on the interpolating operators that are used. Although it is convenient to think of the determinant condition, Eq. (\ref{NNQC}), as a determinant in the $J$-basis, one should expect that for zero CM momentum, this equation splits into five independent quantization conditions corresponding to the five irreps of the cubic group (see table (\ref{groups})). Furthermore, the degeneracy of the energy eigenvalues will reflect the dimension of the corresponding irrep. In general, the FV matrix $\delta \mathcal{G}^{V}$, Eq. (\ref{deltaG}), although being sparse, mixes states corresponding to different irreps of the cubic group. As a result, at least a partial block-diagonalization of this matrix is necessary to unfold different irreps that are present due to the decomposition of a given total angular momentum $J$. When the two-particle system is boosted, the symmetry group of the system is no longer cubic, and the reduction of the determinant condition, Eq. (\ref{NNQC}), takes place according to the irreps of the corresponding point group, table. (\ref{groups}). In the following section, this reduction procedure and the method of block diagonalization will be briefly discussed. In particular, we aim to obtain all the QC satisfied by the phase shifts and mixing parameters of the NN system of channels with $l\leq3$. We constrain this study to the CM momenta $\mathbf{d}=(0,0,0)$, $\mathbf{d}=(0,0,1)$ and $\mathbf{d}=(1,1,0)$, where $\mathbf{d}=\frac{L\mathbf{P}}{2\pi}$, which  provides $47$ independent QCs satisfied by different scattering parameters in these channels. As mentioned earlier, these boost vectors correspond to cubic ($O$), tetragonal ($D_4$) and orthorhombic ($D_2$) point groups, respectively.
\begin{center}
\begin{table} [h]
\label{tab:param1}
\resizebox{15cm}{!}{
\begin{tabular}{|ccccc|}
\hline
$\hspace{.3cm}\mathbf{d}\hspace{.3cm}$&$\hspace{.3cm}$point group$\hspace{.3cm}$&$\hspace{.3cm}$classification$\hspace{.3cm}$&$\hspace{.3cm}N_{\text{elements}}\hspace{.3cm}$&irreps (dimension) \\\hline \hline
$~(0,0,0)~$&${O}$&cubic&$24$&$~A_1(1),A_2(1),E(2),T_1(3),T_2(3)~$\\
$~(0,0,1)~$&$D_{4}$&tetragonal&$8$&$~A_1(1),A_2(1),E(2),B_1(1),B_2(1)~$\\ 
$~(1,1,0)~$&$D_{2}$&orthorhombic&$4$&$~A(1),B_1(1),B_2(1),B_3(1)~$
\\\hline\hline
\end{tabular}
}\caption{The classification of the point groups corresponding to the symmetry groups of the FV calculations with different boost vectors. The forth column shows the number of elements of each group.}
\label{groups}
\end{table}
\end{center}

In order to calculate matrix elements of the FV matrix $\delta \mathcal{G}^{V}$, one can take advantage of the symmetries of the $c_{lm}^{\mathbf{P}}$ functions as defined in Eq. (\ref{clm}). The relations between non-zero $c_{lm}^{\mathbf{P}}$s for any given angular momentum $l$ can be easily deduced from the transformation properties of these functions under symmetry operations of the corresponding point groups
\begin{eqnarray}
c^{\mathbf{P}}_{lm}=\sum_{m'=-l}^{l}\mathcal{D}^{(l)}_{mm'}(R_{\mathcal{X}})~c^{\mathbf{P}}_{lm'},
\label{clm-trans}
\end{eqnarray}
where $R_{\mathcal{X}}$ is the rotation matrix corresponding to each symmetry operation $\mathcal{X}$ of the group, and $\mathcal{D}^{(l)}_{mm'}$ denotes the matrix elements of the Wigner $\mathcal{D}$-matrix \cite{luscher2}. Beside these transformations, one can see that $c^\mathbf{P}_{lm}$s are invariant under inversion as can be easily verified from Eq. (\ref{clm}) for an arbitrary boost, and as a result all $c^\mathbf{P}_{lm}$s with an odd $l$ vanish.\footnote{For systems with non-equal masses, this is no longer true when the system is boosted. Since parity is broken for such systems, even and odd partial-waves mix with each other in the QCs, see Refs. \cite{Bour:2011ef, Davoudi:2011md, Fu:2011xz, Leskovec:2012gb}.} Table (\ref{nonzero-clm}) contains all such relations for non-vanishing $c^\mathbf{P}_{lm}$s up to $l=6$ for $\mathbf{d}=(0,0,0)$, $\mathbf{d}=(0,0,1)$ and $\mathbf{d}=(1,1,0)$ boost vectors.

\begin{table} [h]
\begin{center}
\label{tab:param3}
\resizebox{9cm}{!}{
\begin{tabular}{|c|c|c|}
\hline 
\textbf{d}=(0,0,0) & \textbf{d}=(0,0,1) & \textbf{d}=(1,1,0)\tabularnewline
\hline 
\hline 
$~c_{00}^{P}~$ & $~c_{00}^{P}~$ & $~c_{00}^{P}~$\\
$~c_{40}^{P}~$ & $~c_{20}^{P}~$ & $~c_{20}^{P}~$\\
$~c_{44}^{P}=c_{4,-4}^{P}=\sqrt{\frac{5}{14}}c_{40}^{P}~$ & $~c_{40}^{P}~$ & $~c_{22}^{P}=-c_{2,-2}^{P}~$\\
$~c_{60}^{P}~$ & $~c_{44}^{P}=c_{4,-4}^{P}~$ & $~c_{40}^{P}~$\\
$~c_{64}^{P}=c_{6,-4}^{P}=-\sqrt{\frac{7}{2}}c_{60}^{P}~$ & $~c_{60}^{P}~$ & $~c_{42}^{P}=-c_{4,-2}^{P}~$\\
 & $~c_{64}^{P}=c_{6,-4}^{P}~$ & $~c_{44}^{P}=c_{4,-4}^{P}~$\\
 &  & $~c_{60}^{P}~$\\
 &  & $~c_{62}^{P}=-c_{6,-2}^{P}~$\\
 &  & $~c_{64}^{P}=c_{6,-4}^{P}~$\\
\hline 
\end{tabular} }
\caption{The nonzero $c_{lm}^P$s up to $l=6$ for three different boost vectors $\mathbf{d}$.}
\label{nonzero-clm}
\end{center}
\end{table}

An important point regarding the $c^\mathbf{P}_{lm}$ functions is that they explicitly depend on the direction of the boost vector. In other words, $c^\mathbf{P}_{lm}$s that correspond to different boost vectors with the same magnitude $|\mathbf{d}|=n$  are not equal. As a result the corresponding set of non-zero $c^\mathbf{P}_{lm}$s as well as the relations among them, for permutations of the components of $(0,0,1)$ and $(1,1,0)$ boost vectors are different from those that are listed in Table (\ref{nonzero-clm}). Although this difference in general results in different $\delta \mathcal{G}^{V}$ matrices, e. g. for $(1,0,0)$, $(0,1,0)$ and $(0,0,1)$ boost vectors, as is shown in appendix \ref{app:invariant}, the master equation (\ref{NNQC}) is invariant under a  $\mathbf{P}\rightarrow \mathbf{P}'$ transformation when $\mathbf{P}$ and $\mathbf{P}'$ are related by a cubic rotation and  $|\mathbf{P}|=|\mathbf{P}'|$. The reason is that there exists a unitary transformation that relates $\delta \mathcal{G}^{V,\mathbf{P}}$ to $\delta \mathcal{G}^{V,\mathbf{P}'}$, leaving the determinant condition invariant. 

Since the relations among $c^\mathbf{P}_{lm}$ are simpler when one assumes boost vectors that discriminate the z-axis relative to the other two Cartesian axes, we will present the QCs corresponding to $\mathbf{d}=(0,0,1)$ and $\mathbf{d}=(1,1,0)$ boost vectors only. The lattice practitioner can still use the QCs presented to extract the scattering parameters of the NN system from the energy eigenvalues of lattice calculations with other permutations of these boost vectors. It is however crucial to input the boost vectors that are specified in this paper when calculating the $c^\mathbf{P}_{lm}$ functions in the QCs (instead of the boost vectors that are used in the lattice calculation). In order to increase the precision of the scattering amplitudes obtained, one should perform the lattice calculation with all possible boost vectors of a given magnitude that belong to the same $A_1$ irrep of the cubic group,\footnote{Note that in higher momentum shells, there occurs multiple $A_1$ irreps of the cubic group. This indicates that there are classes of momentum vector that do not transform into each other via a symmetry operation of the cubic group, e. g. $(2,2,1)$ and $(0,0,3)$ vectors in the $\mathbf{n}^2=9$ shell. However, as is discussed, another property of the $c_{lm}^{\mathbf{P}}$ functions for non-relativistic degenerate masses indicates that the value of the FV function is the same for these two boost vectors as they are both of the form $(2n_1,2n_2,2n_3+1)$ with $n_i \in \mathbb{Z}$.} and use the average energy eigenvalues in the QCs presented to determine the scattering parameters; keeping in mind that $c^\mathbf{P}_{lm}$ functions have to be evaluated at the boost vectors considered in this paper.

The other fact that should be pointed out is that due to the symmetries of the $c^\mathbf{P}_{lm}$ function for equal masses, the system at rest with $\mathbf{d}=(0,0,0)$ exhibits the same symmetry transformation as that of the $(2n_1,2n_2,2n_3)$ boost where $n_1,n_2,n_3$ are integers. Similarly, the symmetry group of the calculations with $(0,0,1)$ ($(1,1,0)$) boost is the same as that of $(2n_1,2n_2,2n_3+1)$ ($(2n_1+1,2n_2+1,2n_3)$) boosts. As a result, the quantization conditions presented in appendix \ref{app:QC} can be used with these boost vectors as well. It is worth mentioning that for relativistic two-particle systems with degenerate masses, the above statement is no longer true. This is due to the fact that the boost vector dependence of the relativistic $c_{lm}^{\mathbf{P}}$ function is different from that of the NR counterpart, leading to more distinct point group symmetries for different boosts \cite{movingframe, sharpe1, Christ:2005gi}.

\begin{table}
\begin{centering}
\resizebox{15cm}{!}{

\begin{tabular}{|c||c|c|c|c|c|}
\hline 
J & $0$ & $1$ & $2$ & $3$ & $4$\tabularnewline
\hline 
\hline 
\multirow{4}{*}{$I=0,\; S=1$} & - & $\delta_{1,S},\delta_{1,D},\epsilon_{1,SD}$ & $\delta_{2,D}$ & $\delta_{3,D}$ & -\tabularnewline
\cline{2-6} 
 &  & Eqs.  & Eqs.  & Eqs. & \tabularnewline
 &  & \ref{I000T1}, \ref{I001A2}, \ref{I001E} & \ref{I000E}, \ref{I000T2}, \ref{I001A1}, & \ref{I000A2}, \ref{I000T1}, \ref{I000T2}, \ref{I001B1} & \tabularnewline
 &  & \ref{I110B1}, \ref{I110B2}, \ref{I110B3} & \ref{I001B1}, \ref{I001B2}, \ref{I001E} & \ref{I001B2}, \ref{I001A2}, \ref{I001E}, \ref{I110B1} & \tabularnewline
  &  &  & \ref{I110A} & \ref{I110B2}, \ref{I110B3}, \ref{I110A} & \tabularnewline
\hline
\multirow{5}{*}{$I=1,\; S=0$} & $\delta_{0,S}$ & - & $\delta_{2,D}$ & - & -\tabularnewline
\cline{2-6}
 & Eqs. &  & Eqs.  &  & \tabularnewline
 & \ref{II000A1}, \ref{II001A1}, \ref{II110A} &  & \ref{II000E}, \ref{II000T2}, \ref{II001A1}, &  & \tabularnewline
 &  &  & \ref{II001B1}, \ref{II001B2}, \ref{II001E}, &  & \tabularnewline
 &  &  &  \ref{II110B1}, \ref{II110B2}, \ref{II110B3}, &  & \tabularnewline
 &  &  &   \ref{II110A} &  & \tabularnewline
\hline 
\multirow{5}{*}{$I=0,\; S=0$} & - & $\delta_{1,P}$ & - & $\delta_{3,F}$ & - \tabularnewline
\cline{2-6}
 &  & Eqs.  &  & Eqs.  &  \tabularnewline
 &  & \ref{III000T1}, \ref{III001A2}, \ref{III001E} &  & \ref{III000T1}, \ref{III000A}, \ref{III000T2}, \ref{III001A2}, & \tabularnewline
 &  & \ref{III110B2}, \ref{III110B3}, \ref{III110B1} &  & \ref{III001B1}, \ref{III001B2}, \ref{III001E}, \ref{III110B2}, & \tabularnewline
 &  &  &  & \ref{III110B3}, \ref{III110A}, \ref{III110B1} & \tabularnewline
\hline 
\multirow{5}{*}{$I=1,\; S=1$} & $\delta_{0,P}$ & $\delta_{1,P}$ & $\delta_{2,P},\delta_{2,F},\epsilon_{2,PF}$ & $\delta_{3,F}$ & $\delta_{4,F}$ \tabularnewline
\cline{2-6} 
 & Eqs.  & Eqs.  & Eqs.  & Eqs.  & Eqs.  \tabularnewline
 & \ref{IV000A1}, \ref{IV001A1}, \ref{IV110A} & \ref{IV000T1}, \ref{IV001A2}, \ref{IV001E} & \ref{IV000T2}, \ref{IV000E}, \ref{IV001A1}, & \ref{IV000T1}, \ref{IV000T2}, \ref{IV000A2}, & \ref{IV000T1}, \ref{IV000T2}, \ref{IV000A1}, \tabularnewline
 &  & \ref{IV110A}, \ref{IV110B1}, \ref{IV110B2} & \ref{IV001B1}, \ref{IV001B2}, \ref{IV001E}, & \ref{IV001A2}, \ref{IV001B1}, \ref{IV001B2}, & \ref{IV001A1}, \ref{IV001A2}, \ref{IV001B1}, \tabularnewline
 &  & \ref{IV110B3} & \ref{IV110B1}, \ref{IV110B2}, \ref{IV110B3} & \ref{IV001E}, \ref{IV110A}, \ref{IV110B1}, & \ref{IV001B2}, \ref{IV000E}, \ref{IV001E}, \ref{IV110A},  \tabularnewline
&  &  &  &  \ref{IV110A}, \ref{IV110B2}, \ref{IV110B3} & \ref{IV110B1}, \ref{IV110B2},
\ref{IV110B3}
\tabularnewline
\hline
\end{tabular}}
\par\end{centering}

\caption{The scattering parameters that can be determined from the QCs presented in appendix \ref{app:QC} for all four different spin-isospin channels. The reference to the relevant equations in extracting each parameter is given in the table. These equations are assumed to be used in Eq. (\ref{QC-simplified}). The subscript in each parameter denotes the total $J$ as well as the partial-wave of the channel the parameter corresponds to. }
\label{scatt-param}
\end{table}
%


Back to our main goal, we aim to break the master equations (\ref{NNQC-IS}) into separate QCs corresponding to each irrep of the symmetry group of the problem. In fact, from the transformation law of the $\delta \mathcal{G}^{V}$ function under a symmetry operation of the group, 
\begin{eqnarray}
\left[\delta\mathcal{G}^V\right]_{JM_J,LS;J'M_J',L'S}=\sum_{\bar{M}_J=-J}^{J}\sum_{\bar{M}_J'=-J'}^{J'}\mathcal{D}^{(J)}_{M_J,\bar{M}_J}(R_{\mathcal{X}})\left[\delta\mathcal{G}^V\right]_{J\bar{M}_J,LS;J'\bar{M}_J',L'S}\mathcal{D}^{(J')}_{\bar{M}_J',M_J'}(R_{\mathcal{X}}^{-1}),
\nonumber\\
\label{dG-trans}
\end{eqnarray}
one can deduce that there is a unitary transformation which brings the matrix $\delta \mathcal{G}^{V}$ to a block-diagonal form. Note that we have suppressed the isospin quantum numbers as $\delta\mathcal{G}^V$ is diagonal in the isospin basis. Each of these blocks then can be identified by a given irrep of the symmetry group of the problem. Such transformation eventually breaks the determinant conditions (\ref{NNQC-IS}) to separate determinant conditions corresponding to each irrep of the point group of the system. Explicitly in each spin and isospin sector,
\begin{eqnarray}
\det \left[(\mathcal{M}^{\infty}_{(I,S)})^{-1}+\delta \mathcal{G}^{V}_{(I,S)} \right]=\prod_{\Gamma^i}\det\left[(\mathcal{M}^{\infty-1}_{(I,S)})_{\Gamma^i}+\delta \mathcal{G}^{V,\Gamma^i}_{(I,S)} \right]^{N(\Gamma^i)}=0 .
\label{NNQC-irrep}
\end{eqnarray}
where $\Gamma^i$ denotes each irrep of the corresponding group and $N(\Gamma^i)$ is the dimensionality of each irrep. The dimensionality of each of these smaller determinant conditions is  given by the multiplicity of each irrep in the decomposition of angular momentum channels that are being included in the scattering problem. As is seen in  appendix \ref{app:QC}, although from the master quantization condition, for some of the NN channels with $J\leq4$ and $l\leq 3$, one has to deal with a determinant of $30 \times 30$ matrices, upon such reduction of the master equation, one arrives at QCs that require taking the determinant of at most $9 \times 9$ matrices. We demonstrate this procedure in more detail for one example in appendix \ref{app:red-example}. For the rest of the channels and boosts, only the final result of such reduction will be presented (see appendix \ref{app:QC}).\footnote{Although these QCs are the main results of this chapter, to achieve a better presentation of such long equations, we have tabulated them in an appendix.} It is also shown in appendix \ref{app:red-example} that the QC in Eq. (\ref{NNQC}) is real despite the fact that both $\delta \mathcal{G}^V$ and $\mathcal{M}^{\infty}$ are complex quantities.

In summary, the lattice practitioner may extract the desired scattering parameters of the NN-system by performing the following steps:

\begin{enumerate}
\item For a given irrep $\Gamma$, evaluate the $NN$ correlation function with all possible boost vectors with magnitude ${d}$ that are related to each other via a cubic rotation, $\{C_{NN}^{\Gamma,\textbf{d}_1},\ldots,C_{NN}^{\Gamma,\textbf{d}_{N_d}}\}$.
\item Average the value of the correlation functions over all boost vectors used in the previous step, $C_{NN}^{\Gamma,{d}}=\sum_{i}^{N_d}C_{NN}^{\Gamma,\textbf{d}_i}/N_d$.
\item Obtain the non-relativistic finite volume energy, $E_{NR}^\Gamma=E_{NN}^\Gamma-2m_N$, from the asymptotic behavior of the correlation function and therefore obtain the value of the relative momentum $k^*$ from $k^*=\sqrt{M_NE-(\pi\mathbf{d})^2/L^2}$.
\item Determine scattering parameters from the QCs in appendix \ref{app:QC}:

\begin{enumerate}
\item Use $\mathbf{d}=(0,0,0)$ if $\mathbf{d}$ is a permutation of $(2n_1,2n_2,2n_3)$,
\item Use $\mathbf{d}=(0,0,1)$ if $\mathbf{d}$ is a permutation of $(2n_1,2n_2,2n_3+1)$,
\item Use $\mathbf{d}=(1,1,0)$ if $\mathbf{d}$ is a permutation of $(2n_1+1,2n_2+1,2n_3)$.
\end{enumerate}
\end{enumerate}


\subsubsection{Implication for the $T_1^+$ spectrum at the physical point, $m_\pi\sim140$~MeV}

\begin{figure}[t]
\begin{center} 
\subfigure[]{
\label{swave}
\includegraphics[scale=0.2]{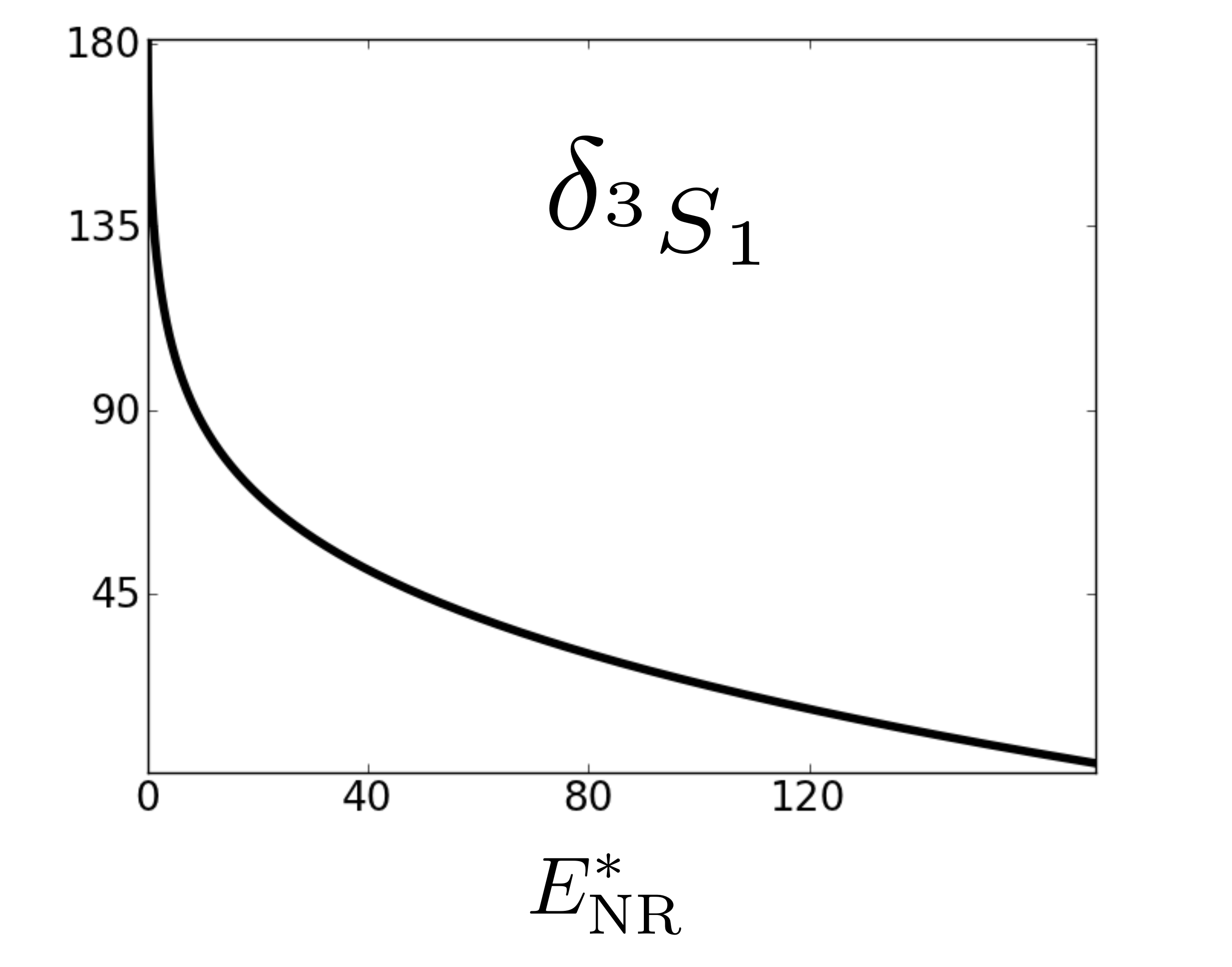}}\\
\subfigure[]{
\label{dwave}
\includegraphics[scale=0.16]{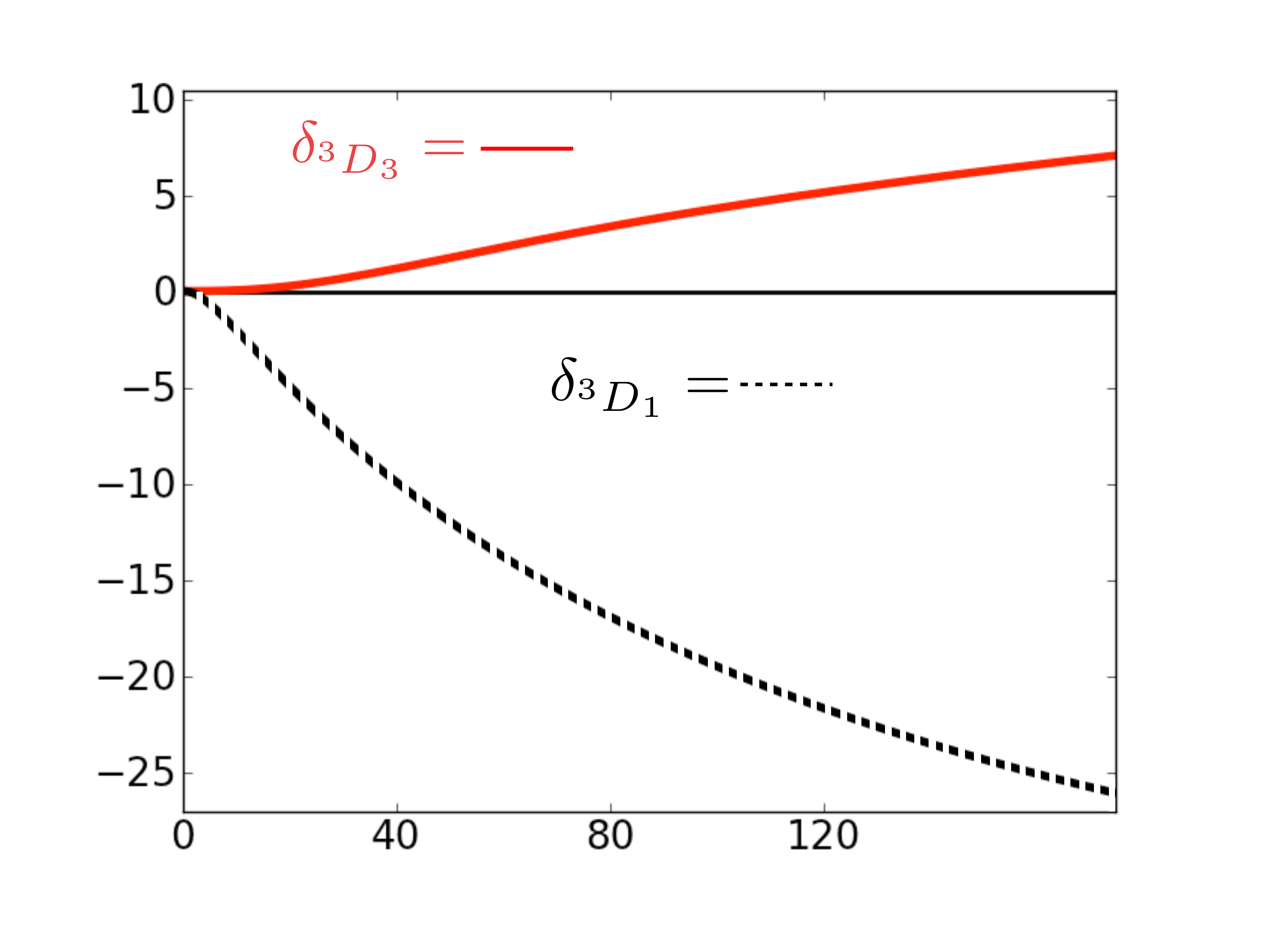}}
\subfigure[]{
\label{eps}
\includegraphics[scale=0.15]{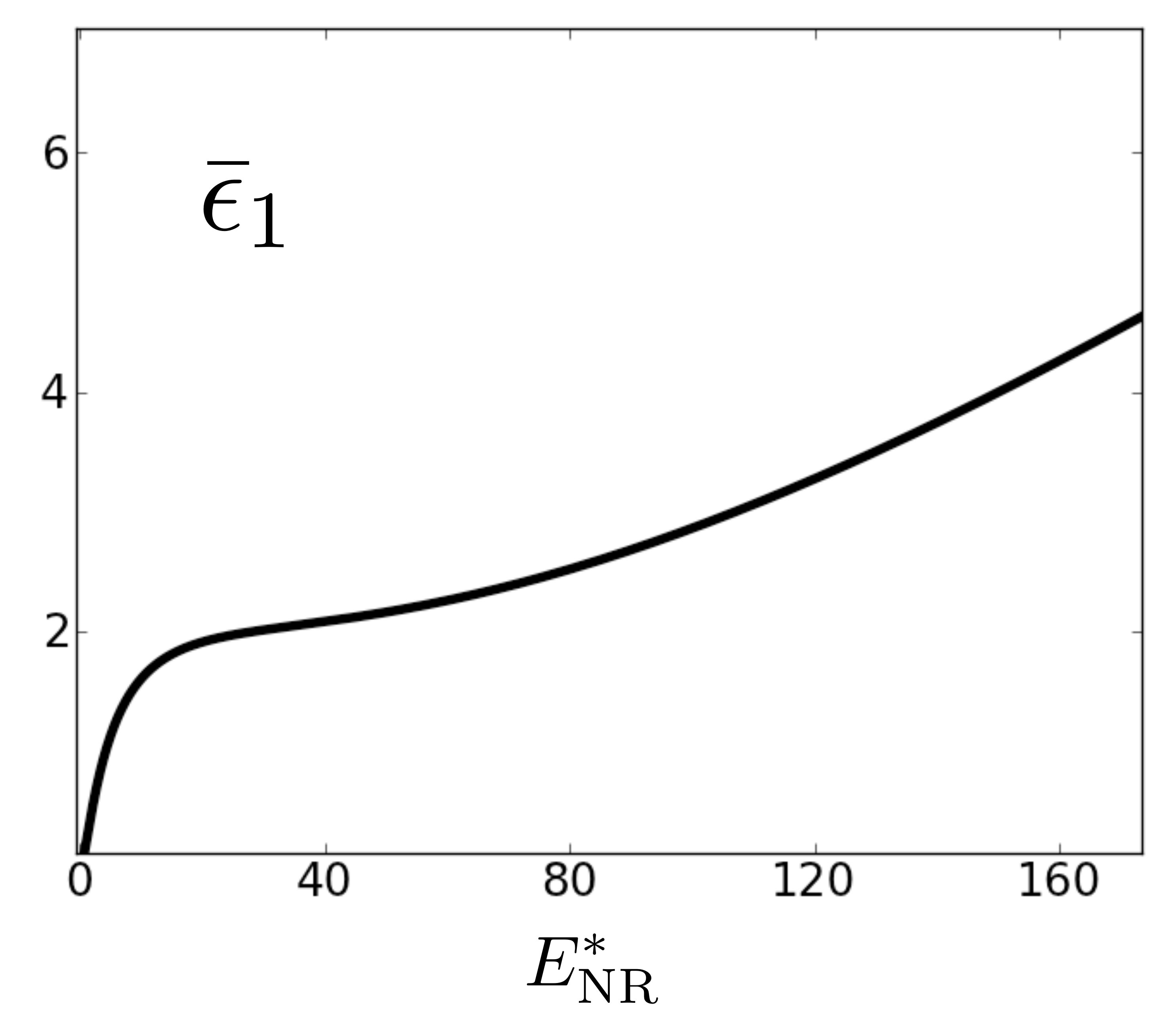}}
\caption[$^3S_1,{^3D_1}$ and ${^3D_3}$ NN scattering phase shifts and mixing angle]{ Shown are the 
of the phase shifts for the (a) $^3S_1$, (b) ${^3D_1}$ and ${^3D_3}$ channels as well as the (c) $J=1$ mixing angle $\bar{\epsilon}_{1}$\cite{T1ps} as a function of the NR CM energy, $E^*_{NR}$.}\label{T1phaseshifts}
\end{center}
\end{figure}

A complete discussion of the implication of these QCs for the forthcoming LQCD calculations requires a rather extensive numerical study using phenomenological phase shifts and exploring $m_\pi$-dependence of these parameters, which is underway. Here we can get a taste of the expectation of the spectrum by considering just the $T_1^+$-irrep in the CM frame. The QC for this channel, written in Eq.~(\ref{T1}), depends on the phase shifts of the $^3S_1,{^3D_1},{^3D_3}$ channels as well as the mixing angle $\bar{\epsilon}_{J=1}$. By inputing the phenomenological scattering parameters obtained from \cite{T1ps} and depicted in Fig.~\ref{T1phaseshifts}, we can predict the spectrum at the physical point. 

\begin{figure}[t]
\begin{center}  
\subfigure[]{
\label{T1_spec}
\includegraphics[scale=0.2]{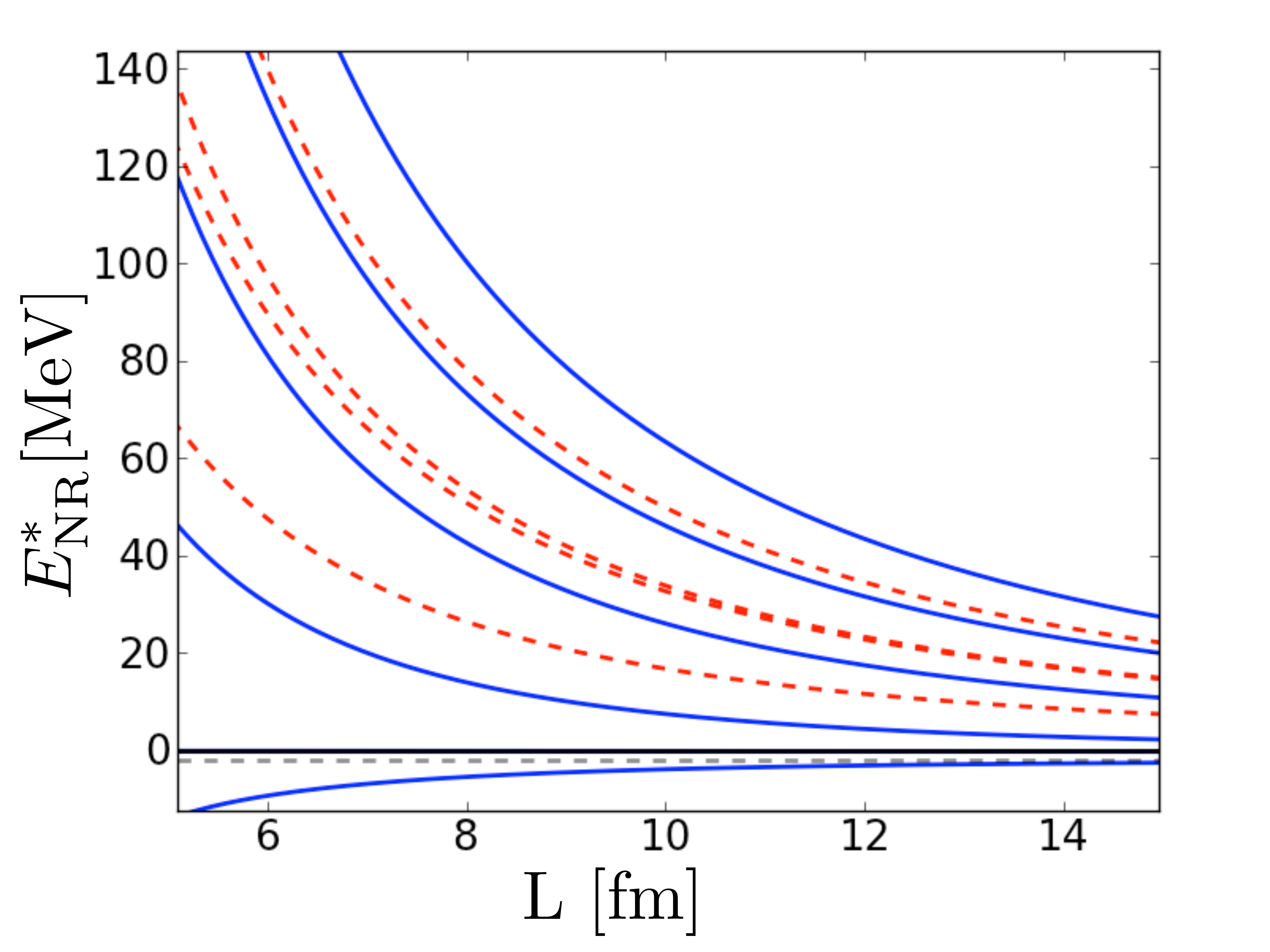}}
\subfigure[]{
\label{T1_qtilde}
\includegraphics[scale=0.2]{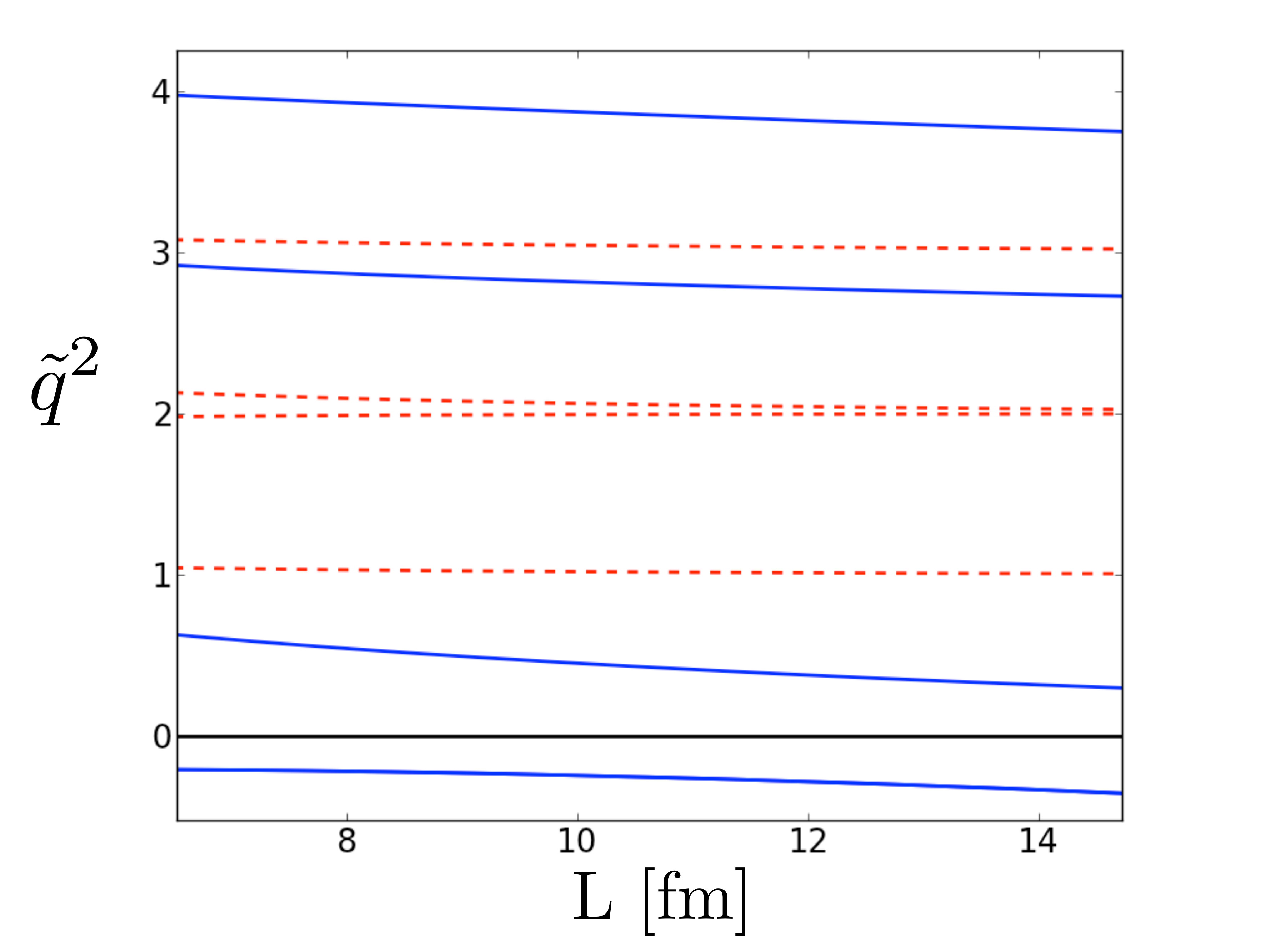}}
\caption[Prediction for the finite-volume spectrum for the $T_1$-irrep at the physical point]{(a) Shown are the nine lowest states satisfying the QC for $T_1$-irrep, Eq.~(\ref{T1}), as a function of the volume $\rm{L}$. The blue-solid lines depict states that are primarily identified as S-wave states, while the red-dashed lines depict states that are primarily identified as D-wave states. The black-dashed line denotes the $-B_d^\infty=2.224644(34)$~MeV line, where $B_d^\infty$ is the infinite volume binding energy of the deuteron. (b) Shown is the dimensionless quantity $\tilde{q}^2=E^*_{NR}{\rm{L}}^2/(4\pi^2)$ plotted as a function of $\rm{L}$.  }\label{T1spec}
\end{center}
\end{figure}

Figure~\ref{T1_spec} shows the nine lowest states, including the bound state, as a function of the volume. The states are identified as either primarily S-wave or D-wave states. This identification is done by comparing the spectrum with the one obtain when setting the mixing angle equal to zero. Immediately one observes that there are two states that are nearly degenerate. It is convenient to plot the dimensionless quantity $\tilde{q}^2=E^*_{NR}{\rm{L}}^2/(4\pi^2)$ as a function of the volume, Fig.~\ref{T1_qtilde}. From Fig.~\ref{T1_qtilde} we observe that in fact the four D-wave states are very close to the free states, which correspond to $\tilde{q}^2=\{0,1,2\ldots\}$. Which one would expect as the D-wave phase shifts are in fact rather small at the physical point, see Fig.~\ref{dwave}. For these states it might be more convenient to use a large-L expansion of the QC around the free-energy solutions since the $c_{lm}$ are divergent at these points. 

Furthermore, it is unclear at this point if technological advancements will ever allow calculations to have a level precision high enough to resolve the two nearly degenerate D-wave states. That being said, it is not physical point that is most interesting. After all, physical NN-scattering parameters are remarkably well constrained. LQCD will have a large impact by obtaining information of the scattering phases as a function of the pion mass.

\section{Proton-proton fusion \label{ppfusion}}

In this section we discuss the weak interaction in the two-nucleon sector. This sector has been previously studied by Detmold and Savage \cite{nnd} in a finite volume. They considered a novel idea of studying electroweak matrix elements using a background field. Since evaluating matrix elements of electroweak currents between NN states, e.g. $\left\langle d\right|\left.A^{\mu}\right.\left|np\right\rangle 
 $,  is naively one or two orders of magnitude more difficult than performing NN-four point functions, they present a procedure for extracting the relevant LECs of the pionless EFT, EFT($\not\hspace{-.08cm}{\pi}$) \cite{pionless}, by calculating four-point functions of nucleons in a finite volume in the presence of a background electroweak field.  This would be a project worth pursuing with great benefits, namely a five-point function is replaced by a four-point function, thereby dramatically reducing the number of propagator contractions. For isovector quantities, this procedure comes at a small computational cost, since for perturbatively small background fields, the QCD generated gauge links get modified by a multiplicative factor that couples the valence quarks to the external field, $U^{QCD}_{\mu}(x)\rightarrow U^{QCD}_{\mu}(x)U_{\mu}^{ext}(x)$. On the other hand, for isoscalar  quantities this approach would require the generation of gauge configurations in the presence of the background field. For both isovector and isoscalar quantities, one would need to perform calculations at a range of background field strengths in order to precisely discern the contribution of the coupling between the background field and the baryonic currents to the NN spectrum. Additionally, the nature of this background field will differ depending on the physics one is interested in.  
Alternatively, one can always evaluate matrix elements of electroweak currents with gauge configuration that solely depend on the QCD action, which is the case considered here. With the improvement in the computational resources available for LQCD calculations, the studies of nucleonic matrix elements will become feasible shortly, and therefore their connection to the physical matrix elements should be properly addressed.

The goal is to explore FV corrections of weak matrix elements in the two-nucleon sector. In particular, we will consider the proton-proton fusion process, $(pp\rightarrow d+e^++\nu_e)$, which couples the $\singlet-\triplet$ channels. In order to do this calculation the mechanism of pionless EFT \cite{pds, pds2, pionless} will be used. The presence of a weak interactions, leads to a contribution to the Lagrangian that couples the axial-vector current $A^{\mu=3}=\frac{1}{2}\left(\bar{u}\gamma^3\gamma^5u-\bar{d}\gamma^3\gamma^5d\right)$ to an external weak current. In terms of the low-energy degrees of freedom, the axial current will receive one-body and two-body contributions. At energies well below the pion-production threshold, the  EFT ($\not\hspace{-.08cm}\pi$) Lagrangian density including weak interactions can be written as \cite{pds,pds2,pionless,Butler:1999sv, Butler:2000zp, Butler:2001jj}
\begin{eqnarray}
\mathcal{L}&=&N^\dag\left(i\partial_0+\frac{\nabla^2}{2m}-\frac{W_3g_A}{2}\sigma^3\tau^3\right)N
-C^{{\left(\singlet\right)}}_0\left(N^TP_1^aN\right)^\dag \left(N^TP_1^aN\right)\nn\\
&&-C^{{\left(\triplet\right)}}_0\left(N^TP_3^jN\right)^\dag \left(N^TP_3^jN\right)
-{L_{1,A}}W_3\left[\left(N^TP_1^3N\right)^\dag \left(N^TP_3^3N\right)+h.c.\right]+\cdots,
\label{pionlag}
\end{eqnarray}
where $N$ is the nucleon annihilation operator with bare mass $m$,  $\{C^{{\left(\singlet\right)}}_0, C^{{\left(\triplet\right)}}_0, g_A,{L_{1,A}}\}$ are the LECs of the theory, $g_A = 1.2701(25)$~\cite{pdg}  is the nucleon axial coupling constant, $W_3$ is the external weak current, and $\{P_1^a, P_3^j\}$ are the standard $\{^1S_0, ^3S_1\}$-projection operators,
\begin{eqnarray}
\label{proj}
P_1^a=\frac{1}{\sqrt{8}}\tau_2\tau^a\sigma_2,\hspace{1cm}
P_3^j=\frac{1}{\sqrt{8}}\tau_2\sigma_2\sigma^j,
\end{eqnarray}
where $\tau$ ($\sigma$) are the Pauli matrices which act in isospin (spin) space. In Eq. (\ref{pionlag}) the ellipsis denotes an infinite tower of higher order operators. The $\mathcal{O}(p^{2n})$-operator for the $\{\singlet, \triplet\}$ state will have  corresponding LECs $\{C^{{\left(\singlet\right)}}_{2n}, C^{{\left(\triplet\right)}}_{2n}\}$, which are included in this calculation. In this section we only consider NN-systems in the S-wave channel, which makes this formalism only applicable near the kinematic threshold. 

At leading order, a weak transition between the isosinglet and isotriplet two-nucleon channels is described by an insertion of the single body current (which is proportional to $g_A$) and the bubble chain of the $C_0^{\triplet}$ and $C_0^{\singlet}$ contact interactions on the corresponding nucleonics legs as discussed in Ref.~\cite{Kong:1999tw}. At NLO, the hadronic matrix element of  $pp\rightarrow d+e^++\nu_e$ receives contributions from one insertion of the $C_2p^2$ operator along with one insertion of the single-body operator proportional to $g_A$ \cite{Kong:1999mp}. At the same order, a single insertion of the two-body current that is proportional to ${L_{1,A}}$ also contributes to the transition amplitude~\cite{Butler:2001jj}. In both of these contributions the dressing of the NN states with the corresponding bubble chain of the LO contact interactions must be assumed. As is discussed in Ref. \cite{Butler:2001jj}, the two-body contribution is estimated to give rise to a few-percent correction to the hadronic matrix element, and its corresponding LEC, $L_{1A}$, is known to contribute to the elastic and inelastic neutrino-deuteron scattering cross sections as well \cite{Butler:1999sv, Butler:2001jj}. Of course, the electromagnetism plays a crucial role in the initial state interactions in the pp-fusion process, but as is shown in Refs. \cite{Kong:1999tw, Kong:1999mp, Butler:2001jj}, the ladder QED diagrams can be summed up to all orders non-perturbatively. Since LQCD calculations of the matrix elements of the axial-vector current involving two-nucleons would allow for a determination of ${L_{1,A}}$, one will achieve tighter theoretical constraints on the cross section of these processes. Furthermore, having obtained the one-body and two-body LECs of the weak sector will allow for the determination of the few-body weak observables.


In the absence of weak interactions, the on-shell scattering amplitude for both channels can be determined exactly in terms of their corresponding LECs by performing a geometric series over all the bubble diagrams \cite{pionless}
\begin{eqnarray}
\mathcal{M}^0=-\frac{\sum_{n=0}^{\infty} C_{2n}q^{*2n}}{1-G^\infty_0\sum_{n=0}^{\infty} C_{2n}q^{*2n}},
\end{eqnarray}
where the on-shell relative momentum in the CM frame is related to the total NR CM energy and momentum of the two-nucleon system via, $q^*=\sqrt{mE^*-\frac{1}{4}{\mathbf{P}^2}}$, and $G^\infty_0$ denotes the loop integral
\begin{eqnarray}
\label{Iinfinity}
G^{\infty}_0&=&\left(\frac{\mu}{2}\right)^{4-D}\int\frac{d^3\mathbf k}{(2\pi)^3}\frac{1}{E-\frac{\mathbf{k}^2}{2m}-\frac{(\mathbf{P}-\mathbf{k})^2}{2m}+i\epsilon}
\end{eqnarray} 
which is linearly divergent. In order to preserve Galilean invariance and maintain a sensible power counting scheme for NR theories with an unnaturally large scattering length, the power-divergence subtraction (PDS)  scheme is used to regularize the  integral \cite{pds,pds2,Beane:2003da}. Using PDS, the integral above becomes
\begin{eqnarray}
\label{IinfinityPDS}
 G^{\infty}_0&=&-\frac{m}{4\pi}\left(\mu+i\sqrt{mE-{P}^2/4}\right)=-\frac{m}{4\pi}\left(\mu+iq^*\right),
\end{eqnarray}
where $\mu$ is the renormalization scale. When the volume is finite, the integral above is replaced by its FV counterpart, $G^{V}_{0}$. It is straightforward to find the relation between the FV correction $\delta G^{V}_{0}=G^{V}_{0}-G^{\infty}_{0}$ and the non-relativistic version of the kinematic function $c_{00}^{P}$ defined in Eq. (\ref{clm}), with $\alpha=\frac{1}{2}$ and $\gamma=1$ for degenerate non-relativistic particles.
One can arrive at the desired relation by adding and subtracting the infinite volume two-particle propagator, Eq. (\ref{Iinfinity}), to $G^{V}_{0}$. One of them can be evaluated using PDS, Eq. (\ref{IinfinityPDS}), and the other one can be written in terms of a regularized principle value integral, leading to 
\begin{eqnarray}
\label{Ifinite}
G^V_0(E,{P})
&=&-\frac{m}{4\pi}\mu
-c_{00}^{P}(q^{*2}),
\end{eqnarray}
therefore arriving at
\begin{eqnarray}
\label{fvloop}
\delta G^V_0(E,{P})&=& 
 =\frac{m}{4\pi}\left(q^*\cot\phi^{P}
 +i q^*\right),
\end{eqnarray}
where we have used the pseudo-phase definition, Eq. (\ref{pseudophase}).

The goal is to find a relation between the FV matrix elements of the axial-vector current and the LECs that parametrize the weak interaction, namely $\{g_A,{L_{1,A}}\}$, following a procedure analogous to section~\ref{EW2B}. The first step is to find the QC satisfied by the energy eigenvalues of the two-nucleon system in  presence of an external weak field as was also considered in Ref.~\cite{nnd}.\footnote{The main distinction between the result that will be obtained here and that of Ref. \cite{nnd} is that we will consider the case where the two-nucleon system has arbitrary momentum below inelastic thresholds, while Ref. \cite{nnd} only considered the two-nucleon system at rest.} After obtaining the QC for this theory, the method by Lellouch and L\"uscher~\cite{LL1} can be utilized to obtain an expression for the FV weak matrix element. The main difference between the problem considered here and the problem discussed in the previous section is that the dominant contribution to the weak processes in the NN-sector comes from the one-body current, namely the term proportional to the axial charge in Lagrangian, Eq.~(\ref{pionlag}). In fact this contribution modifies the nucleon propagator and therefore the on-shell condition. To avoid complications associated with the modification of \emph{external legs} appearing in the FV analogue of the scattering amplitude, $\mathcal{M}^{V}$, we obtain the QC for this system by looking at the pole structure of the NN-correlation function in presence of the weak field. As before a $2\times2$ kernel $\mathcal{K}$ can be formed that incorporates the tree-level $2\rightarrow2$ transitions, 
\begin{eqnarray}
i\mathcal{K}&=&\begin{pmatrix} 
-i\sum\limits_{n} C^{\left(\triplet\right)}_{2n}q^{*2n} &-i{L_{1,A}}\\
-i{L_{1,A}}&-i\sum\limits_{n}  C^{\left(\singlet\right)}_{2n}q^{*2n}\\
\end{pmatrix}.
\end{eqnarray}
The finite-volume function $\mathcal{G}^V$ can be still expressed as a $2\times2$  matrix in the basis of channels, except it will attain off-diagonal elements due to the presence of the single-body operator, in contrast to the scalar sector studied before, 
\begin{eqnarray}
\delta\mathcal{G}^V=\left(\begin{array}{ccc}
G_{+}^{V} &  & G_{-}^{V}\\
\\
G_{-}^{V} &  & G_{+}^{V}
\end{array}\right),
\end{eqnarray}
where FV functions $G^V_+$ and $G^V_-$ are defined as
\begin{eqnarray}
\label{GVpm}
G^{V}_{\pm}=\frac{m}{2L^3}\sum_{\mathbf k}\left[\frac{1}{E-\frac{\mathbf{k}^2}{2m}-\frac{(\mathbf{P}-\mathbf{k})^2}{2}-W_3g_A}\pm
\frac{1}{E-\frac{\mathbf{k}^2}{2m}-\frac{(\mathbf{P}-\mathbf{k})^2}{2m}+W_3g_A}
\right].
\end{eqnarray}
Since we only aim to present the result up to NLO in the EFT expansion according to the power counting discussed above, it suffices to keep only the LO terms in $g_A$ when expanding these FV functions in powers of the weak coupling. Explicitly, $G^V_+=G^V_0(E,{P})+\mathcal{O}(W_3^2g_A^2)$ where $G^V_0$ is defined in Eq.~(\ref{Ifinite}), and $G^V_-=W_3g_A~G^V_1(E,{P})+\mathcal{O}(W_3^3g_A^3)$ with
\begin{eqnarray}
\label{GV1}
G^{V}_1=\frac{m}{L^3}\sum_{\mathbf k}\frac{1}{((\mathbf{k}-\frac{\mathbf{p}}{2})^2-q^{*2})^2}.
\end{eqnarray}

\begin{figure}
\begin{center}
\includegraphics[totalheight=5.0cm]{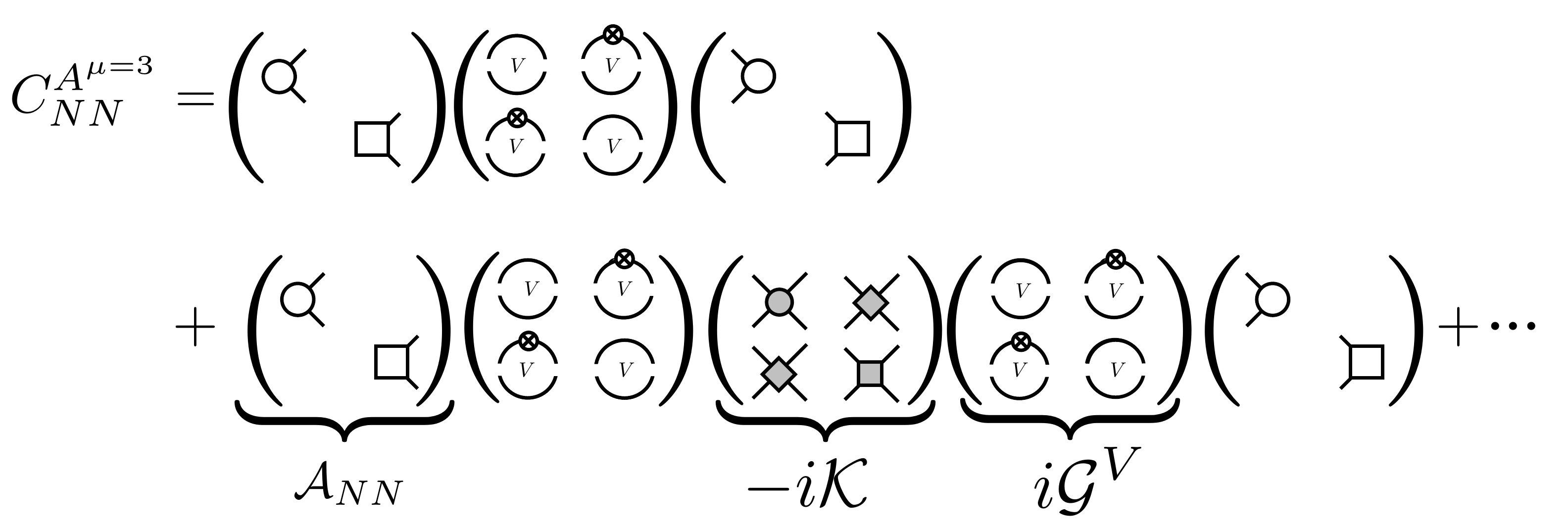}
\end{center}
\caption[NN-correlation function in the isosinglet (isotriplet) channel in the presence of an external weak field]{ Shown is the NN-correlation function in the isosinglet (isotriplet) channel in the presence of an external weak field. $\mathcal{A}_{NN}$ denotes the overlap between the NN-interpolating operators and the vacuum. The two-dimensional kernel is denoted by $\mathcal{K}$. The diagonal terms of the kernel correspond to the strong part of the interactions, while off-diagonal terms depict contributions that arise from the weak interaction, namely $L_{1,A}$. Unlike the scalar sector considered before, the finite-volume function, $\delta\mathcal{G}^V$, has diagonal and off-diagonal contributions due to the presence of the single-body current. }\label{CC_ppfusion}
\par\end{figure}

In order to form the NN correlation function, let us also introduce a diagonal matrix $\mathcal{A}_{NN}$, whose each diagonal element denotes the overlap between the two-nucleon interpolating operators in either isosinglet or isotriplet channels and the vacuum. With theses ingredients, the NN-correlation function in the presence of the external weak field can be easily evaluated, as is diagrammatically presented in Fig.~\ref{CC_ppfusion}. It is important to note that in evaluating the FV loops, one should pay close attention to the pole structure of $G^V_{\pm}$, Eq.~(\ref{GVpm}). In other words, the on-shell condition for the free two-nucleon system is modifies in presence of the single-body weak current, namely, $q^{*2}\rightarrow q^{*2}\pm mW_3g_A$.
Then it is straightforward to show that after keeping only terms up to $\mathcal{O}(C_2q^{*2}W_3g_A,W_3L_{1,A})$, the QC obtained from the pole structure of the NN-correlation function reads
\begin{eqnarray}
\label{1bLONRquant}
\left[q^*\cot\delta_{\singlet}+q^*\cot \phi^P\right]
\left[q^*\cot\delta_{\triplet}+q^*\cot \phi^P\right] 
=\left[\frac{4\pi}{m}W_3{\widetilde{L}_{1,A}}
+\frac{4\pi}{m}W_3g_AG_1^V\right]^2,
\end{eqnarray}
where ${\widetilde{L}_{1,A}}$ that is defined as 
\begin{eqnarray}
{\widetilde{L}_{1,A}}&=&\frac{1}{C^{\left(\triplet\right)}_0C^{\left(\singlet\right)}_0}\left[{{L}_{1,A}}-\frac{g_Am}{2}\left(C^{\left(\triplet\right)}_2+C^{\left(\singlet\right)}_2\right)
\right],
\end{eqnarray}
is a renormalization scale independent quantity~\cite{Butler:1999sv, Butler:2000zp, Butler:2001jj,nnd}
\begin{eqnarray}
\mu \frac{d}{d\mu}{\widetilde{L}_{1,A}}=0.
\end{eqnarray}

Before proceeding let us compare this result with the one presented in Ref.~\cite{nnd}. As discussed, the authors of Ref. \cite{nnd} have evaluated this same quantization condition for two-nucleon systems in presence of an external weak field using a dibaryon formalism. The advantage of this formalism is that the diagrammatic representation of the processes of interest are greatly simplified using an auxiliary field with quantum numbers of two nucleons. In fact, the full di-nucleon propagator sums up all $2\rightarrow2$ interactions non-perturbatively.
In the QC presented in Ref. \cite{nnd}, the contributions of the axial charge current to all orders have been kept, but as the higher order operators that contribute to the weak transition have not been included in their calculation, their result is only valid up to $\mathcal{O}(g_AC_2q^{*2})$~\cite{Butler:1999sv, Butler:2000zp, Butler:2001jj}. In the dibaryon formalism, the two-body weak current is parametrized by $l_{1,A}$ which is related to ${\widetilde{L}_{1,A}}$ in this work via\footnote{Note that Eq.~(31) of Ref.~\cite{nnd}  defines $l_{1,A}$ as $\frac{8\pi}{m}{\widetilde{L}_{1,A}}$, but we suspect this discrepancy is only due to a typo in their result. 
}
\begin{eqnarray}
l_{1,A}=\frac{16\pi}{m}{\widetilde{L}_{1,A}}.
\end{eqnarray}
Using this relation between the LECs of both theories, and keeping in mind the order up to which the resuslt of both calculations are valid, one will find agreement between the result presented in Eq.~(\ref{1bLONRquant}) and that of Ref.~\cite{nnd} after setting the momentum of the CM to zero.

Having obtained the QC for this system, Eq.~(\ref{1bLONRquant}), it is straightforward to obtain the relationship between the FV matrix elements of the Hamiltonian density and the LECs using Lellouch and L\"uscher's trick discussed in section~\ref{EW2B}. In the absence of weak interactions, the two NN-states are assumed to be degenerate with energy $E_0^*$ and on-shell momentum $q^*_0$, satisfying the free quantization condition $\cot(\phi^P)=-\cot(\delta)$. As the weak interaction is turned on, the degeneracy is lifted, leading to a shift in energy equal to $\Delta E^*=V|\mathcal{M}^V_{\singlet-\triplet}|$, where $|\mathcal{M}^V_{\singlet-\triplet}|$ is the FV matrix element of the Hamiltonian density between the $^1S_0$ and $^3S_1$ states. Note that this is matrix element is proportional to $W_3$. Therefore it is convenient to define the purely hadronic matrix element $|\mathcal{M}^V_{W}|=|\mathcal{M}^V_{\singlet-\triplet}|/W_3$ which is in fact what would be calculated using LQCD. Expanding the Eq. (\ref{1bLONRquant}) about the free energy, and keeping LO terms in the weak interaction, one obtains
\begin{eqnarray}
\label{1bNRLL}
\left(\frac{mV}{2}\right)^2\csc^2{\delta_{\singlet}}\csc^2{\delta_{\triplet}}
\left(\phi'_{\singlet}+\delta'_{\singlet}\right)\left(\phi'_{\triplet}+\delta'_{\triplet}\right)|\mathcal{M}^V_{W}|^2
=\left(\frac{4\pi}{m}
{\widetilde{L}_{1,A}}+\frac{4\pi}{m}g_AG_1^V\right)^2.
\end{eqnarray}
This result shows that in order to determine weak matrix elements in the NN-sector,  not only it is necessary to determine the derivatives of the phases shifts in the $\singlet$ and $\triplet$ channels with respect to the on-shell momenta, but also it is necessary to determine the nucleon axial-coupling constant. There is no clear crosscheck for this result, since it is not clear how to implement the density of states approach for this problem. The presence of the one-body operator makes the mixing between the two states non-trivial, therefore one would expect a more complicated relationship between the FV and  infinite volume states than the one predicted via the density of states approach.  Although it would be desirable to obtain a generalization of Lellouch and L\"uscher's result for $2\rightarrow 2$ systems, this example demonstrates that in the two-body sector, one-body currents lead to large finite volume corrections. In principle, the FV matrix elements depend on the nature of the problem that is considered, and each weak hadronic process must be studied separately.

\chapter{Three-boson System in a Finite Volume}{\label{mmmsys}}
Although NN-scattering phase shifts are remarkably well constrained, poor determination of the three-body nuclear force is a large source of systematic error in ab-initio nuclear calculations. Therefore it is desirable to be able to determine the three-body force directly from LQCD calculations. Having studied the two-nucleon sector in chapter~\ref{NNsys} in detail, it would be straightforward to generalize that formalism to include three-body interactions. That being said, the three-body sector is sufficiently complicated on its own that it is practical to first consider the scalar analogue of the three-nucleon problem. With that in mind in this chapter we derive the quantization condition (QC) for the spectrum of a system composed of three identical bosons in a finite volume with periodic boundary conditions. The quantization condition gives a relation between the FV spectrum and infinite volume scattering amplitudes, Eqs. (\ref{STMeq2}), (\ref{QC}). Unlike the two-body sector, this quantization condition in general must be solved numerically, since the relation between the FV energy eigenvalues and three-particle scattering amplitudes is not algebraic. We pay close attention to systems with an attractive two-body force that allows for a two-body bound-state, a \emph{diboson}, and energies below the diboson breakup. For these theories in this energy regime, the quantization condition reduces to the L\"uscher formula with exponential corrections in volume with a length scale that is dictated by the inverse diboson binding momentum. In other words, the boson-diboson scattering phase shifts can be obtained from the three-particle spectrum using the following relation
\begin{eqnarray}
\label{dbQC}
{q}^*_{0}\cot\delta_{Bd} &=&4\pi \ c^P_{00}({q}_{0}^{*})+\eta\frac{e^{-\gamma_d L}}{L} \ ,
\end{eqnarray}
where ${q}_{0}^{*}=\sqrt{\frac{4}{3}\left(mE^*-\bar{q}_{0}^{*2}\right)}$ is the relative momentum in the center of mass (CM) frame of boson-diboson system with $\bar{q}_{0}^{*}$ being the relative momentum of the two bosons in the diboson in the CM frame of the diboson, $m$ is the mass of the three identical particles, $E^*$ is the CM energy, $\gamma_d$ is the binding momentum of the diboson in the infinite volume limit, $\delta_{Bd}$ is the scattering phase shift of the boson-diboson system, $L$ is the spatial extent of the cubic volume, and $\eta$ is an unknown coefficient that must be fitted when extrapolating results to the infinite volume. Given that the diboson is bound, $\bar{q}_{0}^{*2}<0$ and $\bar{q}_{0}^{*2}\rightarrow -\gamma_d^2$ as the volume goes to infinity. $c^{\textbf{P}}_{lm}$ is a kinematic function that is given in Eq.~(\ref{clm}), and for non-relativistic particles $\mathbf{k}^*=\mathbf{k}-\alpha\mathbf{P}$. $\mathbf{P}$ is the total momentum of the boson-diboson system, and $\alpha=\frac{m_1}{m_1+m_2}$ \cite{Bour:2011ef}; so for a diboson that is twice as massive as the boson $\alpha=\frac{1}{3}$. In addition to exponential corrections that are governed by the size of the bound-state wavefunction, there are other exponential volume corrections to the above L\"uscher relation that are arising from the the off-shell states of the 2+1 system. These corrections however are subleading compared to the exponential corrections denoted in Eq. (\ref{dbQC}), and will be discussed in Sec. \ref{recoverL} in more details.

In order to reliably use such an analytical formula, one must necessarily be in the regime where $\gamma_d L$ is at least $4$ so that the infinite volume phase shifts of the bound-state particle scattering can be obtained with a few percent uncertainty. This is an important distinction compared to the the two-body problem where the dominant finite volume corrections to the L\"uscher formula scale like $\sim e^{-m_\pi L}$ where $m_{\pi}$ denotes the mass of the pion (these corrections have been previously calculated for $\pi\pi$~\cite{Bedaque:2006yi} and NN~\cite{Sato:2007ms} systems). So although volumes of the order of $6~\rm{fm}$ or greater would reliably recover, for example, $\pi\pi$ scattering phase shifts at the physical pion mass, in order to accurately recover phase shifts for deuteron-neutron scattering one would naively need $L\gtrsim17~\rm{fm}$.\footnote{Just as L\"uscher's original two-body scalar result~\cite{luscher1, luscher2} can be reliably implemented for studying two-nucleon systems at sufficiently low-energies~\cite{Beane:2003da, Davoudi}, Eq.~(\ref{dbQC}) is expected to hold for near threshold three-nucleon processes. This speculation however remains to be confirmed.} Presumably upon quantifying the coefficients of these exponentials, linear combinations of these exponential corrections can be formed for different boost momenta of the three-particle system so that to cancel out the leading exponential corrections to the above quantization condition, and therefore reduce the size of the volumes needed for a reliable determination of the phase shifts to $L\gtrsim12~\rm{fm}$.\footnote{For a discussion of the improvement of the volume dependence of deuteron binding energy in LQCD calculations see Ref. \cite{Davoudi}.} NLO corrections due to the size of the bound-state scale as $e^{-\sqrt{2}\gamma_d L}/L$. The quantization condition shown in Section \ref{3body} demonstrates that for energies above the diboson breakup Eq. (\ref{dbQC}) gets power-law corrections associated with new possible states that can go on-shell and the quantization condition must be solved numerically.

\section{Three Particles in a Finite Volume: Quantization Condition \label{3body}}
As discussed in detail in section~\ref{dimersec1}, in studying a three-body problem, it is customary to divide the system of three particles to a system of two particles interacting in a given partial-wave $J_{d}$, and a third particle, called the spectator, which interacts with the two-body system with angular momentum $J_{Bd}$. In particular, the dimer formalism is an extremely useful diagrammatic representation of few-body scattering amplitudes which is greatly simplified using the \emph{dimer} field \cite{pionless2, pionless3}. Figure \ref{dimer3} schematically shows how $3\rightarrow 3$ scattering amplitudes are constructed from dimer-boson scattering amplitudes. In this chapter we will simplify the problem even further and truncate the two-particle subsystem to has $J_{d}=0$, and therefore only the S-wave component of the dimer-field discussed in section~\ref{scalardimer} will contribute. This will introduce a systematic error to our calculation that can be corrected.

Consider three bosons with mass $m$ and total energy and momentum $(E,\mathbf{P})$ in the lab frame. The total CM energy of the three-particle system, $E^*$ is then given by $E^{*}=E-\frac{{P}^2}{6m}$. Also the relative momentum of the spectator boson and the dimer in the CM frame of three particles, $\mathbf{q}^{*}$, is related to the momentum of the spectator boson in the lab frame, $\mathbf{q}$ by ${\mathbf{q}}^{*}={\mathbf{q}}-\frac{\textbf{P}}3$. The total CM energy of boson-dimer system can be written as $E^*=\overline{q}^{*2}/m+3{q}^{*2}/4m$, where $\overline{\mathbf{q}}^{*}$ is the relative momentum of the two bosons inside the dimer. 

In section~\ref{scalardimer} we discussed the generalization of the dimer formalism to arbitrary partial waves. By truncating the orbital angular momentum to be equal to zero, the dimer propagator in Eq.~(\ref{D-infinity}) reduces to  
\begin{eqnarray}
\label{dimerprop}
i\mathcal{D}^{\infty}(E_2,{q}^*)=
\frac{-imr/2}{\overline{q}^{*}\cot{\delta_d}-i\overline{q}^{*}+i\epsilon} \ ,
\end{eqnarray}
where $E_2=E^*-q^{*2}/2m$ is the total energy of the dimer. $\delta_d$ denotes the S-wave scattering phase shift of the two-boson system, and $r$ is its effective range. Similarly, by truncating the angular momentum, the FV dimer propagator, Eq.~(\ref{D-FV}) simplifies to 
\begin{eqnarray}
\label{dimerprop}
i\mathcal{D}^{V}(E_2,{q}^*)=
\frac{-imr/2}{\overline{q}^{*} \cot{\delta_d}-4\pi \ c^{{q}^*}_{00}(\overline{q}^{*2}{+i\epsilon}) +i\epsilon} \ ,
\end{eqnarray}
where the kinematic function $c^{\textbf{q}^*}_{lm}$ is defined in Eq. (\ref{clm}).\footnote{It is important to point out that Refs. \cite{Kreuzer:2008bi, Kreuzer:2009jp, Kreuzer:2010ti, Kreuzer:2012sr} used a propagator that corresponds to a dimer at rest. In  future investigation of FV dependence of the Efimov bound states and the triton this needs to be corrected.}
For the case of a dimer composed of identical bosons, one therefore has $\mathbf{k}^*=\mathbf{k}-\frac{\mathbf{q^*}}{2}$.%

The spectrum in a finite volume can be obtained from the poles of the two-particle propagator or equivalently from the poles of the finite volume dimer, Eq. (\ref{clm}),
\begin{eqnarray} 
\label{2QCV}
\overline{q}^{*}_{\kappa} \cot{\delta_d}=4\pi \ c^{{q}^*}_{00}(\overline{q}_{\kappa}^{*2}) \ ,
\end{eqnarray}
where $\overline{q}_{\kappa}$ is the $\kappa^{th}$ solution to the quantization condition for a boosted two particle system \cite{movingframe, sharpe1}. As will be discussed in great details, these poles play an important role in the three-body sector and will be referred to as \emph{L\"uscher} poles. Note that this result is equivalent to the non-relativistic limit of the result obtained in Refs. \cite{movingframe, sharpe1, Christ:2005gi} for the boosted systems of particles with identical masses, and for that of systems with unequal masses~\cite{Davoudi, Fu:2011xz, Leskovec:2012gb}.
 
\begin{figure}[t!]
\begin{center}
\subfigure[]{
\includegraphics[scale=0.475]{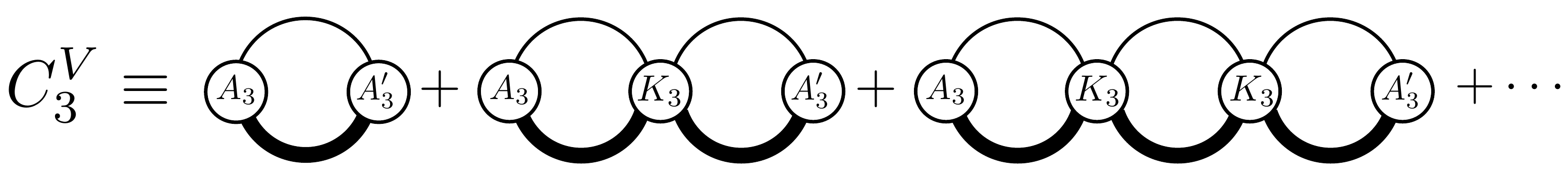}
\label{correlation3}}
\subfigure[]{
\includegraphics[scale=0.35]{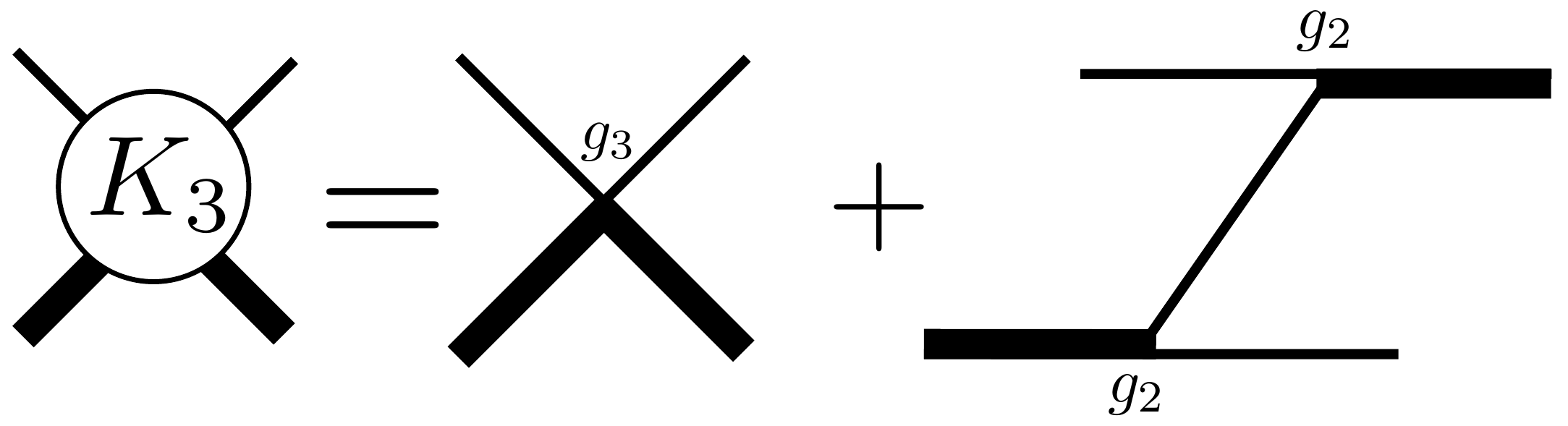}
\label{kernel3}}
\caption[Dimer-boson correlation function in a finite volume]{\label{fig:corrfunc} (a) Diagrammatic expansion of the three-body correlation function $C_3^V$ in the finite volume. $K_3$ denotes the three-body Bethe-Salpeter kernel and $A_3$ $(A'_3)$ is the overlap the creation (annihilation) dimer-boson interpolating operator has with the initial (final) state with total energy $E$ and total momentum $\mathbf{P}$, (b) The effective three-body Bether-Salpeter kernel, $K_3$, is composed of a three-body contact interaction as well as two-body contact interactions via the exchange of a single boson.}
\end{center}
\end{figure}

In order to determine the energy eigenvalues of the three-particle system in a finite volume, one can solve for the poles of the three-particle correlation function as depicted in Fig.~\ref{correlation3}. Algebraically,
\begin{eqnarray}
C^V_3(E,\textbf{P})&=&
\frac{1}{L^3}\sum_{\textbf{q}_1}
A_3\left(\mathbf{q}_1\right){i\mathcal{D}^V(E-\frac{{q}^2_1}{2m},|\textbf{P}-\textbf{q}_1|)} \nn\\
&& \times \left[1+\sum_{n'=2}^\infty  \prod_{n=2}^{n'} \left(\frac{1}{L^3} \sum_{\textbf{q}_n}
iK_3(\textbf{q}_{n-1},\textbf{q}_{n};\mathbf{P},E){i\mathcal{D}^V(E-\frac{{q}^2_n}{2m},|\textbf{P}-\textbf{q}_n|)}\right)\right]A'_3\left(\mathbf{q}_{n'}\right) \ ,\nn\\\label{corr3}
\end{eqnarray}
where $A'_3$ $(A_3)$ is the overlap the annihilation (creation) dimer-boson interpolating operator, $\sigma_3$ $(\sigma_3^\dag)$, has with the initial (final) state with total energy $E$ and total momentum $\mathbf{P}$. Note that we have suppressed the total energy and momentum dependence of the overlap factors in our notation. The interactions between three bosons are incorporated in an effective three-body Bethe-Salpeter kernel, $K_3$, Fig.~\ref{kernel3},
\begin{eqnarray}
iK_3(\textbf{p},\textbf{k};{\mathbf{P}},E)&\equiv&-{ig_3}-\frac{i{g_2^2}}{E-\frac{\textbf{p}^2}{2m}-\frac{\textbf{k}^2}{2m}-\frac{(\textbf{P}-\textbf{p}-\textbf{k})^2}{2m}+i\epsilon} \ ,
\label{Kernel}
\end{eqnarray}
where the incoming (outgoing) boson has momentum $\mathbf{p}$ $(\mathbf{k})$ and the incoming (outgoing) dimer has momentum $\mathbf{P}-\mathbf{p}$ $(\mathbf{P}-\mathbf{k})$, and $(E,\mathbf{P})$ denote the total energy and momentum of the three-particle system as before. The first term in the kernel, Eq. (\ref{Kernel}), is the three-body contact interaction, while the second term describes the interaction of three particles via exchange of an intermediate particle through two-body contact interactions.

The finite volume contribution to the first term in the expansion of the three-body correlation function, Eq. (\ref{corr3}), can be evaluated easily using the Poisson resummation formula and the kinematic relations between the CM and lab frame momenta as presented earlier,
\begin{eqnarray}
\left\{\frac{1}{L^3}\sum_{\textbf{q}_1}-\int\frac{d^3\textbf{q}_1}{(2\pi)^3}\right\}
A_3\left(\mathbf{q}_1\right){i\mathcal{D}^V(E-\frac{{q}^2_1}{2m},|\textbf{P}-\textbf{q}_1|)}A'_3\left(\mathbf{q}_1\right) =
\sum^{N_{E^*}}_{\kappa}A_3({q}_{\kappa}^{*}) \ i \delta\tilde{\mathcal{G}}^V_\kappa({q}_{\kappa}^{*})A'_3({q}_{\kappa}^{*}) \  .\nn\\\label{firstCorr}
\end{eqnarray}
While $A_3$ and $A'_3$ in the LHS of Eq. (\ref{firstCorr}) are functions of the relative coordinate $\mathbf{q}_1$, they are represented as vectors in the space of the boson-dimer angular momentum, $J_{Bd}$, in the RHS, and are evaluated at the poles of the dimer propagator, ${q}^{*}_{\kappa}$, and the sum runs over these poles. $\delta \tilde{\mathcal{G}}^{V}_{\kappa}$ is a matrix in the same angular momentum basis whose elements are defined by 
\begin{eqnarray}
\label{gtilde}
(\delta\tilde{\mathcal{G}}^V_\kappa)_{l_1m_1,l_2m_2}&\equiv&
\frac{R^V_\kappa}{m}(\delta\mathcal{G}^V_\kappa)_{l_1m_1,l_2m_2} \ ,
\end{eqnarray}
with
\begin{eqnarray}
(\delta{\mathcal{G}}^V_\kappa)_{l_1m_1,l_2m_2}&=&i\frac{m~{q}_{\kappa}^{*}}{3\pi} \left(\delta_{l_1,l_2}\delta_{m_1,m_2}+i\sum_{lm}\frac{\sqrt{4\pi}}{{q}_{\kappa}^{*l+1}}c^{\textbf{P}}_{lm}({q}_{\kappa}^{*})\int d\Omega~Y^*_{l_1,m_1}Y^*_{l,m}Y_{l_2,m_2}\right) \ .\nn\\\label{dgv} 
\end{eqnarray}
The kinematic function $c^{\textbf{P}}_{lm}$ is defined in Eq. (\ref{clm}) with $\alpha=\frac{1}{3}$, since the dimer is twice as massive as the boson. The on-shell CM momentum of the boson-dimer system, ${q}^{*}_{\kappa}$, is defined by $\kappa^{th}$ pole of the FV dimer propagator, $\overline{q}^{*2}_{\kappa}=mE^*-\frac{3}{4}{q}^{*2}_{\kappa}$, and $R^{V}_{\kappa}$ is its residue at the $\kappa^{th}$ pole. Explicitly,
\begin{eqnarray}
\label{residue}
\lim_{\overline{q}^{*2}\rightarrow {\overline{q}}_{\kappa}^{*2}}i\mathcal{D}^V(E-\frac{{q}^2}{2m},|\textbf{P}-\textbf{q}|)
\approx \frac{iR^V_\kappa}{\overline{q}^{*2}-{\overline{q}}_{\kappa}^{*2}+i\epsilon}
=
-\frac{4}{3}\frac{iR^V_\kappa}{{q}^{*2}-{q}_{\kappa}^{*2}-i\epsilon} \ ,
\end{eqnarray}
where
\begin{eqnarray}
\label{residuedef}
R^{V}_{\kappa}=-\frac{mr}{2} \left[\left. \frac{\partial}{\partial \overline{q}^{*2}}\left(\overline{q}^{*}\cot\delta_d-{4\pi \ {c^{|\mathbf{P}-\mathbf{q}|}_{00}\left(\overline{q}^{*2}\right)}}\right)\right|_{\overline{q}^{*2}={\overline{q}}_{\kappa}^{*2}}\right]^{-1} \ .
\end{eqnarray}
Note that the poles of the FV dimer propagator correspond to the energy eigenvalues of the boosted two-particle system in the finite volume, Eq. (\ref{2QCV}), i.e. the L\"uscher poles.

\begin{figure}[t]
\begin{center}
\includegraphics[scale=0.375]{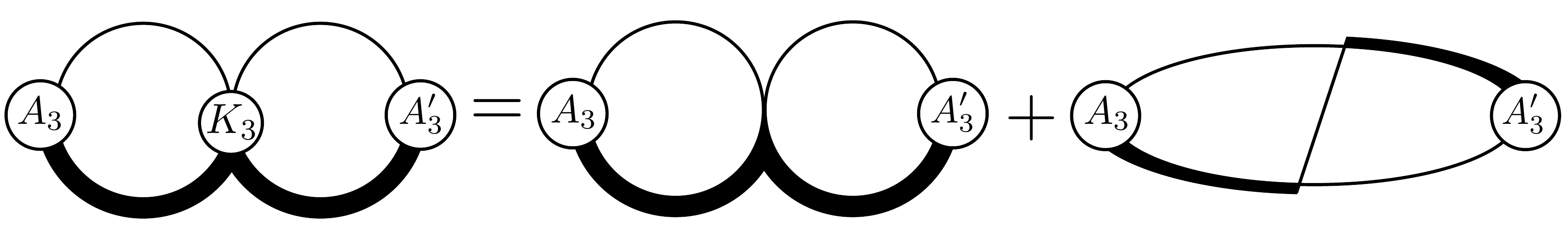}
\caption[NLO contribution to the dimer-boson correlation function]{\label{fig:twoloop} The NLO contribution to the three-body correlation function with one insertion of the three-body kernel.}
\end{center}
\end{figure}

Equation (\ref{firstCorr}) reflects the fact that, unlike the two-body case, a single on-shell condition does not simultaneously fix the relative momenta of the boson-dimer pair as well as that of the bosons inside the dimer. What is important to observe here is that for a given CM energy $E^*$, there is only a finite number of ``channels" $N_{E^*}$ that can go on-shell, each being identified by a particular configuration of the boson-dimer relative momentum and the relative momentum of the two bosons (of the dimer),
{\small
\begin{eqnarray}
\label{conf}
\left\{\overline{q}^{*}_{\kappa},q^{*}_{\kappa}\right\}
=\left\{\left(\overline{q}^{*}_{0},\sqrt{\frac{4}{3}(mE^*-\overline{q}^{*2}_{0})}\right),\left(\overline{q}^{*}_{1},\sqrt{\frac{4}{3}(mE^*-\overline{q}^{*2}_{1})}\right),\dots,\left(\overline{q}^{*}_{N_{E^*}},\sqrt{\frac{4}{3}(mE^*-\overline{q}^{*2}_{N_{E^*}})}\right)\right\} \ .
\end{eqnarray}}
These channels contribute to the quantization condition since $\overline{q}^{*2}_{\kappa}/m<E^*$. This observation makes the analogy to the ``coupled-channel" systems self-evident. For CM energies that are below the dimer energy, $E^*<\overline{q}^{*2}_{\kappa}/m$, the energy is not sufficient to allow the three-particle system to go on-shell. Subsequently, these states can be neglected as they give rise to exponential corrections in volume rather than power-law. Furthermore, similar to the two-body case, the on-shell condition does not fix the directional degrees of freedom of the relative momentum of the $2+1$ system, and therefore it is convenient to upgrade all finite volume quantities into infinite-dimensional matrices in angular momentum.

The calculation of the second term in the expansion of the correlation function, Eq. (\ref{corr3}), is more involved as it comes with one insertion of the three-body kernel, Fig.~\ref{fig:twoloop}, and due to the one boson exchange contribution couples the momenta running into the loops,
\begin{eqnarray}
C^V_{3,1}(E)&=&
\frac{1}{L^6}\sum_{\textbf{q}_1,\textbf{q}_2}
A_3\left(\mathbf{q}_1\right){i\mathcal{D}^V(E-\frac{{q}^2_1}{2m},|\textbf{P}-\textbf{q}_1|)} \nn\\
\nn
& & \times \left[-ig_{3}-\frac{i{g_2^2}}{E-\frac{\textbf{q}_1^2}{2m}-\frac{\textbf{q}_2^2}{2m}-\frac{(\textbf{P}-\textbf{q}_1-\textbf{q}_2)^2}{2m}+i\epsilon}\right]
{i\mathcal{D}^V(E-\frac{{q}^2_2}{2m},|\textbf{P}-\textbf{q}_2|)}A'_3\left(\mathbf{q}_2\right).\\\label{Cexd0}
\end{eqnarray}
Although at the first glance, there appears to be poles arising from the exchange boson propagator, one can verify that the poles of the three-body kernel are exactly canceled by the zeros of the full finite volume dimer propagator\footnote{This important observation was first pointed out to us by Michael D\"oring and Akaki Rusetsky for the relativistic three-particle systems \cite{RusetskyDoring}.}. As a result, the only power law volume dependence of such diagrams arise from the poles of the dimer propagator only. Given this observation, it is straightforward to show that
{\small
\begin{eqnarray}
C^V_{3,1}(E)&=& 
\int\frac{d^3\textbf{q}_1}{(2\pi)^3}\frac{d^3\textbf{q}_2}{(2\pi)^3}
A_3\left(\mathbf{q}_1\right){i\mathcal{D}^V(E-\frac{{q}^2_1}{2m},|\textbf{P}-\textbf{q}_1|)}
iK_3(\textbf{q}_1,\textbf{q}_2;\mathbf{P},E)
{i\mathcal{D}^V(E-\frac{{q}^2_2}{2m},|\textbf{P}-\textbf{q}_2|)}A'_3\left(\mathbf{q}_2\right)
\nn\\
& - & 2\int\frac{d^3\textbf{q}_1}{(2\pi)^3} \ A_3\left(\mathbf{q}_1\right) {i\mathcal{D}^V(E-\frac{{q}_1^2}{2m},|\textbf{P}-\textbf{q}_1|)} 
 \sum^{N_{E^*}}_{\kappa}\left[K_{3} (\mathbf{q}_1,{q}^*_{\kappa};E^*)\delta\tilde{\mathcal{G}}^V_\kappa({q}_{\kappa}^{*})A'_3(q^*_{\kappa})\right]
\nn\\
\label{Cexd}
& - &
\sum^{N_{E^*}}_{\kappa,\kappa'}
\left[A_3(q^*_{\kappa'})\delta\tilde{\mathcal{G}}^V_{\kappa'}({q}_{\kappa'}^{*})iK_{3} ({q}^*_{\kappa'},{q}^*_{\kappa};E^*)\delta\tilde{\mathcal{G}}^V_\kappa({q}_{\kappa}^{*})A'_3(q^*_{\kappa})\right] \ ,
\end{eqnarray}}
where a summation over angular momentum is understood for the terms inside the brackets. The summation over the two-body L\"uscher poles is left explicit.
The result of Eq. (\ref{Cexd}), along with the fact that the dimer propagator can be decomposed in a series over its poles,
\begin{eqnarray}
\label{dimerdecomp}
i\mathcal{D}^{V}(mE-3{q}^2/4,{q})
=\sum^{N_{E^*}}_{\kappa}\frac{iR^V_\kappa}{\overline{q}^{*2}-{\overline{q}}_{\kappa}^{*2}+i\epsilon} \ ,
\label{decomp}
\end{eqnarray}
suggests that the dimer propagator can be upgraded unto a diagonal matrix in the space of $N_{E^*}$ available FV states which is a useful representation when performing the sum over all diagrams contributing to the correlation function. Each element of this matrix is then effectively a single particle propagator with the corresponding FV pole and residue that contain finite volume dependence of the propagators,
 \begin{eqnarray}
 \label{dimerM}
 \left[i\mathcal{D}^{V}(mE-3{q}^2/4,{q})\right]_{\kappa \kappa'}
=\frac{iR^V_\kappa}{\overline{q}^{*2}-{\overline{q}}_{\kappa}^{*2}+i\epsilon} \ \delta _{\kappa \kappa'} \ .
\label{propmatx}
\end{eqnarray}
\begin{figure}[t]
\begin{center}
\subfigure[]{
\includegraphics[scale=0.325]{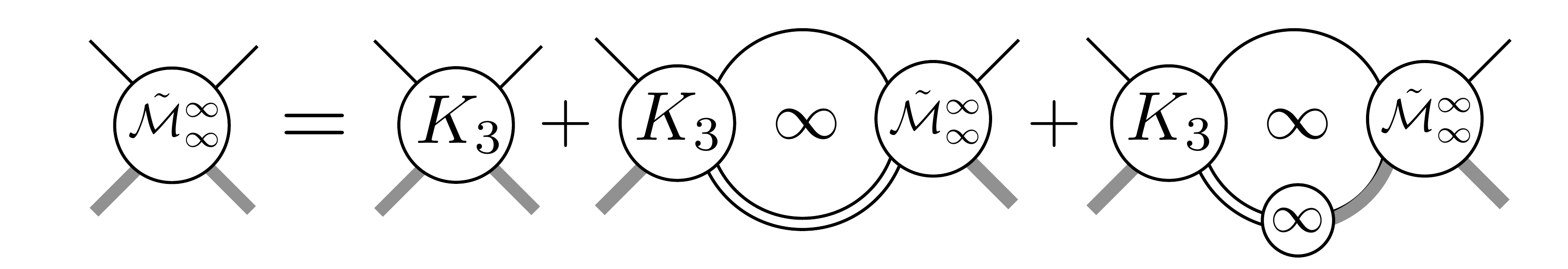}
\label{MII}}
\subfigure[]{
\includegraphics[scale=0.325]{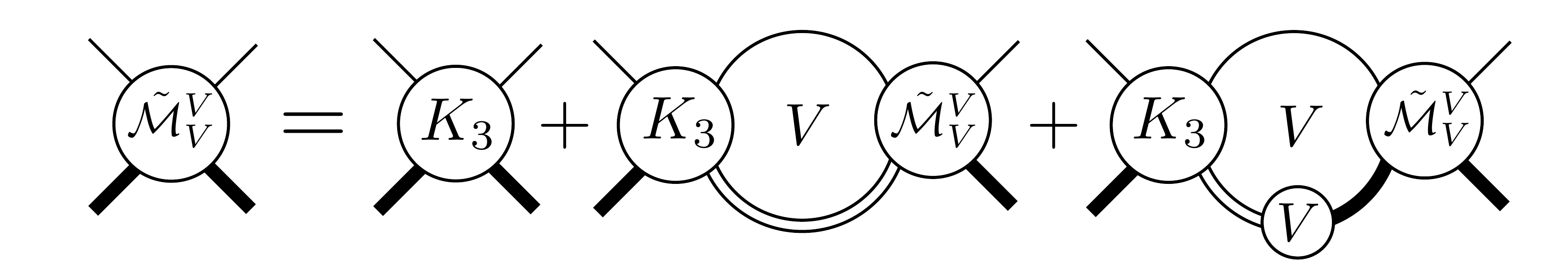}
\label{MVV}}
\subfigure[]{
\includegraphics[scale=0.325]{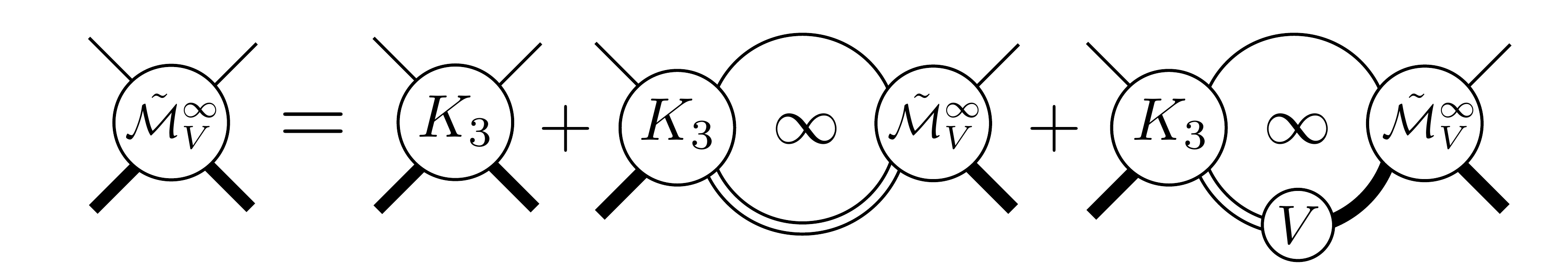}
\label{MVI}}
\caption[STM equation in finite and infinite volumes]{\label{fig:scattamp} (a) Diagrammatic representation of the inhomogenous integral equation satisfied by the three-body scattering amplitude in the infinite volume, (b) The corresponding sum equation satisfied by the FV scattering amplitude, (c) The diagrammatic representation of the integral equation satisfied by $\tilde{\mathcal{M}}^{\infty}_{V}$, Eq. (\ref{STMeq2}).}
\end{center}
\end{figure}

Given the simplifying feature of the FV loop sums as is evident from Eq. (\ref{Cexd}), and the representation of the FV dimer a matrix over available channels, Eq. (\ref{dimerM}), it is straightforward to sum over the infinite number of terms appearing in the boson-dimer correlation function, Eq. (\ref{corr3}). Denoting the boson-FV dimer propagator as $\mathcal{G}^{\infty}_{V}$, one can show that Eq. (\ref{corr3}) is equal to
\begin{eqnarray}
C^V_3(E,\textbf{P})-C^{\infty}_{3,V}(E,\textbf{P})&=&iA_3\left[(1-\tilde{\mathcal{G}}^{\infty}_{V}\tilde{\mathcal{M}}^{\infty}_{V}) \delta \tilde{\mathcal{G}}^V\frac{1}{1+\tilde{\mathcal{M}}^{\infty}_{V}\delta\tilde{\mathcal{G}}^V}(1-\tilde{\mathcal{M}}^{\infty}_{V}\tilde{\mathcal{G}}^{\infty}_{V})\right]A'_3 \nn,\\
\label{kernelexp}
\end{eqnarray}
where $C^{\infty}_{3,V}(E,\textbf{P}) \equiv iA_3\tilde{\mathcal{G}}^{\infty}_{V}{(1-\tilde{\mathcal{M}}^{\infty}_{V}\tilde{\mathcal{G}}^{\infty}_{V})}A'_3$. $\tilde{\mathcal{M}}^{\infty}_{V}$ is defined as the sum over all infinite volume diagrams containing a boson and a finite volume dimer, Fig.~\ref{MVI}, and can be interpreted as the non-renormalized infinite volume scattering amplitude between a boson and a FV dimer.\footnote{The difference between renormalized and non-renormalized scattering amplitudes will be explained shortly.} That is to say, while the relative momentum between the dimer and boson is continuous in $\tilde{\mathcal{M}}^{\infty}_{V}$, the relative momentum of the two bosons inside the dimer remains discretized,
{\small
\begin{eqnarray}
\tilde{\mathcal{M}}^{\infty}_{V}\left(\mathbf{p},\mathbf{k};\mathbf{P},E\right)& = & K_3(\textbf{p},\textbf{k};{\mathbf{P}},E) -
\int\frac{d^3q}{(2\pi)^3}K_3(\textbf{p},\textbf{q};{\mathbf{P}},E){\mathcal{D}^{V}(E-\frac{q^2}{2m},|\mathbf{P}-\mathbf{q}|)} \tilde{\mathcal{M}}^{\infty}_{V}\left(\mathbf{q},\mathbf{k};\mathbf{P},E\right) \nn\\
& = & 
 \tilde{\mathcal{M}}^{\infty}_{\infty}(\textbf{p},\textbf{k};{\mathbf{P}},E) -
\int\frac{d^3q}{(2\pi)^3}\tilde{\mathcal{M}}^{\infty}_{\infty}\textbf{p},\textbf{q};{\mathbf{P}},E)\delta{\mathcal{D}^{V}(E-\frac{q^2}{2m},|\mathbf{P}-\mathbf{q}|)} \tilde{\mathcal{M}}^{\infty}_{V}\left(\mathbf{q},\mathbf{k};\mathbf{P},E\right) \nn\\
\label{STMeq2}
\end{eqnarray}}
where $\delta\mathcal{D}^{V}=\mathcal{D}^{V}-\mathcal{D}^{\infty}$ and $\tilde{\mathcal{M}}^{\infty}_{\infty}$ is the non-renormalized infinite volume scattering amplitude, Fig.~\ref{MII}. For comparison, the full FV scattering amplitude of the dimer-boson system is also depicted in Fig.~\ref{MVV}, where all relative momenta between the three-particles are necessarily discretized. Note that the boson-dimer propagator and $\tilde{\mathcal{M}}^{\infty}_{V}$ are evaluated on-shell where the scattering energies of each boson-dimer channel is given by ${q}^{*2}_{\kappa}/m$.

The poles of the FV correlation function, Eq. (\ref{kernelexp}), determine the spectrum, 
\begin{eqnarray}
\label{QC}
\rm{Det}(1+\tilde{\mathcal{M}}^{\infty}_{V}\delta\tilde{\mathcal{G}}^V)=0 \ ,
\end{eqnarray}
where the determinant $\rm{det}_{\rm{oc}}$ is over the $N_{E^*}$ open channels and the determinant $\rm{det}_{\rm{pw}}$ is over the boson-dimer relative angular momentum. In practice it is necessary to perform a truncation over the partial-waves and choose a maximal angular momentum. This quantization condition however incorporate the partial-wave mixing due to the reduced symmetry of the boson-dimer wavefunction in the finite cubic volume as will be discussed in more details in the next section.

The reason that the scattering amplitude quantities introduced above are not renormalized is that unlike single particle operators in a non-relativistic field theory, the dimer field corresponds to an interpolating operator that has overlap with two-particle states, and as a result must be renormalized \cite{Chen:1999tn, Bedaque:1998km}. The renormalization factor in the finite volume can be obtained from the residue of the FV dimer propagator,
\begin{eqnarray}
\label{ZV}
(\mathcal{Z}^{V}_\kappa)^{-1}=i\left.\frac{\partial}{\partial E^*}\frac{1}{i\mathcal{D}^V\left(E-\frac{{q}^2}{2m},|\textbf{P}-\textbf{q}|\right)}\right|_{E^*=\frac{\bar{q}^{*2}_{\kappa}}m+\frac{3q^{*2}_{\kappa}}{4m}}
=\frac{m}{R^V_\kappa} \ .
\end{eqnarray}
Upon renormalizing the dimer field, therefore, one arrives at the normalized scattering amplitudes in the finite volume, e.g. $(\mathcal{M}^{\infty}_{V})_{\kappa \kappa'}=(\mathcal{Z}^{V}_\kappa)^{1/2} (\tilde{\mathcal{M}}^{\infty}_{V})_{\kappa \kappa'} (\mathcal{Z}^{V}_{\kappa'})^{1/2}$.

The quantization condition, Eq. (\ref{QC}), resembles that of the two-body coupled-channel systems as presented in Refs. \cite{Hansen:2012tf, Briceno:2012yi}. As discussed earlier this illustrates that a single on-shell condition does not fix the magnitude of both relative momenta and there is a freedom in scattering in any of \emph{finite} number of available channels. The other characteristic of Eq. (\ref{QC}) is that it does not still provide a algebraic relation between the infinite volume scattering amplitude and the energy eigenvalues of the boson-diboson system, simply because $\tilde{\mathcal{M}}^{\infty}_{V}$ still has possibly large volume corrections arising from FV dimer propagator, Eq. (\ref{STMeq2}). Despite all these complexities, this quantization condition not only gives better insight into the three-body problem in a finite volume, it automatically reduces to the L\"uscher quantization condition for the bound-state-particle scattering below the bound-state breakup, up to exponential corrections that are due to the size of the bound-state wave-function. This will be discussed in the next section in more details.

\section{Boson-diboson scattering below the breakup threshold \label{recoverL}}
The formalism developed in the previous section does not assume any specific form for the interactions in the three-body system. Therefore the result presented is universal regardless of the nature of the interactions or whether the theory contains any number of two-body or three-body bound-states. In this section, though, we consider a theory with an attractive two-body force which allows a two-body bound-state, a \emph{diboson}. We will show how Eq. (\ref{QC}) reduces to the well-known two-body result below the diboson breakup.

For such energies there is only one state that can go on-shell and introduce power-law volume corrections, the boson-diboson state. By restricting to $l_{\rm max}=0$, the low-energy parametrization of the scattering amplitude becomes that of a two-particle system in an S-wave with masses $m_1=\tfrac{m_2}{2}=m$,
\begin{eqnarray}
\label{Bdscattamp}
{\mathcal{M}}_{Bd}&=&\frac{3\pi}{m}\frac{1}{{q}^*_{0}\cot\delta_{Bd}-i{q}^*_{0}},
\end{eqnarray}
where ${q}_{0}^{*2}/m\equiv \frac{4}{3}(E^*-\bar{q}^{*2}_0/m)$ is the boson-diboson scattering energy in the CM frame, $\bar{q}^{*2}_0/m$ is the boosted diboson FV binding energy, and $\delta_{Bd}$ denotes the boson-diboson scattering phase-shift. However, this is the non-renormalized quantity $\tilde{\mathcal{M}}^{\infty}_{V}$ that appears in the QC, Eq. (\ref{QC}), and not the physical scattering amplitude. Here we argue that by introducing a systematic error of the order of $e^{-\gamma_d L}/L$ to the final result, Eq. (\ref{dbQC}), the scattering phase shifts can be derived from the QC, Eq. (\ref{QC}), after replacing $\delta\tilde{\mathcal{G}}^V\tilde{\mathcal{M}}^{\infty}_{V}$ with $\delta\mathcal{G}^V\mathcal{M}^{\infty}_{\infty}\equiv\delta\mathcal{G}^V{\mathcal{M}}_{Bd}$. $\gamma_d$ denotes the infinite volume binding momentum of the diboson which satisfies,
\begin{eqnarray}
\label{2QCI}
\left. \left(\overline{q}^{*}\cot{\delta_d}-i\overline{q}^{*}\right)\right|_{\overline{q}^{*}=i\gamma_d}=0 \ .
\end{eqnarray}

The first step to prove this claim is to note that the bound-state pole of the FV dimer propagator is exponentially close to the bound-state pole of the infinite volume dimer, $\bar{q}^{*}_0=i\gamma_d+\mathcal{O}(e^{-\gamma_d L}/L)$,
which is evident from Eq. (\ref{2QCV}) after analytically continuing the momentum $\tilde{q}^*_0$ to the imaginary axis. These exponential corrections have been previously calculated for two-body bound states in the CM frame \cite{luscher1, luscher2, Beane:2003da, Sasaki:2006jn} as well as moving frames \cite{Bour:2011ef, Davoudi}. Now in evaluating $\tilde{\mathcal{M}}^{\infty}_{V}$, one needs to perform a series of coupled integrals as is given in the first line of Eq. (\ref{STMeq2}). For negatives energies, the only singularity of the integrands in the range of integration occurs when the diboson pole of the FV dimer propagators is reached. The contribution to the integrals due to this singularity is proportional to the residue of the FV dimer at that pole. Since the residue of the infinite volume dimer propagator at the diboson pole,
\begin{eqnarray}
\label{residuedef}
R^{\infty}_{d}=-\frac{mr}{2} \left[\frac{\partial}{\partial \overline{q}^{*2}}\left(\overline{q}^{*}\cot\delta_d-i\overline{q}^{*}\right)|_{\overline{q}^{*2}=-\gamma_d^2}\right]^{-1} \ ,
\end{eqnarray}
is exponentially close to its FV counterpart, Eq. (\ref{residuedef}),\footnote{There is another correction to the residue function at the diboson pole that occurs at $\mathcal{O}\left(e^{-\gamma_d L} / \gamma_d L\right)$. Since for $\gamma_dL\sim 1$ the diboson does not fit in the volume, and the finite volume formalism is no longer valid, one must make sure to use sufficiently large volumes for shallow bound-states so that $\gamma_dL \gg 1$. It then follows that these corrections are subleading compared to the $\mathcal{O}(e^{-\gamma_d L})$ corrections in Eq. (\ref{RVRI}) and could be ignored.}
\begin{eqnarray}
\label{RVRI}
R^{V}_{d}=R^{\infty}_{d}\left[1+\mathcal{O}(e^{-\gamma_d L})\right] \ ,
\end{eqnarray}
one can replace $\mathcal{D}^V$ with $\mathcal{D}^{\infty}$ up to the exponential corrections that scale by the size of the bound-state wave-function. Consequently, from Eq. (\ref{STMeq2}) one observes that $\tilde{\mathcal{M}}^{\infty}_{\infty}$ is equal to $\tilde{\mathcal{M}}^{\infty}_{V}$ up to exponentially small corrections. Note that $\tilde{\mathcal{M}}^{\infty}_{V}$ and $\tilde{\mathcal{M}}^{\infty}_{\infty}$ are renormalized differently, however, the finite volume dimer field renormalization factor $\mathcal{Z}^V$, Eq. (\ref{ZV}) is exponentially close to the renormalization factor of the infinite volume diboson field $\mathcal{Z}^{\infty}$ around the bound-state pole, which is defined as
\begin{eqnarray}
\label{ZI}
(\mathcal{Z}^{\infty}_d)^{-1}=i\left.\frac{\partial}{\partial E^*}\frac{1}{i\mathcal{D}^{\infty}\left(E-\frac{{q}^2}{2m},|\textbf{P}-\textbf{q}|\right)}\right|_{E^*=-\frac{\gamma_d^2}m+\frac{3q^{*2}_{d}}{4m}}
=\frac{m}{R^{\infty}_d} \ .
\end{eqnarray}
Therefore one can approximate $\delta\tilde{\mathcal{G}}^V\tilde{\mathcal{M}}^{\infty}_{V}=\delta\mathcal{G}^V\mathcal{M}^{\infty}_{V}$ in Eq. (\ref{QC}), with $\delta\mathcal{G}^V\mathcal{M}^{\infty}_{\infty}$ for elastic processes that occur in this energy regime up to these exponential corrections as stated.

Keeping in mind these exponential corrections, one can now apply the expression for the scattering amplitude in Eq. (\ref{Bdscattamp}), to Eq. (\ref{QC}). Using Eqs. (\ref{gtilde}), (\ref{dgv}), one can recover the two-particle quantization condition for two-particle systems up to the exponential corrections explained above as shown in Eq. (\ref{dbQC}) and reiterated here for clarity
\begin{eqnarray}
\label{dbQC2}
{q}^*_{0}\cot\delta_{Bd} &=&4\pi \ c^P_{00}({q}_{0}^{*})+\eta\frac{e^{-\gamma_d L}}{L}.
\end{eqnarray}
This result confirms the postulate and numerical verification made by Bour \emph{et al.}~\cite{Bour:2012hn} that upon subtracting off the FV binding energy of the bound-state from the total energy of the three-particle system, the scattering energy eigenvalues of the bound-state-particle system can be reliably related to the scattering phase shift of the system through the use of L\"uscher formula for two-body systems after extrapolating to the infinite volume limit. This offers the practitioner a reliable method to extract the infinite volume phase shift of elastic bound-state-particle scattering by fitting to an exponential form.

This result also illustrates that, in order to obtain the boson-diboson scattering phase shift, not only does one need to determine the boosted three-particle energy spectrum, but also needs to obtain the scattering parameters of the boosted two-particle system. It is also evident that if the interactions support a boson-diboson bound-state, a \emph{triboson}, after analytically continuing the momentum in Eq. (\ref{dbQC}) to the imaginary plane, ${q}^*_{0}=i\gamma_{Bd}$, the binding energy of the three-particle system, $B_3=\frac{3\gamma_{Bd} ^2}{4m}$, can be obtained easily via Eq. (\ref{dbQC}), as is well-known for bound-states appearing in the two-body sector~\cite{luscher1, luscher2, Beane:2003da}. Alternatively, one can also solve for the triboson poles of the FV scattering amplitude from the FV counterpart of the Skorniakov and Ter-Martirosian (STM) equation, Fig.~\ref{MVV}, as is pursued in Refs.~\cite{Kreuzer:2008bi, Kreuzer:2009jp, Kreuzer:2010ti, Kreuzer:2012sr}. 

The boson-diboson QC, Eq. (\ref{dbQC}), is a low-energy approximation of Eq. (\ref{QC}), which at NLO has two sources of exponential corrections. First, the QC receive corrections associated with the finite volume binding momentum of the diboson which scale like $\mathcal{O}(e^{-\sqrt{2}\gamma_d L}/L)$ at next to leading order. It also acquires exponential corrections associated with the truncation of off-shell states appearing in the decomposition of the dimer propagator, Eq. (\ref{dimerdecomp}), as mentioned before. More explicitly, the next excited state of the boson-diboson system corresponds to a CM scattering energy of ${q}_{1}^{*2}/m\equiv\frac{4}{3}(E^*-\bar{q}^{*2}_1/m)$, where $\bar{q}^{*}_1$ is the boosted momentum for an unbound two-boson system. For $E^*<\bar{q}^{*2}_1$/m, the three-boson scattering energy, ${q}_{1}^{*2}/m$, is negative which leads to exponential corrections of $\mathcal{O}\left(e^{-{{q}_{1}^{*}}L}/L\right)$ to the single-channel QC, Eq. (\ref{dbQC}), which however are subleading compared to $\mathcal{O}\left(e^{-{{q}_{0}^{*}}L}/L\right)\sim\mathcal{O}(e^{-\gamma_d L}/L)$ corrections. For sufficiently high energies, these exponential corrections become power-law in the volume, and one necessarily has to study a coupled-channel system made up of a boson-diboson state and a three-boson state. For energies just above the diboson breakup Eq. (\ref{QC}) can be written as  
\begin{eqnarray}
\label{QCbreakup}
\left(1+\tilde{\mathcal{M}}^{\infty}_{V,Bd-Bd}~\delta\tilde{\mathcal{G}}^V_{Bd}\right)\left(1+\tilde{\mathcal{M}}^{\infty}_{V,BBB-BBB}~\delta\tilde{\mathcal{G}}^V_{BBB}\right)=|\tilde{\mathcal{M}}^{\infty}_{V,Bd-BBB}|^2~\delta\tilde{\mathcal{G}}^V_{Bd}~\delta\tilde{\mathcal{G}}^V_{BBB}\ ,
\end{eqnarray}
where $\delta\tilde{\mathcal{G}}^V_{Bd}$ and $\delta\tilde{\mathcal{G}}^V_{BBB}$ are respectively the boson-diboson and three boson propagators, $\tilde{\mathcal{M}}^{\infty}_{V,\kappa-\kappa'}$ denotes the elements of $\tilde{\mathcal{M}}^{\infty}_{V}$ for the $\kappa^{th}$ ($\kappa'^{th}$) initial (final) state. For such energies, the approximations made before are no longer valid and determination of infinite volume scattering cross sections from the finite volume spectrum requires numerically solving an integral equation for $\tilde{\mathcal{M}}^{\infty}_{V}$, Eq. (\ref{STMeq2}).

Lastly we comment on the systematic uncertainties associated with the dimer formalism and partial-wave mixing. Assume, for example, that both the dimer and the boson-dimer wavefunctions are projected onto the $A_1^+$ irreducible representation of the cubic group and that the three particles are degenerate. Then in the boson-dimer CM frame, the system has an overlap with $(J_{d},J_{Bd})=(0,0)$ as well as $(J_{d},J_{Bd})=\{(2,0),(4,0),(0,4),(2,4),(2,6),\ldots\}$ angular momentum states, with the leading contamination arising from the D-wave dimer. As discussed in Sec. \ref{3body}, this is due to the the fact that the dimer in this $2+1$ body set-up is boosted and its symmetry group in its CM frame is reduced compared to the original cubic group \cite{movingframe}. If one then proceeds to consider a reference frame where the dimer-boson system has non-zero momentum, then the ground state will have overlap with $(J_{d},J_{Bd})=(0,0)$ as well as $(J_{d},J_{Bd})=\{(0,1),(2,0),(2,1),(0,2),(2,2),\ldots\}$ angular momentum states. This is because the dimer-boson is effectively a two-particle system where one of the particles is twice as massive as the other, and therefore S and P-wave mixing is unavoidable \cite{Fu:2011xz}. As a result one needs to simultaneously determine S and P-wave scattering parameters. Note that although the dimer field used in this paper is an S-wave field which does not lead to inclusion of higher partial-waves in the two-body QC, the boson-diboson scattering QC, Eq. (\ref{dbQC}) fully incorporates the partial-wave mixing in the space of the boson-diboson angular momentum states.

  

 \chapter{Conclusion}{\label{conclusion}}

Determining nuclear properties directly from quantum chromodynamics (QCD) will impact our understanding of a wide range of phenomena. Given the non-perturbative nature of QCD, currently LQCD is the only reliable way to carry out such an ambitious program. As discussed extensively throughout this work, LQCD calculations are necessarily performed in a finite Euclidean spacetime. Therefore, it is necessary to construct formalism that connects the finite-volume observables determined via LQCD to the infinite-volume quantities of interest. 

\begin{figure}[t]
\begin{center}   
\includegraphics[scale=.8]{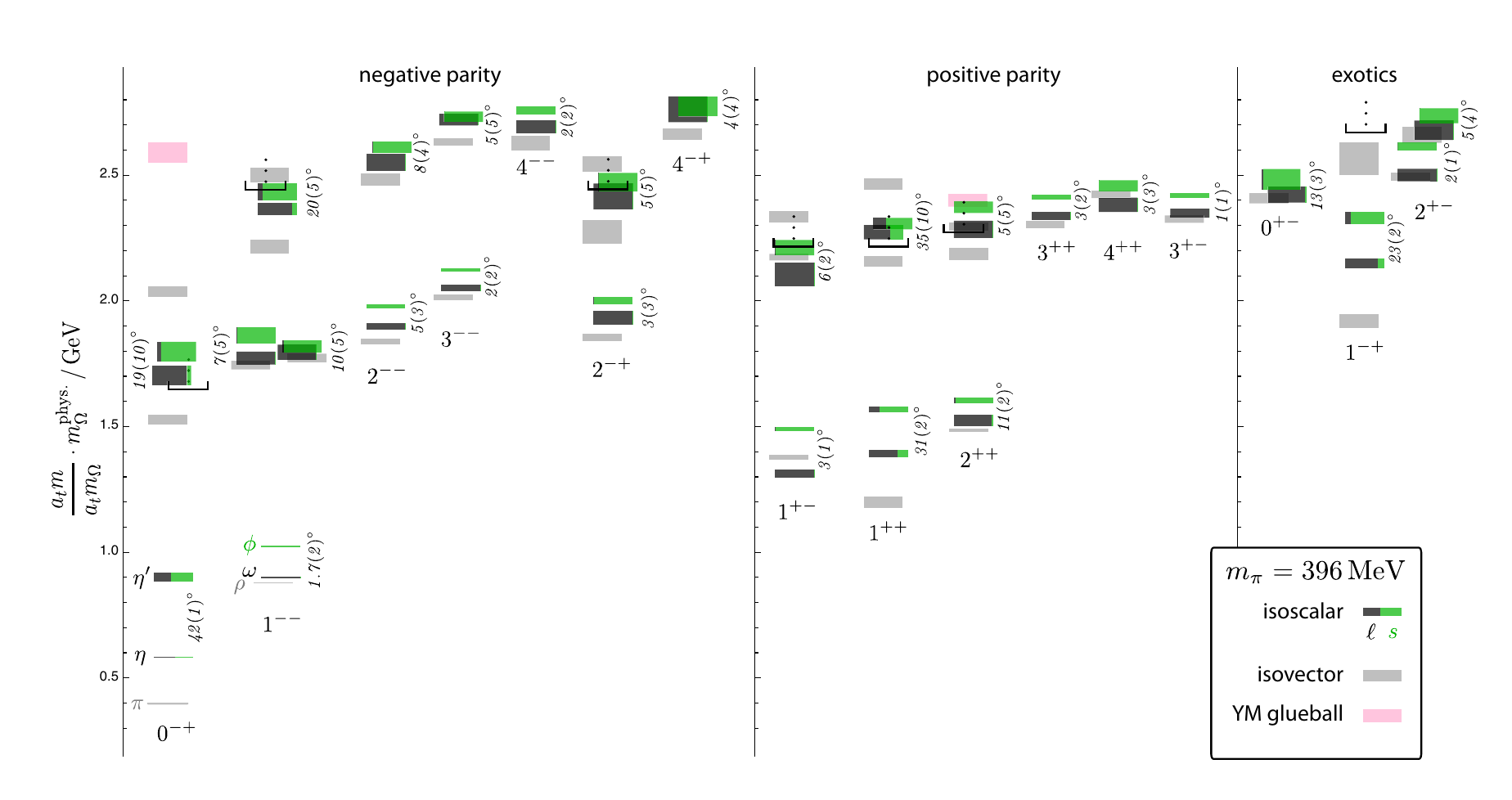}
\caption[Isoscalar meson spectrum calculated by the Hadron Spectrum Collaboration with $m_\pi\approx 396~\rm{MeV}$]{Shown is the isoscalar meson spectrum calculated by the Hadron Spectrum Collaboration with $m_\pi\approx 396~\rm{MeV}$ \cite{Dudek:2011tt}. States are labeled by $J^{PC}$, the size of the boxes corresponds to the one sigma statistical uncertainty, and fraction of the state corresponding to light(strange) content is depicted in grey(green). Light grey boxes denote states obtained from Ref.~\cite{Dudek:2010wm}, and the pink boxes depict glueball states obtained in the quarkless Yang-Mills theory \cite{Morningstar:1999rf}. The figure is reproduced with permission from the Hadron Spectrum Collaboration.}\label{JLab}
\end{center}
\end{figure}

In chapter~\ref{mmsys} we reviewed L\"uscher's seminal work \cite{luscher1, luscher2}, which allows for the extraction of meson-meson scattering phase shifts from the FV spectrum below inelastic threshold. Although this formalism has allowed for the study of multiple scattering channels (e.g. $\pi^+\pi^+$ see section~\ref{pipiscat}), improvement in algorithms and increase in computational resources have allowed modern day LQCD to extract energies well above inelastic thresholds. For example, in 2011 the Hadron Spectrum Collaboration determined the isoscalar meson spectrum for light quark masses corresponding to $m_\pi\approx 396~\rm{MeV}$ \cite{Dudek:2011tt} up to energies of approximately 2800~MeV, which is well above several inelastic thresholds (see Fig.~\ref{JLab}). Although this is an impressive computational achievement, it surpasses our current theoretical understanding of the FV spectrum and therefore our understanding of the physical implication of such calculation is limited at this point. With this class of challenges in mind, in section~\ref{coudpledchannels} we derived the generalization of L\"uscher formalism for multiple channels composed of two-mesons with nonzero total momentum (first derived in Refs.~\cite{Briceno:2012yi, Hansen:2012tf}) and discussed the implication for the $\pi\pi-K\bar{K}$ isosinglet spectrum for $m_\pi\approx310$~MeV.

\begin{figure}[t]
\begin{center}   
\includegraphics[scale=.8]{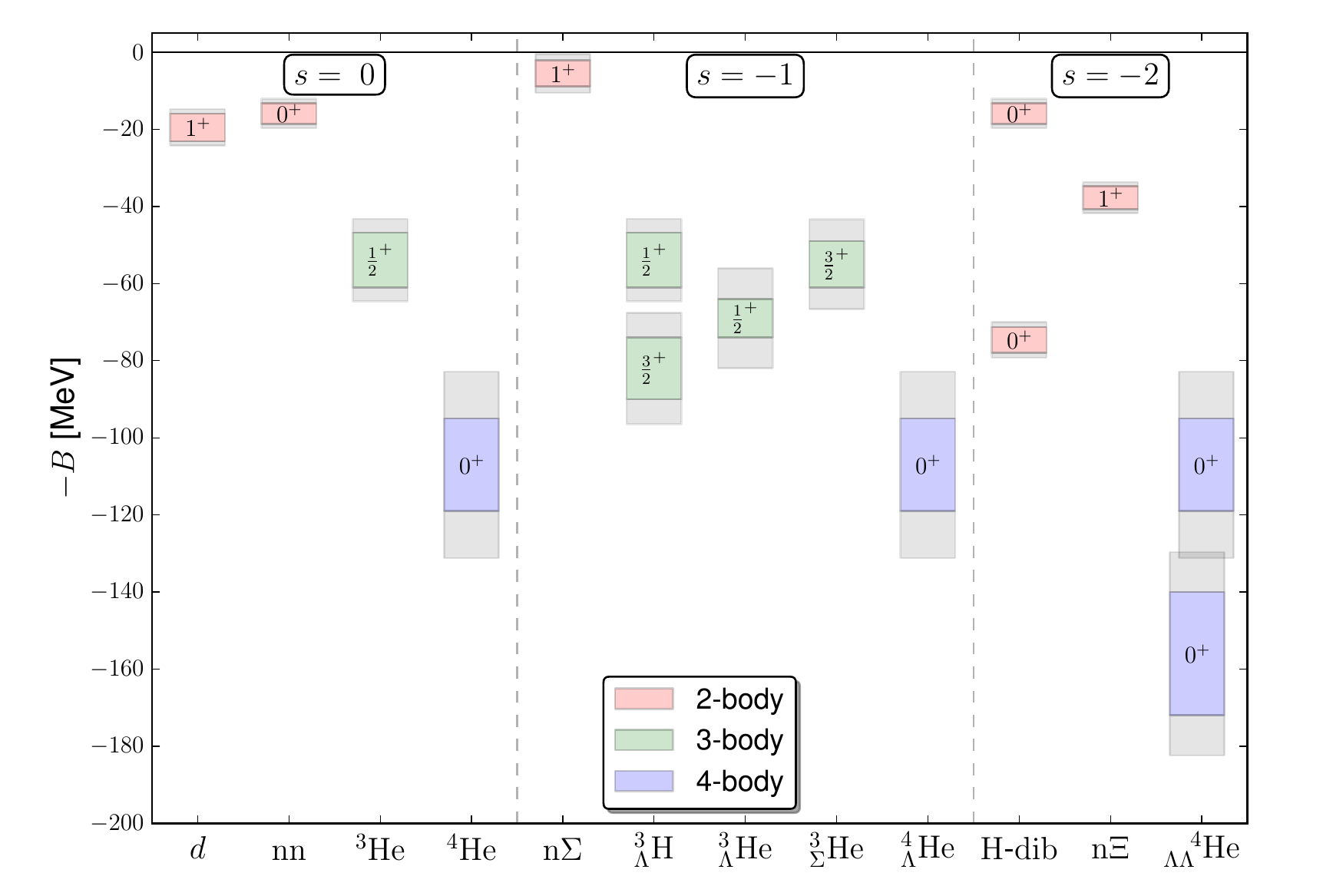}
\caption[Light nuclei and hypernuclei binding energies calculated by the NPLQCD Collaboration~\cite{Beane:2012vq}]{ Shown is the binding energies calculated by the NPLQCD Collaboration~\cite{Beane:2012vq} for nuclei and hypernuclei with four baryons or less, including states with strangenes $s=\{0,1,2\}$. These are calculated at the $SU(3)$ flavor symmetry point, which correspond to $m_\pi=m_K\approx{800}$~MeV. The figure is reproduced with permission from the NPLQCD Collaboration.}\label{NPLQCD}
\end{center}
\end{figure}

Due to a poor  signal-to-ratio ratio~\cite{Lepage89}, numerical calculations involving baryonic systems are significantly more challenging than their mesonic counterparts. As a result, it has not been until recently that it has been shown that calculations of few-nucleon systems are possible \cite{Beane:2013br, Beane:2012vq, Beane:2011iw, Inoue:2011ai, Yamazaki:2011nd, Yamazaki:2012hi, Murano:2013xxa, Doi:2012xd, Orginos:2012js}. Figure~(\ref{NPLQCD}) shows the binding energies calculated by the NPLQCD Collaboration~\cite{Beane:2012vq} for systems including up to four baryons in the limit of exact $SU(3)$ flavor symmetry. It is not unrealistic to expect these calculations to be performed near the physical point in the upcoming years. Although it is desirable to use the formalism developed by L\"uscher to study NN-systems, this is only suitable when the NN-system is projected into an S-wave~\cite{Beane:2003da}. That being said, in chapter~\ref{NNsys} we saw that this formalism can be generalized for two-nucleon systems with arbitrary momentum. 

In studying the two nucleon system, it was convenient to first generalize the auxiliary field formalism to arbitrary partial-waves in both the scalar~\ref{scalardimer} and nuclear sectors~\ref{nucleardimer} in infinite and finite volumes. This formalism was used to derive a master equation that relates the FV two-nucleon energies and the scattering parameters of the two-nucleon systems with arbitrary spin, isospin and angular momentum. This master equation, Eq.~(\ref{NNQC}) is valid for arbitrary total momentum up to inelastic thresholds.

The quantization condition (QC) is a determinant over an infinite-dimensional matrix in the basis of angular momentum, and in practice it is necessary to truncate the number of partial-waves that contribute to the scattering. By taking advantage of the symmetries of the problem, we show how the master equation can be reduced to finite-size blocks that relate particular partial-wave channels (and their mixing) to different spin-isospin channels and different irreps of the corresponding point group of the system. By truncating the matrices at $l\leq 3$, this procedure requires block-diagonalizing matrices as large as $30\times30$. The resulting QCs are determinant conditions involving matrices that are at most $9\times9$, and are therefore practical to be used in future LQCD calculations of NN systems. We have provided one explicit example of this reduction for the scattering in the positive parity isosinglet channel for zero CM momentum in appendix \ref{app:red-example}.  All other QCs for different CM boosts, parity, isospin, spin, and angular momentum are enumerated in appendix \ref{app:QC}. Having studied the zero CM boost as well as $(0,0,1)$ and $(1,1,0)$ boosts, we arrive at $47$ independent QCs for four different spin and isospin channels giving access to all 16 phase shifts and mixing parameters in these channels. Table \ref{scatt-param} summarizes all such  scattering parameters and the corresponding equations that give access to each parameter as presented in this paper. Given the fact that NN-systems couple different partial-waves, in order to reliably extract scattering parameters from LQCD calculations, these calculations must be necessarily performed  in multiple boosts and various irreps of the corresponding symmetry group.

 A sector where LQCD will certainly have a big impact is in electroweak processes involving multi-nucleon systems. With this in mind, in section~\ref{ppfusion} we have used EFT(${\not\hspace{-.09cm}{\pi}}$)  \cite{pionless,pds, pds2, pionless2, pionless3, pionless3} to determine the FV expression for NN-matrix element of the axial-vector current that mixes the $\singlet-\triplet$ NN channels. This is pertinent for performing calculations of proton-proton fusion, among other interesting processes, directly from LQCD~\cite{Kong:1999tw, Kong:1999mp, Butler:2000zp, nnd}. The channels in this system are mixed not only by a two-body operator but also by a one-body operator. As it is shown, FV effects arising from the insertion of a one-body operator are sizable and therefore must be included. Unlike the scalar analogue discussed in section~\ref{EW2B}, the FV and infinite volume weak matrix elements are not simply proportional to each other. The result demonstrates that in fact the FV matrix element is proportional to a linear combination of the LO and NLO LECs that parametrize the weak interactions in the NN-sector.

Finally, in chapter~\ref{mmmsys} we have determined a model-independent representation of the quantization condition for energy eigenvalues of three identical bosons in a finite volume with the periodic boundary conditions. Using a non-relativistic EFT, the FV three-particle spectrum has been shown to be related to the infinite volume S-matrix elements. For arbitrary energies, this correspondence requires solving an integral equation. With nuclear systems in mind, close attention is paid to scalar theories that support a two-body bound-state. It is shown that for energies below the diboson breakup, the quantization condition reduces to the well known L\"uscher result for two particles with unequal masses, with exponential corrections dictated by the size of the diboson, Eq. (\ref{dbQC}). Although physically intuitive, this is a non-trivial observation that will require the Lattice practitioner to extrapolate the phase shifts obtained from the L\"uscher formula for the particle bound-state elastic scattering to the infinite volume.

In deriving the quantization condition we limited the dimer field to have the same quantum numbers as of the two-boson system in an S-wave. This simplification comes at the cost of neglecting the partial-wave mixing associated with a cubic finite volume with the periodic boundary conditions. Given the formalism presented in Sect.~\ref{scalardimer}, this is an approximation that will no longer be needed. By evaluating poles of the full three-boson correlation function in the finite volume, the quantization condition for the three-particle system with non-zero total momentum is derived. The poles are given by a determinant condition, Eq. (\ref{QC}), where the determinant is taken over dimer-boson relative angular momentum states as well as $N_{E^*}$ available boson-dimer eigenstates for each CM energy, $E^*$. As is shown, the corresponding quantization condition has strong parallels with two-body coupled-channel systems discussed in setcion~\ref{coudpledchannels}. However one has to be careful that this is not the physical scattering amplitude that directly shows up in the QC, but rather the scattering amplitude of a boson-FV dimer system. These two quantities are related to each other through an integral equation, Eq. (\ref{STMeq2}).

Furthermore, as is explained in detail, the exponential volume corrections from the off-shell excited states of the dimer are accounted for in the full quantization condition. For sufficiently high energies, these exponential corrections become power-law in volume and can no longer be neglected. Then one would have to consider a coupled-channel system where the number of channels are determined by the total CM energy of the three-particle system, as shown in Eq. (\ref{QCbreakup}) for energies just above the diboson breakup. The formalism presented considers three-particle with non-zero total momentum which eventually allows for more independent measurements at a given energy. This however leads to a practical complexity as the symmetry of the system is reduced, and the ground state of the system is expected to mix with the P-wave scattering state \cite{Bour:2011ef, Davoudi,Fu:2011xz}. The quantization condition derived predicts this mixing between S and P partial-waves, and indicates that the truncation of the determinant condition at S-wave could, in practice, introduce large systematics to the calculation.

With these observations at hand, it is argued that future LQCD studies of nuclear reactions and resonances involving three-particle states will require the following steps. First, one needs to reliably determine scattering phase shifts for the two-particle sector from which one can obtain the boosted L\"uscher poles as a function of the boost momenta and energy. From there, one would proceed to obtained the three-particle spectrum. This requires high statistics to obtain multiple states with clean signals. Also in order to disentangle the coupled-channel nature of the three particle system, these calculations need to be performed with different boosts and in different volumes. In addition one has to simultaneously determine energy eigenvalues of three-particle states in different irreps of the cubic group to correctly deal with the partial wave mixing which is more severe than the two-particle sector. All of this information should be simultaneously fit to numerically solve the quantization condition presented. This would lead to an accurate determination of the three-body Bethe-Salpeter kernel (or equivalently the LECs describing the systems at hand), which encodes all of the infinite volume physics up to the four-particle inelastic threshold.

We are entering an important era for nuclear physics. As has been discussed throughout this work, nuclear physics spans a wide range of scales from the cosmological to the subatomic. Nuclear reactions were responsible for the formation of light nuclei in the early stages of the universe, and continue to be responsible for fueling the evolution of stars. Also nuclear physics exhibits an extremely rich spectrum with some seemingly fined tune systems, such as the deuteron which is nearly unbound. We observe a complex structure of resonances and rare isotopes that continue to be studied to this day, both experimentally and theoretically. At the core of all of this structure lies QCD, and for the very first time in history we will soon be able to connect the complexity observed in nature with the standard model of particle of physics. This thesis outlined key steps needed to carry this ambitious program of unifying our understanding of the nuclear force.

 \printendnotes

%
%

\bibliographystyle{jhep}
\bibliography{bibi}
%
%

\appendix
\raggedbottom\sloppy
 
\chapter{Scattering Theory}{\label{scattheory}}
The defining object behind scattering theory is the $S$-matrix, which is the unitary matrix connecting states of the Hilbert space. More formally, ``in" and ``out" states are asymptotic states of particles at $T=-\infty$ and $T=\infty$, respectively. Consider the case where the initial(final) state is compose of $n(n')$-states with four-momenta $\{{q}_1,{q}_2,\ldots, {q}_n\}(\{{p}_1,{p}_2,\ldots, {p}_{n'}\})$. Then, the overlap between these two states can be written in terms of unitary time transformation
\begin{eqnarray}
_{out}\langle \{{p}_1,{p}_2,\ldots, {p}_{n'} |\{{q}_1,{q}_2,\ldots, {q}_n\}\rangle_{\rm{in}}&=&
\lim_{T\rightarrow \infty}\langle \{{p}_1,{p}_2,\ldots, {p}_{n'} |e^{-i2HT}|\{{q}_1,{q}_2,\ldots, {q}_n\}\rangle\nn\\
&\equiv&\langle \{{p}_1,{p}_2,\ldots, {p}_{n'} |S|\{{q}_1,{q}_2,\ldots, {q}_n\}\rangle,
\end{eqnarray}
where $H$ denotes the Hamiltonian of the theory and the two states on the right of the equality are defined at a common time reference frame. It is in this limit case that the $S$-matrix is defined. If the initial particles do not interact, then $S$ is the equal to the identity. Therefore, it is convenient to isolate the interactive piece of the $S$-matrix into the $T$-matrix 
\begin{eqnarray}
iT\equiv S-\bf{1}.
\end{eqnarray}
Furthermore, due to momentum-conservation the expectation value of the $T$-matrix is proportional to a four-dimensional delta-function, so all of the physics  can be encapsulated in the scattering amplitude, $\mathcal{M}$, defined by
\begin{eqnarray}
\langle \{{p}_1,{p}_2,\ldots, {p}_{n'} |iT|\{{q}_1,{q}_2,\ldots, {q}_n\}\rangle=(2\pi)^4\delta^{(4)}({p}_{out}- {q}_{in})~i\mathcal{M}.
\end{eqnarray}
It is the scattering amplitude that is typically calculated. For example, in the two-body sector, one can determine the scattering amplitude by evaluating the sum of all $2\rightarrow 2$ amputated Feynman diagrams. 

By definition, the representation of the scattering amplitude will depend on the nature of the states. For example, because the normalization of relativistic and non-relativistic states are different, the kinematic factors in the scattering amplitude will differ. 

For systems with a single two-particle channel, due to unitarity constraints the $S$-matrix must be equal to $S_{1}=e^{i2\delta}$, where $\delta$ is a real phase which depends on the relative momentum, $\bf{q}^*$, between the particle and the total energy 	center of mass energy $E^*$. Since total angular momentum is a good quantum number in the infinite volume limit, the $S$-matrix can be written as a infinite-dimensional diagonal matrix over all angular momentum channels, $S_1=diag(e^{i2\delta^{(0)}},e^{i2\delta^{(1)}},\ldots,e^{i2\delta^{(l)}},\ldots)$. The relativistic and non-relativistic scattering amplitudes can then be written in terms of the scattering phase shift

\begin{eqnarray}
\label{scatrel}
(\mathcal{M}^{(l)})_{rel}&=&\frac{8\pi E^*}{nq^*}\frac{e^{2i\delta^{(l)} }-1}{2i}=
\frac{8\pi E^*}{n}\frac{q^{*2l+1}}{q^{*2l+1}\cot \delta^{(l)}-iq^{*2l+1}},
\\
\label{scatNR}
(\mathcal{M}^{(l)})_{NR}&=&\frac{2\pi}{n \mu  q}\frac{e^{2i\delta^{(l)} }-1}{2i}=\frac{2\pi}{n \mu  }\frac{q^{*2l}}{q^{*2l+1}\cot \delta^{(l)}-iq^{*2l+1}},
\end{eqnarray}
where $n=1/2$ for identical particles and $n=1$ otherwise, and $\mu=(m_1^{-1}+m_2^{-1})^{-1}$ is the reduced mass of the two-particle system. At low-energies $q^{*2l+1}\cot \delta^{(l)}$ can be expanded a function of $q^{*2}$, this is the \emph{effective range expansion} (ERE)
\begin{eqnarray}
\label{EFE}
{q}^{*2l+1}\cot\delta^{(l)}=-\frac{1}{a_l}+\frac{r_{l}{q}^{*2}}{2}+\sum_{n=2}^\infty\frac{\rho_{n,l}}{2n!}~{q}^{*2n},
\end{eqnarray}
where $a_l$, $r_{l}$ and $\rho_{n,l}$ are referred to as shape parameters. For S-wave channels $a_0$ and $r_{0}$ are known as the scattering lengths and effective range, respectively.

When there are two open channels in the system, the $l^{th}$ component of the $S$-matrix can be written using the ``barred" parameterization \cite{Smatrix}  
\begin{eqnarray}
\label{smatrix2}
S_2^{(l)}=\begin{pmatrix} 
e^{i2\delta_I^{(l)}}\cos{2\overline{\epsilon}^{(l)}}&ie^{i(\delta_I^{(l)}+\delta_{II}^{(l)})}\sin{2\overline{\epsilon}^{(l)}}\\
ie^{i(\delta_I^{(l)}+\delta_{II}^{(l)})}\sin{2\overline{\epsilon}^{(l)}}&e^{i2\delta_{II}^{(l)}}\cos{2\overline{\epsilon}^{(l)}} \\
\end{pmatrix},
\end{eqnarray}
where the mixing angle $\overline{\epsilon}^{(l)}$ parametrizes mixing between the $I$ and $II$ channels in the $l^{th}$-partial wave. When there an N coupled-channels, the S-matrix becomes and N-dimensional matrix. In general the scattering amplitude can be written in terms of the S-matrix elements,
\begin{eqnarray}
\label{scatrel2}
(\mathcal{M}_{i,j}^{(l)})_{rel}&=&\frac{8\pi E^*}{\sqrt{n_in_{j}q^*_iq^*_{j}}}\frac{(S^{(l_1)})_{i,j}-\delta_{i,j}}{2i}\\
\label{scatNR2}
(\mathcal{M}_{i,j}^{(l)})_{NR}&=&\frac{2\pi  }{\sqrt{n_in_{j}q^*_iq^*_{j}\mu_{i}\mu_{j}}}\frac{(S^{(l_1)})_{i,j}-\delta_{i,j}}{2i}.
\end{eqnarray}

\chapter{Quantization Conditions under $\mathbf{P}\rightarrow \mathbf{P}'$ transformation when $\mathbf{P}$ and $\mathbf{P}'$ are related by a cubic rotation and $|\mathbf{P}|=|\mathbf{P}'|$} \label{app:invariant}
We aim to show that the master QC, Eq. (\ref{NNQC}), is invariant under a $\mathbf{P}\rightarrow \mathbf{P}'$ transformation where $\mathbf{P}$ and $\mathbf{P}'$ are two boost vectors that are related by a cubic rotation. Denoting such rotation by $R$, it is straightforward to show that
\begin{eqnarray}
c^{\mathbf{P}'}_{lm}=\sum_{m'=-l}^{l}\mathcal{D}^{(l)}_{mm'}(R)~c^{\mathbf{P}}_{lm'}.
\label{clm-trans-P}
\end{eqnarray}
Note that for $\mathbf{P}=0$ this relation reduces to Eq. (\ref{clm-trans}), while for a general non-zero boost vector, it only holds if the rotation $R$ corresponds to the symmetry operation of the cube. For example, such a transformation can take the $c^{\mathbf{P}}_{lm}$ function evaluated with $\textbf{d}=(0,0,1)$ to a $c^{\mathbf{P}'}_{lm}$ evaluated with $\textbf{d}=(1,0,0)$. To proceed let us rewrite the $\delta \mathcal{G}^{V}$ matrix elements as given in Eq. (\ref{deltaG}) in terms of the matrix elements of the $\mathcal{F}^{FV}$ that is defined in Eq. (\ref{F}) for the scalar sector,
{\small
\begin{eqnarray}
&& \left[\delta\mathcal{G}^{V,\mathbf{P}}\right]_{JM_J,IM_I,LS;J'M_J',I'M_I',L'S'}=\frac{iM}{4\pi}\delta_{II'}\delta_{M_IM_I'}\delta_{SS'}\times
\nonumber\\
&& \qquad \qquad \qquad ~ \times \left[k^*\delta_{JJ'}\delta_{M_JM_J'}\delta_{LL'} +i\sum_{M_L,M_L',M_S}\langle JM_J|LM_L,SM_S\rangle \langle L'M_L',SM_S|J'M_J'\rangle \mathcal{F}^{FV,\mathbf{P}}_{LM_L,L'M_L'} \right].
\nonumber\\
\label{deltaG-F}
\end{eqnarray}}
Superscript $\mathbf{P}$ on $\delta\mathcal{G}^{V}$ and $\mathcal{F}^{FV}$ reflects the fact that they depend on both magnitude and direction of the boost vector. Now given the transformation of $c^{\mathbf{P}}_{lm}$ under a cubic rotation of the boost vector, Eq. (\ref{clm-trans-P}), one can write $\mathcal{F}^{FV,\mathbf{P}'}$ as following
\begin{eqnarray}
\left[\mathcal{F}^{FV,\mathbf{P}'}\right]_{LM_L,L'M_L'}&=&\sum_{l,m}\sum_{m'=-l}^{l}\mathcal{D}^{(l)}_{mm'}(R)\frac{(4\pi)^{3/2}}{k^{*l}}c^{\mathbf{P}}_{lm'}(k^{*2})\int d\Omega~Y^*_{L,M_L}Y^*_{l,m}Y_{L',M_L'}
\nonumber\\
\nonumber\\
&=& \sum_{\bar{M}_L=-L}^{L}\sum_{\bar{M}_L'=-L'}^{L'}\mathcal{D}^{(L)}_{\bar{M}_LM_L}(R^{-1})\left[\mathcal{F}^{FV,\mathbf{P}}\right]_{L\bar{M}_L,L'\bar{M}_L'}
\mathcal{D}^{(L')}_{M_L'\bar{M}_L'}(R),
\label{F-trans}
\end{eqnarray}
where in the last equality we have used the fact that under rotation
\begin{eqnarray}
\sum_{M'=-L}^{L}\mathcal{D}^{(L)}_{MM'}(R)~Y_{LM'}(\hat{\mathbf{r}})=Y_{LM}(R\hat{\mathbf{r}}).
\label{deltaG}
\end{eqnarray}

Now one can obtain the relation between $\delta\mathcal{G}^{V,\mathbf{P}'}$ and $\delta\mathcal{G}^{V,\mathbf{P}}$ using Eqs. (\ref{deltaG-F}, \ref{F-trans}),
{ \small 
\begin{eqnarray}
\left[\delta\mathcal{G}^{V,\mathbf{P}'}\right]_{JM_J,L;J'M_J',L'}&=&\frac{iM}{4\pi}\times \left[k^*\delta_{JJ'}\delta_{M_JM_J'}\delta_{LL'} +i \sum_{M_L,M_L',M_S}\langle JM_J|LM_L,SM_S\rangle \langle L'M_L',SM_S|J'M_J'\rangle\right.
\nonumber\\
&& \qquad \qquad \qquad  \left. \times  \sum_{\bar{M}_L=-L}^{L}\sum_{\bar{M}_L'=-L'}^{L'}\mathcal{D}^{(L)}_{\bar{M}_LM_L}(R^{-1})\left[\mathcal{F}^{FV,\mathbf{P}}\right]_{L\bar{M}_L,L'\bar{M}_L'}
\mathcal{D}^{(L')}_{M_L'\bar{M}_L'}(R) \right],
\nonumber\\
\label{G-trans1}
\end{eqnarray}}
where we have suppressed spin and isospin indices for the sake of compactness. Using the fact that
{ \small 
\begin{eqnarray}
\langle JM_J|LM_L,SM_S\rangle=\sum_{\widetilde{M}_J=-J}^{J}\sum_{\widetilde{M}_L=-L}^{L}\sum_{\widetilde{M}_S=-S}^{S}\mathcal{D}^{(J)}_{M_J\widetilde{M}_J}(R^{-1})\mathcal{D}^{(L)}_{M_L\widetilde{M}_L}(R)\mathcal{D}^{(S)}_{M_S\widetilde{M}_S}(R)
\langle J\widetilde{M}_J|L\widetilde{M}_L,S\widetilde{M}_S\rangle,
\nonumber\\
\label{CG-trans}
\end{eqnarray}}
and given that Wigner $\mathcal{D}$-matrices are unitary, one easily arrives at
\begin{eqnarray}
\left[\delta\mathcal{G}^{V,\mathbf{P}'}\right]_{JM_J,L;J'M_J',L'}=\sum_{\bar{M}_J=-J}^{J}\sum_{\bar{M}_J'=-J'}^{J'}\mathcal{D}^{(J)}_{\bar{M}_JM_J}(R^{-1})\left[\delta\mathcal{G}^{V,\mathbf{P}}\right]_{J\bar{M}_J,L;J'\bar{M}_J',L'}\mathcal{D}^{(J')}_{M_J'\bar{M}_J'}(R),
\nonumber\\
\label{G-trans2}
\end{eqnarray}
or in the matrix notation, $\delta\mathcal{G}^{V,\mathbf{P}'}=\mathcal{D}^*(R)\delta\mathcal{G}^{V,\mathbf{P}}\mathcal{D}^T(R)$. Given that the scattering amplitude is diagonal in the $|J,M_J\rangle$ basis, and that the quantization condition Eq. (\ref{NNQC}) is a determinant condition, one obtains
\begin{eqnarray}
\det\left[{(\mathcal{M}^{\infty})^{-1}+\delta\mathcal{G}^{V,\mathbf{P}'}}\right]&=&\det\left[\mathcal{D}^*(R)\left({(\mathcal{M}^{\infty})^{-1}+\delta\mathcal{G}^{V,\mathbf{P}}}\right)\mathcal{D}^T(R)\right]
\nonumber\\
&=&\det\left[{(\mathcal{M}^{\infty})^{-1}+\delta\mathcal{G}^{V,\mathbf{P}}}\right]=0.
\label{QC-trans}
\end{eqnarray}
As one would expect, although the FV functions are in general different for different boosts with the same magnitude within a given $A_1$ irrep of the cubic group, the spectrum does not depend on the choice of the direction of the boost vector. As discussed in Sec. \ref{sec: Reduction}, in order to extract the scattering parameters of NN systems from the QCs presented in this paper, one needs to use the specific boost vectors that are studied in this paper. However, the value of energy eigenvalues can be taken from the LQCD calculations that are performed with any other boost vector that is a cubic rotation of the boost vectors presented here.


\chapter{Reduction Procedure for Positive Parity Isosinglet Channel with $\mathbf{P}=\mathbf{0}$} \label{app:red-example}

\noindent Consider the NN system in the positive parity isosinglet channel where the ground state in the infinite volume is known to be a shallow bound state, the deuteron, whose wave-function is an admixture of both S-wave and D-wave. In order to obtain the phase shifts and mixing parameter in this channel from the energy eigenvalues of the two-nucleon system at rest from a LQCD calculation, one must first construct sources and sinks that transform according to a given irrep of the cubic group, e.g. $T_1$ when $\textbf{P}=0$. The extracted energies then needs to be put in the determinant condition for this channel in the corresponding irrep of the cubic group, Eq. (\ref{NNQC-irrep}), and subsequently solve for the scattering parameters. If one assumes the contributions from scattering channels with $J>4$ and $l\geq4$ to be negligible, the scattering amplitude matrix in the LHS of Eq. (\ref{NNQC-irrep}) can be written as
\begin{eqnarray}
\mathcal{M}^{\infty}_{(0,1)}=\left(\begin{array}{cccc}
\mathcal{M}_{1;0}^{(0,0;1)} & \mathcal{M}_{1;0}^{(0,2;1)} & 0 & 0\\
\mathcal{M}_{1;0}^{(2,0;1)} & \mathcal{M}_{1;0}^{(2,2;1)} & 0 & 0\\
0 & 0 & \mathcal{M}_{2;0}^{(2,2;1)} & 0\\
0 & 0 & 0 & \mathcal{M}_{3;0}^{(2,2;1)}
\end{array}\right),
\end{eqnarray}
where each element, $ \mathcal{M}_{J;I}^{(L,L';S)}$, is a diagonal $(2J+1)^2\times(2I+1)^2$-matrix. Note that this is an $18\times18$ matrix which is parametrized by two phase shifts and a mixing angle in the $J=1$ channel, and two D-wave phase shifts in the $J=2$ and $J=3$ channels. Although there is a mixing between D-wave and G-wave channels in the $J=3$ sector, due to the assumption of a negligible $G$-wave scattering, the scattering amplitude in this channel is truncated to the D-wave.

The elements of the FV matrix $\delta \mathcal{G}^V$ in the LHS of Eq. (\ref{NNQC-irrep}) for this channel can be evaluated from Eq. (\ref{deltaG}). The result reads
\begin{eqnarray}
\delta \mathcal{G}^V_{(0,1)}=\left(\begin{array}{cccc}
\delta\mathcal{G}{}_{1,1;0}^{V,(0,0;1)} & \delta\mathcal{G}_{1,1;0}^{V,(0,2;1)} & \delta\mathcal{G}_{12;0}^{V,(0,2;1)} & \delta\mathcal{G}_{1,3;0}^{V,(0,2;1)}\\
\\
\delta\mathcal{G}_{1,1;0}^{V,(2,0;1)} & \delta\mathcal{G}_{1,1;0}^{V,(2,2;1)} & \delta\mathcal{G}_{1,2;0}^{V,(2,2;1)} & \delta\mathcal{G}_{1,3;0}^{V,(2,2;1)}\\
\\
\delta\mathcal{G}_{2,1;0}^{V,(2,0;1)} & \delta\mathcal{G}_{2,1;0}^{V,(2,2;1)} & \delta\mathcal{G}_{2,2;0}^{V,(2,2;1)} & \delta\mathcal{G}_{2,3;0}^{V,(2,2;1)}\\
\\
\delta\mathcal{G}_{3,1;0}^{V,(2,0;1)} & \delta\mathcal{G}_{3,1;0}^{V,(2,2;1)} & \delta\mathcal{G}_{3,2;0}^{V,(2,2;1)} & \delta\mathcal{G}_{3,3;0}^{V,(2,2;1)}
\end{array}\right),
\end{eqnarray}
where each element still represents a matrix $\delta\mathcal{G}_{J,J';I}^{V,(L,L';S)}$
in the $|J,M_J\rangle$ basis and whose explicit forms are as following\footnote{We will drop the superscript $\mathbf{P}$ on the $c_{lm}$s in this example as they are evaluated for $\mathbf{P}=0$.}
\begin{eqnarray}
\delta\mathcal{G}_{1,1;0}^{V,(0,0;1)}&=&\delta\mathcal{G}_{1,1;0}^{V,(2,2;1)}=M(-c_{00}+\frac{i k^*}{4 \pi })~\mathbf{I}_3,
\end{eqnarray}
\begin{eqnarray}
 \delta\mathcal{G}_{1,3;0}^{V,(2,2;1)}&=&\left[\delta\mathcal{G}_{3,1;0}^{V,(2,2;1)}\right]^T=\frac{M}{k^{*4}}c_{40}\left(
\begin{array}{ccccccc}
 0 & 0 & -\frac{3  }{7} & 0 & 0 & 0 & -\frac{\sqrt{15}}{7} \\
 0 & 0 & 0 & \frac{2 \sqrt{6}}{7} & 0 & 0 & 0 \\
 -\frac{\sqrt{15}}{7} & 0 & 0 & 0 & -\frac{3}{7} & 0 & 0
\end{array}
\right),
\end{eqnarray}
\begin{eqnarray}
\delta \mathcal{G}^{V,(2,2;1)}_{(2,2;0)}&=& M(-c_{00}+\frac{i k^*}{4 \pi})~\mathbf{I}_5+\frac{M}{k^{*4}}c_{40}\left(\begin{array}{ccccc}
 \frac{2 }{21} & 0 & 0 & 0 & \frac{10 }{21} \\
 0 &-\frac{8 }{21} & 0 & 0 & 0 \\
 0 & 0 & \frac{4 }{7} & 0 & 0 \\
 0 & 0 & 0 &-\frac{8 }{21} & 0 \\
 \frac{10 }{21} & 0 & 0 & 0 & \frac{2 }{21}
\end{array}\right),
\end{eqnarray}
\begin{eqnarray}
\delta \mathcal{G}^{V,(2,2;1)}_{(2,3;0)}&=&\left[\delta \mathcal{G}^{V,(2,2;1)}_{(3,2;0)}\right]^T=\frac{M}{k^{*4}}c_{40}\left(
\begin{array}{ccccccc}
 0 & \frac{5 \sqrt{2}}{21} & 0 & 0 & 0 & \frac{5 \sqrt{2}}{21} & 0 \\
 0 & 0 & -\frac{5 \sqrt{5}}{21} & 0 & 0 & 0 & \frac{5}{7 \sqrt{3}} \\
 0 & 0 & 0 & 0 & 0 & 0 & 0 \\
 -\frac{5}{7 \sqrt{3}} & 0 & 0 & 0 & \frac{5 \sqrt{5}}{21} & 0 & 0 \\
 0 & -\frac{5 \sqrt{2}}{21} & 0 & 0 & 0 & -\frac{5 \sqrt{2}}{21} & 0
\end{array}
\right),
\end{eqnarray}
\begin{eqnarray}
\delta \mathcal{G}^{V,(2,2;1)}_{(3,3;0)}&=&M(-c_{00}+\frac{i k^*}{4 \pi})~\mathbf{I}_7-
\frac{M}{k^{*4}}c_{40}\left(\begin{array}{ccccccc}
 \frac{1}{7} & 0 & 0 & 0 & \frac{\sqrt{\frac{5}{3}}}{7} & 0 & 0 \\
 0 & -\frac{1}{3} & 0 & 0 & 0 & \frac{5}{21} & 0 \\
 0 & 0 & \frac{1}{21} & 0 & 0 & 0 & \frac{\sqrt{\frac{5}{3}}}{7} \\
 0 & 0 & 0 & \frac{2}{7} & 0 & 0 & 0 \\
 \frac{\sqrt{\frac{5}{3}}}{7} & 0 & 0 & 0 & \frac{1}{21} & 0 & 0 \\
 0 & \frac{5}{21} & 0 & 0 & 0 & -\frac{1}{3} & 0 \\
 0 & 0 & \frac{\sqrt{\frac{5}{3}}}{7} & 0 & 0 & 0 & \frac{1}{7}
\end{array}\right),
\nonumber\\
\end{eqnarray}
where $\mathbf{I}_n$ is the $n \times n$ identity matrix, and the rest of the blocks are zero. As is suggested in Ref. \cite{Luu:2011ep}, a unitary matrix, that can bring the $\delta \mathcal{G}^V$ matrix into a block-diagonalized form, can be found by diagonalizing the blocks that are located on the diagonal of the $\delta \mathcal{G}^V$ matrix, $\delta \mathcal{G}^{V,(L,L';1)}_{(J,J;0)}$. It turns out that when there are multiple occurrences of a given irrep in each angular momentum $J$ (see table \ref{irreps}), the procedure of block diagonalization becomes more cumbersome, and a systematic procedure must be taken which is based on the knowledge of the basis functions corresponding to each occurrence of any given irrep. Such basis functions for the irreps of the point groups considered in this paper have been previously worked out (see Table II in Ref.~\cite{Thomas:2011rh}). These basis functions correspond to each occurrence of the irreps in the decomposition of the angular momentum states into the irreps of the $O$, $D_4$ and $D_2$ point groups up to $J=4$. For this channel, however, such a unitary matrix can be found easily based on the method described in Ref. \cite{Luu:2011ep}. One finds
\begin{eqnarray}
S=\left(\begin{array}{ccc}
S_{11} & 0 & 0\\
0 & S_{22} & 0\\
0 & 0 & S_{33}
\end{array}\right),
\end{eqnarray}
where the zero elements denote sub-blocks of appropriate dimension with all elements equal to zero, and the non-trivial blocks are the following matrices
\begin{eqnarray}
S_{11}=\mathbf{I}_6,
~S_{22}=\left(\begin{array}{ccccc}
0 & 0 & 0 & 1 & 0\\
0 & 1 & 0 & 0 & 0\\
0 & 0 & 1 & 0 & 0\\
-\frac{1}{\sqrt{2}} & 0 & 0 & 0 & \frac{1}{\sqrt{2}}\\
\frac{1}{\sqrt{2}} & 0 & 0 & 0 & \frac{1}{\sqrt{2}}
\end{array}\right), 
~S_{33}=\left(\begin{array}{ccccccc}
0 & 0 & \sqrt{\frac{3}{8}} & 0 & 0 & 0 & \sqrt{\frac{5}{8}}\\
\sqrt{\frac{5}{8}} & 0 & 0 & 0 & \sqrt{\frac{3}{8}} & 0 & 0\\
0 & 0 & 0 & 1 & 0 & 0 & 0\\
0 & 0 & -\sqrt{\frac{5}{8}} & 0 & 0 & 0 & \sqrt{\frac{3}{8}}\\
0 & \frac{1}{\sqrt{2}} & 0 & 0 & 0 & \frac{1}{\sqrt{2}} & 0\\
-\sqrt{\frac{3}{8}} & 0 & 0 & 0 & \sqrt{\frac{5}{8}} & 0 & 0\\
0 & -\frac{1}{\sqrt{2}} & 0 & 0 & 0 & \frac{1}{\sqrt{2}} & 0
\end{array}\right).
\nonumber\\
\end{eqnarray}
The resultant partially block-diagonalized matrix can then be obtained by,
{ \small 
\begin{eqnarray}
&& S[(\mathcal{M}^{\infty}_{(0,1)})^{-1}+\delta \mathcal{G}^{V}_{(0,1)}]S^T=
\nonumber\\
&& \qquad ~~~~ \left(\begin{array}{cccccccccccccccccc}
x_{1} & 0 & 0 & y_{1} & 0 & 0 & 0 & 0 & 0 & 0 & 0 & 0 & 0 & 0 & 0 & 0 & 0 & 0\\
0 & x_{1} & 0 & 0 & y_{1} & 0 & 0 & 0 & 0 & 0 & 0 & 0 & 0 & 0 & 0 & 0 & 0 & 0\\
0 & 0 & x_{1} & 0 & 0 & y_{1} & 0 & 0 & 0 & 0 & 0 & 0 & 0 & 0 & 0 & 0 & 0 & 0\\
y_{1} & 0 & 0 & x_{2} & 0 & 0 & 0 & 0 & 0 & 0 & 0 & -y_{2} & 0 & 0 & 0 & 0 & 0 & 0\\
0 & y_{1} & 0 & 0 & x_{2} & 0 & 0 & 0 & 0 & 0 & 0 & 0 & 0 & y_{2} & 0 & 0 & 0 & 0\\
0 & 0 & y_{1} & 0 & 0 & x_{2} & 0 & 0 & 0 & 0 & 0 & 0 & -y_{2} & 0 & 0 & 0 & 0 & 0\\
0 & 0 & 0 & 0 & 0 & 0 & x_{3} & 0 & 0 & 0 & 0 & 0 & 0 & 0 & 0 & 0 & y_{3} & 0\\
0 & 0 & 0 & 0 & 0 & 0 & 0 & x_{3} & 0 & 0 & 0 & 0 & 0 & 0 & y_{3} & 0 & 0 & 0\\
0 & 0 & 0 & 0 & 0 & 0 & 0 & 0 & x_{4} & 0 & 0 & 0 & 0 & 0 & 0 & 0 & 0 & 0\\
0 & 0 & 0 & 0 & 0 & 0 & 0 & 0 & 0 & x_{3} & 0 & 0 & 0 & 0 & 0 & -y_{3} & 0 & 0\\
0 & 0 & 0 & 0 & 0 & 0 & 0 & 0 & 0 & 0 & x_{4} & 0 & 0 & 0 & 0 & 0 & 0 & 0\\
0 & 0 & 0 & -y_{2} & 0 & 0 & 0 & 0 & 0 & 0 & 0 & x_{5} & 0 & 0 & 0 & 0 & 0 & 0\\
0 & 0 & 0 & 0 & 0 & -y_{2} & 0 & 0 & 0 & 0 & 0 & 0 & x_{5} & 0 & 0 & 0 & 0 & 0\\
0 & 0 & 0 & 0 & y_{2} & 0 & 0 & 0 & 0 & 0 & 0 & 0 & 0 & x_{5} & 0 & 0 & 0 & 0\\
0 & 0 & 0 & 0 & 0 & 0 & 0 & y_{3} & 0 & 0 & 0 & 0 & 0 & 0 & x_{6} & 0 & 0 & 0\\
0 & 0 & 0 & 0 & 0 & 0 & 0 & 0 & 0 & -y_{3} & 0 & 0 & 0 & 0 & 0 & x_{6} & 0 & 0\\
0 & 0 & 0 & 0 & 0 & 0 & y_{3} & 0 & 0 & 0 & 0 & 0 & 0 & 0 & 0 & 0 & x_{6} & 0\\
0 & 0 & 0 & 0 & 0 & 0 & 0 & 0 & 0 & 0 & 0 & 0 & 0 & 0 & 0 & 0 & 0 & x_{7}
\end{array}\right),\nn\\
\label{BD-form}
\end{eqnarray}}
where
{ \small 
\begin{eqnarray}
x_{1}&=&-M c_{00}+\frac{iMk^*}{4\pi}+\frac{\mathcal{M}_{1;0}^{(2,2;1)}}{\det(\mathcal{M}^{SD})},~
x_{2}=-M c_{00}+\frac{iMk^*}{4\pi}+\frac{\mathcal{M}_{1;0}^{(0,0;1)}}{\det(\mathcal{M}^{SD})},
\nonumber\\
x_{3}&=&-M c_{00}-\frac{8}{21}\frac{M}{k^{*4}}c_{40}+\frac{iMk^*}{4\pi}+\frac{1}{\mathcal{M}^{(22;1)}_{2;0}}, ~
x_{4}=-M c_{00}+\frac{4}{7}\frac{M}{k^{*4}}c_{40}+\frac{iMk^*}{4\pi}+\frac{1}{\mathcal{M}^{(22;1)}_{2;0}},
\nonumber\\
x_{5}&=&-M c_{00}-\frac{2}{7}\frac{M}{k^{*4}}c_{40}+\frac{iMk^*}{4\pi}+\frac{1}{\mathcal{M}^{(22;1)}_{3;0}},~
x_{6}=-M c_{00}+\frac{2}{21}\frac{M}{k^{*4}}c_{40}+\frac{iMk^*}{4\pi}+\frac{1}{\mathcal{M}^{(22;1)}_{3;0}},
\nonumber\\
x_{7}&=&-M c_{00}+\frac{4}{7}\frac{M}{k^{*4}}c_{40}+\frac{iMk^*}{4\pi}+\frac{1}{\mathcal{M}^{(22;1)}_{3;0}},
y_{1}=-\frac{\mathcal{M}_{1;0}^{(0,2;1)}}{\det(\mathcal{M}^{SD})},
y_{2}=\frac{2\sqrt{6}}{7}\frac{M}{k^{*4}}c_{40},
y_{3}=\frac{10\sqrt{2}}{21}\frac{M}{k^{*4}}c_{40},
\nonumber\\
\end{eqnarray}}
and $\det(\mathcal{M}^{SD})$ in these relations denotes the determinant of the $J=1$ sub-block of the scattering amplitude, $\det(\mathcal{M}^{SD})=\mathcal{M}_{1;0}^{(0,0;1)} \mathcal{M}_{1;0}^{(2,2;1)}- (\mathcal{M}_{1;0}^{(0,2;1)})^2$. This matrix can now clearly be broken to 4 independent blocks corresponding to 4 irreps of the cubic group. The degeneracy of the diagonal elements of this matrix, as well as the coupling between different rows and columns, indicate which irrep of the cubic group each block corresponds to. According to table \ref{irreps}, the one-dimensional irrep $A_2$ only occurs in the decomposition of $J=3$ angular momentum. As is seen from Eq. (\ref{BD-form}), the element $x_7$ belongs to the $J=3$ sector and has a one-fold degeneracy. Also it does not mix with other angular momentum channels, therefore it must correspond to the $A_2$ irrep. So the one-dimensional QC corresponding to the $A_2$ irrep is
\begin{eqnarray}
A_2: ~ \frac{1}{\mathcal{M}^{(22;1)}_{3;0}}-M c_{00}+\frac{4}{7}\frac{M}{k^{*4}}c_{40}+\frac{iMk^*}{4\pi}=0.
\label{A2}
\end{eqnarray}

The QC corresponding to the two-dimensional irrep $E$ can be also deduced easily as it only has overlap with the $J=2$ channel. Clearly the element corresponding to this irrep is $x_4$ with two-fold degeneracy and the corresponding QC reads
\begin{eqnarray}
E: ~ \frac{1}{\mathcal{M}^{(22;1)}_{2;0}}-M c_{00}+\frac{4}{7}\frac{M}{k^{*4}}c_{40}+\frac{iMk^*}{4\pi}=0.
\label{E}
\end{eqnarray}

The three-dimensional irrep $T_2$ appears in the decomposition of both $J=2$ and $J=3$ angular momentum, and as is seen from Eq. (\ref{BD-form}) mixes the $x_3$, $x_6$ and $y_3$ elements through the following QC
\begin{eqnarray}
T_2: ~ \det\left(\begin{array}{cc}
\frac{1}{\mathcal{M}^{(22;1)}_{2;0}}-M c_{00}-\frac{8}{21}\frac{M}{k^{*4}}c_{40}+\frac{iMk^*}{4\pi} & \frac{10\sqrt{2}}{21}\frac{M}{k^{*4}}c_{40} \\
\frac{10\sqrt{2}}{21}\frac{M}{k^{*4}}c_{40} & \frac{1}{\mathcal{M}^{(22;1)}_{3;0}}-M c_{00}+\frac{2}{21}\frac{M}{k^{*4}}c_{40}+\frac{iMk^*}{4\pi}
\end{array}\right)=0.
\nonumber\\
\label{T2}
\end{eqnarray}
As is clear, the energy eigenvalues in this irrep have a three-fold degeneracy (there are three copies of this QC) that is consistent with the dimensionality of the irrep. The remaining irrep is $T_1$ which is a three-dimensional irrep and contribute to both $J=1$ and $J=3$ channels. As there are two $J=1$ sectors corresponding to S-wave and D-wave scatterings, the QC must be the determinant of a $3 \times 3$ matrix. This is in fact the case by looking closely at the partially block-diagonalized matrix in Eq. (\ref{BD-form}). One finds explicitly
{\small
\begin{eqnarray}
T_1: &~& \det\left(\begin{array}{ccc}
\frac{\mathcal{M}_{1;0}^{(2,2;1)}}{\det(\mathcal{M}^{SD})}-M c_{00}+\frac{iMk^*}{4\pi} & -\frac{\mathcal{M}_{1;0}^{(0,2;1)}}{\det(\mathcal{M}^{SD})} & 0\\
-\frac{\mathcal{M}_{1;0}^{(0,2;1)}}{\det(\mathcal{M}^{SD})} & \frac{\mathcal{M}_{1;0}^{(0,0;1)}}{\det(\mathcal{M}^{SD})}-M c_{00}+\frac{iMk^*}{4\pi} & -\frac{2\sqrt{6}}{7}\frac{M}{k^{*4}}c_{40}\\
0 & -\frac{2\sqrt{6}}{7}\frac{M}{k^{*4}}c_{40} & \frac{1}{\mathcal{M}^{(22;1)}_{3;0}}-M c_{00}-\frac{2}{7}\frac{M}{k^{*4}}c_{40}+\frac{iMk^*}{4\pi}
\end{array}\right)
\nonumber\\
&& \qquad \qquad \qquad \qquad \qquad \qquad
\qquad \qquad \qquad \qquad \qquad \qquad 
\qquad \qquad \qquad \qquad \qquad \qquad \qquad ~ =0.
\nonumber\\
\label{T1}
\end{eqnarray}}
Again there is a three-fold degeneracy for the energy-eigenvalues as there are three copies of this QC for this irrep. This is an important QC as it gives access to the mixing angle between S- and D-partial-waves. Note that the QC for $A_2$ irrep, Eq. (\ref{A2}), by its own determines the phase-shift in the $J=3$ channel, which can then be used in Eq. (\ref{T1}) for the $T_1$ irrep to determine the phase-shifts and mixing angle in the $J=1$ channel. Eq. (\ref{E}) for the $E$ irrep gives access to the phase shift in the $J=2$ channel, and finally Eq. (\ref{T2}) provides another relation for the phase-shifts in the $J=2$ and $J=3$ channels. In practice, one needs multiple energy levels in order to be able to reliably extract these parameters from the QCs presented. This is specially a challenging task when it comes to the determination of the scattering parameters in the channels with physical mixing, e. g. S-D mixing, since there are at least three unknown parameters to be determined from the QC, e. g. see Eq. (\ref{T1}). By performing the LQCD calculations of boosted two-nucleon systems, one will obtain more energy levels corresponding to other QCs. These QCs then provide a set of equations that the same scattering parameters satisfy, and therefore better constraints can be put on these quantities. Without going into the detail of the reduction procedure that leads to such QCs for boosted systems, as well as QCs for the other three spin-isospin channels, we tabulate these QCs in the next appendix. Before presenting the rest of QCs though, let us show that the QCs are all real conditions.

\begin{center}
\begin{table}
\label{tab:param2}
\resizebox{15cm}{!}{
\begin{tabular}{|c|c|c|cc|}
\hline
$\hspace{.3cm}J\hspace{.3cm}$&$O$&${D}_{4}$&${D}_{2}$&
 \\\hline \hline
0&${A}_1$&${A}_1$&${A}$& 
\\\hline
1&${T}_1$&${A}_2\oplus E$&${B}_1\oplus {B}_2 \oplus {B}_3$& \\\hline
2&$E\oplus T_2$&
$A_1\oplus E \oplus B_1 \oplus B_2$&$A\oplus A \oplus B_1 \oplus B_2 \oplus B_3$&
\\
\hline
3&$A_2\oplus T_1 \oplus T_2$
&$A_2\oplus E\oplus E\oplus B_1\oplus B_2$
&$B_1\oplus B_1 \oplus B_2 \oplus B_2 \oplus B_3 \oplus B_3 \oplus A$
&
\\
\hline4&
$A_1\oplus E \oplus T_1 \oplus T_2$
&
$A_1\oplus A_1 \oplus A_2 \oplus E \oplus E \oplus B_1 \oplus B_2$
&
$A\oplus A \oplus A \oplus B_1 \oplus B_1 \oplus B_2 \oplus B_2 \oplus B_3 \oplus B_3$&\\
\hline
\end{tabular}}
\caption{The decomposition of the irreps of the rotational group up to $J=4$ in terms of the irreps of the cubic ($O$), tetragonal ($D_4$) and orthorhombic ($D_2$) groups, see Refs. \cite{luscher2, Feng:2004ua, Dresselhaus, Thomas:2011rh, Gockeler:2012yj}. The corresponding basis functions  of each irrep are given in Ref.~\cite{Thomas:2011rh} for both spin-zero and spin-one systems. These basis functions become useful in reducing the full determinant condition, Eq. (\ref{NNQC}) into separate QCs corresponding to each irrep of the point group considered, see Sec. \ref{sec: Reduction}.
}
\label{irreps}
\end{table}
\end{center}

As all the $c_{lm}^{\mathbf{P}}$ functions are real, the only imaginary part of the FV matrix $\delta \mathcal{G}^V$ shows up in the diagonal elements of this matrix.\footnote{This is not always the case as for example, the $\delta \mathcal{G}^V$ matrix for the $\mathbf{d}=(1,1,0)$ boost contains off-diagonal complex elements as well. For all of those case, it can be checked that although the elements of the matrix $(\mathcal{M}^{\infty})^{-1}+\delta\mathcal{G}^{V}$ are complex, the determinant of the matrix remains real, see QCs in appendix \ref{app:QC}.} For the angular momentum channels $J$ where there is no coupling between different partial-waves, the inverse scattering amplitude matrix has only diagonal elements, whose imaginary part exactly cancels that of the $\delta \mathcal{G}^V$ matrix, see Eq. (\ref{M-single}). Explicitly,
\begin{eqnarray}
\Im [(\mathcal{M}^{LL;S}_{JM_J;IM_I})^{-1}+\delta \mathcal{G}^{V,(LL;S)}_{JM_J,JM_J;IM_I}]=-\frac{iMk^*}{4\pi}+\frac{iMk^*}{4\pi}=0.
\label{Im-diagonal}
\end{eqnarray}
For the angular momentum channels where there are off-diagonal terms due to the partial-wave mixing, one can still write the inverse of the scattering amplitude in that sector, Eqs. (\ref{M-coupled1}, \ref{M-coupled2}), as following
{ \small 
\begin{eqnarray}
(\mathcal{M}^{LL';1})^{-1}&=&
\left( \begin{array}{cc}
-\frac{Mk^*}{4\pi}
\frac{\cos{2\epsilon}~\sin({\delta '-\delta})
+\sin({\delta '+\delta})}{\cos({\delta '+\delta})-\cos({\delta '-\delta})\cos({2\epsilon})}-\frac{iMk^*}{4\pi}&
\frac{Mk^*}{2\pi}
\frac{\cos(\epsilon)\sin(\epsilon)}{\cos(\delta '+\delta)-\cos(\delta '-\delta)\cos(2\epsilon)}
\\
\frac{Mk^*}{2\pi}
\frac{\cos({\epsilon})\sin({\epsilon})}{\cos({\delta '+\delta})-\cos({\delta '-\delta})\cos({2\epsilon})}
&
-\frac{Mk^*}{4\pi}
\frac{\cos(2\epsilon)~\sin({\delta-\delta '})
+\sin(\delta '+\delta)}{\cos(\delta '+\delta)-\cos(\delta '-\delta)\cos(2\epsilon)}-\frac{iMk^*}{4\pi}
\\
\end{array} \right),
\nonumber\\
\label{Minverse-coupled}
\end{eqnarray}}
where $L=J \pm 1$ ($L'=J \mp 1$) and $\delta$ ($\delta '$) denotes the phase shift corresponding to the $L$ ($L'$) partial-wave. The off-diagonal elements of this matrix are real. Given that the FV function $\delta \mathcal{G}^V$ has real off-diagonal terms, these terms in the QC lead to a real off-diagonal element. For the diagonal elements, the imaginary part of the inverse scattering amplitude is isolated and has the same form as the imaginary part of the $\delta \mathcal{G}^V$ matrix, so a similar cancellation as that given in Eq. (\ref{Im-diagonal}) occurs in this case as well.

\chapter{List of Quantization Conditions for NN-Systems}
\label{app:QC}
In order to make the presentation of the QCs clear and compact, we will introduce a simpler notation in this section as following. Let us introduce a new FV function $\mathcal{F}^{(\Gamma),{\textbf{P}}}$ that is projected to a particular irrep of the point group of the problem, $\Gamma_i$,
\begin{eqnarray}
\mathcal{F}^{(\Gamma_i),{\textbf{P}}}(k^{*2};L)& \equiv &-\left[\delta\mathcal{G}^{V}(k^{*2};L)-\frac{iM k^*}{4\pi}\right]_{\Gamma_i}
\nonumber\\
&=&{M}\sum_{l,m}\frac{1}{k^{*l}}~{\mathbb{F}
}_{lm}^{(\Gamma_i),{\textbf{P}}}~{c_{lm}^{\textbf{P}}(k^{*2};L)},
\label{def-F}
\end{eqnarray}
where the volume dependence of the FV functions has been made explicit, while the reference to each $(I,S)$ channel is implicit. In this form, all the detail of the corresponding projected FV functions are embedded in purely numerical matrices, ${\mathbb{F}
}_{lm}^{(\Gamma_i),{\textbf{P}}}$. Similarly, the projection of the inverse of the scattering amplitude in each spin-isospin channel unto a particular irreducible representation, ${\mathbb{M}}^{(\Gamma_i)}$, is defined as
\begin{eqnarray}
{\mathbb{M}}_{(I,S)}^{(\Gamma_i)}\equiv\left(\mathcal{M}^{\infty-1}_{(I,S)}\right)_{\Gamma_i}.
\label{def-M}
\end{eqnarray}
With this notation, the quantization condition for the irreducible representation $\Gamma_i$ can be simply written as
\begin{eqnarray}
\det\left({\mathbb{M}}_{(I,S)}^{(\Gamma_i)}+\frac{iMk^{*}}{4\pi }-\mathcal{F}^{(\Gamma_i),{\textbf{P}}}_{(I,S)}\right)=0.
\label{QC-simplified}
\end{eqnarray}
Since we aim to present the QCs for each spin-isospin channel in separate subsections, the $(I,S)$ subscripts can be dropped in the following presentation. Although the $(I,S)$ index of the scattering amplitudes is assumed implicitly, one should keep the $(J,L)$ quantum numbers of the elements of the scattering amplitude matrix explicit. In order to simplify the notation, however, the diagonal elements of the scattering amplitude in the $L$-basis in each spin-isospin channel will be denoted by
\begin{eqnarray}
\mathcal{M}_{J,L} \equiv \mathcal{M}^{(LL;S)}_{JM_J;IM_{I}},
\end{eqnarray}
while the off-diagonal elements will be defined as
\begin{eqnarray}
\mathcal{M}_{J,LL'} \equiv \mathcal{M}^{(LL';S)}_{JM_J;IM_{I}}.
\end{eqnarray}
The determinant of the $2\times2$ sub-sector that presents the mixing between partial-waves in the $J$ sector is denoted by $\det \mathcal{M}_J$. Explicitly,
\begin{eqnarray}
\det\mathcal{M}_{J}=\det \left( \begin{array}{cc}
\mathcal{M}_{J,L}&\mathcal{M}_{J,LL'}\\
\mathcal{M}_{J,L'L}&\mathcal{M}_{J,L}\\
\end{array} \right)\delta_{L,J-1}\delta_{L',J+1}.
\end{eqnarray}
Instead of using numerical values for the partial-wave $L$, we have used the  conventional spectroscopic notations for $L=0,1,2,3$ as S, P, D and F waves, respectively.

\subsection{Positive parity isosinglet channel}
The scattering amplitude matrix in this channel, after truncating the scatterings at $J=4$ and $L=3$, reads
\begin{eqnarray}
\mathcal{M}_{(0,1)}^{\infty}=\left(\begin{array}{cccc}
\mathcal{M}_{1,S} & \mathcal{M}_{1,SD} & 0 & 0\\
\mathcal{M}_{1,DS} & \mathcal{M}_{1;D} & 0 & 0\\
0 & 0 & \mathcal{M}_{2,D} & 0\\
0 & 0 & 0 & \mathcal{M}_{3,D}
\end{array}\right).
\end{eqnarray}
As is clear, each element is still a $(2J+1)^2$ matrix due to the $M_J$ quantum number. As a result, the truncated scattering amplitude is a $16\times16$ matrix that will be used in the master QC for this spin-isospin channel. In the following, the elements of matrices $\mathbb{F}$ and $\mathbb{M}$ as defied above, Eqs. (\ref{def-F}, \ref{def-M}), will be given for this channel for different irreps of the cubic, tetragonal and orthorhombic point groups.
{\small
\subsubsection{$\mathbf{d}=(0,0,0)$}
\begin{align}
& E: \hspace{.5cm}
\mathbb{F}_{00}^{(E)}=1, \hspace{.5cm}
\mathbb{F}_{40}^{(E)}=-4/7,\hspace{.5cm}
{\mathbb{M}}^{(E)}=\mathcal{M}_{2,D}^{-1}. \label{I000E}
\\
& A_2: \hspace{.5cm}
\mathbb{F}_{00}^{(A_2)}= 1, \hspace{.5cm}
\mathbb{F}_{40}^{(A_2)}= -4/7, \hspace{.5cm}
{\mathbb{M}}^{(A_2)}=
\mathcal{M}_{3,D}^{-1}. \label{I000A2}
\\
& T_1: \hspace{.5cm}
\mathbb{F}_{00}^{(T_1)}=\textbf{I}_{3},\hspace{.5cm}
\mathbb{F}_{40}^{(T_1)}=
\left(
\begin{array}{ccc}
 0 & 0 & 0 \\
 0 & 0 & \frac{2 \sqrt{6}}{7} \\
 0 & \frac{2 \sqrt{6}}{7} & \frac{2}{7} \\
\end{array}
\right),\hspace{.5cm}
{\mathbb{M}}^{(T_1)}=\left(
\begin{array}{ccc}
 \frac{\mathcal{M}_{1,D}}{{\det\mathcal{M}_{1}}} & -\frac{\mathcal{M}_{1,SD}}{\det\mathcal{M}_{1}} & 0 \\
 -\frac{\mathcal{M}_{1,SD}}{\det\mathcal{M}_{1}} & \frac{\mathcal{M}_{1,S}}{\det\mathcal{M}_{1}} & 0 \\
 0 & 0 &\mathcal{M}_{3,D}^{-1} \\
\end{array}
\right). \label{I000T1}
\end{align}
\begin{align}
& T_2: \hspace{.5cm}
\mathbb{F}_{00}^{(T_2)}=\textbf{I}_{2},\hspace{.5cm}
\mathbb{F}_{40}^{(T_2)}=\left(
\begin{array}{cc}
 \frac{8}{21} & -\frac{10 \sqrt{2}}{21} \\
 -\frac{10 \sqrt{2}}{21} & -\frac{2}{21} \\
\end{array}
\right),\hspace{.5cm}
{\mathbb{M}}^{(T_2)}=
\left(
\begin{array}{ccc}
\mathcal{M}_{2,D}^{-1}&0  \\
0 &\mathcal{M}_{3,D}^{-1} \\ 
\end{array}
\right). \label{I000T2}
\end{align}
%
\subsubsection{$\mathbf{d}=(0,0,1)$}
\begin{align}
& A_1:\hspace{.5cm}\mathbb{F}_{00}^{(A_1)}= 1,\hspace{.5cm}
\mathbb{F}_{20}^{(A_1)}=\frac{\sqrt{5}}{7},\hspace{.5cm}
\mathbb{F}_{40}^{(A_1)}= -4/7,\hspace{.5cm}
{\mathbb{M}}^{(A_1)}=
\mathcal{M}_{2,D}^{-1}.
\label{I001A1}
\\
& B_1:\hspace{.5cm}
\mathbb{F}_{00}^{(B_1)}=\textbf{I}_{2},\hspace{.5cm}
\mathbb{F}_{20}^{(B_1)}=\left(
\begin{array}{cc}
 -\frac{\sqrt{5}}{7} & -\frac{\sqrt{10}}{7} \\
 -\frac{\sqrt{10}}{7} & 0 \\
\end{array}
\right),\hspace{.5cm}
\mathbb{F}_{40}^{(B_1)}=
\left(
\begin{array}{cc}
 -\frac{2}{21} & \frac{5 \sqrt{2}}{21} \\
 \frac{5 \sqrt{2}}{21} & -\frac{1}{3} \\
\end{array}
\right),
\nonumber\\
&\hspace{1.2cm}
\mathbb{F}_{44}^{(B_1)}=\frac{1}{3}
\left(
\begin{array}{cc}
 - 2 \sqrt{\frac{10}{7}} & -2 \sqrt{\frac{5}{7}}  \\
 -2 \sqrt{\frac{5}{7}} & -{\sqrt{\frac{10}{7}}} \\
\end{array}
\right),\hspace{.5cm}
{\mathbb{M}}^{(B_1)}=
\left(
\begin{array}{ccc}
\mathcal{M}_{2,D}^{-1}&0  \\
0 &\mathcal{M}_{3,D}^{-1} \\ 
\end{array}
\right).
\label{I001B1}
\\
& B_2:\hspace{.5cm}
\mathbb{F}_{00}^{(B_2)}=\textbf{I}_{2},\hspace{.5cm}
\mathbb{F}_{20}^{(B_2)}=\left(
\begin{array}{cc}
 -\frac{\sqrt{5}}{7} & -\frac{\sqrt{10}}{7} \\
 -\frac{\sqrt{10}}{7} & 0 \\
\end{array}
\right),\hspace{.5cm}
\mathbb{F}_{40}^{(B_2)}=
\left(
\begin{array}{cc}
 -\frac{2}{21} & \frac{5 \sqrt{2}}{21} \\
 \frac{5 \sqrt{2}}{21} & -\frac{1}{3} \\
\end{array}
\right),
\nonumber\\
&\hspace{1.25cm}
\mathbb{F}_{44}^{(B_2)}=\frac{1}{3}
\left(
\begin{array}{cc}
  2 \sqrt{\frac{10}{7}} & 2 \sqrt{\frac{5}{7}}  \\
 2 \sqrt{\frac{5}{7}} & {\sqrt{\frac{10}{7}}} \\
\end{array}
\right),\hspace{.5cm}
{\mathbb{M}}^{(B_2)}=
\left(
\begin{array}{ccc}
\mathcal{M}_{2,D}^{-1}&0  \\
0 &\mathcal{M}_{3,D}^{-1} \\ 
\end{array}
\right).
\label{I001B2}
\end{align}
\begin{align}
& A_2:\hspace{.5cm}
\mathbb{F}_{00}^{(A_2)}=\textbf{I}_{3},\hspace{.5cm}
\mathbb{F}_{20}^{(A_2)}=
\left(
\begin{array}{ccc}
 \frac{2}{\sqrt{5}} & 0 & -\frac{9}{7 \sqrt{5}} \\
 0 & -\frac{1}{\sqrt{5}} & \frac{6 }{7}\sqrt{\frac{2}{5}} \\
 -\frac{9}{7 \sqrt{5}} & \frac{6 }{7}\sqrt{\frac{2}{5}} & \frac{8}{7 \sqrt{5}} \\
\end{array}
\right),\hspace{.5cm}
\mathbb{F}_{40}^{(A_2)}=\left(
\begin{array}{ccc}
 0 & 0 & -\frac{4}{7} \\
 0 & 0 & -\frac{2 \sqrt{2}}{7}  \\
 -\frac{4}{7} & -\frac{2 \sqrt{2}}{7} & \frac{2}{7} \\
\end{array}
\right),
\nonumber\\
&\hspace{1.2cm}
{\mathbb{M}}^{(A_2)}= \left(
\begin{array}{ccc}
 \frac{2 \mathcal{M}_{1,S}+2 \sqrt{2}\mathcal{M}_{1,SD}+\mathcal{M}_{1,D}}{3 \det\mathcal{M}_{1}} & \frac{\sqrt{2} \mathcal{M}_{1,S}-\mathcal{M}_{1,SD}-\sqrt{2}\mathcal{M}_{1,D}}{3 \det\mathcal{M}_{1}} & 0 \\
 \frac{\sqrt{2} \mathcal{M}_{1,S}-\mathcal{M}_{1,SD}-\sqrt{2}\mathcal{M}_{1,D}}{3 \det\mathcal{M}_{1}} & \frac{\mathcal{M}_{1,S}-2 \sqrt{2}\mathcal{M}_{1,SD}+2\mathcal{M}_{1,D}}{3 \det\mathcal{M}_{1}} & 0 \\
 0 & 0 & \mathcal{M}_{3,D}^{-1}  \\
\end{array}
\right).
\label{I001A2}
\\
& E:\hspace{.5cm}\mathbb{F}_{00}^{(E)}=\textbf{I}_{4},\hspace{.5cm}
\mathbb{F}_{20}^{(E)}=
\left(
\begin{array}{ccccc}
 \frac{1}{2 \sqrt{5}} & 0 & -\frac{\sqrt{3}}{2} & 0 & \frac{4 \sqrt{\frac{3}{5}}}{7} \\
 0 & -\frac{1}{\sqrt{5}} & 0 & 0 & -\frac{3}{7}   \sqrt{\frac{6}{5}} \\
 -\frac{\sqrt{3}}{2} & 0 & \frac{\sqrt{5}}{14} & 0 & -\frac{2}{7} \\
 0 & 0 & 0 & -\frac{2 }{7} \sqrt{5}  & 0 \\
 \frac{4 \sqrt{\frac{3}{5}}}{7} & -\frac{3}{7}   \sqrt{\frac{6}{5}} & -\frac{2}{7} & 0 & \frac{6}{7 \sqrt{5}} \\
\end{array}
\right),
\nonumber\\
&\hspace{1cm}\mathbb{F}_{40}^{(E)}=\left(
\begin{array}{ccccc}
 0 & 0 & 0 & 0 & \frac{\sqrt{3}}{7} \\
 0 & 0 & 0 & 0 & \frac{\sqrt{6}}{7} \\
 0 & 0 & \frac{8}{21} & 0 & -\frac{5 \sqrt{5}}{21} \\
 0 & 0 & 0 & \frac{1}{7} & 0 \\
 \frac{\sqrt{3}}{7} & \frac{\sqrt{6}}{7} & -\frac{5 \sqrt{5}}{21} & 0 & \frac{1}{21}
\end{array}
\right),
\hspace{.5cm}
\mathbb{F}_{44}^{(E)}=
\left(
\begin{array}{ccccc}
 0 & 0 & 0 & \sqrt{\frac{2}{7}} & 0 \\
 0 & 0 & 0 & \frac{2}{\sqrt{7}} & 0 \\
 0 & 0 & 0 & \sqrt{\frac{10}{21}} & 0 \\
 \sqrt{\frac{2}{7}} & \frac{2}{\sqrt{7}} & \sqrt{\frac{10}{21}} & 0 & \sqrt{\frac{2}{21}} \\
 0 & 0 & 0 & \sqrt{\frac{2}{21}} & 0
\end{array}
\right),
\nonumber
\end{align}
\begin{align}
&\hspace{1cm}{\mathbb{M}}^{(E)}=
 \left(
\begin{array}{ccccc}
 \frac{\mathcal{M}_{1,S}-2 \sqrt{2}\mathcal{M}_{1,SD}+2\mathcal{M}_{1,D}}{3 \det\mathcal{M}_{1}} & \frac{\sqrt{2} \mathcal{M}_{1,S}-\mathcal{M}_{1,SD}-\sqrt{2}\mathcal{M}_{1,D}}{3 \det\mathcal{M}_{1}} & 0 & 0 & 0 \\
 \frac{\sqrt{2} \mathcal{M}_{1,S}-\mathcal{M}_{1,SD}-\sqrt{2}\mathcal{M}_{1,D}}{3 \det\mathcal{M}_{1}} & \frac{2 \mathcal{M}_{1,S}+2 \sqrt{2}\mathcal{M}_{1,SD}+\mathcal{M}_{1,D}}{3 \det\mathcal{M}_{1}} & 0 & 0 & 0 \\
 0 & 0 & \mathcal{M}_{2,D}^{-1} & 0 & 0 \\
 0 & 0 & 0 & \mathcal{M}_{3,D}^{-1} & 0 \\
 0 & 0 & 0 & 0 & \mathcal{M}_{3,D}^{-1} \\
\end{array}
\right).
\label{I001E}
\end{align}
%
\subsubsection{$\mathbf{d}=(1,1,0)$}
\begin{align}
& B_1: \hspace{.5cm}
\mathbb{F}_{00}^{(B_1)}=\textbf{I}_{5},\hspace{.5cm}
\mathbb{F}_{20}^{(B_1)}=\left(
\begin{array}{ccccc}
 \frac{2}{\sqrt{5}} & 0 & 0 & -\frac{9}{7 \sqrt{5}} & 0 \\
 0 & -\frac{1}{\sqrt{5}} & 0 & \frac{6}{7}\sqrt{\frac{2}{5}} & 0 \\
 0 & 0 & -\frac{\sqrt{5}}{7} & 0 & -\frac{\sqrt{10}}{7} \\
 -\frac{9}{7 \sqrt{5}} & \frac{6}{7}\sqrt{\frac{2}{5}} & 0 & \frac{8}{7 \sqrt{5}} & 0 \\
 0 & 0 & -\frac{\sqrt{10}}{7} & 0 & 0 \\
\end{array}
\right),
\nonumber\\
&\hspace{1.2cm}
\mathbb{F}_{22}^{(B_1)}=\left(
\begin{array}{ccccc}
 0 & 0 & 0 & 0 & -\frac{3 \sqrt{2}}{7}   \\
 0 & 0 & \sqrt{2} & 0 & \frac{4}{7} \\
 0 & -\sqrt{2} & 0 & \frac{2}{7} & 0 \\
 0 & 0 & -\frac{2}{7} & 0 & -\frac{4 \sqrt{2}}{7}   \\
 \frac{3 \sqrt{2}}{7} & -\frac{4}{7} & 0 & \frac{4 \sqrt{2}}{7} & 0 \\
\end{array}
\right)
,\hspace{.5cm}
\mathbb{F}_{40}^{(B_1)}=\left(
\begin{array}{ccccc}
 0 & 0 & 0 & -\frac{4}{7} & 0 \\
 0 & 0 & 0 & -\frac{2 \sqrt{2}}{7}   & 0 \\
 0 & 0 & -\frac{2}{21} & 0 & \frac{5 \sqrt{2}}{21} \\
 -\frac{4}{7} & -\frac{2 \sqrt{2}}{7}   & 0 & \frac{2}{7} & 0 \\
 0 & 0 & \frac{5 \sqrt{2}}{21} & 0 & -\frac{1}{3} \\
\end{array}
\right),
\nonumber
\end{align}
\begin{align}
&\hspace{1.2cm}\mathbb{F}_{42}^{(B_1)}=\left(
\begin{array}{ccccc}
 0 & 0 & 0 & 0 & -\frac{2 \sqrt{6}}{7}   \\
 0 & 0 & 0 & 0 & -\frac{2 \sqrt{3}}{7}   \\
 0 & 0 & 0 & -\frac{10}{7 \sqrt{3}} & 0 \\
 0 & 0 & \frac{10}{7 \sqrt{3}} & 0 & -\frac{\sqrt{\frac{2}{3}}}{7} \\
 \frac{2 \sqrt{6}}{7} & \frac{2 \sqrt{3}}{7} & 0 & \frac{\sqrt{\frac{2}{3}}}{7} & 0 \\
\end{array}
\right),\hspace{.5cm}
\mathbb{F}_{44}^{(B_1)}=\left(
\begin{array}{ccccc}
 0 & 0 & 0 & 0 & 0 \\
 0 & 0 & 0 & 0 & 0 \\
 0 & 0 & -\frac{2}{3}\sqrt{\frac{10}{7}} & 0 & -\frac{2}{3}  \sqrt{\frac{5}{7}} \\
 0 & 0 & 0 & 0 & 0 \\
 0 & 0 & -\frac{2}{3}  \sqrt{\frac{5}{7}} & 0 & -\frac{\sqrt{\frac{10}{7}}}{3} \\
\end{array}
\right),
\nonumber\\
&\hspace{1.2cm}{\mathbb{M}}^{(B_1)}=
\left(
\begin{array}{ccccc}
 \frac{2 \mathcal{M}_{1,S}+2 \sqrt{2}\mathcal{M}_{1,SD}+\mathcal{M}_{1,D}}{3\det\mathcal{M}_{1}} & \frac{\sqrt{2} \mathcal{M}_{1,S}-\mathcal{M}_{1,SD}-\sqrt{2}\mathcal{M}_{1,D}}{3 \det\mathcal{M}_{1}} & 0 & 0 & 0 \\
 \frac{\sqrt{2} \mathcal{M}_{1,S}-\mathcal{M}_{1,SD}-\sqrt{2}\mathcal{M}_{1,D}}{3 \det\mathcal{M}_{1}} & \frac{\mathcal{M}_{1,S}-2 \sqrt{2}\mathcal{M}_{1,SD}+2\mathcal{M}_{1,D}}{3 \det\mathcal{M}_{1}} & 0 & 0 & 0 \\
 0 & 0 & \mathcal{M}_{2,D}^{-1} & 0 & 0 \\
 0 & 0 & 0 & \mathcal{M}_{3,D}^{-1} & 0 \\
 0 & 0 & 0 & 0 & \mathcal{M}_{3,D}^{-1} \\
\end{array}
\right).
\label{I110B1}
\end{align}
\begin{align}
& B_2:\hspace{.5cm}\mathbb{F}_{00}^{(B_2)}=\textbf{I}_{5},\hspace{.5cm}
\mathbb{F}_{20}^{(B_2)}=\left(
\begin{array}{ccccc}
 -\frac{1}{\sqrt{5}} & 0 & 0 & 0 & -\frac{3}{7}  \sqrt{\frac{6}{5}} \\
 0 & \frac{1}{2 \sqrt{5}} & -\frac{\sqrt{3}}{2} & 0 & \frac{4 \sqrt{\frac{3}{5}}}{7} \\
 0 & -\frac{\sqrt{3}}{2} & \frac{\sqrt{5}}{14} & 0 & -\frac{2}{7} \\
 0 & 0 & 0 & -\frac{2\sqrt{5}}{7}   & 0 \\
 -\frac{3}{7}  \sqrt{\frac{6}{5}} & \frac{4 \sqrt{\frac{3}{5}}}{7} & -\frac{2}{7} & 0 & \frac{6}{7 \sqrt{5}} \\
\end{array}
\right),
\nonumber\\
&\hspace{1.2cm}\mathbb{F}_{22}^{(B_2)}=\left(
\begin{array}{ccccc}
 -i \sqrt{\frac{6}{5}} & 0 & 0 & -\frac{3 \sqrt{3}}{7}  & -\frac{3 i}{7 \sqrt{5}} \\
 0 & i \sqrt{\frac{3}{10}} & \frac{i}{\sqrt{2}} & \frac{2 \sqrt{6}}{7} & \frac{2}{7} i \sqrt{\frac{2}{5}} \\
 0 & \frac{i}{\sqrt{2}} & \frac{i}{7}  \sqrt{\frac{15}{2}} & \frac{\sqrt{10}}{7} & -\frac{i \sqrt{6}}{7}  \\
 \frac{3 \sqrt{3}}{7} & -\frac{2 \sqrt{6}}{7}   & -\frac{\sqrt{10}}{7} & 0 & \frac{2 \sqrt{2}}{7} \\
 -\frac{3 i}{7 \sqrt{5}} & \frac{2}{7} i \sqrt{\frac{2}{5}} & -\frac{i \sqrt{6}}{7}  & -\frac{2 \sqrt{2}}{7}   & \frac{-4i}{7}   \sqrt{\frac{6}{5}} \\
\end{array}
\right),\hspace{.5cm}
\mathbb{F}_{40}^{(B_2)}=\left(
\begin{array}{ccccc}
 0 & 0 & 0 & 0 & \frac{\sqrt{6}}{7} \\
 0 & 0 & 0 & 0 & \frac{\sqrt{3}}{7} \\
 0 & 0 & \frac{8}{21} & 0 & -\frac{5 \sqrt{5}}{21}  \\
 0 & 0 & 0 & \frac{1}{7} & 0 \\
 \frac{\sqrt{6}}{7} & \frac{\sqrt{3}}{7} & -\frac{5 \sqrt{5}}{21}  & 0 & \frac{1}{21} \\
\end{array}
\right),
\nonumber\\
&\hspace{1.2cm}
\mathbb{F}_{42}^{(B_2)}=\left(
\begin{array}{ccccc}
 0 & 0 & 0 & \frac{1}{7} & \frac{i \sqrt{15}}{7} \\
 0 & 0 & 0 & \frac{1}{7 \sqrt{2}} & \frac{i}{7}  \sqrt{\frac{15}{2}} \\
 0 & 0 & -i\frac{4  \sqrt{10}}{21}  & -\frac{\sqrt{\frac{15}{2}}}{7} & -\frac{5 i}{21 \sqrt{2}} \\
 -\frac{1}{7} & -\frac{1}{7 \sqrt{2}} & \frac{\sqrt{\frac{15}{2}}}{7} & 0 & -\frac{\sqrt{6}}{7} \\
 \frac{i \sqrt{15}}{7} & \frac{i}{7}  \sqrt{\frac{15}{2}} & -\frac{5 i}{21 \sqrt{2}} & \frac{\sqrt{6}}{7} & - i \frac{2\sqrt{10}}{21}  \\
\end{array}
\right)
,\nonumber\\
&\hspace{1.2cm}
\mathbb{F}_{44}^{(B_2)}=\left(
\begin{array}{ccccc}
 0 & 0 & 0 & \frac{2 i}{\sqrt{7}} & 0 \\
 0 & 0 & 0 & i \sqrt{\frac{2}{7}} & 0 \\
 0 & 0 & 0 & i \sqrt{\frac{10}{21}} & 0 \\
 -\frac{2 i}{\sqrt{7}} & -i \sqrt{\frac{2}{7}} & -i \sqrt{\frac{10}{21}} & 0 & -i \sqrt{\frac{2}{21}} \\
 0 & 0 & 0 & i \sqrt{\frac{2}{21}} & 0 \\
\end{array}
\right),
\nonumber\\
&\hspace{1.2cm}
{\mathbb{M}}^{(B_2)}=\left(
\begin{array}{ccccc}
 \frac{2 \mathcal{M}_{1,S}+2 \sqrt{2}\mathcal{M}_{1,SD}+\mathcal{M}_{1,D}}{3\det\mathcal{M}_{1}} & \frac{\sqrt{2} \mathcal{M}_{1,S}-\mathcal{M}_{1,SD}-\sqrt{2}\mathcal{M}_{1,D}}{3 \det\mathcal{M}_{1}} & 0 & 0 & 0 \\
 \frac{\sqrt{2} \mathcal{M}_{1,S}-\mathcal{M}_{1,SD}-\sqrt{2}\mathcal{M}_{1,D}}{3 \det\mathcal{M}_{1}} & \frac{\mathcal{M}_{1,S}-2 \sqrt{2}\mathcal{M}_{1,SD}+2\mathcal{M}_{1,D}}{3 \det\mathcal{M}_{1}} & 0 & 0 & 0 \\
 0 & 0 & \mathcal{M}_{2,D}^{-1} & 0 & 0 \\
 0 & 0 & 0 & \mathcal{M}_{3,D}^{-1} & 0 \\
 0 & 0 & 0 & 0 & \mathcal{M}_{3,D}^{-1} \\
\end{array}
\right).
\label{I110B2}
\end{align}
\begin{align}
&B_3:\hspace{.5cm}
\mathbb{F}_{00}^{(B_3)}=\textbf{I}_{5},\hspace{.5cm}
\mathbb{F}_{20}^{(B_3)}=\left(
\begin{array}{ccccc}
 \frac{1}{2 \sqrt{5}} & 0 & -\frac{\sqrt{3}}{2} & \frac{4 \sqrt{\frac{3}{5}}}{7} & 0 \\
 0 & -\frac{1}{\sqrt{5}} & 0 & -\frac{3}{7} \sqrt{\frac{6}{5}} & 0 \\
 -\frac{\sqrt{3}}{2} & 0 & \frac{\sqrt{5}}{14} & -\frac{2}{7} & 0 \\
 \frac{4 \sqrt{\frac{3}{5}}}{7} & -\frac{3}{7} \sqrt{\frac{6}{5}} & -\frac{2}{7} & \frac{6}{7 \sqrt{5}} & 0 \\
 0 & 0 & 0 & 0 & -\frac{2 \sqrt{5}}{7}  \\
\end{array}
\right),
\nonumber\\
&\hspace{1.2cm}
\mathbb{F}_{22}^{(B_3)}=\left(
\begin{array}{ccccc}
 -i \sqrt{\frac{3}{10}} & 0 & -\frac{i}{\sqrt{2}} & \frac{-2i}{7}   \sqrt{\frac{2}{5}} & \frac{2 \sqrt{6}}{7} \\
 0 & i \sqrt{\frac{6}{5}} & 0 & \frac{3 i}{7 \sqrt{5}} & -\frac{3 \sqrt{3}}{7}   \\
 -\frac{i}{\sqrt{2}} & 0 & -\frac{i}{7}  \sqrt{\frac{15}{2}} & \frac{i \sqrt{6}}{7} & \frac{\sqrt{10}}{7} \\
 \frac{-2i}{7}   \sqrt{\frac{2}{5}} & \frac{3 i}{7 \sqrt{5}} & \frac{i \sqrt{6}}{7} & \frac{4}{7} i \sqrt{\frac{6}{5}} & -\frac{2 \sqrt{2}}{7}   \\
 -\frac{2 \sqrt{6}}{7}   & \frac{3 \sqrt{3}}{7} & -\frac{\sqrt{10}}{7} & \frac{2 \sqrt{2}}{7} & 0 \\
\end{array}
\right)
,\hspace{.5cm}
\mathbb{F}_{40}^{(B_3)}=\left(
\begin{array}{ccccc}
 0 & 0 & 0 & \frac{\sqrt{3}}{7} & 0 \\
 0 & 0 & 0 & \frac{\sqrt{6}}{7} & 0 \\
 0 & 0 & \frac{8}{21} & -\frac{5 \sqrt{5}}{21}  & 0 \\
 \frac{\sqrt{3}}{7} & \frac{\sqrt{6}}{7} & -\frac{5 \sqrt{5}}{21}  & \frac{1}{21} & 0 \\
 0 & 0 & 0 & 0 & \frac{1}{7} \\
\end{array}
\right),
\nonumber\\
&\hspace{1.2cm}
\mathbb{F}_{42}^{(B_3)}=\left(
\begin{array}{ccccc}
 0 & 0 & 0 & -\frac{i}{7}  \sqrt{\frac{15}{2}} & \frac{1}{7 \sqrt{2}} \\
 0 & 0 & 0 & -\frac{i \sqrt{15}}{7}   & \frac{1}{7} \\
 0 & 0 & \frac{4 i \sqrt{10}}{21} & \frac{5 i}{21 \sqrt{2}} & -\frac{\sqrt{\frac{15}{2}}}{7} \\
 -\frac{i}{7}  \sqrt{\frac{15}{2}} & -\frac{i \sqrt{15}}{7}   & \frac{5 i}{21 \sqrt{2}} & \frac{2 i \sqrt{10}}{21} & \frac{\sqrt{6}}{7} \\
 -\frac{1}{7 \sqrt{2}} & -\frac{1}{7} & \frac{\sqrt{\frac{15}{2}}}{7} & -\frac{\sqrt{6}}{7} & 0 \\
\end{array}
\right)
,\nonumber\\
&\hspace{1.2cm}
\mathbb{F}_{44}^{(B_3)}=\left(
\begin{array}{ccccc}
 0 & 0 & 0 & 0 & -i \sqrt{\frac{2}{7}} \\
 0 & 0 & 0 & 0 & -\frac{2 i}{\sqrt{7}} \\
 0 & 0 & 0 & 0 & -i \sqrt{\frac{10}{21}} \\
 0 & 0 & 0 & 0 & -i \sqrt{\frac{2}{21}} \\
 i \sqrt{\frac{2}{7}} & \frac{2 i}{\sqrt{7}} & i \sqrt{\frac{10}{21}} & i \sqrt{\frac{2}{21}} & 0 \\
\end{array}
\right),
\nonumber\\
&\hspace{1.2cm}
{\mathbb{M}}^{(B_3)}=
\left(
\begin{array}{ccccc}
 \frac{\mathcal{M}_{1,S}-2 \sqrt{2}\mathcal{M}_{1,SD}+2\mathcal{M}_{1,D}}{3 \det\mathcal{M}_{1}} & \frac{\sqrt{2} \mathcal{M}_{1,S}-\mathcal{M}_{1,SD}-\sqrt{2}\mathcal{M}_{1,D}}{3 \det\mathcal{M}_{1}} & 0 & 0 & 0 \\
 \frac{\sqrt{2} \mathcal{M}_{1,S}-\mathcal{M}_{1,SD}-\sqrt{2}\mathcal{M}_{1,D}}{3 \det\mathcal{M}_{1}} & \frac{2 \mathcal{M}_{1,S}+2 \sqrt{2}\mathcal{M}_{1,SD}+\mathcal{M}_{1,D}}{3 \det\mathcal{M}_{1}} & 0 & 0 & 0 \\
 0 & 0 & \mathcal{M}_{2,D}^{-1} & 0 & 0 \\
 0 & 0 & 0 & \mathcal{M}_{3,D}^{-1} & 0 \\
 0 & 0 & 0 & 0 & \mathcal{M}_{3,D}^{-1} \\
\end{array}
\right).
\label{I110B3}\\
& A: \hspace{.5cm}\mathbb{F}_{00}^{(A)}=\textbf{I}_{3},\hspace{.5cm}
\mathbb{F}_{20}^{(A)}=\left(
\begin{array}{ccc}
 \frac{\sqrt{5}}{7} & 0 & 0 \\
 0 & -\frac{\sqrt{5}}{7} & -\frac{\sqrt{10}}{7} \\
 0 & -\frac{\sqrt{10}}{7} & 0 \\
\end{array}
\right),\hspace{.5cm}
\mathbb{F}_{22}^{(A)}=\left(
\begin{array}{ccc}
 0 & -\frac{\sqrt{10}}{7} & \frac{2 \sqrt{5}}{7} \\
 \frac{\sqrt{10}}{7} & 0 & 0 \\
 -\frac{2 \sqrt{5}}{7}  & 0 & 0 \\
\end{array}
\right),
\nonumber\\
&\hspace{1.2cm}
\mathbb{F}_{40}^{(A)}=\left(
\begin{array}{ccc}
 -\frac{4}{7} & 0 & 0 \\
 0 & -\frac{2}{21} & \frac{5 \sqrt{2}}{21} \\
 0 & \frac{5 \sqrt{2}}{21} & -\frac{1}{3} \\
\end{array}
\right),\hspace{.5cm}
\nonumber
\mathbb{F}_{42}^{(A)}=\left(
\begin{array}{ccc}
 0 & -\frac{2 \sqrt{\frac{10}{3}}}{7}   & \frac{4 \sqrt{\frac{5}{3}}}{7} \\
 \frac{2 \sqrt{\frac{10}{3}}}{7} & 0 & 0 \\
 -\frac{4}{7} \sqrt{\frac{5}{3}} & 0 & 0 \\
\end{array}
\right),
\nonumber\\
&\hspace{1.2cm}
\mathbb{F}_{44}^{(A)}=\left(
\begin{array}{ccc}
 0 & 0 & 0 \\
 0 & \frac{2 \sqrt{\frac{10}{7}}}{3} & \frac{2 \sqrt{\frac{5}{7}}}{3} \\
 0 & \frac{2 \sqrt{\frac{5}{7}}}{3} & \frac{\sqrt{\frac{10}{7}}}{3} \\
\end{array}
\right),\hspace{.5cm}
{\mathbb{M}}^{(A)}=\left(
\begin{array}{ccc}
 \mathcal{M}_{2,D}^{-1} & 0 & 0 \\
 0 & \mathcal{M}_{2,D}^{-1} & 0 \\
 0 & 0 & \mathcal{M}_{3,D}^{-1} \\
\end{array}
\right).
\label{I110A}
\end{align}
%

\subsection{Positive parity isotriplet channel}
The scattering amplitude matrix for this channel is only a $6\times6$ matrix as following
\begin{eqnarray}
\mathcal{M}^{\infty}_{(1,0)} = \left( \begin{array}{cccccccccc}
\mathcal{M}_{0,S}&0\\
0&\mathcal{M}_{2,D}\\
\end{array} \right),
\end{eqnarray}
}
%

\subsection{Positive parity isotriplet channel}
The scattering amplitude matrix for this channel is only a $6\times6$ matrix as following
\begin{eqnarray}
\mathcal{M}^{\infty}_{(1,0)} = \left( \begin{array}{cccccccccc}
\mathcal{M}_{0,S}&0\\
0&\mathcal{M}_{2,D}\\
\end{array} \right),
\end{eqnarray}
where each element is still a diagonal matrix in the $M_J$ basis. The QC in Eq. (\ref{QC-simplified}) for each irrep of the corresponding point group should be understood with the matrices that are given below.
{\small
\subsubsection{$\mathbf{d}=(0,0,0)$}
\begin{align}
&A_1:\hspace{.5cm}\mathbb{F}_{00}^{(A_1)}=1,\hspace{.5cm}
{\mathbb{M}}^{(A_1)}=\mathcal{M}_{0,S}^{-1}.
\label{II000A1}
\\
&E:\hspace{.5cm}
\mathbb{F}_{00}^{(E)}=1,\hspace{.5cm}
\mathbb{F}_{40}^{(E)}=\frac{6}{7},\hspace{.5cm}
{\mathbb{M}}^{(E)}=\mathcal{M}_{2,D}^{-1}.
\label{II000E}
\\
&T_2:\hspace{.5cm}
\mathbb{F}_{00}^{(T_2)}=1,\hspace{.5cm}
\mathbb{F}_{40}^{(T_2)}=-\frac{4}{7},\hspace{.5cm}
{\mathbb{M}}^{(T_2)}=\mathcal{M}_{2,D}^{-1}.
\label{II000T2}
\end{align}
%
\subsubsection{$\mathbf{d}=(0,0,1)$}
\begin{align}
&A_1:\hspace{.5cm}\mathbb{F}_{00}^{(A_1)}=\textbf{I}_{2},\hspace{.5cm}
\mathbb{F}_{20}^{(A_1)}=\left(
\begin{array}{cc}
 0 & 1 \\
 1 & \frac{2 \sqrt{5}}{7} \\
\end{array}
\right),\hspace{.5cm}
\mathbb{F}_{40}^{(A_1)}=\left(
\begin{array}{cc}
 0 & 0 \\
 0 & \frac{6}{7} \\
\end{array}
\right),
\hspace{.5cm} {\mathbb{M}}^{(A_1)}=\left(
\begin{array}{cc}
 \mathcal{M}_{0,S}^{-1} & 0 \\
 0 & \mathcal{M}_{2,D}^{-1} \\
\end{array}
\right).
\label{II001A1}
\\
&B_1:\hspace{.5cm}\mathbb{F}_{00}^{(B_1)}=1,\hspace{.5cm}
\mathbb{F}_{20}^{(B_1)}=-\frac{2 \sqrt{5}}{7} ,\hspace{.5cm}
\mathbb{F}_{40}^{(B_1)}=\frac{1}{7},\hspace{.5cm}
\mathbb{F}_{44}^{(B_1)}=\sqrt{\frac{10}{7}},\hspace{.5cm}
{\mathbb{M}}^{(B_1)}=\mathcal{M}_{2,D}^{-1}.
\label{II001B1}
\\
&B_1:\hspace{.5cm}\mathbb{F}_{00}^{(B_1)}=1,\hspace{.5cm}
\mathbb{F}_{20}^{(B_1)}=-\frac{2 \sqrt{5}}{7} ,\hspace{.5cm}
\mathbb{F}_{40}^{(B_1)}=\frac{1}{7},\hspace{.5cm}
\mathbb{F}_{44}^{(B_1)}=-\sqrt{\frac{10}{7}},\hspace{.5cm}
{\mathbb{M}}^{(B_1)}=\mathcal{M}_{2,D}^{-1}.
\label{II001B2}
\\
&E:\hspace{.5cm}\mathbb{F}_{00}^{(E)}=1,\hspace{.5cm}
\mathbb{F}_{20}^{(E)}=\frac{\sqrt{5}}{7},\hspace{.5cm}
\mathbb{F}_{40}^{(E)}=-\frac{4}{7},\hspace{.5cm}
{\mathbb{M}}^{(E)}=\mathcal{M}_{2,D}^{-1}.
\label{II001E}
\end{align}
%
\subsubsection{$\mathbf{d}=(1,1,0)$}
\begin{align}
&B_1:\hspace{.5cm}\mathbb{F}_{00}^{(B_1)}=1,\hspace{.5cm}
\mathbb{F}_{20}^{(B_1)}=-\frac{2 \sqrt{5}}{7},\hspace{.5cm}
\mathbb{F}_{40}^{(B_1)}=\frac{1}{7},\hspace{.5cm}
\mathbb{F}_{44}^{(B_1)}=\sqrt{\frac{10}{7}},\hspace{.5cm}
{\mathbb{M}}^{(B_1)}=\mathcal{M}_{2,D}^{-1}.
\label{II110B1}
\\
&B_2:\hspace{.5cm}\mathbb{F}_{00}^{(B_2)}=1,\hspace{.5cm}
\mathbb{F}_{20}^{(B_2)}=\frac{\sqrt{5}}{7},\hspace{.5cm}
\mathbb{F}_{22}^{(B_2)}=-\frac{i \sqrt{30}}{7} ,\hspace{.5cm}
\mathbb{F}_{40}^{(B_2)}=-\frac{4}{7},\hspace{.5cm}
\mathbb{F}_{42}^{(B_2)}=-\frac{2 i \sqrt{10}}{7}  ,
\nonumber\\
&\hspace{1.2cm}
{\mathbb{M}}^{(B_2)}=\mathcal{M}_{2,D}^{-1}.
\label{II110B2}
\\
&B_3:\hspace{.5cm}\mathbb{F}_{00}^{(B_3)}=1,\hspace{.5cm}
\mathbb{F}_{20}^{(B_3)}=\frac{\sqrt{5}}{7},\hspace{.5cm}
\mathbb{F}_{22}^{(B_3)}=\frac{i \sqrt{30}}{7},\hspace{.5cm}
\mathbb{F}_{40}^{(B_3)}=-\frac{4}{7},
\hspace{.5cm}
\mathbb{F}_{42}^{(B_3)}=\frac{2 i \sqrt{10}}{7},
\nonumber\\
&\hspace{1.2cm}
{\mathbb{M}}^{(B_3)}=\mathcal{M}_{2,D}^{-1}.
\label{II110B3}
\end{align}
\begin{align}
&A:\hspace{.5cm}
\mathbb{F}_{00}^{(A)}=\textbf{I}_{3},\hspace{.5cm}
\mathbb{F}_{20}^{(A)}=\left(
\begin{array}{ccc}
 0 & 0 & 1 \\
 0 & -\frac{2 \sqrt{5}}{7} & 0 \\
 1 & 0 & \frac{2 \sqrt{5}}{7}
\end{array}
\right),\hspace{.5cm}
\mathbb{F}_{22}^{(A)}=\left(
\begin{array}{ccc}
 0 & \sqrt{2} & 0 \\
 -\sqrt{2} & 0 & \frac{2 \sqrt{10}}{7} \\
 0 & -\frac{2 \sqrt{10}}{7} & 0
\end{array}
\right),
\nonumber\\
&\hspace{1.2cm}
\mathbb{F}_{40}^{(A)}=\left(
\begin{array}{ccc}
 0 & 0 & 0 \\
 0 & \frac{1}{7} & 0 \\
 0 & 0 & \frac{6}{7}
\end{array}
\right),\hspace{.5cm}
\mathbb{F}_{42}^{(A)}=\left(
\begin{array}{ccc}
 0 & 0 & 0 \\
 0 & 0 & -\frac{\sqrt{30}}{7} \\
 0 & \frac{\sqrt{30}}{7} & 0
\end{array}
\right),\hspace{.5cm}
\mathbb{F}_{44}^{(A)}=\left(
\begin{array}{ccc}
 0 & 0 & 0 \\
 0 & -\sqrt{\frac{10}{7}} & 0 \\
 0 & 0 & 0
\end{array}
\right),
\nonumber\\
&\hspace{1.2cm}
{\mathbb{M}}^{(A)}=\left(
\begin{array}{ccc}
 \mathcal{M}_{0,S}^{-1} & 0 & 0 \\
 0 & \mathcal{M}_{2,D}^{-1} & 0 \\
 0 & 0 & \mathcal{M}_{2,D}^{-1} \\
\end{array}
\right).
\label{II110A}
\end{align}
}
\subsection{Negative parity isosinglet channel}
Given the truncation made on the angular momentum in the master QC, the scattering amplitude matrix for this channel is a $10\times10$ matrix, and is given by
\begin{eqnarray}
\mathcal{M}^{\infty}_{(0,0)} = \left( \begin{array}{cccccccccc}
\mathcal{M}_{1,P}&0\\
0&\mathcal{M}_{3,F}\\
\end{array} \right),
\end{eqnarray}
where each element is still a diagonal matrix in the $M_J$ basis. The following matrices should be used in the QC in Eq. (\ref{QC-simplified}) for this channel. 

{\small
\subsubsection{$\mathbf{d}=(0,0,0)$}
\begin{align}
&T_1:\hspace{.5cm}
\mathbb{F}_{00}^{(T_1)}=\textbf{I}_{2},\hspace{.5cm}
\mathbb{F}_{40}^{(T_1)}=\left(
\begin{array}{cc}
 0 & -\frac{4}{\sqrt{21}} \\
 -\frac{4}{\sqrt{21}} & \frac{6}{11} \\
\end{array}
\right),\hspace{.5cm}
\mathbb{F}_{60}^{(T_1)}=\left(
\begin{array}{cc}
 0 & 0 \\
 0 & \frac{100}{33 \sqrt{13}} \\
\end{array}
\right),
\nonumber\\
&\hspace{1.2cm}{\mathbb{M}}^{(T_1)}=
\left(
\begin{array}{cc}
\mathcal{M}_{1,P}^{-1} & 0 \\
 0 & \mathcal{M}_{3,F}^{-1} \\
\end{array}
\right).
\label{III000T1}
\\
&A_2:\hspace{.5cm}
\mathbb{F}_{00}^{(A_2)}=1,\hspace{.5cm}
\mathbb{F}_{40}^{(A_2)}=-\frac{12}{11},\hspace{.5cm}
\mathbb{F}_{60}^{(A_2)}=\frac{80}{11 \sqrt{13}},\hspace{.5cm}
{\mathbb{M}}^{(A_2)}=\mathcal{M}_{3,F}^{-1}.
\label{III000A}
\\
&T_2:\hspace{.5cm}
\mathbb{F}_{00}^{(T_2)}=1,\hspace{.5cm}
\mathbb{F}_{40}^{(T_2)}=-\frac{2}{11},\hspace{.5cm}
\mathbb{F}_{60}^{(T_2)}=-\frac{60}{11 \sqrt{13}},\hspace{.5cm}
{\mathbb{M}}^{(T_2)}=\mathcal{M}_{3,F}^{-1}.
\label{III000T2}
\end{align}
%

\subsubsection{$\mathbf{d}=(0,0,1)$}
\begin{align}
&A_2:\hspace{.5cm}
\mathbb{F}_{00}^{(A_2)}=\textbf{I}_{2},\hspace{.5cm}
\mathbb{F}_{20}^{(A_2)}=\left(
\begin{array}{cc}
 \frac{2}{\sqrt{5}} & 3 \sqrt{\frac{3}{35}} \\
 3 \sqrt{\frac{3}{35}} & \frac{4}{3 \sqrt{5}} \\
\end{array}
\right),\hspace{.5cm}
\mathbb{F}_{40}^{(A_2)}=\left(
\begin{array}{cc}
 0 & \frac{4}{\sqrt{21}} \\
 \frac{4}{\sqrt{21}} & \frac{6}{11} \\
\end{array}
\right),
\nonumber\\
&\hspace{1.25cm}\mathbb{F}_{60}^{(A_2)}=\left(
\begin{array}{cc}
 0 & 0 \\
 0 & \frac{100}{33 \sqrt{13}} \\
\end{array}
\right),\hspace{.5cm}
{\mathbb{M}}^{(A_2)}=\left(
\begin{array}{cc}
\mathcal{M}_{1,P}^{-1} & 0 \\
 0 & \mathcal{M}_{3,F}^{-1} \\
\end{array}
\right).
\label{III001A2}
\\
&B_1:\hspace{.5cm}
\mathbb{F}_{00}^{(B_1)}=1,\hspace{.5cm}
\mathbb{F}_{40}^{(B_1)}=-\frac{7}{11},\hspace{.5cm}
\mathbb{F}_{44}^{(B_1)}=-\frac{\sqrt{70}}{11},\hspace{.5cm}
\mathbb{F}_{60}^{(B_1)}=\frac{10}{11 \sqrt{13}},\hspace{.5cm}
\nonumber\\
&\hspace{1.25cm}
\mathbb{F}_{64}^{(B_1)}=-\frac{10 \sqrt{\frac{14}{13}}}{11},\hspace{.5cm}
{\mathbb{M}}^{(B_1)}=\mathcal{M}_{3,F}^{-1}.
\label{III001B1}
\end{align}
\begin{align}
&B_2:\hspace{.5cm}
\mathbb{F}_{00}^{(B_2)}=1,\hspace{.5cm}
\mathbb{F}_{40}^{(B_2)}=-\frac{7}{11},\hspace{.5cm}
\mathbb{F}_{44}^{(B_2)}=\frac{\sqrt{70}}{11},\hspace{.5cm}
\mathbb{F}_{60}^{(B_2)}=\frac{10}{11 \sqrt{13}},\hspace{.5cm}
\nonumber\\
&\hspace{1.2cm}
\mathbb{F}_{64}^{(B_2)}=\frac{10 \sqrt{\frac{14}{13}}}{11},\hspace{.5cm}
{\mathbb{M}}^{(B_2)}=\mathcal{M}_{3,F}^{-1}.
\label{III001B2}
\\
&E:\hspace{.5cm}
\mathbb{F}_{00}^{(E)}=\textbf{I}_{3},\hspace{.5cm}
\mathbb{F}_{20}^{(E)}=\left(
\begin{array}{ccc}
 -\frac{1}{\sqrt{5}} & 3 \sqrt{\frac{2}{35}} & 0 \\
 3 \sqrt{\frac{2}{35}} & \frac{1}{\sqrt{5}} & 0 \\
 0 & 0 & -\frac{\sqrt{5}}{3} \\
\end{array}
\right),\hspace{.5cm}
\mathbb{F}_{40}^{(E)}=\left(
\begin{array}{ccc}
 0 & -\sqrt{\frac{2}{7}} & 0 \\
 -\sqrt{\frac{2}{7}} & \frac{1}{11} & 0 \\
 0 & 0 & \frac{3}{11} \\
\end{array}
\right),
\nonumber\\
&\hspace{1.1cm}\mathbb{F}_{44}^{(E)}=\left(
\begin{array}{ccc}
 0 & 0 & -\frac{2}{\sqrt{3}} \\
 0 & 0 & \frac{\sqrt{42}}{11} \\
 -\frac{2}{\sqrt{3}} & \frac{\sqrt{42}}{11} & 0 \\
\end{array}
\right),\hspace{.5cm}
\mathbb{F}_{60}^{(E)}=
\left(
\begin{array}{ccc}
 0 & 0 & 0 \\
 0 & -\frac{25}{11 \sqrt{13}} & 0 \\
 0 & 0 & -\frac{5}{33 \sqrt{13}} \\
\end{array}
\right),
\nonumber\\
&\hspace{1.2cm}
\mathbb{F}_{64}^{(E)}=\left(
\begin{array}{ccc}
 0 & 0 & 0 \\
 0 & 0 & -\frac{5}{11}  \sqrt{\frac{70}{39}} \\
 0 & -\frac{5}{11}  \sqrt{\frac{70}{39}} & 0 \\
\end{array}
\right),
\hspace{.5cm}
{\mathbb{M}}^{(E)}=\left(
\begin{array}{ccc}
\mathcal{M}_{1,P}^{-1} & 0 & 0 \\
 0 & \mathcal{M}_{3,F}^{-1} & 0 \\
 0 & 0 & \mathcal{M}_{3,F}^{-1} \\
\end{array}
\right).
\label{III001E}
\end{align}
%

\subsubsection{$\mathbf{d}=(1,1,0)$}
\begin{align}
&B_2:\hspace{.5cm}\mathbb{F}_{00}^{(B_2)}=\textbf{I}_{3},\hspace{.5cm}
\mathbb{F}_{20}^{(B_2)}=\left(
\begin{array}{ccc}
 -\frac{1}{\sqrt{5}} & 3 \sqrt{\frac{2}{35}} & 0 \\
 3 \sqrt{\frac{2}{35}} & \frac{1}{\sqrt{5}} & 0 \\
 0 & 0 & -\frac{\sqrt{5}}{3} \\
\end{array}
\right),\hspace{.5cm}
\mathbb{F}_{22}^{(B_2)}=\left(
\begin{array}{ccc}
 i \sqrt{\frac{6}{5}} & -i \sqrt{\frac{3}{35}} & \frac{3}{\sqrt{7}} \\
 -i \sqrt{\frac{3}{35}} & 2 i \sqrt{\frac{2}{15}} & -\frac{\sqrt{2}}{3} \\
 -\frac{3}{\sqrt{7}} & \frac{\sqrt{2}}{3} & 0 \\
\end{array}
\right),
\nonumber
\\
&\hspace{1.2cm}
\mathbb{F}_{40}^{(B_2)}=\left(
\begin{array}{ccc}
 0 & -\sqrt{\frac{2}{7}} & 0 \\
 -\sqrt{\frac{2}{7}} & \frac{1}{11} & 0 \\
 0 & 0 & \frac{3}{11} \\
\end{array}
\right),\hspace{.5cm}
\mathbb{F}_{42}^{(B_2)}=\left(
\begin{array}{ccc}
 0 & i \sqrt{\frac{5}{7}} & -\frac{1}{\sqrt{21}} \\
 i \sqrt{\frac{5}{7}} & \frac{2 i \sqrt{10}}{11} & \frac{3 \sqrt{6}}{11} \\
 \frac{1}{\sqrt{21}} & -\frac{3 \sqrt{6}}{11} & 0 \\
\end{array}
\right),
\nonumber
\\
&\hspace{1.2cm}\mathbb{F}_{44}^{(B_2)}=\left(
\begin{array}{ccc}
 0 & 0 & \frac{2 i}{\sqrt{3}} \\
 0 & 0 & -\frac{i \sqrt{42}}{11} \\
 -\frac{2 i}{\sqrt{3}} & \frac{i \sqrt{42}}{11} & 0 \\
\end{array}
\right),\hspace{.5cm}
\mathbb{F}_{60}^{(B_2)}=\left(
\begin{array}{ccc}
 0 & 0 & 0 \\
 0 & -\frac{25}{11 \sqrt{13}} & 0 \\
 0 & 0 & -\frac{5}{33 \sqrt{13}} \\
\end{array}
\right),
\nonumber
\\
&\hspace{1.2cm}
\mathbb{F}_{62}^{(B_2)}=\left(
\begin{array}{ccc}
 0 & 0 & 0 \\
 0 & \frac{10}{11} i \sqrt{\frac{35}{39}} & -\frac{10}{33} \sqrt{\frac{7}{13}} \\
 0 & \frac{10 \sqrt{\frac{7}{13}}}{33} & 0 \\
\end{array}
\right),\hspace{.5cm}
\mathbb{F}_{64}^{(B_2)}=\left(
\begin{array}{ccc}
 0 & 0 & 0 \\
 0 & 0 & \frac{5}{11} i \sqrt{\frac{70}{39}} \\
 0 & -\frac{1}{11} 5 i \sqrt{\frac{70}{39}} & 0 \\
\end{array}
\right),
\nonumber
\\
&\hspace{1.2cm}
\mathbb{F}_{66}^{(B_2)}=\left(
\begin{array}{ccc}
 0 & 0 & 0 \\
 0 & 0 & 0 \\
 0 & 0 & -10 i \sqrt{\frac{7}{429}} \\
\end{array}
\right),\hspace{.5cm}
{\mathbb{M}}^{(B_2)}=\left(
\begin{array}{ccc}
\mathcal{M}_{1,P}^{-1} & 0 & 0 \\
 0 & \mathcal{M}_{3,F}^{-1} & 0 \\
 0 & 0 & \mathcal{M}_{3,F}^{-1} \\
\end{array}
\right).
\label{III110B2}
\end{align}
\begin{align}
&B_3:\hspace{.5cm}\mathbb{F}_{00}^{(B_3)}=\textbf{I}_{3},\hspace{.5cm}
\mathbb{F}_{20}^{(B_3)}=\left(
\begin{array}{ccc}
 -\frac{1}{\sqrt{5}} & 0 & 3 \sqrt{\frac{2}{35}} \\
 0 & -\frac{\sqrt{5}}{3} & 0 \\
 3 \sqrt{\frac{2}{35}} & 0 & \frac{1}{\sqrt{5}} \\
\end{array}
\right),\hspace{.5cm}
\mathbb{F}_{22}^{(B_3)}=\left(
\begin{array}{ccc}
 -i \sqrt{\frac{6}{5}} & \frac{3}{\sqrt{7}} & i \sqrt{\frac{3}{35}} \\
 -\frac{3}{\sqrt{7}} & 0 & \frac{\sqrt{2}}{3} \\
 i \sqrt{\frac{3}{35}} & -\frac{\sqrt{2}}{3} & -2 i \sqrt{\frac{2}{15}} \\
\end{array}
\right),
\nonumber
\\
&\hspace{1.2cm}\mathbb{F}_{40}^{(B_3)}=\left(
\begin{array}{ccc}
 0 & 0 & -\sqrt{\frac{2}{7}} \\
 0 & \frac{3}{11} & 0 \\
 -\sqrt{\frac{2}{7}} & 0 & \frac{1}{11} \\
\end{array}
\right),\hspace{.5cm}
\mathbb{F}_{42}^{(B_3)}=\left(
\begin{array}{ccc}
 0 & -\frac{1}{\sqrt{21}} & -i \sqrt{\frac{5}{7}} \\
 \frac{1}{\sqrt{21}} & 0 & -\frac{3 \sqrt{6}}{11} \\
 -i \sqrt{\frac{5}{7}} & \frac{3 \sqrt{6}}{11} & -\frac{i2 \sqrt{10} }{11} \\
\end{array}
\right),
\nonumber
\\
&\hspace{1.2cm}
\mathbb{F}_{44}^{(B_3)}=\left(
\begin{array}{ccc}
 0 & -\frac{2 i}{\sqrt{3}} & 0 \\
 \frac{2 i}{\sqrt{3}} & 0 & - \frac{i \sqrt{42}}{11} \\
 0 & \frac{i \sqrt{42}}{11} & 0 \\
\end{array}
\right),\hspace{.5cm}
\mathbb{F}_{60}^{(B_3)}=\left(
\begin{array}{ccc}
 0 & 0 & 0 \\
 0 & -\frac{5}{33 \sqrt{13}} & 0 \\
 0 & 0 & -\frac{25}{11 \sqrt{13}} \\
\end{array}
\right),
\nonumber
\\
&\hspace{1.2cm}
\mathbb{F}_{62}^{(B_3)}=\left(
\begin{array}{ccc}
 0 & 0 & 0 \\
 0 & 0 & \frac{10 \sqrt{\frac{7}{13}}}{33} \\
 0 & -\frac{10}{33}  \sqrt{\frac{7}{13}} & -\frac{10i}{11}   \sqrt{\frac{35}{39}} \\
\end{array}
\right),\hspace{.5cm}
\mathbb{F}_{64}^{(B_3)}=\left(
\begin{array}{ccc}
 0 & 0 & 0 \\
 0 & 0 & \frac{5}{11} i \sqrt{\frac{70}{39}} \\
 0 & -\frac{1}{11} 5 i \sqrt{\frac{70}{39}} & 0 \\
\end{array}
\right),
\nonumber
\\
&\hspace{1.2cm}
\mathbb{F}_{66}^{(B_3)}=\left(
\begin{array}{ccc}
 0 & 0 & 0 \\
 0 & 10 i \sqrt{\frac{7}{429}} & 0 \\
 0 & 0 & 0 \\
\end{array}
\right),\hspace{.5cm}
{\mathbb{M}}^{(B_3)}=\left(
\begin{array}{ccc}
\mathcal{M}_{1,P}^{-1} & 0 & 0 \\
 0 & \mathcal{M}_{3,F}^{-1} & 0 \\
 0 & 0 & \mathcal{M}_{3,F}^{-1} \\
\end{array}
\right).
\label{III110B3}
\end{align}
\begin{align}
&A:\hspace{.5cm}\mathbb{F}_{00}^{(A)}=1,\hspace{.5cm}
\mathbb{F}_{40}^{(A)}=-\frac{7}{11},\hspace{.5cm}
\mathbb{F}_{44}^{(A)}=\frac{\sqrt{70}}{11},\hspace{.5cm}\mathbb{F}_{60}^{(A)}=\frac{10}{11 \sqrt{13}},\hspace{.5cm}
\nonumber\\
&\hspace{1.2cm}
\mathbb{F}_{64}^{(A)}=\frac{10 \sqrt{\frac{14}{13}}}{11},\hspace{.5cm}
{\mathbb{M}}^{(A)}=\mathcal{M}_{3,F}^{-1}.
\label{III110A}
\\
&B_1:\hspace{.5cm}
\mathbb{F}_{00}^{(B_1)}=\textbf{I}_{3},\hspace{.5cm}
\mathbb{F}_{20}^{(B_1)}=\left(
\begin{array}{ccc}
 \frac{2}{\sqrt{5}} & 3 \sqrt{\frac{3}{35}} & 0 \\
 3 \sqrt{\frac{3}{35}} & \frac{4}{3 \sqrt{5}} & 0 \\
 0 & 0 & 0 \\
\end{array}
\right),\hspace{.5cm}
\mathbb{F}_{22}^{(B_1)}=\left(
\begin{array}{ccc}
 0 & 0 & \sqrt{\frac{6}{7}} \\
 0 & 0 & -\frac{2 \sqrt{2}}{3}  \\
 -\sqrt{\frac{6}{7}} & \frac{2 \sqrt{2}}{3} & 0 \\
\end{array}
\right),
\nonumber
\\
&\hspace{1.2cm}
\mathbb{F}_{40}^{(B_1)}=\left(
\begin{array}{ccc}
 0 & \frac{4}{\sqrt{21}} & 0 \\
 \frac{4}{\sqrt{21}} & \frac{6}{11} & 0 \\
 0 & 0 & -\frac{7}{11} \\
\end{array}
\right),\hspace{.5cm}
\mathbb{F}_{42}^{(B_1)}=\left(
\begin{array}{ccc}
 0 & 0 & 2 \sqrt{\frac{2}{7}} \\
 0 & 0 & -\frac{\sqrt{6}}{11} \\
 -2 \sqrt{\frac{2}{7}} & \frac{\sqrt{6}}{11} & 0 \\
\end{array}
\right),\hspace{.5cm}
\mathbb{F}_{44}^{(B_1)}=\left(
\begin{array}{ccc}
 0 & 0 & 0 \\
 0 & 0 & 0 \\
 0 & 0 & -\frac{\sqrt{70}}{11} \\
\end{array}
\right),
\nonumber\\
&\hspace{1.2cm}
\mathbb{F}_{60}^{(B_1)}=\left(
\begin{array}{ccc}
 0 & 0 & 0 \\
 0 & \frac{100}{33 \sqrt{13}} & 0 \\
 0 & 0 & \frac{10}{11 \sqrt{13}} \\
\end{array}
\right),\hspace{.5cm}
\mathbb{F}_{62}^{(B_1)}=\left(
\begin{array}{ccc}
 0 & 0 & 0 \\
 0 & 0 & \frac{40 \sqrt{\frac{7}{13}}}{33} \\
 0 & -\frac{40}{33} \sqrt{\frac{7}{13}} & 0 \\
\end{array}
\right),
\nonumber
\\
&\hspace{1.2cm}\mathbb{F}_{64}^{(B_1)}=\left(
\begin{array}{ccc}
 0 & 0 & 0 \\
 0 & 0 & 0 \\
 0 & 0 & -\frac{10}{11} \sqrt{\frac{14}{13}} \\
\end{array}
\right),\hspace{.5cm}
{\mathbb{M}}^{(B_1)}=\left(
\begin{array}{ccc}
\mathcal{M}_{1,P}^{-1} & 0 & 0 \\
 0 & \mathcal{M}_{3,F}^{-1} & 0 \\
 0 & 0 & \mathcal{M}_{3,F}^{-1} \\
\end{array}
\right).
\label{III110B1}
\end{align}
}
%

\subsection{Negative parity isotriplet channel}
The scattering amplitude in this channel is a $30 \times 30$ matrix with the following elements
 \begin{eqnarray}
\mathcal{M}_{(1,1)}^{\infty}=\left(\begin{array}{cccccc}
\mathcal{M}_{0,P} & 0 & 0 & 0 & 0 & 0\\
0 & \mathcal{M}_{1,P} & 0 & 0 & 0 & 0\\
0 & 0 & \mathcal{M}_{2,P} & \mathcal{M}_{2,PF} & 0 & 0\\
0 & 0 & \mathcal{M}_{2,FP} & \mathcal{M}_{2,F} & 0 & 0\\
0 & 0 & 0 & 0 & \mathcal{M}_{3,F} & 0\\
0 & 0 & 0 & 0 & 0 & \mathcal{M}_{4,F}
\end{array}\right)
 \end{eqnarray}
As is seen, various QCs in this channel will give access to the P- F-wave mixing parameter in the NN negative parity sector. The QCs are obtained via inputing the following matrices in Eq. (\ref{QC-simplified}).

{\small
\subsubsection{$\mathbf{d}=(0,0,0)$}
\begin{align}
&T_1\hspace{.5cm}\mathbb{F}_{00}^{(T_1)}=\textbf{I}_{3},\hspace{.5cm}
\mathbb{F}_{40}^{(T_1)}=\left(
\begin{array}{ccc}
 0 & \sqrt{\frac{2}{7}} & \sqrt{\frac{10}{21}} \\
 \sqrt{\frac{2}{7}} & \frac{1}{11} & \frac{\sqrt{15}}{11} \\
 \sqrt{\frac{10}{21}} & \frac{\sqrt{15}}{11} & \frac{3}{11} \\
\end{array}
\right),\hspace{.5cm}
\mathbb{F}_{60}^{(T_1)}=\left(
\begin{array}{ccc}
 0 & 0 & 0 \\
 0 & -\frac{25}{11 \sqrt{13}} & \frac{35 \sqrt{\frac{5}{39}}}{11} \\
 0 & \frac{35 \sqrt{\frac{5}{39}}}{11} & -\frac{5}{33 \sqrt{13}} \\
\end{array}
\right),
\nonumber
\\
&\hspace{.9cm}{\mathbb{M}}^{(T_1)}=\left(
\begin{array}{ccc}
\mathcal{M}_{1,P}^{-1} & 0 & 0 \\
 0 & \mathcal{M}_{3,F}^{-1} & 0 \\
 0 & 0 & \mathcal{M}_{4,F}^{-1} \\
\end{array}
\right).
\label{IV000T1}
\end{align}
\begin{align}
&T_2:\hspace{.5cm}\mathbb{F}_{00}^{(T_2)}=\textbf{I}_{4},\hspace{.5cm}
\mathbb{F}_{40}^{(T_2)}=\left(
\begin{array}{cccc}
 0 & \frac{4 \sqrt{\frac{2}{3}}}{7} & \sqrt{\frac{10}{21}} & -\frac{\sqrt{\frac{10}{3}}}{7} \\
 \frac{4 \sqrt{\frac{2}{3}}}{7} & -\frac{4}{21} & \frac{2 \sqrt{\frac{5}{7}}}{3} & \frac{6 \sqrt{5}}{77} \\
 \sqrt{\frac{10}{21}} & \frac{2 \sqrt{\frac{5}{7}}}{3} & -\frac{1}{33} & -\frac{15}{11 \sqrt{7}} \\
 -\frac{\sqrt{\frac{10}{3}}}{7} & \frac{6 \sqrt{5}}{77} & -\frac{15}{11 \sqrt{7}} & -\frac{39}{77} \\
\end{array}
\right),
\nonumber
\\
&\hspace{1.2cm}
\mathbb{F}_{60}^{(T_2)}=\left(
\begin{array}{cccc}
 0 & 0 & 0 & 0 \\
 0 & 0 & 0 & -\frac{80}{33} \sqrt{\frac{5}{13}} \\
 0 & 0 & \frac{45}{11 \sqrt{13}} & \frac{35 \sqrt{\frac{7}{13}}}{33} \\
 0 & -\frac{80}{33} \sqrt{\frac{5}{13}} & \frac{35 \sqrt{\frac{7}{13}}}{33} & \frac{25}{33 \sqrt{13}} \\
\end{array}
\right),\hspace{.5cm}
{\mathbb{M}}^{(T_2)}= \left(
\begin{array}{cccc}
 \frac{\mathcal{M}_{2,F}}{\det\mathcal{M}_{2}} & -\frac{\mathcal{M}_{2,PF}}{\det\mathcal{M}_{2}} & 0 & 0 \\
 -\frac{\mathcal{M}_{2,PF}}{\det\mathcal{M}_{2}} & \frac{\mathcal{M}_{2,P}}{\det\mathcal{M}_{2}} & 0 & 0 \\
 0 & 0 & \mathcal{M}_{3,F}^{-1} & 0 \\
 0 & 0 & 0 & \mathcal{M}_{4,F}^{-1} \\
\end{array}
\right).
\label{IV000T2}
\\
&A_2:\hspace{.5cm}\mathbb{F}_{00}^{(A_2)}=1,\hspace{.5cm}
\mathbb{F}_{40}^{(A_2)}=-\frac{2}{11},\hspace{.5cm}
\mathbb{F}_{60}^{(A_2)}=-\frac{60}{11 \sqrt{13}}\hspace{.5cm}
{\mathbb{M}}^{(A_2)}=\mathcal{M}_{3,F}^{-1}.
\label{IV000A2}
\\
&A_1:\hspace{.5cm}\mathbb{F}_{00}^{(A_1)}=\textbf{I}_{2},\hspace{.5cm}
\mathbb{F}_{40}^{(A_1)}=\left(
\begin{array}{cc}
 0 & -\frac{4}{\sqrt{21}} \\
 -\frac{4}{\sqrt{21}} & \frac{6}{11} \\
\end{array}
\right), \hspace{.5cm}\mathbb{F}_{60}^{(A_1)}=\left(
\begin{array}{cc}
 0 & 0 \\
 0 & \frac{100}{33 \sqrt{13}} \\
\end{array}
\right),
\nonumber
\\
&\hspace{1.2cm}
{\mathbb{M}}^{(A_1)}= \left(
\begin{array}{cc}
 \mathcal{M}_{0,P}^{-1} & 0 \\
 0 & \mathcal{M}_{4,F}^{-1} \\
\end{array}
\right).
\label{IV000A1}
\end{align}
\begin{align}
&E:\hspace{.5cm}
\mathbb{F}_{00}^{(E)}=\textbf{I}_{2},\hspace{.5cm}
\mathbb{F}_{40}^{(E)}=\left(
\begin{array}{ccc}
 0 & -\frac{2 \sqrt{6}}{7}   & -\frac{2 \sqrt{\frac{10}{3}}}{7}   \\
 -\frac{2 \sqrt{6}}{7}   & \frac{2}{7} & \frac{12 \sqrt{5}}{77} \\
 -\frac{2 \sqrt{\frac{10}{3}}}{7}   & \frac{12 \sqrt{5}}{77} & \frac{6}{77} \\
\end{array}
\right),\hspace{.5cm}
\mathbb{F}_{60}^{(E)}=\left(
\begin{array}{ccc}
 0 & 0 & 0 \\
 0 & 0 & \frac{20 \sqrt{\frac{5}{13}}}{11} \\
 0 & \frac{20 \sqrt{\frac{5}{13}}}{11} & -\frac{80}{33 \sqrt{13}} \\
\end{array}
\right),
\nonumber
\\
&\hspace{1.2cm}
{\mathbb{M}}^{(E)}=
\left(
\begin{array}{ccc}
 \frac{\mathcal{M}_{2,F}}{\det\mathcal{M}_{2}} & -\frac{\mathcal{M}_{2,PF}}{\det\mathcal{M}_{2}} & 0 \\
 -\frac{\mathcal{M}_{2,PF}}{\det\mathcal{M}_{2}} & \frac{\mathcal{M}_{2,P}}{\det\mathcal{M}_{2}} & 0 \\
 0 & 0 & \mathcal{M}_{4,F}^{-1} \\
\end{array}
\right).
\label{IV000E}
\end{align}
%

\subsubsection{$\mathbf{d}=(0,0,1)$}
\begin{align}
&A_1:\hspace{.5cm}\mathbb{F}_{00}^{(A_1)}=\textbf{I}_{4},\hspace{.5cm}
\mathbb{F}_{20}^{(A_1)}=\left(
\begin{array}{ccccc}
 0 & -\sqrt{\frac{2}{5}} & \sqrt{\frac{3}{5}} & 0 & 0 \\
 -\sqrt{\frac{2}{5}} & \frac{1}{\sqrt{5}} & -\frac{\sqrt{\frac{6}{5}}}{7} & 0 & \frac{9 \sqrt{\frac{2}{5}}}{7} \\
 \sqrt{\frac{3}{5}} & -\frac{\sqrt{\frac{6}{5}}}{7} & \frac{8}{7 \sqrt{5}} & 0 & -\frac{2}{7 \sqrt{15}} \\
 0 & 0 & 0 & -\frac{\sqrt{5}}{3} & 0 \\
 0 & \frac{9 \sqrt{\frac{2}{5}}}{7} & -\frac{2}{7 \sqrt{15}} & 0 & \frac{5 \sqrt{5}}{21} \\
\end{array}
\right),
\nonumber
\\
&\hspace{1.2cm}
\mathbb{F}_{40}^{(A_1)}=\left(
\begin{array}{ccccc}
 0 & 0 & 0 & 0 & -\frac{2}{3} \\
 0 & 0 & -\frac{2 \sqrt{6}}{7}   & 0 & \frac{5 \sqrt{2}}{21} \\
 0 & -\frac{2 \sqrt{6}}{7}   & \frac{2}{7} & 0 & -\frac{10}{77} \sqrt{3} \\
 0 & 0 & 0 & \frac{3}{11} & 0 \\
 -\frac{2}{3} & \frac{5 \sqrt{2}}{21} & -\frac{10}{77} \sqrt{3} & 0 & \frac{27}{77} \\
\end{array}
\right),\hspace{.5cm}
\mathbb{F}_{44}^{(A_1)}=\left(
\begin{array}{ccccc}
 0 & 0 & 0 & -\frac{2 \sqrt{2}}{3}  & 0 \\
 0 & 0 & 0 & -\frac{2}{3} & 0 \\
 0 & 0 & 0 & \frac{2 \sqrt{6}}{11} & 0 \\
 -\frac{2 \sqrt{2}}{3}  & -\frac{2}{3} & \frac{2 \sqrt{6}}{11} & 0 & \frac{3 \sqrt{2}}{11} \\
 0 & 0 & 0 & \frac{3 \sqrt{2}}{11} & 0 \\
\end{array}
\right),
\nonumber
\\
&\hspace{1.2cm}
\mathbb{F}_{60}^{(A_1)}=\left(
\begin{array}{ccccc}
 0 & 0 & 0 & 0 & 0 \\
 0 & 0 & 0 & 0 & 0 \\
 0 & 0 & 0 & 0 & -\frac{50}{11 \sqrt{39}} \\
 0 & 0 & 0 & -\frac{5}{33 \sqrt{13}} & 0 \\
 0 & 0 & -\frac{50}{11 \sqrt{39}} & 0 & \frac{25}{33 \sqrt{13}} \\
\end{array}
\right),\hspace{.5cm}
\mathbb{F}_{64}^{(A_1)}=\left(
\begin{array}{ccccc}
 0 & 0 & 0 & 0 & 0 \\
 0 & 0 & 0 & 0 & 0 \\
 0 & 0 & 0 & -\frac{10 }{11} \sqrt{\frac{10}{39}} & 0 \\
 0 & 0 & -\frac{10 }{11} \sqrt{\frac{10}{39}} & 0 & -\frac{5}{11}  \sqrt{\frac{10}{13}} \\
 0 & 0 & 0 & -\frac{5}{11}  \sqrt{\frac{10}{13}} & 0 \\
\end{array}
\right),
\nonumber\\
&\hspace{1.2cm}{\mathbb{M}}^{(A_1)}=\left(
\begin{array}{ccccc}
 \mathcal{M}_{0,P}^{-1} & 0 & 0 & 0 & 0 \\
 0 & \frac{\mathcal{M}_{2,F}}{\det\mathcal{M}_{2}} & -\frac{\mathcal{M}_{2,PF}}{\det\mathcal{M}_{2}} & 0 & 0 \\
 0 & -\frac{\mathcal{M}_{2,PF}}{\det\mathcal{M}_{2}} & \frac{\mathcal{M}_{2,P}}{\det\mathcal{M}_{2}} & 0 & 0 \\
 0 & 0 & 0 & \mathcal{M}_{4,F}^{-1} & 0 \\
 0 & 0 & 0 & 0 & \mathcal{M}_{4,F}^{-1} \\
\end{array}
\right).
\label{IV001A1}
\\
&A_2:\hspace{.5cm}\mathbb{F}_{00}^{(A_2)}=\textbf{I}_{3},\hspace{.5cm}
\mathbb{F}_{20}^{(A_2)}=\left(
\begin{array}{ccc}
 -\frac{1}{\sqrt{5}} & \sqrt{\frac{18}{35}} & 0 \\
 \sqrt{\frac{18}{35}} & \frac{1}{\sqrt{5}} & 0 \\
 0 & 0 & -\frac{\sqrt{5}}{3} \\
\end{array}
\right),\hspace{.5cm}
\mathbb{F}_{40}^{(A_2)}=\left(
\begin{array}{ccc}
 0 & -\sqrt{\frac{2}{7}} & 0 \\
 -\sqrt{\frac{2}{7}} & \frac{1}{11} & 0 \\
 0 & 0 & \frac{3}{11} \\
\end{array}
\right),
\nonumber
\\
&\hspace{1.2cm}\mathbb{F}_{44}^{(A_2)}=\left(
\begin{array}{ccc}
 0 & 0 & \frac{2}{\sqrt{3}} \\
 0 & 0 & -\frac{\sqrt{42}}{11} \\
 \frac{2}{\sqrt{3}} & -\frac{\sqrt{42}}{11} & 0 \\
\end{array}
\right),\hspace{.5cm}
\mathbb{F}_{60}^{(A_2)}=\left(
\begin{array}{ccc}
 0 & 0 & 0 \\
 0 & -\frac{25}{11 \sqrt{13}} & 0 \\
 0 & 0 & -\frac{5}{33 \sqrt{13}} \\
\end{array}
\right),
\nonumber
\\
&\hspace{1.2cm}
\mathbb{F}_{64}^{(A_2)}=\left(
\begin{array}{ccc}
 0 & 0 & 0 \\
 0 & 0 & \frac{5 \sqrt{\frac{70}{39}}}{11} \\
 0 & \frac{5 \sqrt{\frac{70}{39}}}{11} & 0 \\
\end{array}
\right),\hspace{.5cm}
{\mathbb{M}}^{(A_2)}=
\left(
\begin{array}{ccc}
\mathcal{M}_{1,P}^{-1} & 0 & 0 \\
 0 & \mathcal{M}_{3,F}^{-1} & 0 \\
 0 & 0 & \mathcal{M}_{4,F}^{-1} \\
\end{array}
\right).
\label{IV001A2}
\end{align}
\begin{align}
&B_1:\hspace{.5cm}\mathbb{F}_{00}^{(B_1)}=\textbf{I}_{4},\hspace{.5cm}
\mathbb{F}_{20}^{(B_1)}=\left(
\begin{array}{cccc}
 -\frac{1}{\sqrt{5}} & \frac{\sqrt{\frac{6}{5}}}{7} & -\sqrt{\frac{3}{14}} & \frac{3 \sqrt{\frac{3}{2}}}{7} \\
 \frac{\sqrt{\frac{6}{5}}}{7} & -\frac{8}{7 \sqrt{5}} & -\frac{1}{\sqrt{7}} & -\frac{1}{21} \\
 -\sqrt{\frac{3}{14}} & -\frac{1}{\sqrt{7}} & 0 & -\frac{\sqrt{\frac{5}{7}}}{3} \\
 \frac{3 \sqrt{\frac{3}{2}}}{7} & -\frac{1}{21} & -\frac{\sqrt{\frac{5}{7}}}{3} & \frac{2 \sqrt{5}}{21} \\
\end{array}
\right),
\nonumber
\\
&\hspace{1.2cm}
\mathbb{F}_{40}^{(B_1)}=\left(
\begin{array}{cccc}
 0 & -\frac{\sqrt{\frac{2}{3}}}{7} & \sqrt{\frac{5}{42}} & -\frac{\sqrt{\frac{15}{2}}}{7} \\
 -\frac{\sqrt{\frac{2}{3}}}{7} & \frac{1}{21} & \frac{\sqrt{\frac{5}{7}}}{3} & \frac{9 \sqrt{5}}{77} \\
 \sqrt{\frac{5}{42}} & \frac{\sqrt{\frac{5}{7}}}{3} & -\frac{7}{66} & -\frac{15}{22 \sqrt{7}} \\
 -\frac{\sqrt{\frac{15}{2}}}{7} & \frac{9 \sqrt{5}}{77} & -\frac{15}{22 \sqrt{7}} & -\frac{3}{14} \\
\end{array}
\right),\hspace{.5cm}
\mathbb{F}_{44}^{(B_1)}=\left(
\begin{array}{cccc}
 0 & -\sqrt{\frac{20}{21}} & -\frac{1}{\sqrt{3}} & -\frac{1}{\sqrt{21}} \\
 -\sqrt{\frac{20}{21}} & \frac{\sqrt{\frac{10}{7}}}{3} & -\frac{\sqrt{2}}{3} & \frac{\sqrt{\frac{18}{7}}}{11} \\
 -\frac{1}{\sqrt{3}} & -\frac{\sqrt{2}}{3} & -\frac{\sqrt{\frac{35}{2}}}{33} & \frac{\sqrt{\frac{45}{2}}}{11} \\
 -\frac{1}{\sqrt{21}} & \frac{\sqrt{\frac{18}{7}}}{11} & \frac{\sqrt{\frac{45}{2}}}{11} & \frac{9 \sqrt{\frac{5}{14}}}{11} \\
\end{array}
\right),
\nonumber
\\
&\hspace{1.2cm}\mathbb{F}_{60}^{(B_1)}=\left(
\begin{array}{cccc}
 0 & 0 & 0 & 0 \\
 0 & 0 & 0 & -\frac{10}{33} \sqrt{\frac{5}{13}} \\
 0 & 0 & -\frac{15}{22 \sqrt{13}} & \frac{35 \sqrt{\frac{7}{13}}}{66} \\
 0 & -\frac{10}{33} \sqrt{\frac{5}{13}} & \frac{35 \sqrt{\frac{7}{13}}}{66} & -\frac{5}{6 \sqrt{13}} \\
\end{array}
\right),
\hspace{.5cm}
\mathbb{F}_{64}^{(B_1)}=\left(
\begin{array}{cccc}
 0 & 0 & 0 & 0 \\
 0 & 0 & 0 & -\frac{10}{33} \sqrt{\frac{70}{13}} \\
 0 & 0 & \frac{15 \sqrt{\frac{7}{26}}}{11} & \frac{35}{33 \sqrt{26}} \\
 0 & -\frac{10}{33} \sqrt{\frac{70}{13}} & \frac{35}{33 \sqrt{26}} & \frac{5 \sqrt{\frac{7}{26}}}{11} \\
\end{array}
\right),
\nonumber
\\
&\hspace{1.2cm}
{\mathbb{M}}^{(B_1)}=\left(
\begin{array}{cccc}
 \frac{\mathcal{M}_{2,F}}{\det\mathcal{M}_{2}} & -\frac{\mathcal{M}_{2,PF}}{\det\mathcal{M}_{2}} & 0 & 0 \\
 -\frac{\mathcal{M}_{2,PF}}{\det\mathcal{M}_{2}} & \frac{\mathcal{M}_{2,P}}{\det\mathcal{M}_{2}} & 0 & 0 \\
 0 & 0 & \mathcal{M}_{3,F}^{-1} & 0 \\
 0 & 0 & 0 & \mathcal{M}_{4,F}^{-1} \\
\end{array}
\right).
\label{IV001B1}
\\
&B_2:\hspace{.5cm}
\mathbb{F}_{00}^{(B_2)}=\textbf{I}_{4},\hspace{.5cm}
\mathbb{F}_{20}^{(B_1)}=\left(
\begin{array}{cccc}
 -\frac{1}{\sqrt{5}} & \frac{\sqrt{\frac{6}{5}}}{7} & -\sqrt{\frac{3}{14}} & \frac{3 \sqrt{\frac{3}{2}}}{7} \\
 \frac{\sqrt{\frac{6}{5}}}{7} & -\frac{8}{7 \sqrt{5}} & -\frac{1}{\sqrt{7}} & -\frac{1}{21} \\
 -\sqrt{\frac{3}{14}} & -\frac{1}{\sqrt{7}} & 0 & -\frac{\sqrt{\frac{5}{7}}}{3} \\
 \frac{3 \sqrt{\frac{3}{2}}}{7} & -\frac{1}{21} & -\frac{\sqrt{\frac{5}{7}}}{3} & \frac{2 \sqrt{5}}{21} \\
\end{array}
\right),
\nonumber
\\
&\hspace{1.2cm}\mathbb{F}_{40}^{(B_2)}=\left(
\begin{array}{cccc}
 0 & -\frac{\sqrt{\frac{2}{3}}}{7} & \sqrt{\frac{5}{42}} & -\frac{\sqrt{\frac{15}{2}}}{7} \\
 -\frac{\sqrt{\frac{2}{3}}}{7} & \frac{1}{21} & \frac{\sqrt{\frac{5}{7}}}{3} & \frac{9 \sqrt{5}}{77} \\
 \sqrt{\frac{5}{42}} & \frac{\sqrt{\frac{5}{7}}}{3} & -\frac{7}{66} & -\frac{15}{22 \sqrt{7}} \\
 -\frac{\sqrt{\frac{15}{2}}}{7} & \frac{9 \sqrt{5}}{77} & -\frac{15}{22 \sqrt{7}} & -\frac{3}{14} \\
\end{array}
\right),
\hspace{.5cm}
\mathbb{F}_{44}^{(B_2)}=\left(
\begin{array}{cccc}
 0 & 2 \sqrt{\frac{5}{21}} & \frac{1}{\sqrt{3}} & \frac{1}{\sqrt{21}} \\
 2 \sqrt{\frac{5}{21}} & -\frac{\sqrt{\frac{10}{7}}}{3} & \frac{\sqrt{2}}{3} & -\frac{3}{11}  \sqrt{\frac{2}{7}} \\
 \frac{1}{\sqrt{3}} & \frac{\sqrt{2}}{3} & \frac{\sqrt{\frac{35}{2}}}{33} & -\frac{3}{11} \sqrt{\frac{5}{2}} \\
 \frac{1}{\sqrt{21}} & -\frac{3}{11}  \sqrt{\frac{2}{7}} & -\frac{3}{11} \sqrt{\frac{5}{2}} & -\frac{9}{11}  \sqrt{\frac{5}{14}} \\
\end{array}
\right),
\nonumber
\\
&\hspace{1.2cm}\mathbb{F}_{60}^{(B_2)}=\left(
\begin{array}{cccc}
 0 & 0 & 0 & 0 \\
 0 & 0 & 0 & -\frac{10}{33} \sqrt{\frac{5}{13}} \\
 0 & 0 & -\frac{15}{22 \sqrt{13}} & \frac{35 \sqrt{\frac{7}{13}}}{66} \\
 0 & -\frac{10}{33} \sqrt{\frac{5}{13}} & \frac{35 \sqrt{\frac{7}{13}}}{66} & -\frac{5}{6 \sqrt{13}} \\
\end{array}
\right),
\hspace{.5cm}
\mathbb{F}_{64}^{(B_2)}=\left(
\begin{array}{cccc}
 0 & 0 & 0 & 0 \\
 0 & 0 & 0 & \frac{10 \sqrt{\frac{70}{13}}}{33} \\
 0 & 0 & -\frac{15}{11} \sqrt{\frac{7}{26}} & -\frac{35}{33 \sqrt{26}} \\
 0 & \frac{10 \sqrt{\frac{70}{13}}}{33} & -\frac{35}{33 \sqrt{26}} & -\frac{5}{11} \sqrt{\frac{7}{26}} \\
\end{array}
\right),
\nonumber
\\
&\hspace{1.1cm}
{\mathbb{M}}^{(B_2)}=\left(
\begin{array}{cccc}
 \frac{\mathcal{M}_{2,F}}{\det\mathcal{M}_{2}} & -\frac{\mathcal{M}_{2,PF}}{\det\mathcal{M}_{2}} & 0 & 0 \\
 -\frac{\mathcal{M}_{2,PF}}{\det\mathcal{M}_{2}} & \frac{\mathcal{M}_{2,P}}{\det\mathcal{M}_{2}} & 0 & 0 \\
 0 & 0 & \mathcal{M}_{3,F}^{-1} & 0 \\
 0 & 0 & 0 & \mathcal{M}_{4,F}^{-1} \\
\end{array}
\right).
\label{IV001B2}
\end{align}
\begin{align}
&E:\hspace{.5cm}
\mathbb{F}_{00}^{(E)}=\textbf{I}_{7},
\hspace{.5cm} \mathbb{F}_{20}^{(E)}=\left(
\begin{array}{ccccccc}
 \frac{1}{2 \sqrt{5}} & \frac{3}{2 \sqrt{5}} & \sqrt{\frac{3}{10}} & 2 \sqrt{\frac{3}{35}} & 0 & 0 & 0 \\
 \frac{3}{2 \sqrt{5}} & \frac{1}{2 \sqrt{5}} & -\frac{\sqrt{\frac{3}{10}}}{7} & \sqrt{\frac{3}{35}} & 0 & \frac{3 \sqrt{3}}{7} & 0 \\
 \sqrt{\frac{3}{10}} & -\frac{\sqrt{\frac{3}{10}}}{7} & \frac{4}{7 \sqrt{5}} & \sqrt{\frac{2}{35}} & 0 & -\frac{\sqrt{2}}{21} & 0 \\
 2 \sqrt{\frac{3}{35}} & \sqrt{\frac{3}{35}} & \sqrt{\frac{2}{35}} & \frac{3}{4 \sqrt{5}} & 0 & \frac{5}{12 \sqrt{7}} & 0 \\
 0 & 0 & 0 & 0 & -\frac{\sqrt{5}}{4} & 0 & -\frac{\sqrt{\frac{5}{3}}}{4} \\
 0 & \frac{3 \sqrt{3}}{7} & -\frac{\sqrt{2}}{21} & \frac{5}{12 \sqrt{7}} & 0 & \frac{17 \sqrt{5}}{84} & 0 \\
 0 & 0 & 0 & 0 & -\frac{\sqrt{\frac{5}{3}}}{4} & 0 & -\frac{\sqrt{5}}{12}
\end{array}
\right),
\nonumber
\end{align}
\begin{align}
&\hspace{1.2cm}\mathbb{F}_{40}^{(E)}=\left(
\begin{array}{ccccccc}
 0 & 0 & 0 & \frac{\sqrt{\frac{3}{7}}}{2} & 0 & \frac{\sqrt{\frac{5}{3}}}{2} & 0 \\
 0 & 0 & \frac{4 \sqrt{\frac{2}{3}}}{7} & \frac{5}{2 \sqrt{21}} & 0 & \frac{\sqrt{\frac{5}{3}}}{14} & 0 \\
 0 & \frac{4 \sqrt{\frac{2}{3}}}{7} & -\frac{4}{21} & \frac{5}{3 \sqrt{14}} & 0 & -\frac{3 \sqrt{\frac{5}{2}}}{77} & 0 \\
 \frac{\sqrt{\frac{3}{7}}}{2} & \frac{5}{2 \sqrt{21}} & \frac{5}{3 \sqrt{14}} & \frac{1}{66} & 0 & \frac{9 \sqrt{\frac{5}{7}}}{22} & 0 \\
 0 & 0 & 0 & 0 & \frac{1}{22} & 0 & \frac{5 \sqrt{3}}{22} \\
 \frac{\sqrt{\frac{5}{3}}}{2} & \frac{\sqrt{\frac{5}{3}}}{14} & -\frac{3 \sqrt{\frac{5}{2}}}{77} & \frac{9 \sqrt{\frac{5}{7}}}{22} & 0 & \frac{27}{154} & 0 \\
 0 & 0 & 0 & 0 & \frac{5 \sqrt{3}}{22} & 0 & -\frac{9}{22}
\end{array}
\right),
\nonumber\\
&\hspace{1.2cm}\mathbb{F}_{44}^{(E)}=\left(
\begin{array}{ccccccc}
 0 & 0 & 0 & 0 & \frac{1}{\sqrt{2}} & 0 & \frac{1}{\sqrt{6}} \\
 0 & 0 & 0 & 0 & -\frac{1}{\sqrt{2}} & 0 & -\frac{1}{\sqrt{6}} \\
 0 & 0 & 0 & 0 & -\frac{1}{\sqrt{3}} & 0 & \frac{3}{11} \\
 0 & 0 & 0 & 0 & \frac{\sqrt{\frac{7}{6}}}{11} & 0 & -\frac{3 \sqrt{\frac{7}{2}}}{11} \\
 \frac{1}{\sqrt{2}} & -\frac{1}{\sqrt{2}} & -\frac{1}{\sqrt{3}} & \frac{\sqrt{\frac{7}{6}}}{11} & 0 & \frac{\sqrt{\frac{15}{2}}}{11} & 0 \\
 0 & 0 & 0 & 0 & \frac{\sqrt{\frac{15}{2}}}{11} & 0 & \frac{3 \sqrt{\frac{5}{2}}}{11} \\
 \frac{1}{\sqrt{6}} & -\frac{1}{\sqrt{6}} & \frac{3}{11} & -\frac{3 \sqrt{\frac{7}{2}}}{11} & 0 & \frac{3 \sqrt{\frac{5}{2}}}{11} & 0
\end{array}
\right),
\nonumber\\
&\hspace{1.2cm}\mathbb{F}_{60}^{(E)}=\left(
\begin{array}{ccccccc}
 0 & 0 & 0 & 0 & 0 & 0 & 0 \\
 0 & 0 & 0 & 0 & 0 & 0 & 0 \\
 0 & 0 & 0 & 0 & 0 & \frac{20 \sqrt{\frac{10}{13}}}{33} & 0 \\
 0 & 0 & 0 & \frac{75}{44 \sqrt{13}} & 0 & \frac{35 \sqrt{\frac{35}{13}}}{132} & 0 \\
 0 & 0 & 0 & 0 & \frac{5}{44 \sqrt{13}} & 0 & -\frac{35}{44 \sqrt{39}} \\
 0 & 0 & \frac{20 \sqrt{\frac{10}{13}}}{33} & \frac{35 \sqrt{\frac{35}{13}}}{132} & 0 & -\frac{5}{132 \sqrt{13}} & 0 \\
 0 & 0 & 0 & 0 & -\frac{35}{44 \sqrt{39}} & 0 & \frac{85}{132 \sqrt{13}}
\end{array}
\right),
\nonumber\end{align}
\begin{align}
&\mathbb{F}_{64}^{(E)}=\left(
\begin{array}{ccccccc}
 0 & 0 & 0 & 0 & 0 & 0 & 0 \\
 0 & 0 & 0 & 0 & 0 & 0 & 0 \\
 0 & 0 & 0 & 0 & 0 & 0 & \frac{40 \sqrt{\frac{5}{13}}}{33} \\
 0 & 0 & 0 & 0 & \frac{5 \sqrt{\frac{105}{26}}}{22} & 0 & -\frac{25 \sqrt{\frac{35}{26}}}{66} \\
 0 & 0 & 0 & \frac{5 \sqrt{\frac{105}{26}}}{22} & 0 & -\frac{35 \sqrt{\frac{3}{26}}}{22} & 0 \\
 0 & 0 & 0 & 0 & -\frac{35 \sqrt{\frac{3}{26}}}{22} & 0 & \frac{5}{22 \sqrt{26}} \\
 0 & 0 & \frac{40 \sqrt{\frac{5}{13}}}{33} & -\frac{25 \sqrt{\frac{35}{26}}}{66} & 0 & \frac{5}{22 \sqrt{26}} & 0
\end{array}
\right),
\nonumber
\\
&{\mathbb{M}}^{(E)}=\left(
\begin{array}{ccccccc}
\mathcal{M}_{1,P}^{-1} & 0 & 0 & 0 & 0 & 0 & 0 \\
 0 & \frac{\mathcal{M}_{2,F}}{\det\mathcal{M}_{2}} & -\frac{\mathcal{M}_{2,PF}}{\det\mathcal{M}_{2}} & 0 & 0 & 0 & 0 \\
 0 & -\frac{\mathcal{M}_{2,PF}}{\det\mathcal{M}_{2}} & \frac{\mathcal{M}_{2,P}}{\det\mathcal{M}_{2}} & 0 & 0 & 0 & 0 \\
 0 & 0 & 0 & \mathcal{M}_{3,F}^{-1} & 0 & 0 & 0 \\
 0 & 0 & 0 & 0 & \mathcal{M}_{3,F}^{-1} & 0 & 0 \\
 0 & 0 & 0 & 0 & 0 & \mathcal{M}_{4,F}^{-1} & 0 \\
 0 & 0 & 0 & 0 & 0 & 0 & \mathcal{M}_{4,F}^{-1} \\
\end{array}
\right).
\label{IV001E}
\end{align}
%

\subsubsection{$\mathbf{d}=(1,1,0)$}
\begin{align}
&A:\hspace{.5cm}\mathbb{F}_{00}^{(A)}=\textbf{I}_{9},
\nonumber\\
&\hspace{1.2cm}
\mathbb{F}_{20}^{(A)}=\left(
\begin{array}{ccccccccc}
 0 & -\sqrt{\frac{2}{5}} & 0 & \sqrt{\frac{3}{5}} & 0 & 0 & 0 & 0 & 0 \\
 -\sqrt{\frac{2}{5}} & \frac{1}{\sqrt{5}} & 0 & -\frac{\sqrt{\frac{6}{5}}}{7} & 0 & 0 & 0 & \frac{9 \sqrt{\frac{2}{5}}}{7} & 0 \\
 0 & 0 & -\frac{1}{\sqrt{5}} & 0 & \frac{\sqrt{\frac{6}{5}}}{7} & -\sqrt{\frac{3}{14}} & \frac{3 \sqrt{\frac{3}{2}}}{7} & 0 & 0 \\
 \sqrt{\frac{3}{5}} & -\frac{\sqrt{\frac{6}{5}}}{7} & 0 & \frac{8}{7 \sqrt{5}} & 0 & 0 & 0 & -\frac{2}{7 \sqrt{15}} & 0 \\
 0 & 0 & \frac{\sqrt{\frac{6}{5}}}{7} & 0 & -\frac{8}{7 \sqrt{5}} & -\frac{1}{\sqrt{7}} & -\frac{1}{21} & 0 & 0 \\
 0 & 0 & -\sqrt{\frac{3}{14}} & 0 & -\frac{1}{\sqrt{7}} & 0 & -\frac{\sqrt{\frac{5}{7}}}{3} & 0 & 0 \\
 0 & 0 & \frac{3 \sqrt{\frac{3}{2}}}{7} & 0 & -\frac{1}{21} & -\frac{\sqrt{\frac{5}{7}}}{3} & \frac{2 \sqrt{5}}{21} & 0 & 0 \\
 0 & \frac{9 \sqrt{\frac{2}{5}}}{7} & 0 & -\frac{2}{7 \sqrt{15}} & 0 & 0 & 0 & \frac{5 \sqrt{5}}{21} & 0 \\
 0 & 0 & 0 & 0 & 0 & 0 & 0 & 0 & -\frac{\sqrt{5}}{3} \\
\end{array}
\right),
\nonumber
\end{align}
\begin{align}
&\nonumber\hspace{1cm}\mathbb{F}_{22}^{(A)}=\left(
\begin{array}{ccccccccc}
 0 & 0 & -\frac{2}{\sqrt{5}} & 0 & \sqrt{\frac{6}{5}} & 0 & 0 & 0 & 0 \\
 0 & 0 & -\sqrt{\frac{2}{5}} & 0 & \frac{2 \sqrt{\frac{3}{5}}}{7} & \sqrt{\frac{3}{7}} & \frac{3 \sqrt{3}}{7} & 0 & 0 \\
 \frac{2}{\sqrt{5}} & \sqrt{\frac{2}{5}} & 0 & -\frac{2}{7} \sqrt{\frac{3}{5}} & 0 & 0 & 0 & -\frac{3}{7 \sqrt{5}} & \frac{3}{\sqrt{7}} \\
 0 & 0 & \frac{2 \sqrt{\frac{3}{5}}}{7} & 0 & -\frac{8}{7}\sqrt{\frac{2}{5}} & \sqrt{\frac{2}{7}} & -\frac{\sqrt{2}}{21} & 0 & 0 \\
 -\sqrt{\frac{6}{5}} & -\frac{2}{7} \sqrt{\frac{3}{5}} & 0 & \frac{8 \sqrt{\frac{2}{5}}}{7} & 0 & 0 & 0 & \frac{\sqrt{\frac{2}{15}}}{21} & -\frac{\sqrt{\frac{2}{21}}}{3} \\
 0 & -\sqrt{\frac{3}{7}} & 0 & -\sqrt{\frac{2}{7}} & 0 & 0 & 0 & \frac{5}{3 \sqrt{42}} & \frac{\sqrt{\frac{5}{6}}}{3} \\
 0 & -\frac{3 \sqrt{3}}{7}   & 0 & \frac{\sqrt{2}}{21} & 0 & 0 & 0 & \frac{5 \sqrt{\frac{3}{2}}}{7} & -\sqrt{\frac{5}{42}} \\
 0 & 0 & \frac{3}{7 \sqrt{5}} & 0 & -\frac{\sqrt{\frac{2}{15}}}{21} & -\frac{5}{3 \sqrt{42}} & -\frac{5}{7}  \sqrt{\frac{3}{2}} & 0 & 0 \\
 0 & 0 & -\frac{3}{\sqrt{7}} & 0 & \frac{\sqrt{\frac{2}{21}}}{3} & -\frac{\sqrt{\frac{5}{6}}}{3} & \sqrt{\frac{5}{42}} & 0 & 0 \\
\end{array}
\right),
\nonumber
\\
&\hspace{1.2cm}
\mathbb{F}_{40}^{(A)}=\left(
\begin{array}{ccccccccc}
 0 & 0 & 0 & 0 & 0 & 0 & 0 & -\frac{2}{3} & 0 \\
 0 & 0 & 0 & -\frac{2 \sqrt{6}}{7}   & 0 & 0 & 0 & \frac{5 \sqrt{2}}{21} & 0 \\
 0 & 0 & 0 & 0 & -\frac{\sqrt{\frac{2}{3}}}{7} & \sqrt{\frac{5}{42}} & -\frac{\sqrt{\frac{15}{2}}}{7} & 0 & 0 \\
 0 & -\frac{2 \sqrt{6}}{7}   & 0 & \frac{2}{7} & 0 & 0 & 0 & -\frac{10}{77} \sqrt{3} & 0 \\
 0 & 0 & -\frac{\sqrt{\frac{2}{3}}}{7} & 0 & \frac{1}{21} & \frac{\sqrt{\frac{5}{7}}}{3} & \frac{9 \sqrt{5}}{77} & 0 & 0 \\
 0 & 0 & \sqrt{\frac{5}{42}} & 0 & \frac{\sqrt{\frac{5}{7}}}{3} & -\frac{7}{66} & -\frac{15}{22 \sqrt{7}} & 0 & 0 \\
 0 & 0 & -\frac{\sqrt{\frac{15}{2}}}{7} & 0 & \frac{9 \sqrt{5}}{77} & -\frac{15}{22 \sqrt{7}} & -\frac{3}{14} & 0 & 0 \\
 -\frac{2}{3} & \frac{5 \sqrt{2}}{21} & 0 & -\frac{10}{77} \sqrt{3} & 0 & 0 & 0 & \frac{27}{77} & 0 \\
 0 & 0 & 0 & 0 & 0 & 0 & 0 & 0 & \frac{3}{11} \\
\end{array}
\right),
\nonumber
\end{align}
\begin{align}
&\hspace{1.2cm}
\mathbb{F}_{42}^{(A)}=\left(
\begin{array}{ccccccccc}
 0 & 0 & 0 & 0 & 0 & 0 & -\frac{2 \sqrt{2}}{3}  & 0 & 0 \\
 0 & 0 & 0 & 0 & -\frac{2 \sqrt{5}}{7}  & \frac{2}{\sqrt{7}} & \frac{4}{21} & 0 & 0 \\
 0 & 0 & 0 & \frac{2 \sqrt{5}}{7} & 0 & 0 & 0 & \frac{\sqrt{15}}{7} & -\frac{1}{\sqrt{21}} \\
 0 & 0 & -\frac{2 \sqrt{5}}{7}  & 0 & \frac{\sqrt{\frac{10}{3}}}{7} & 2 \sqrt{\frac{2}{21}} & -\frac{4\sqrt{6}}{77}  & 0 & 0 \\
 0 & \frac{2 \sqrt{5}}{7} & 0 & -\frac{\sqrt{\frac{10}{3}}}{7} & 0 & 0 & 0 & -\frac{9 \sqrt{10}}{77}  & \frac{3 \sqrt{\frac{2}{7}}}{11} \\
 0 & -\frac{2}{\sqrt{7}} & 0 & -2 \sqrt{\frac{2}{21}} & 0 & 0 & 0 & -\frac{15}{11 \sqrt{14}} & -\frac{3}{11} \sqrt{\frac{5}{2}} \\
 \frac{2 \sqrt{2}}{3} & -\frac{4}{21} & 0 & \frac{4 \sqrt{6}}{77} & 0 & 0 & 0 & \frac{3}{7 \sqrt{2}} & \frac{9 \sqrt{\frac{5}{14}}}{11} \\
 0 & 0 & -\frac{\sqrt{15}}{7} & 0 & \frac{9 \sqrt{10}}{77} & \frac{15}{11 \sqrt{14}} & -\frac{3}{7 \sqrt{2}} & 0 & 0 \\
 0 & 0 & \frac{1}{\sqrt{21}} & 0 & -\frac{3}{11}  \sqrt{\frac{2}{7}} & \frac{3 \sqrt{\frac{5}{2}}}{11} & -\frac{9}{11}  \sqrt{\frac{5}{14}} & 0 & 0 \\
\end{array}
\right),
\nonumber
\end{align}
\begin{align}
&\mathbb{F}_{44}^{(A)}=\left(
\begin{array}{ccccccccc}
 0 & 0 & 0 & 0 & 0 & 0 & 0 & 0 & -\frac{2 \sqrt{2}}{3}  \\
 0 & 0 & 0 & 0 & 0 & 0 & 0 & 0 & -\frac{2}{3} \\
 0 & 0 & 0 & 0 & 2 \sqrt{\frac{5}{21}} & \frac{1}{\sqrt{3}} & \frac{1}{\sqrt{21}} & 0 & 0 \\
 0 & 0 & 0 & 0 & 0 & 0 & 0 & 0 & \frac{2 \sqrt{6}}{11} \\
 0 & 0 & 2 \sqrt{\frac{5}{21}} & 0 & -\frac{\sqrt{\frac{10}{7}}}{3} & \frac{\sqrt{2}}{3} & -\frac{3}{11}  \sqrt{\frac{2}{7}} & 0 & 0 \\
 0 & 0 & \frac{1}{\sqrt{3}} & 0 & \frac{\sqrt{2}}{3} & \frac{\sqrt{\frac{35}{2}}}{33} & -\frac{3}{11} \sqrt{\frac{5}{2}} & 0 & 0 \\
 0 & 0 & \frac{1}{\sqrt{21}} & 0 & -\frac{3}{11}  \sqrt{\frac{2}{7}} & -\frac{3}{11} \sqrt{\frac{5}{2}} & -\frac{9}{11}  \sqrt{\frac{5}{14}} & 0 & 0 \\
 0 & 0 & 0 & 0 & 0 & 0 & 0 & 0 & \frac{3 \sqrt{2}}{11} \\
 -\frac{2 \sqrt{2}}{3}  & -\frac{2}{3} & 0 & \frac{2 \sqrt{6}}{11} & 0 & 0 & 0 & \frac{3 \sqrt{2}}{11} & 0 \\
\end{array}
\right),
\nonumber
\\
&\mathbb{F}_{60}^{(A)}=\left(
\begin{array}{ccccccccc}
 0 & 0 & 0 & 0 & 0 & 0 & 0 & 0 & 0 \\
 0 & 0 & 0 & 0 & 0 & 0 & 0 & 0 & 0 \\
 0 & 0 & 0 & 0 & 0 & 0 & 0 & 0 & 0 \\
 0 & 0 & 0 & 0 & 0 & 0 & 0 & -\frac{50}{11 \sqrt{39}} & 0 \\
 0 & 0 & 0 & 0 & 0 & 0 & -\frac{10}{33} \sqrt{\frac{5}{13}} & 0 & 0 \\
 0 & 0 & 0 & 0 & 0 & -\frac{15}{22 \sqrt{13}} & \frac{35 \sqrt{\frac{7}{13}}}{66} & 0 & 0 \\
 0 & 0 & 0 & 0 & -\frac{10}{33} \sqrt{\frac{5}{13}} & \frac{35 \sqrt{\frac{7}{13}}}{66} & -\frac{5}{6 \sqrt{13}} & 0 & 0 \\
 0 & 0 & 0 & -\frac{50}{11 \sqrt{39}} & 0 & 0 & 0 & \frac{25}{33 \sqrt{13}} & 0 \\
 0 & 0 & 0 & 0 & 0 & 0 & 0 & 0 & -\frac{5}{33 \sqrt{13}} \\
\end{array}
\right),
\nonumber
\\
&\mathbb{F}_{62}^{(A)}=\left(
\begin{array}{ccccccccc}
 0 & 0 & 0 & 0 & 0 & 0 & 0 & 0 & 0 \\
 0 & 0 & 0 & 0 & 0 & 0 & 0 & 0 & 0 \\
 0 & 0 & 0 & 0 & 0 & 0 & 0 & 0 & 0 \\
 0 & 0 & 0 & 0 & 0 & 0 & -\frac{40}{33}  \sqrt{\frac{7}{13}} & 0 & 0 \\
 0 & 0 & 0 & 0 & 0 & 0 & 0 & \frac{20 \sqrt{\frac{35}{39}}}{33} & -\frac{10}{33 \sqrt{39}} \\
 0 & 0 & 0 & 0 & 0 & 0 & 0 & -\frac{140}{33 \sqrt{39}} & \frac{5 \sqrt{\frac{35}{39}}}{33} \\
 0 & 0 & 0 & \frac{40 \sqrt{\frac{7}{13}}}{33} & 0 & 0 & 0 & 0 & -\frac{5}{11}  \sqrt{\frac{5}{39}} \\
 0 & 0 & 0 & 0 & -\frac{20}{33} \sqrt{\frac{35}{39}} & \frac{140}{33 \sqrt{39}} & 0 & 0 & 0 \\
 0 & 0 & 0 & 0 & \frac{10}{33 \sqrt{39}} & -\frac{5}{33}  \sqrt{\frac{35}{39}} & \frac{5 \sqrt{\frac{5}{39}}}{11} & 0 & 0 \\
\end{array}
\right),
\nonumber
\end{align}
\begin{align}
&\hspace{1.1cm}\mathbb{F}_{64}^{(A)}=\left(
\begin{array}{ccccccccc}
 0 & 0 & 0 & 0 & 0 & 0 & 0 & 0 & 0 \\
 0 & 0 & 0 & 0 & 0 & 0 & 0 & 0 & 0 \\
 0 & 0 & 0 & 0 & 0 & 0 & 0 & 0 & 0 \\
 0 & 0 & 0 & 0 & 0 & 0 & 0 & 0 & -\frac{10 }{11} \sqrt{\frac{10}{39}} \\
 0 & 0 & 0 & 0 & 0 & 0 & \frac{10 \sqrt{\frac{70}{13}}}{33} & 0 & 0 \\
 0 & 0 & 0 & 0 & 0 & -\frac{15}{11} \sqrt{\frac{7}{26}} & -\frac{35}{33 \sqrt{26}} & 0 & 0 \\
 0 & 0 & 0 & 0 & \frac{10 \sqrt{\frac{70}{13}}}{33} & -\frac{35}{33 \sqrt{26}} & -\frac{5}{11} \sqrt{\frac{7}{26}} & 0 & 0 \\
 0 & 0 & 0 & 0 & 0 & 0 & 0 & 0 & -\frac{5}{11}  \sqrt{\frac{10}{13}} \\
 0 & 0 & 0 & -\frac{10 }{11} \sqrt{\frac{10}{39}} & 0 & 0 & 0 & -\frac{5}{11}  \sqrt{\frac{10}{13}} & 0 \\
\end{array}
\right),
\nonumber
\\
&\hspace{1.1cm}\mathbb{F}_{66}^{(A)}=\left(
\begin{array}{ccccccccc}
 0 & 0 & 0 & 0 & 0 & 0 & 0 & 0 & 0 \\
 0 & 0 & 0 & 0 & 0 & 0 & 0 & 0 & 0 \\
 0 & 0 & 0 & 0 & 0 & 0 & 0 & 0 & 0 \\
 0 & 0 & 0 & 0 & 0 & 0 & 0 & 0 & 0 \\
 0 & 0 & 0 & 0 & 0 & 0 & 0 & 0 & 10 \sqrt{\frac{5}{429}} \\
 0 & 0 & 0 & 0 & 0 & 0 & 0 & 0 & 5 \sqrt{\frac{7}{429}} \\
 0 & 0 & 0 & 0 & 0 & 0 & 0 & 0 & \frac{5}{\sqrt{429}} \\
 0 & 0 & 0 & 0 & 0 & 0 & 0 & 0 & 0 \\
 0 & 0 & 0 & 0 & -10 \sqrt{\frac{5}{429}} & -5 \sqrt{\frac{7}{429}} & -\frac{5}{\sqrt{429}} & 0 & 0 \\
\end{array}
\right),
\nonumber
\\
&\hspace{1.2cm}
{\mathbb{M}}^{(A)}=
\left(
\begin{array}{ccccccccc}
 \mathcal{M}_{0,P}^{-1} & 0 & 0 & 0 & 0 & 0 & 0 & 0 & 0 \\
 0 & \frac{\mathcal{M}_{2,F}}{\det\mathcal{M}_{2}} & 0 & -\frac{\mathcal{M}_{2,PF}}{\det\mathcal{M}_{2}} & 0 & 0 & 0 & 0 & 0 \\
 0 & 0 & \frac{\mathcal{M}_{2,F}}{\det\mathcal{M}_{2}} & 0 & -\frac{\mathcal{M}_{2,PF}}{\det\mathcal{M}_{2}} & 0 & 0 & 0 & 0 \\
 0 & -\frac{\mathcal{M}_{2,PF}}{\det\mathcal{M}_{2}} & 0 & \frac{\mathcal{M}_{2,P}}{\det\mathcal{M}_{2}} & 0 & 0 & 0 & 0 & 0 \\
 0 & 0 & -\frac{\mathcal{M}_{2,PF}}{\det\mathcal{M}_{2}} & 0 & \frac{\mathcal{M}_{2,P}}{\det\mathcal{M}_{2}} & 0 & 0 & 0 & 0 \\
 0 & 0 & 0 & 0 & 0 & \mathcal{M}_{3,F}^{-1} & 0 & 0 & 0 \\
 0 & 0 & 0 & 0 & 0 & 0 & \mathcal{M}_{4,F}^{-1} & 0 & 0 \\
 0 & 0 & 0 & 0 & 0 & 0 & 0 & \mathcal{M}_{4,F}^{-1} & 0 \\
 0 & 0 & 0 & 0 & 0 & 0 & 0 & 0 & \mathcal{M}_{4,F}^{-1} \\
\end{array}
\right).
\label{IV110A}
\\
&B_1:\hspace{.5cm}\mathbb{F}_{00}^{(B_1)}=\textbf{I}_{7},\hspace{.5cm}
\mathbb{F}_{20}^{(B_1)}=\left(
\begin{array}{ccccccc}
 -\frac{1}{\sqrt{5}} & 0 & 0 & 3 \sqrt{\frac{2}{35}} & 0 & 0 & 0 \\
 0 & -\frac{1}{\sqrt{5}} & \frac{\sqrt{\frac{6}{5}}}{7} & 0 & -\sqrt{\frac{3}{14}} & 0 & \frac{3 \sqrt{\frac{3}{2}}}{7} \\
 0 & \frac{\sqrt{\frac{6}{5}}}{7} & -\frac{8}{7 \sqrt{5}} & 0 & -\frac{1}{\sqrt{7}} & 0 & -\frac{1}{21} \\
 3 \sqrt{\frac{2}{35}} & 0 & 0 & \frac{1}{\sqrt{5}} & 0 & 0 & 0 \\
 0 & -\sqrt{\frac{3}{14}} & -\frac{1}{\sqrt{7}} & 0 & 0 & 0 & -\frac{\sqrt{\frac{5}{7}}}{3} \\
 0 & 0 & 0 & 0 & 0 & -\frac{\sqrt{5}}{3} & 0 \\
 0 & \frac{3 \sqrt{\frac{3}{2}}}{7} & -\frac{1}{21} & 0 & -\frac{\sqrt{\frac{5}{7}}}{3} & 0 & \frac{2 \sqrt{5}}{21} \\
\end{array}
\right),
\nonumber
\end{align}
\begin{align}
&\hspace{1.2cm}
\mathbb{F}_{22}^{(B_1)}=\left(
\begin{array}{ccccccc}
 0 & \sqrt{\frac{6}{5}} & \frac{2}{\sqrt{5}} & 0 & \frac{2}{\sqrt{7}} & 0 & 0 \\
 -\sqrt{\frac{6}{5}} & 0 & 0 & \sqrt{\frac{3}{35}} & 0 & \frac{3}{\sqrt{7}} & 0 \\
 -\frac{2}{\sqrt{5}} & 0 & 0 & \sqrt{\frac{2}{35}} & 0 & -\frac{\sqrt{\frac{2}{21}}}{3} & 0 \\
 0 & -\sqrt{\frac{3}{35}} & -\sqrt{\frac{2}{35}} & 0 & -\frac{1}{\sqrt{2}} & 0 & \frac{5}{3 \sqrt{14}} \\
 -\frac{2}{\sqrt{7}} & 0 & 0 & \frac{1}{\sqrt{2}} & 0 & \frac{\sqrt{\frac{5}{6}}}{3} & 0 \\
 0 & -\frac{3}{\sqrt{7}} & \frac{\sqrt{\frac{2}{21}}}{3} & 0 & -\frac{\sqrt{\frac{5}{6}}}{3} & 0 & \sqrt{\frac{5}{42}} \\
 0 & 0 & 0 & -\frac{5}{3 \sqrt{14}} & 0 & -\sqrt{\frac{5}{42}} & 0 \\
\end{array}
\right),
\nonumber
\\
&\hspace{1.2cm}
\mathbb{F}_{40}^{(B_1)}=\left(
\begin{array}{ccccccc}
 0 & 0 & 0 & -\sqrt{\frac{2}{7}} & 0 & 0 & 0 \\
 0 & 0 & -\frac{\sqrt{\frac{2}{3}}}{7} & 0 & \sqrt{\frac{5}{42}} & 0 & -\frac{\sqrt{\frac{15}{2}}}{7} \\
 0 & -\frac{\sqrt{\frac{2}{3}}}{7} & \frac{1}{21} & 0 & \frac{\sqrt{\frac{5}{7}}}{3} & 0 & \frac{9 \sqrt{5}}{77} \\
 -\sqrt{\frac{2}{7}} & 0 & 0 & \frac{1}{11} & 0 & 0 & 0 \\
 0 & \sqrt{\frac{5}{42}} & \frac{\sqrt{\frac{5}{7}}}{3} & 0 & -\frac{7}{66} & 0 & -\frac{15}{22 \sqrt{7}} \\
 0 & 0 & 0 & 0 & 0 & \frac{3}{11} & 0 \\
 0 & -\frac{\sqrt{\frac{15}{2}}}{7} & \frac{9 \sqrt{5}}{77} & 0 & -\frac{15}{22 \sqrt{7}} & 0 & -\frac{3}{14} \\
\end{array}
\right),
\nonumber
\\
&\hspace{1.2cm}
\mathbb{F}_{42}^{(B_1)}=\left(
\begin{array}{ccccccc}
 0 & 0 & 0 & 0 & -\sqrt{\frac{3}{7}} & 0 & \frac{1}{\sqrt{3}} \\
 0 & 0 & 0 & -\sqrt{\frac{5}{7}} & 0 & -\frac{1}{\sqrt{21}} & 0 \\
 0 & 0 & 0 & -\sqrt{\frac{10}{21}} & 0 & \frac{3 \sqrt{\frac{2}{7}}}{11} & 0 \\
 0 & \sqrt{\frac{5}{7}} & \sqrt{\frac{10}{21}} & 0 & -\frac{1}{11 \sqrt{6}} & 0 & \frac{13 \sqrt{\frac{3}{14}}}{11} \\
 \sqrt{\frac{3}{7}} & 0 & 0 & \frac{1}{11 \sqrt{6}} & 0 & -\frac{3}{11} \sqrt{\frac{5}{2}} & 0 \\
 0 & \frac{1}{\sqrt{21}} & -\frac{3}{11}  \sqrt{\frac{2}{7}} & 0 & \frac{3 \sqrt{\frac{5}{2}}}{11} & 0 & -\frac{9}{11}  \sqrt{\frac{5}{14}} \\
 -\frac{1}{\sqrt{3}} & 0 & 0 & -\frac{13}{11} \sqrt{\frac{3}{14}} & 0 & \frac{9 \sqrt{\frac{5}{14}}}{11} & 0 \\
\end{array}
\right),
\nonumber
\\
&\hspace{1.2cm}
\mathbb{F}_{44}^{(B_1)}=\left(
\begin{array}{ccccccc}
 0 & 0 & 0 & 0 & 0 & \frac{2}{\sqrt{3}} & 0 \\
 0 & 0 & -2 \sqrt{\frac{5}{21}} & 0 & -\frac{1}{\sqrt{3}} & 0 & -\frac{1}{\sqrt{21}} \\
 0 & -2 \sqrt{\frac{5}{21}} & \frac{\sqrt{\frac{10}{7}}}{3} & 0 & -\frac{\sqrt{2}}{3} & 0 & \frac{3 \sqrt{\frac{2}{7}}}{11} \\
 0 & 0 & 0 & 0 & 0 & -\frac{\sqrt{42}}{11} & 0 \\
 0 & -\frac{1}{\sqrt{3}} & -\frac{\sqrt{2}}{3} & 0 & -\frac{\sqrt{\frac{35}{2}}}{33} & 0 & \frac{3 \sqrt{\frac{5}{2}}}{11} \\
 \frac{2}{\sqrt{3}} & 0 & 0 & -\frac{\sqrt{42}}{11} & 0 & 0 & 0 \\
 0 & -\frac{1}{\sqrt{21}} & \frac{3 \sqrt{\frac{2}{7}}}{11} & 0 & \frac{3 \sqrt{\frac{5}{2}}}{11} & 0 & \frac{9 \sqrt{\frac{5}{14}}}{11} \\
\end{array}
\right),
\nonumber\\
&\hspace{1.2cm}
\mathbb{F}_{60}^{(B_1)}=\left(
\begin{array}{ccccccc}
 0 & 0 & 0 & 0 & 0 & 0 & 0 \\
 0 & 0 & 0 & 0 & 0 & 0 & 0 \\
 0 & 0 & 0 & 0 & 0 & 0 & -\frac{10}{33} \sqrt{\frac{5}{13}} \\
 0 & 0 & 0 & -\frac{25}{11 \sqrt{13}} & 0 & 0 & 0 \\
 0 & 0 & 0 & 0 & -\frac{15}{22 \sqrt{13}} & 0 & \frac{35 \sqrt{\frac{7}{13}}}{66} \\
 0 & 0 & 0 & 0 & 0 & -\frac{5}{33 \sqrt{13}} & 0 \\
 0 & 0 & -\frac{10}{33} \sqrt{\frac{5}{13}} & 0 & \frac{35 \sqrt{\frac{7}{13}}}{66} & 0 & -\frac{5}{6 \sqrt{13}} \\
\end{array}
\right),
\nonumber
\end{align}
\begin{align}
&\hspace{1.2cm}
\mathbb{F}_{62}^{(B_1)}=\left(
\begin{array}{ccccccc}
 0 & 0 & 0 & 0 & 0 & 0 & 0 \\
 0 & 0 & 0 & 0 & 0 & 0 & 0 \\
 0 & 0 & 0 & 0 & 0 & -\frac{10}{33 \sqrt{39}} & 0 \\
 0 & 0 & 0 & 0 & -\frac{10}{11} \sqrt{\frac{7}{13}} & 0 & \frac{70}{33 \sqrt{13}} \\
 0 & 0 & 0 & \frac{10 \sqrt{\frac{7}{13}}}{11} & 0 & \frac{5 \sqrt{\frac{35}{39}}}{33} & 0 \\
 0 & 0 & \frac{10}{33 \sqrt{39}} & 0 & -\frac{5}{33}  \sqrt{\frac{35}{39}} & 0 & \frac{5 \sqrt{\frac{5}{39}}}{11} \\
 0 & 0 & 0 & -\frac{70}{33 \sqrt{13}} & 0 & -\frac{5}{11}  \sqrt{\frac{5}{39}} & 0 \\
\end{array}
\right),
\nonumber
\\
&\hspace{1.2cm}
\mathbb{F}_{64}^{(B_1)}=\left(
\begin{array}{ccccccc}
 0 & 0 & 0 & 0 & 0 & 0 & 0 \\
 0 & 0 & 0 & 0 & 0 & 0 & 0 \\
 0 & 0 & 0 & 0 & 0 & 0 & -\frac{10}{33} \sqrt{\frac{70}{13}} \\
 0 & 0 & 0 & 0 & 0 & \frac{5 \sqrt{\frac{70}{39}}}{11} & 0 \\
 0 & 0 & 0 & 0 & \frac{15 \sqrt{\frac{7}{26}}}{11} & 0 & \frac{35}{33 \sqrt{26}} \\
 0 & 0 & 0 & \frac{5 \sqrt{\frac{70}{39}}}{11} & 0 & 0 & 0 \\
 0 & 0 & -\frac{10}{33} \sqrt{\frac{70}{13}} & 0 & \frac{35}{33 \sqrt{26}} & 0 & \frac{5 \sqrt{\frac{7}{26}}}{11} \\
\end{array}
\right),
\nonumber\
\\
&\hspace{1.2cm}
\mathbb{F}_{66}^{(B_1)}=\left(
\begin{array}{ccccccc}
 0 & 0 & 0 & 0 & 0 & 0 & 0 \\
 0 & 0 & 0 & 0 & 0 & 0 & 0 \\
 0 & 0 & 0 & 0 & 0 & -10 \sqrt{\frac{5}{429}} & 0 \\
 0 & 0 & 0 & 0 & 0 & 0 & 0 \\
 0 & 0 & 0 & 0 & 0 & -5 \sqrt{\frac{7}{429}} & 0 \\
 0 & 0 & 10 \sqrt{\frac{5}{429}} & 0 & 5 \sqrt{\frac{7}{429}} & 0 & \frac{5}{\sqrt{429}} \\
 0 & 0 & 0 & 0 & 0 & -\frac{5}{\sqrt{429}} & 0 \\
\end{array}
\right),
\nonumber\\
&\hspace{1.2cm}
{\mathbb{M}}^{(B_1)}=
\left(
\begin{array}{ccccccc}
\mathcal{M}_{1,P}^{-1} & 0 & 0 & 0 & 0 & 0 & 0 \\
 0 & \frac{\mathcal{M}_{2,F}}{\det\mathcal{M}_{2}} & -\frac{\mathcal{M}_{2,PF}}{\det\mathcal{M}_{2}} & 0 & 0 & 0 & 0 \\
 0 & -\frac{\mathcal{M}_{2,PF}}{\det\mathcal{M}_{2}} & \frac{\mathcal{M}_{2,P}}{\det\mathcal{M}_{2}} & 0 & 0 & 0 & 0 \\
 0 & 0 & 0 & \mathcal{M}_{3,F}^{-1} & 0 & 0 & 0 \\
 0 & 0 & 0 & 0 & \mathcal{M}_{3,F}^{-1} & 0 & 0 \\
 0 & 0 & 0 & 0 & 0 & \mathcal{M}_{4,F}^{-1} & 0 \\
 0 & 0 & 0 & 0 & 0 & 0 & \mathcal{M}_{4,F}^{-1} \\
\end{array}
\right).
\label{IV110B1}
\\
&B_2:\hspace{.5cm}\mathbb{F}_{00}^{(B_2)}=\textbf{I}_{7},\hspace{.5cm}
\mathbb{F}_{20}^{(B_2)}=\left(
\begin{array}{ccccccc}
 \frac{1}{2 \sqrt{5}} & -\frac{3}{2 \sqrt{5}} & -\sqrt{\frac{3}{10}} & 2 \sqrt{\frac{3}{35}} & 0 & 0 & 0 \\
 -\frac{3}{2 \sqrt{5}} & \frac{1}{2 \sqrt{5}} & -\frac{\sqrt{\frac{3}{10}}}{7} & -\sqrt{\frac{3}{35}} & 0 & 0 & \frac{3 \sqrt{3}}{7} \\
 -\sqrt{\frac{3}{10}} & -\frac{\sqrt{\frac{3}{10}}}{7} & \frac{4}{7 \sqrt{5}} & -\sqrt{\frac{2}{35}} & 0 & 0 & -\frac{\sqrt{2}}{21} \\
 2 \sqrt{\frac{3}{35}} & -\sqrt{\frac{3}{35}} & -\sqrt{\frac{2}{35}} & \frac{3}{4 \sqrt{5}} & 0 & 0 & -\frac{5}{12 \sqrt{7}} \\
 0 & 0 & 0 & 0 & -\frac{\sqrt{5}}{4} & -\frac{\sqrt{\frac{5}{3}}}{4} & 0 \\
 0 & 0 & 0 & 0 & -\frac{\sqrt{\frac{5}{3}}}{4} & -\frac{\sqrt{5}}{12} & 0 \\
 0 & \frac{3 \sqrt{3}}{7} & -\frac{\sqrt{2}}{21} & -\frac{5}{12 \sqrt{7}} & 0 & 0 & \frac{17 \sqrt{5}}{84} \\
\end{array}
\right),
\nonumber
\end{align}
\begin{align}
&\hspace{1.2cm}
\mathbb{F}_{22}^{(B_2)}=\left(
\begin{array}{ccccccc}
 -i \sqrt{\frac{3}{10}} & -i \sqrt{\frac{3}{10}} & -\frac{i}{\sqrt{5}} & -i \sqrt{\frac{2}{35}} & \sqrt{\frac{6}{7}} & 0 & 0 \\
 -i \sqrt{\frac{3}{10}} & -i \sqrt{\frac{3}{10}} & \frac{3 i}{7 \sqrt{5}} & \frac{3 i}{\sqrt{70}} & \sqrt{\frac{3}{14}} & \frac{3}{\sqrt{14}} & \frac{3 i}{7 \sqrt{2}} \\
 -\frac{i}{\sqrt{5}} & \frac{3 i}{7 \sqrt{5}} & \frac{-4i}{7}  \sqrt{\frac{6}{5}} & i \sqrt{\frac{3}{35}} & \frac{1}{\sqrt{7}} & -\frac{1}{3 \sqrt{21}} & -\frac{i}{21 \sqrt{3}} \\
 -i \sqrt{\frac{2}{35}} & \frac{3 i}{\sqrt{70}} & i \sqrt{\frac{3}{35}} & i \sqrt{\frac{3}{10}} & -\frac{1}{2 \sqrt{2}} & \frac{5}{6 \sqrt{6}} & -\frac{5 i}{3 \sqrt{42}} \\
 -\sqrt{\frac{6}{7}} & -\sqrt{\frac{3}{14}} & -\frac{1}{\sqrt{7}} & \frac{1}{2 \sqrt{2}} & 0 & 0 & \frac{\sqrt{\frac{5}{14}}}{6} \\
 0 & -\frac{3}{\sqrt{14}} & \frac{1}{3 \sqrt{21}} & -\frac{5}{6 \sqrt{6}} & 0 & 0 & \frac{\sqrt{\frac{15}{14}}}{2} \\
 0 & \frac{3 i}{7 \sqrt{2}} & -\frac{i}{21 \sqrt{3}} & -\frac{5 i}{3 \sqrt{42}} & -\frac{\sqrt{\frac{5}{14}}}{6} & -\frac{\sqrt{\frac{15}{14}}}{2} & \frac{-5i}{7}  \sqrt{\frac{5}{6}} \\
\end{array}
\right),
\nonumber
\\
&\hspace{1.2cm}
\mathbb{F}_{40}^{(B_2)}=\left(
\begin{array}{ccccccc}
 0 & 0 & 0 & \frac{\sqrt{\frac{3}{7}}}{2} & 0 & 0 & -\frac{\sqrt{\frac{5}{3}}}{2} \\
 0 & 0 & \frac{4 \sqrt{\frac{2}{3}}}{7} & -\frac{5}{2 \sqrt{21}} & 0 & 0 & \frac{\sqrt{\frac{5}{3}}}{14} \\
 0 & \frac{4 \sqrt{\frac{2}{3}}}{7} & -\frac{4}{21} & -\frac{5}{3 \sqrt{14}} & 0 & 0 & -\frac{3}{77}  \sqrt{\frac{5}{2}} \\
 \frac{\sqrt{\frac{3}{7}}}{2} & -\frac{5}{2 \sqrt{21}} & -\frac{5}{3 \sqrt{14}} & \frac{1}{66} & 0 & 0 & -\frac{9}{22} \sqrt{\frac{5}{7}} \\
 0 & 0 & 0 & 0 & \frac{1}{22} & \frac{5 \sqrt{3}}{22} & 0 \\
 0 & 0 & 0 & 0 & \frac{5 \sqrt{3}}{22} & -\frac{9}{22} & 0 \\
 -\frac{\sqrt{\frac{5}{3}}}{2} & \frac{\sqrt{\frac{5}{3}}}{14} & -\frac{3}{77}  \sqrt{\frac{5}{2}} & -\frac{9}{22} \sqrt{\frac{5}{7}} & 0 & 0 & \frac{27}{154} \\
\end{array}
\right),
\nonumber
\end{align}
\begin{align}
&\mathbb{F}_{42}^{(B_2)}=\left(
\begin{array}{ccccccc}
 0 & 0 & 0 & -\frac{1}{2} i \sqrt{\frac{15}{14}} & \frac{1}{2 \sqrt{14}} & -\frac{\sqrt{\frac{7}{6}}}{2} & -\frac{1}{2} i \sqrt{\frac{3}{2}} \\
 0 & 0 & \frac{4}{7} i \sqrt{\frac{5}{3}} & \frac{1}{2} i \sqrt{\frac{5}{42}} & -\frac{3}{2 \sqrt{14}} & \frac{5}{2 \sqrt{42}} & \frac{-3}{14} i \sqrt{\frac{3}{2}} \\
 0 & \frac{4}{7} i \sqrt{\frac{5}{3}} & -i\frac{2  \sqrt{10}}{21}  & \frac{1}{6} i \sqrt{\frac{5}{7}} & -\frac{\sqrt{\frac{3}{7}}}{2} & -\frac{15}{22 \sqrt{7}} & \frac{27 i}{154} \\
 -\frac{1}{2} i \sqrt{\frac{15}{14}} & \frac{1}{2} i \sqrt{\frac{5}{42}} & \frac{1}{6} i \sqrt{\frac{5}{7}} & \frac{i \sqrt{10}}{33} & \frac{\sqrt{\frac{3}{2}}}{11} & \frac{3}{11 \sqrt{2}} & \frac{-i3}{11}  \sqrt{\frac{2}{7}} \\
 -\frac{1}{2 \sqrt{14}} & \frac{3}{2 \sqrt{14}} & \frac{\sqrt{\frac{3}{7}}}{2} & -\frac{\sqrt{\frac{3}{2}}}{11} & 0 & 0 & -\frac{5}{11} \sqrt{\frac{15}{14}} \\
 \frac{\sqrt{\frac{7}{6}}}{2} & -\frac{5}{2 \sqrt{42}} & \frac{15}{22 \sqrt{7}} & -\frac{3}{11 \sqrt{2}} & 0 & 0 & -\frac{3}{11} \sqrt{\frac{5}{14}} \\
 -\frac{1}{2} i \sqrt{\frac{3}{2}} & \frac{-3}{14} i \sqrt{\frac{3}{2}} & \frac{27 i}{154} & \frac{-3i}{11}  \sqrt{\frac{2}{7}} & \frac{5 \sqrt{\frac{15}{14}}}{11} & \frac{3 \sqrt{\frac{5}{14}}}{11} & -i\frac{9}{77}  \sqrt{10} \\
\end{array}
\right),
\nonumber
\\
&\mathbb{F}_{44}^{(B_2)}=\left(
\begin{array}{ccccccc}
 0 & 0 & 0 & 0 & -\frac{i}{\sqrt{2}} & -\frac{i}{\sqrt{6}} & 0 \\
 0 & 0 & 0 & 0 & -\frac{i}{\sqrt{2}} & -\frac{i}{\sqrt{6}} & 0 \\
 0 & 0 & 0 & 0 & -\frac{i}{\sqrt{3}} & \frac{3 i}{11} & 0 \\
 0 & 0 & 0 & 0 & -\frac{1}{11} i \sqrt{\frac{7}{6}} & \frac{3}{11} i \sqrt{\frac{7}{2}} & 0 \\
 \frac{i}{\sqrt{2}} & \frac{i}{\sqrt{2}} & \frac{i}{\sqrt{3}} & \frac{1}{11} i \sqrt{\frac{7}{6}} & 0 & 0 & -\frac{1}{11} i \sqrt{\frac{15}{2}} \\
 \frac{i}{\sqrt{6}} & \frac{i}{\sqrt{6}} & -\frac{3 i}{11} & \frac{-3i}{11}  \sqrt{\frac{7}{2}} & 0 & 0 & \frac{-3i}{11}  \sqrt{\frac{5}{2}} \\
 0 & 0 & 0 & 0 & \frac{1}{11} i \sqrt{\frac{15}{2}} & \frac{3}{11} i \sqrt{\frac{5}{2}} & 0 \\
\end{array}
\right),
\nonumber
\end{align}
\begin{align}
&\mathbb{F}_{60}^{(B_2)}=\left(
\begin{array}{ccccccc}
 0 & 0 & 0 & 0 & 0 & 0 & 0 \\
 0 & 0 & 0 & 0 & 0 & 0 & 0 \\
 0 & 0 & 0 & 0 & 0 & 0 & \frac{20 \sqrt{\frac{10}{13}}}{33} \\
 0 & 0 & 0 & \frac{75}{44 \sqrt{13}} & 0 & 0 & -\frac{35}{132} \sqrt{\frac{35}{13}} \\
 0 & 0 & 0 & 0 & \frac{5}{44 \sqrt{13}} & -\frac{35}{44 \sqrt{39}} & 0 \\
 0 & 0 & 0 & 0 & -\frac{35}{44 \sqrt{39}} & \frac{85}{132 \sqrt{13}} & 0 \\
 0 & 0 & \frac{20 \sqrt{\frac{10}{13}}}{33} & -\frac{35}{132} \sqrt{\frac{35}{13}} & 0 & 0 & -\frac{5}{132 \sqrt{13}} \\
\end{array}
\right),
\nonumber
\\
&\mathbb{F}_{62}^{(B_2)}=\left(
\begin{array}{ccccccc}
 0 & 0 & 0 & 0 & 0 & 0 & 0 \\
 0 & 0 & 0 & 0 & 0 & 0 & 0 \\
 0 & 0 & 0 & 0 & 0 & \frac{40 \sqrt{\frac{2}{39}}}{33} & \frac{40}{33} i \sqrt{\frac{14}{39}} \\
 0 & 0 & 0 & \frac{-5i}{22}  \sqrt{\frac{105}{13}} & \frac{5 \sqrt{\frac{7}{13}}}{22} & -\frac{5}{66} \sqrt{\frac{91}{3}} & -\frac{35 i}{66 \sqrt{39}} \\
 0 & 0 & 0 & -\frac{5}{22}  \sqrt{\frac{7}{13}} & 0 & 0 & \frac{35 \sqrt{\frac{5}{13}}}{66} \\
 0 & 0 & -\frac{40}{33} \sqrt{\frac{2}{39}} & \frac{5 \sqrt{\frac{91}{3}}}{66} & 0 & 0 & -\frac{5}{22} \sqrt{\frac{15}{13}} \\
 0 & 0 & \frac{40}{33} i \sqrt{\frac{14}{39}} & -\frac{35 i}{66 \sqrt{39}} & -\frac{35}{66} \sqrt{\frac{5}{13}} & \frac{5 \sqrt{\frac{15}{13}}}{22} & \frac{-5i}{22}  \sqrt{\frac{35}{39}} \\
\end{array}
\right),
\nonumber
\end{align}
\begin{align}
&\hspace{1.2cm}\mathbb{F}_{64}^{(B_2)}=\left(
\begin{array}{ccccccc}
 0 & 0 & 0 & 0 & 0 & 0 & 0 \\
 0 & 0 & 0 & 0 & 0 & 0 & 0 \\
 0 & 0 & 0 & 0 & 0 & \frac{40}{33} i \sqrt{\frac{5}{13}} & 0 \\
 0 & 0 & 0 & 0 & \frac{-5i}{22}  \sqrt{\frac{105}{26}} & \frac{25}{66} i \sqrt{\frac{35}{26}} & 0 \\
 0 & 0 & 0 & \frac{5}{22} i \sqrt{\frac{105}{26}} & 0 & 0 & \frac{35}{22} i \sqrt{\frac{3}{26}} \\
 0 & 0 & -i\frac{40}{33}  \sqrt{\frac{5}{13}} & \frac{-25i}{66}  \sqrt{\frac{35}{26}} & 0 & 0 & -\frac{5 i}{22 \sqrt{26}} \\
 0 & 0 & 0 & 0 & \frac{-35i}{22} \sqrt{\frac{3}{26}} & \frac{5 i}{22 \sqrt{26}} & 0 \\
\end{array}
\right),
\nonumber
\\
&\hspace{1.2cm}\mathbb{F}_{66}^{(B_2)}=\left(
\begin{array}{ccccccc}
 0 & 0 & 0 & 0 & 0 & 0 & 0 \\
 0 & 0 & 0 & 0 & 0 & 0 & 0 \\
 0 & 0 & 0 & 0 & 0 & 0 & 0 \\
 0 & 0 & 0 & 0 & 0 & 0 & 0 \\
 0 & 0 & 0 & 0 & \frac{5}{2} i \sqrt{\frac{21}{143}} & \frac{5}{2} i \sqrt{\frac{7}{143}} & 0 \\
 0 & 0 & 0 & 0 & \frac{5}{2} i \sqrt{\frac{7}{143}} & \frac{5}{2} i \sqrt{\frac{7}{429}} & 0 \\
 0 & 0 & 0 & 0 & 0 & 0 & 0 \\
\end{array}
\right),
\nonumber
\\
&\hspace{1.2cm}{\mathbb{M}}^{(B_2)}=\left(
\begin{array}{ccccccc}
\mathcal{M}_{1,P}^{-1} & 0 & 0 & 0 & 0 & 0 & 0 \\
 0 & \frac{\mathcal{M}_{2,F}}{\det\mathcal{M}_{2}} & -\frac{\mathcal{M}_{2,PF}}{\det\mathcal{M}_{2}} & 0 & 0 & 0 & 0 \\
 0 & -\frac{\mathcal{M}_{2,PF}}{\det\mathcal{M}_{2}} & \frac{\mathcal{M}_{2,P}}{\det\mathcal{M}_{2}} & 0 & 0 & 0 & 0 \\
 0 & 0 & 0 & \mathcal{M}_{3,F}^{-1} & 0 & 0 & 0 \\
 0 & 0 & 0 & 0 & \mathcal{M}_{3,F}^{-1} & 0 & 0 \\
 0 & 0 & 0 & 0 & 0 & \mathcal{M}_{4,F}^{-1} & 0 \\
 0 & 0 & 0 & 0 & 0 & 0 & \mathcal{M}_{4,F}^{-1} \\
\end{array}
\right).
\label{IV110B2}
\end{align}
\begin{align}
&B_3:\hspace{.5cm}\mathbb{F}_{00}^{(B_3)}=\textbf{I}_{7},\hspace{.5cm}
\mathbb{F}_{20}^{(B_3)}=\left(
\begin{array}{ccccccc}
 \frac{1}{2 \sqrt{5}} & -\frac{3}{2 \sqrt{5}} & -\sqrt{\frac{3}{10}} & 0 & 2 \sqrt{\frac{3}{35}} & 0 & 0 \\
 -\frac{3}{2 \sqrt{5}} & \frac{1}{2 \sqrt{5}} & -\frac{\sqrt{\frac{3}{10}}}{7} & 0 & -\sqrt{\frac{3}{35}} & \frac{3 \sqrt{3}}{7} & 0 \\
 -\sqrt{\frac{3}{10}} & -\frac{\sqrt{\frac{3}{10}}}{7} & \frac{4}{7 \sqrt{5}} & 0 & -\sqrt{\frac{2}{35}} & -\frac{\sqrt{2}}{21} & 0 \\
 0 & 0 & 0 & -\frac{\sqrt{5}}{4} & 0 & 0 & -\frac{\sqrt{\frac{5}{3}}}{4} \\
 2 \sqrt{\frac{3}{35}} & -\sqrt{\frac{3}{35}} & -\sqrt{\frac{2}{35}} & 0 & \frac{3}{4 \sqrt{5}} & -\frac{5}{12 \sqrt{7}} & 0 \\
 0 & \frac{3 \sqrt{3}}{7} & -\frac{\sqrt{2}}{21} & 0 & -\frac{5}{12 \sqrt{7}} & \frac{17 \sqrt{5}}{84} & 0 \\
 0 & 0 & 0 & -\frac{\sqrt{\frac{5}{3}}}{4} & 0 & 0 & -\frac{\sqrt{5}}{12} \\
\end{array}
\right),
\nonumber\\
&\hspace{1.2cm}
\mathbb{F}_{22}^{(B_3)}=\left(
\begin{array}{ccccccc}
 i \sqrt{\frac{3}{10}} & i \sqrt{\frac{3}{10}} & \frac{i}{\sqrt{5}} & \sqrt{\frac{6}{7}} & i \sqrt{\frac{2}{35}} & 0 & 0 \\
 i \sqrt{\frac{3}{10}} & i \sqrt{\frac{3}{10}} & -\frac{3 i}{7 \sqrt{5}} & \sqrt{\frac{3}{14}} & -\frac{3 i}{\sqrt{70}} & -\frac{3 i}{7 \sqrt{2}} & \frac{3}{\sqrt{14}} \\
 \frac{i}{\sqrt{5}} & -\frac{3 i}{7 \sqrt{5}} & \frac{4}{7} i \sqrt{\frac{6}{5}} & \frac{1}{\sqrt{7}} & -i \sqrt{\frac{3}{35}} & \frac{i}{21 \sqrt{3}} & -\frac{1}{3 \sqrt{21}} \\
 -\sqrt{\frac{6}{7}} & -\sqrt{\frac{3}{14}} & -\frac{1}{\sqrt{7}} & 0 & \frac{1}{2 \sqrt{2}} & \frac{\sqrt{\frac{5}{14}}}{6} & 0 \\
 i \sqrt{\frac{2}{35}} & -\frac{3 i}{\sqrt{70}} & -i \sqrt{\frac{3}{35}} & -\frac{1}{2 \sqrt{2}} & -i \sqrt{\frac{3}{10}} & \frac{5 i}{3 \sqrt{42}} & \frac{5}{6 \sqrt{6}} \\
 0 & -\frac{3 i}{7 \sqrt{2}} & \frac{i}{21 \sqrt{3}} & -\frac{\sqrt{\frac{5}{14}}}{6} & \frac{5 i}{3 \sqrt{42}} & \frac{5}{7} i \sqrt{\frac{5}{6}} & -\frac{\sqrt{\frac{15}{14}}}{2} \\
 0 & -\frac{3}{\sqrt{14}} & \frac{1}{3 \sqrt{21}} & 0 & -\frac{5}{6 \sqrt{6}} & \frac{\sqrt{\frac{15}{14}}}{2} & 0 \\
\end{array}
\right),
\nonumber\\
&\hspace{1.2cm}\mathbb{F}_{40}^{(B_3)}=\left(
\begin{array}{ccccccc}
 0 & 0 & 0 & 0 & \frac{\sqrt{\frac{3}{7}}}{2} & -\frac{\sqrt{\frac{5}{3}}}{2} & 0 \\
 0 & 0 & \frac{4 \sqrt{\frac{2}{3}}}{7} & 0 & -\frac{5}{2 \sqrt{21}} & \frac{\sqrt{\frac{5}{3}}}{14} & 0 \\
 0 & \frac{4 \sqrt{\frac{2}{3}}}{7} & -\frac{4}{21} & 0 & -\frac{5}{3 \sqrt{14}} & -\frac{3}{77}  \sqrt{\frac{5}{2}} & 0 \\
 0 & 0 & 0 & \frac{1}{22} & 0 & 0 & \frac{5 \sqrt{3}}{22} \\
 \frac{\sqrt{\frac{3}{7}}}{2} & -\frac{5}{2 \sqrt{21}} & -\frac{5}{3 \sqrt{14}} & 0 & \frac{1}{66} & -\frac{9}{22} \sqrt{\frac{5}{7}} & 0 \\
 -\frac{\sqrt{\frac{5}{3}}}{2} & \frac{\sqrt{\frac{5}{3}}}{14} & -\frac{3}{77}  \sqrt{\frac{5}{2}} & 0 & -\frac{9}{22} \sqrt{\frac{5}{7}} & \frac{27}{154} & 0 \\
 0 & 0 & 0 & \frac{5 \sqrt{3}}{22} & 0 & 0 & -\frac{9}{22} \\
\end{array}
\right),
\nonumber
\\
&\hspace{1.2cm}\mathbb{F}_{42}^{(B_3)}=\left(
\begin{array}{ccccccc}
 0 & 0 & 0 & \frac{1}{2 \sqrt{14}} & \frac{1}{2} i \sqrt{\frac{15}{14}} & \frac{1}{2} i \sqrt{\frac{3}{2}} & -\frac{\sqrt{\frac{7}{6}}}{2} \\
 0 & 0 & \frac{-4i}{7}  \sqrt{\frac{5}{3}} & -\frac{3}{2 \sqrt{14}} & -\frac{1}{2} i \sqrt{\frac{5}{42}} & \frac{3}{14} i \sqrt{\frac{3}{2}} & \frac{5}{2 \sqrt{42}} \\
 0 & \frac{-4i}{7}  \sqrt{\frac{5}{3}} & \frac{2 i \sqrt{10}}{21} & -\frac{\sqrt{\frac{3}{7}}}{2} & -\frac{1}{6} i \sqrt{\frac{5}{7}} & -i\frac{27}{154}  & -\frac{15}{22 \sqrt{7}} \\
 -\frac{1}{2 \sqrt{14}} & \frac{3}{2 \sqrt{14}} & \frac{\sqrt{\frac{3}{7}}}{2} & 0 & -\frac{\sqrt{\frac{3}{2}}}{11} & -\frac{5}{11} \sqrt{\frac{15}{14}} & 0 \\
 \frac{1}{2} i \sqrt{\frac{15}{14}} & -\frac{1}{2} i \sqrt{\frac{5}{42}} & -\frac{1}{6} i \sqrt{\frac{5}{7}} & \frac{\sqrt{\frac{3}{2}}}{11} & -\frac{i \sqrt{10}}{33} & \frac{3}{11} i \sqrt{\frac{2}{7}} & \frac{3}{11 \sqrt{2}} \\
 \frac{1}{2} i \sqrt{\frac{3}{2}} & \frac{3}{14} i \sqrt{\frac{3}{2}} & -i\frac{27}{154}  & \frac{5 \sqrt{\frac{15}{14}}}{11} & \frac{3}{11} i \sqrt{\frac{2}{7}} & \frac{9 i \sqrt{10}}{77} & \frac{3 \sqrt{\frac{5}{14}}}{11} \\
 \frac{\sqrt{\frac{7}{6}}}{2} & -\frac{5}{2 \sqrt{42}} & \frac{15}{22 \sqrt{7}} & 0 & -\frac{3}{11 \sqrt{2}} & -\frac{3}{11} \sqrt{\frac{5}{14}} & 0 \\
\end{array}
\right),
\nonumber\\
&\hspace{1.2cm}\mathbb{F}_{44}^{(B_3)}=\left(
\begin{array}{ccccccc}
 0 & 0 & 0 & \frac{i}{\sqrt{2}} & 0 & 0 & \frac{i}{\sqrt{6}} \\
 0 & 0 & 0 & \frac{i}{\sqrt{2}} & 0 & 0 & \frac{i}{\sqrt{6}} \\
 0 & 0 & 0 & \frac{i}{\sqrt{3}} & 0 & 0 & -i\frac{3}{11} \\
 -\frac{i}{\sqrt{2}} & -\frac{i}{\sqrt{2}} & -\frac{i}{\sqrt{3}} & 0 & -\frac{1}{11} i \sqrt{\frac{7}{6}} & \frac{1}{11} i \sqrt{\frac{15}{2}} & 0 \\
 0 & 0 & 0 & \frac{1}{11} i \sqrt{\frac{7}{6}} & 0 & 0 & \frac{-3i}{11}  \sqrt{\frac{7}{2}} \\
 0 & 0 & 0 & -\frac{1}{11} i \sqrt{\frac{15}{2}} & 0 & 0 & \frac{-3i}{11}  \sqrt{\frac{5}{2}} \\
 -\frac{i}{\sqrt{6}} & -\frac{i}{\sqrt{6}} & \frac{3 i}{11} & 0 & \frac{3}{11} i \sqrt{\frac{7}{2}} & \frac{3}{11} i \sqrt{\frac{5}{2}} & 0 \\
\end{array}
\right)
\nonumber
\end{align}
\begin{align}
&\mathbb{F}_{60}^{(B_3)}=\left(
\begin{array}{ccccccc}
 0 & 0 & 0 & 0 & 0 & 0 & 0 \\
 0 & 0 & 0 & 0 & 0 & 0 & 0 \\
 0 & 0 & 0 & 0 & 0 & \frac{20 \sqrt{\frac{10}{13}}}{33} & 0 \\
 0 & 0 & 0 & \frac{5}{44 \sqrt{13}} & 0 & 0 & -\frac{35}{44 \sqrt{39}} \\
 0 & 0 & 0 & 0 & \frac{75}{44 \sqrt{13}} & -\frac{35}{132} \sqrt{\frac{35}{13}} & 0 \\
 0 & 0 & \frac{20 \sqrt{\frac{10}{13}}}{33} & 0 & -\frac{35}{132} \sqrt{\frac{35}{13}} & -\frac{5}{132 \sqrt{13}} & 0 \\
 0 & 0 & 0 & -\frac{35}{44 \sqrt{39}} & 0 & 0 & \frac{85}{132 \sqrt{13}} \\
\end{array}
\right),
\nonumber\\
&\mathbb{F}_{62}^{(B_3)}=\left(
\begin{array}{ccccccc}
 0 & 0 & 0 & 0 & 0 & 0 & 0 \\
 0 & 0 & 0 & 0 & 0 & 0 & 0 \\
 0 & 0 & 0 & 0 & 0 & -i\frac{40}{33}  \sqrt{\frac{14}{39}} & \frac{40 \sqrt{\frac{2}{39}}}{33} \\
 0 & 0 & 0 & 0 & -\frac{5}{22}  \sqrt{\frac{7}{13}} & \frac{35 \sqrt{\frac{5}{13}}}{66} & 0 \\
 0 & 0 & 0 & \frac{5 \sqrt{\frac{7}{13}}}{22} & \frac{5}{22} i \sqrt{\frac{105}{13}} & \frac{35 i}{66 \sqrt{39}} & -\frac{5}{66} \sqrt{\frac{91}{3}} \\
 0 & 0 & -i\frac{40}{33}  \sqrt{\frac{14}{39}} & -\frac{35}{66}  \sqrt{\frac{5}{13}}& \frac{35 i}{66 \sqrt{39}} & \frac{5}{22} i \sqrt{\frac{35}{39}} & \frac{5 \sqrt{\frac{15}{13}}}{22} \\
 0 & 0 & -\frac{40 }{33} \sqrt{\frac{2}{39}} & 0 & \frac{5 \sqrt{\frac{91}{3}}}{66} & -\frac{5}{22}   \sqrt{\frac{15}{13}} & 0 \\
\end{array}
\right),
\nonumber
\end{align}
\begin{align}
&\mathbb{F}_{64}^{(B_3)}=\left(
\begin{array}{ccccccc}
 0 & 0 & 0 & 0 & 0 & 0 & 0 \\
 0 & 0 & 0 & 0 & 0 & 0 & 0 \\
 0 & 0 & 0 & 0 & 0 & 0 & -i\frac{40}{33}  \sqrt{\frac{5}{13}} \\
 0 & 0 & 0 & 0 & \frac{-5i}{22}  \sqrt{\frac{105}{26}} & \frac{-35i}{22} \sqrt{\frac{3}{26}} & 0 \\
 0 & 0 & 0 & \frac{5}{22} i \sqrt{\frac{105}{26}} & 0 & 0 & \frac{-25i}{66}  \sqrt{\frac{35}{26}} \\
 0 & 0 & 0 & \frac{35}{22} i \sqrt{\frac{3}{26}} & 0 & 0 & -\frac{5 i}{22 \sqrt{26}} \\
 0 & 0 & \frac{40}{33} i \sqrt{\frac{5}{13}} & 0 & \frac{25}{66} i \sqrt{\frac{35}{26}} & \frac{5 i}{22 \sqrt{26}} & 0 \\
\end{array}%
\right).
\label{IV110B3}
\end{align}
}
%
 
 \raggedbottom\sloppy
  
  \vita{ Ra\'ul was born in Caracas, Venezuela, in 1984. Raised as the only man in a Catholic household full of short and bossy Hispanic women. Life was good until he moved to Bethesda, Maryland in fifth grade along with his mom and grandma. Despite the fact that nobody in his \emph{cookie-cutter} classroom could point out Italy in the map, he quickly adjusted to people talking to him as though he was mentally challenged since he knew no English and was the only foreigner in the whole school. 
  
  From there, he moved every other year, back and forth between Venezuela and the States. It wasn't until the grace of God put Chavez into power and Ra\'ul's dad and stepmom ran back to the states in fear of the ``socialist revolution", that he set roots in the states. Again, highschool kids in Florida left a lot to desire, and it would be a safe bet that they also couldn't point Italy in the map. This made studying and painting very appealing. After a couple of years, he got a full ride to attend college at New College of Florida where he would meet lots of equally weird people that would restore his faith in his generation.   
  
  When debating whether to pursue a career in science or in fine arts, he figured that if he would spend his days contemplating esoteric thoughts he might as well make a living at it. Also, there was no reason to not continue painting on the side. Little did he know that aside from being completely consumed by research, the average physicist makes $\epsilon$ more than a starving artist. After 4 years of forgettable/unforgettable moments at NCF he graduated with a BA in Physics in 2007.

In the fall of 2007, he began attending UW for graduate school. He joined the nuclear theory group at the beginning of his third year. In March of 2013, he co-organized along with Zohreh Davoudi (UW) and Tom Luu (LLN) the first Institute for Nuclear Theory (INT) workshop on ``Nuclear Reactions from Lattice QCD". This was also the first INT workshop to have graduate student organizers. After graduating from UW with his PhD Ra\'ul will start a postdoc at the Thomas Jefferson National Accelerator Facility (JLab) where he will continue his work on Lattice QCD. He is eager to move to Virginia with his pup and join the theory group at JLab, but will definitely miss Seattle and all the great people he's met here through the years. 
}

\end{document}